\documentstyle[12pt]{article}
\def\ll{\label}
\def\re{\ref}
\def\c{\cite}
\def\b{\begin}
\def\La{\Lambda}
\def\r1{(\ref{$1})}
\def\ot{\otimes}

\def\sn{\rm sn}
\def\pa{\partial}
\def\kap{\kappa}

\def\cR{\cal R}
\def\cF{\cal F}
\def\cP{\cal P}
\def\ti{\tilde}
\def\cn{\rm cn}
\def\dn{\rm dn}
\def\ga{\gamma}
\def\ep{\epsilon}
\def\th{\theta}
\def\ba{\begin{array}{c}}
\def\e{\end}
\def\sk{\smallskip}
\def\ea{\end{array}}
\def\pr{\prod}
\def\ni{\noindent}
\def\si{\sigma}
\def\da{\dagger}
\def\De{\Delta}
\def\de{\delta}
\def\bet{\beta}
\def\ov{\over}
\def\ha{{1\over 2}}
\def\qr{{1\over 4}}
\def\l{\left}
\def\l({\left(}
\def\r){\right)}
\def\r{\right}
\def\rw{\rightarrow}
\def\om{\omega}
\def\la{\lambda}
\def\al{\alpha}

\def\be{\begin{equation}}
\def\bc{\begin{center}}
\def\ec{\end{center}}
\def\bit{\begin{itemize}}
\def\eit{\end{itemize}}
\def\ee{\end{equation}}
\def\ed{\end{document}}
\def\bea{\begin{eqnarray}}
\def\eea{\end{eqnarray}}
\def\uqg{{\cal U}_q(g)}

\def\Rt{\widetilde R}
\def\St{\widetilde S}
\def\Ri{R^{-1}}

\def\K#1#2{K^{#1}_{#2}}
\def\L#1#2{L^{#1}_{#2}}

\def\bfl{\begin{flushleft}}
\def\efl{\end{flushleft}}
\def\bfr{\begin{flushright}}
\def\efr{\end{flushright}}

\begin{document}
%\eqnsection
\bibliographystyle{unsrt}

%_________________________________\input {preport0.tex}
%=========================== titlepage =================================
%\begin{titlepage}
%\rightline{SINP/TNP/yy-nn}
%\rightline{hep-th/9612046} \vspace{12pt}
%\begin{center}\LARGE{\bf{\sf papername }}\\[24pt]
%\Large{\sf authors }\\
%\normalsize{\em Saha Institute of Nuclear Physics\\
%1/AF Bidhannagar, 700 064 Calcutta, INDIA }\\[12pt]
%\today\\[48pt]
%{\Large Abstract}\\[12pt]
%\end{center}
%\end{titlepage}
%\markboth{Anjan Kundu:Quantum Integrable Systems}
%{Anjan Kundu:Quantum Integrable Systems}
%\begin{document}
%\eqnsection
%\rightline{hep-th/9612046} \vspace{12pt}
\title{ Quantum Integrable Systems:
Construction, Solution, Algebraic Aspect }
\author{
Anjan Kundu  \\
 Saha Institute of Nuclear Physics  \\
 Theory Group\\ 1/AF Bidhan Nagar,Calcutta 700 064,India.
 }
\date{}
\maketitle
\vskip 1 cm
%Running title: {\it Quantum integrabe systems}
%\vskip 4 cm
%\e{itemize}
\begin{abstract}
%------------------------------------------------------------
Various aspects of  the theory of quantum integrable systems are reviewed.
Basic ideas behind the construction of integrable
ultralocal and nonultralocal quantum   models are explored by 
exploiting the underlying algebraic structures related to 
the  Yang--Baxter equations.
Physical meaning of abstract mathematical notions
like  universal ${\cal R}$-matrix,
 quantized algebras, Sklyanin algebra, braided algebra, Hopf algebra
 etc. and the   role played by them in integrable  systems
are highlighted. Systematic construction of quantum integrable
lattice as well as  field  models and their exact excitation spectra
   are 
presented through examples. The coordinate and algebraic formulations
of the  Bethe
ansatz are  illustrated with    comparison,  along with the description of
nested and functional Bethe ansatzes. The techniques for deriving quantum
Hamiltonians from the Lax operators are demonstrated on concrete models.  
The exposition of this review is kept in a fairly
 elementary level with emphasis on the
physical contents.

\end{abstract}

\newpage

\tableofcontents
\newpage

%_____________________________________\input {preport1.tex}
\section{Introduction}
\subsection{Background and major concepts}
\setcounter{equation}{0}
Search for unifying theories continues in all scientific fields.
 In recent years we have   witnessed  again the
 emergence of  deep interconnections
 between seemingly diverse subjects like quantum and statistical
 systems, knot and braid theories,  Yang-Baxter equation,
 quantized and braided algebras
   etc. centered around the theory of integrable systems
 in $(1+1)$ dimensions.
 The linkage between integrable systems and    the
 conformal field theory, string theory
 as well as with  topological models
 have also been revealed  and a wide varieties of review articles are
  dedicated to these aspects \c{deVega,jimbo,selzub,wadati,alvarez,
moor,
kauffman}.
  The  aim of the present  review work
 however, is to focus on  the subject of
 the integrable systems itself, concentrating  mainly
on the role played by the  recently discovered   algebraic structures
underlying this theory and their
applications in constructing and solving  physical models.
We attempt to  present this beautiful
interplay between the abstract  mathematical objects in one hand and
 the physical problems   in  integrable quantum models in the other,
 in an elementary  way.

%_____________________________________-Classical Integrability_______

  The  theory   and  applications  of nonlinear  Integrable  systems,
   an immensely  important  subject in  mathematical physics of today,
 has made
profound  influence  in  many  branches  of  physics  as well as in  other
disciplines.
The seed of its development
 can   possibly  be traced back to    1934,
when on  an  August day     of 
a British engineer--historian named Scott Russell had a chance
encounter with  a strange
stable  wave in the Union  canal of Edinburgh \c{russel}.
Next evidence of such {\em paradoxically} stable solution was found
in the famous computer experiment of Fermi, Ulam and Pasta \c{fup}.
However only   in the  mid-sixties
such  phenomena
were understood  to be the   exact localized solutions
 of integrable  equations like Korteweg-de Vries (KdV) equation
\be
u_t(x,t)+ u_{xxx}(x,t)+ u(x,t) u(x,t)_x=0,
\ll{kdv}\ee
solvable through inverse scattering method (ISM) \c {ggkm} 
and were named  as {\it Solitons} \c{krus65}.
Subsequently,
this  was revealed to be  an universal feature 
of all nonlinear integrable  systems  in $(1+1)$ dimensions \c
{zakh72,soliton,ism}.

Mathematical basis of classical integrable systems was laid down
mainly through the
works of Sofia Kawalewskaya, Fuchs, Painlev\'e, Liuoville and others
\c{lakh93}.
There exist in fact various  definitions of integrability \c{lakh93}.
 We however,
for subsequent
 generalization to the quantum models,
 shall adopt the notion of  integrability
in the Liuoville sense, where integrability means the existence of
{\it action-angle} variables. That is,  if in a Hamiltonian system
$\ H [p(x,t), q(x,t)] \ $ given by the nonlinear equation
\be \dot {p}= - {\delta H  \over \delta q}, \quad
\dot {q}=  {\delta H  \over \delta p}, \ll{nle}\ee
it is possible to find a canonical transformation $\quad \left(p(x,t),
q(x,t)\right) \rightarrow \left( a(\la), b(\la,t)\right), \quad$
such that the new Hamiltonian  becomes dependent only on the action
variables, i.e.  $\ H[a(\la)] \ $,
then the  system may be called {\em completely integrable}.
In this case the dynamical
equations
$ \quad \dot {a}= - {\delta H  \over \delta b} =0, \quad
\dot {b}=  {\delta H  \over \delta a} = \om , \quad  $
can be trivially solved and as a bonus we also get $a(\la)$ as the generator
of  the  conserved quantities 
$\ln (a(\la))=\sum_{j=0}^\infty {c_j  \la^j} \ ,$ or $~~
\ln (a(\la))=\sum_{j=0}^\infty {c_{(-j)}}  \la^{-j} \ $ with $~
\ \  \dot c_{(\pm j)}=0 \ . $
 The number of such independent set of conserved quantities 
$\ \{ c_j \} \ $  coincides with the degree
of freedom   of the system, which  in case of field models with infinite
degrees of freedom also becomes 
infinite.  One of these conserved quantities 
 may  be considered as the Hamiltonian of the 
system.
For example, for the  KdV equation $c_{(-3)}$ is usually taken  
as  the Hamiltonian.
 For discrete systems with finite  number of particles,
only $N$ number of conserved quantities (usually first $N$)
among an infinite set turns out to be 
independent. Therefore  the existence of infinite set of 
 conserved quantities itself is not enough, their  
   involution (commutativity) 
must also be established for  concluding the complete integrability.
 We shall
see later that the validity of the famous
Yang--Baxter equation usually guarantees such involution,
even in the quantum case.

The major development in the theory of integrable systems
has been achieved in  seventies and eighties  with the invention 
and application of ISM \c{soliton,ism, russel}.
The important feature of this method is that, instead of attacking the
nonlinear equation (\re{nle}) directly, one constructs 
corresponding linear
scattering problem
\be
{\cal T}_x(x,\lambda)= L(q(x,t),p(x,t), \lambda)~ {\cal T}(x,\lambda)
,\ll{laxeq} \ee
where  the Lax operator $L(q,p, \lambda)~$ with the fields $q, p$ 
contains all  information
about the original nonlinear system. Some concrete examples
of Lax operators are   given with the list of integrable models in 
sect. 2. The parameter $\lambda,$
known as the
{\it spectral} parameter, acts like the {  momentum} or the rapidity 
of the scattering waves, while  the original field $q$ becomes the scattering
potential.
Interestingly, the function  $a(\la)$ related to the {\it transition }
 and $b(\la,t)$ to the {\it reflection}  coefficient
 of the scattering process
 are linked with  the action and the angle variables, respectively.
 The aim of
the  ISM is  presizely to find  the canonical mapping between 
 the original field
and the action-angle variables and  using this
to construct  the exact solutions for the
 original field. Soliton solution
 is a special solution, which corresponds
 to the reflectionless ($b=0$) potential and the form of $a(\la)$ as 
$~~~
a(\la)=\pr_i^m  {{\la-\la_i }\ov {\la-\la_i^* }},
~~~$ 
with zeroes (poles) at discrete points $\la_i~(\la^*_i)$
 in the complex spectral parameter plane.
We shall see in sect. 7  that even in  quantum problems some bound states
(called string solutions) exhibit such solitonic structure.
 The Hamiltonian is identified from the infinite set of
 conserved quantities $\{ c_j \} $,  
generated by $\ln a(\la)$, by knowing the   Poisson bracket 
and  the  dynamical equation.
The  Lax operator $L(q,p, \lambda)~$ therefore is the true 
representative of an   integrable system
   and together  with the Poisson bracket can generate 
 the whole integrable hierarchy. This fact
makes the systematic construction of the   Lax operators, to which 
a major  portion  of the present review is devoted, so important.
%%%%%%%%%%%%%%%%%%%%%%%%%%%%%%%%%%%%%%%%%%%%%%%%%%%%%%%%%%%%%%%%%%
%-------------------------------------- Quantum intyegrability__________

\medskip

%222
A new era
started with the formulation of quantum
inverse scattering method (QISM) by  extending 
   the notion of integrability to the quantum domain
\c {fadrev,qism,kulskly}.
 To highlight   various aspects of quantum integrable systems 
 is the main purpose  of the 
current   review.
The action--angle variables in the quantum case  become 
 operators with generalized
commutation relation
\be
A(\la) B(\mu)= f(\la-\mu) B(\mu)A(\la) \ll{a-b}\ee
 For  concluding complete  integrability of a quantum system,
 the  infinite set of conserved quantities, which  now  become  
operators must be in involution: $~ [ C_n, C_m]=0 .$ 
This is guaranteed by $~[\tau(\la), \tau (\mu)]=0~$, since 
the transfer matrix  $\tau (\la)$ 
 generates the conserved operators : $\ln \tau (\la)=\sum_j C_j
\la^j$. The existence  of  commuting transfer matrices in turn is assured by  
 the 
matrix relation known as the Quantum Yang--Baxter equation (QYBE)
$~~~ R(\lambda , \mu)~ { T_1} (\lambda)~ { T_2}(\mu )
~ = ~  { T_2}(\mu )~ { T_1 }(\lambda)~~ R(\lambda , \mu),
~~T_1=T \ot I,~ T_2 =I \ot T, ~~$
involving the quantum $ R(\la, \mu)$-matrix and the monodromy matrix 
$~~T(\la)=\prod_{i=1}^N L_i(\la)~$, where $L_i(\la)$ is the Lax operator
 of the corresponding   lattice
model at point $i$ and $~\tau(\la)=tr T(\la)~$. 
The QYBE associated with  the quantum integrable systems
and  ensuring  the commutativity of action variables 
for different $\la$, exhibits 
 amazingly rich algebraic properties.
It represents in fact a  set of  generalized commutation relations 
like ({\re{a-b})
between
action  and angle like variables  given in the matrix form.

For the commonly discussed  models
known as {\it ultralocal} models,
 the Lax operators commute  at different points:
$~~[L_{1i},L_{2j}]=0~~$ aand 
the above QYBE  can be obtained as a consequence of its local version
\begin {equation}
R(\lambda , \mu)~ L_{j1}(\lambda)~ L_{j2}(\mu )
~ = ~  L_{j2}(\mu )~ L_{j1}(\lambda)~ R(\lambda , \mu),~~~j=1,\ldots ,N.
\ll{lqybe}\end {equation}
On the other hand, for rest of  the models constituting another
 important class, known as   {\em nonultralocal }
models \c{Mailletc,maillet, khijmp96} we have 
 $~~[L_{1i},L_{2j}] \neq 0~,~$ and 
  the QYBE (\re{lqybe}) has to  be extended to its braided version
(BQYBE)
\begin{equation}
{R}_{12}(\la-\mu)Z_{21}^{-1}L_{1j}(\la)\tilde Z_{21}L_{2j}(\mu)
= Z_{12}^{-1}L_{2j}(\mu) \tilde Z_{12}L_{1j}(\la){R}_{12}(\la-\mu).
\ll{lbqybe}\end{equation}
with additional matrices $Z, \ti Z$. For describing the integrability 
one has to complement (\re{lbqybe}) also with commutation  relations
between $~L_{1i},~L_{2j} ~$
at different lattice points $i \neq j$, known as braiding relations.
As a result the local BQYBE can be raised to the global level giving 
braided QYBE for the monodromy matrix similar to (\re{lbqybe}).
\subsection{ Guide to the contents of the review} 

 We  detail the   basic idea behind the 
quantum and statistical integrable systems 
 in sect. 2, by  introducing quantum $R$-matrices,
Lax operators,
 Yang--Baxter equations and their extensions \c{baxter,xyz,fadrev,khijmp96}.
 We also present here a list of
well known quantum  integrable models of both ultralocal and nonultralocal
type along with their explicit Lax operators, belonging to the trigonometric
 $R_{trig}(\la)$ as well as to  the rational $R_{rat}(\la)$ matrices.
%2222

%3333
We focus on the algebraic aspects of the quantum integrable systems
\c{drinfeld,fad95, kbmpl92,kbmpl95,dobrev}
in sect. 3. The 
 significance of   
Hopf algebras in integrable systems is stressed 
 by showing how  the local and global   QYBE relations are
linked together through the  related coproduct properties and how and why
they
differ for  ultralocal and nonultralocal systems, depending on their
 underlying algebraic structures \c{khijmp96}.
The origin of quantum algebras \c{qa,drinfeld} in integrable theory 
is traced out   and the  fascinating   
 structures like quadratic Sklyanin algebra \c{skalg}, its   extensions
\c{kbjp92}, 
  the Yangian algebra 
\c{drinfeld,cheri} and the braided \c{majid} and quantized braided algebra 
\c {majidq,hlavqb} are explored through 
quantum Yang-Baxter like equations. The notions of quantum group \c{ncgeom}, 
Faddeev--Reshitikhin--Takhtajan (FRT) algebra \c{frt}, Hecke and
Birman-Wenzl-Murakami
algebra \c{wadati} are also introduced in this section.
Abstraction of the quantum $R$-matrix  can also be raised to another
level of algebraic sophistication by defining an Universal ${\cal R}$-matrix
intertwining between two different coproducts in an
quasi-triangular Hopf algebra \c{drinfeld,urslq2}.
 The  Universal ${\cal R}$-matrix
 for the general reductive Lie algebras \c{reductive,ktjp95}
 is constructed  here with the    
  intention of its subsequent application in the next  section.

Such abstract algebraic notions have  not only 
 enriched the  integrable systems alone, but also the   algebraic 
 aspects of the quantum integrable theory  have   made profound
 influence 
 on  pure mathematics,  by introducing examples of noncocommuting
and quasi-triangular Hopf algebras \c{drinfeld}, giving 
FRT formulation  of dual  algebras \c{frt}
 and introducing various knot and link
polynomials through Yang-Baxter algebra \c{wadati}.
%3333
%4444-------

For systematic construction of the key objects involved in the quantum
integrable ultralocal theory,  a Yang-Baxterization
scheme  \c{jones,ktjp95} is formulated  in sect. 4 by   exploiting
 the  universal ${\cal R}$-matrix.
 The richness of this abstract object is shown 
from the  application point of view,
by deriving from it 
the FRT relations, QYBE, YBE  and
most importantly 
the spectral parameter dependent  quantum $R$-matrices and the Lax
 operators.
The  models thus generated  through realizations 
of the underlying quantum algebra  and its undeformed limit 
include the well known sine-Gordon (SG)  equation, 
nonlinear Schr\"odinger (NLS) equation   and 
the  $XXZ, XXX$ spin-$\ha$ models   as well as the  multicomponent 
Toda field equations \c{babelontoda}. 
  {\em Twisting  } transformation of the  Hopf algebra \c{twist}
also allows to construct from  the   ${\cal R}$
the  {\em colored} and {\em twisted} generalizations of the models.

%4444
%555------

For covering wider class of models
an extension of the trigonometric Sklyanin   algebra (ETSA)
   \c{kbmpl92} is considered 
 in  sect. 5, 
which through Yang-Baxterization  
  leads to  the construction   of some integrable ancestor model.
 Different realizations of the underlying ETSA generate
 variety of  quantum integrable   models  in a systematic way.
Such models include 
Liuoville model \c{liuFad}, massive Thirring model \c{kulskly}, relativistic
 Toda chains \c{qrtoda}, Ablowitz--Ladik model \c{ALM},
quantum derivative NLS \c{kb-dnls} , $q$-oscillator models etc.
apart from  the quantum algebra related models 
found in sect. 4. 
All these descendant
models 
 are associated with the same
 quantum $R_{trig}$ matrix of trigonometric type.   
 At  $q\rw 1$ limit linked  to  the Yangian algebra, one can
construct similarly  the related 
 ancestor model for generating  its  integrable family of descendant  models 
associated with the rational $R_{rat}$-matrix solution.
An added  advantage of this approach  is that, one can get
directly the exact lattice regularized 
 Lax operators, which are otherwise difficult to
guess from the known field models. Such discrete Lax operators, 
apart from their own interest,   
yield also  the corresponding field models 
at   vanishing  lattice spacing.
In this   approach one starts 
from a
 known  $R$-matrix solutions 
and constructs subsequently  the  Lax operator of
an ancestor
model involving generators of  quantized algebra,
  determined  by the QYBE.
Different reductions of this ancestor
model through suitable realizations in physical variables
like bosons, spin operators or $q$-oscillators \c{qoscl}
   generate in turn 
  wide range of  integrable  descendant  models.
 The result also 
 brings out
 the hidden interrelations between  diverse 
 models and classifies them
 according to the associated $R$-matrix.
Various realizations of the extended quadratic algebra  provides 
 important  criteria for recognizing {\em integrable nonlinearities}.
To focus on such   application aspects of  abstract algebraic objects in
quantum integrable  theory  is the main objective   of this section.

%555

%666666
The monodromy matrix is formed by multiplying local
 Lax operators 
and the conserved quantities are found
by expanding the trace of the monodromy matrix in the spectral parameter.
However,
the explicit construction of the conserved operators including the 
Hamiltonian is often a difficult task in quantum models. 
We take up this issue in sect. 6. For the fundamental  models \c{sglsg},
which include
spin chains and Hubberd model, 
 we 
  derive  the Hamiltonians in explicit form
starting  from the associated Lax operators.
The construction of twisted, colored, impurity  as well as
inhomogeneous models are also discussed here.
The generalized spin model and the  projector methods  for constructing 
the Hamiltonian of nonfundamental models \c{Coker,sglsg}, e.g.   bosonic
models are described on the example of lattice NLS and 
the related difficulties  for the lattice SG model are pointed out.
 Some  special examples of 
nonfundamental models like Toda chains,  a simpler lattice NLS \c{krjp94}   
allowing explicit construction of Hamiltonian are highlighted.
%66666

%7777---
A problem of significant physical interest in   
 quantum models  is to find the energy spectrum by 
solving   the 
eigenvalue problem
$~~  H\mid m> =  E_m\mid m>  , ~~$
where $ H$ is a  nonlinear  Hamiltonian operator 
and $\mid m>$ is the
$m$-particle eigenstate.
In case of   integrable models this problem is usually    solvable
  in an exact and nonperturbative way. Moreover, 
 one can even find  the eigenvalues  $\Lambda(\la)$ for all
other conserved quantities from       the 
transfer matrix: 
$ \ \ln \tau(\la)\mid m>= \ln \Lambda (\la)\mid m>~~$ 
by  expanding  $~~ \ln \Lambda (\la)=
\sum_{j=0}^\infty { c_j  \la^j }
~~,$
similar to the classical case.
Interestingly, in this formulation the { angle} variable $B(\la)$ acts
like  the { creation} operator and the $m$-particle state $\mid m>$
can be obtained by applying $m$ times the  operators $B(\la_i), i\in [1,m]$
  on the  pseudovacuum $\mid 0>$. Therefore using the generalized
commutation relation (\re{a-b})
and the properties of the pseudovacuum, one can calculate 
the required eigenvalues $ \Lambda (\la)$.
We  discuss in sect. 7
 the  basic concepts  of 
the Bethe ansatz method,
 a powerful method for solving   the
eigenvalue  problem exactly.
 Its algebraic formulation \c{baxter,fadrev,izumov,xyz,deVega} is explained
on  examples of spin models like $XXZ$ and $XXX$ chains
and  bosonic models like the  SG and the NLS.
We consider  in parallel the coordinate formulation \c{cba,qism,baxter,korbook} 
 introduced by Bethe way back in 1931 \c{Bethe},
 on the same examples, mainly 
 to bring out the linkage  
between these two approaches. In considering more general algebraic
 structures, the nested Bethe ansatz  method \c{nba} is described 
for the models like the Hubberd  and the vector NLS 
 and the functional Bethe ansatz method
\c{fba,toda} is demonstrated on the examples   of relativistic and
nonrelativistic Toda chains.
%7777
%888---corrected above

Our next concern in sect. 8 is another major and  important
class of integrable models, namely quantum  { nonultralocal} models, discussions of which
are  usually avoided in general reviews, due to the nonavailability
 of their
well framed
theories apart from the studies in \c{maillet}.
 Such systems, examples of which are well known models like
 KdV and modified KdV \c{soliton}, nonlinear $\sigma$-model \c{simodel},
 WZWN model \c{wzwn},
 complex SG model \c {comsg} etc. are difficult to
tackle, mainly due to the
 presence of  { derivative  $\delta$-}function type terms in their
 commutation relations. This makes the transition from local to the global
 QYBE difficult and the application of the standard 
 Yang-Baxter relations related to   the
 quantum algebra fails for such systems. However,
   based on the concept of  quantized braided
 algebra \c{majidq,hlavqb}, the integrability of nonultralocal models
 may be formulated \c{khijmp96}, much in parallel to the ultralocal case,
through braided extensions of the QYBE  and a trace
 factorization  \c{skly-r,hlavopsc}. The applicability  of this theory 
 is demonstrated  on  nonultralocal models like 
 modified KdV \c{kmpl95}, WZWN model \c{wzwn},
nonabelian Toda chain \c{natoda},
quantum mapping  \c{Nijhof}, Coulomb gas picture of CFT
\cite {babelon} etc.
It is remarkable that the supersymmetric \c{SUSY}
  and anyonic \c{anyon1} integrable models
as well as the systems related to the reflection equation \c{skly-r}
 also fit well
in this formulation.
%888
%999---corrected above

Finally we discuss in sect. 9 
some   recent developments in this subject and
few open problems, the starting concepts of which are mostly contained in the
present review.
 Our  discussions are
  focused on the  multicomponent extension
of the ETSA \c{kbjp92} with possible applications,
 Cologero-Sutherland  spin model \c {cs}, its 
 possible   $R$-matrix formulation \c{csavan} and
 deformed Yangian symmetry \c{csyan},
 finite size correction and conformal properties of the integrable systems
\c{karow},
quantum mKdV and its  application for an alternative formulation of 
CFT \c{BLZ},
 reaction-diffusion equation as
integrable system \c{reacrit}. We discuss further 
 some newly discovered
ladder symmetric models \c{frahm,ladder},    recent development in
 { elliptic quantum group} \c{eqg},  the possibility of
generating  coupled spin chains and extension of Hubberd model 
 through twisting transformation \c{kplan} 
and finally  the problem in  constructing  quantum KdV,
nonlinear $\sigma$-model and complex SG model as quantum nonultralocal
systems.
%999
%101010

In  sect. 10
we conclude with a few  remarks.

The arrangement of this paper is  outlined in the contents.

%____________________________________\input {preport2.tex}
\setcounter{section}{1}

\section { Quantum
 integrable systems: basic features and properties}
The  Lax operator, a key object in classical
  integrable systems \c{soliton} is generalized   
to the quantum case, where it   continues   to play  a major role. 
Its matrix elements, classically   being  functions 
of  field variables, become now quantum operators. Examples of such
quantum Lax operators as $2\times 2$ matrices     representing 
a variety of integrable models are listed  in sect. 2.4. 
 The  explicit form  of the  Lax operator determines the associated 
quantum model.  We show here how  the integrability of such models 
are ensured by certain relations, known as the
 quantum Yang-Baxter equation (QYBE), satisfied by their Lax operators. 
As a result, using the Lax operator one can   
 construct  the commuting family 
of conserved quantities including the Hamiltonian.   

Quantum integrable systems  (QIS)
 can be divided into two broad classes:
  {\it ultralocal} and {\it nonultralocal} depending on an important 
property of their representative Lax operators $L_{ai}(\la)$, where 
$a$ and $i$ denote  
the auxiliary  and the quantum spaces, respectively and $\la$ 
is the spectral
parameter.  
For  ultralocal QIS, which include well known models like
Nonlinear Schr\"odinger equation, sine-Gordon  model, Toda chain,
etc. the Lax operators exhibit a common  { ultralocality} property, i.e.
they  commute  at different lattice points $i \not =j$: $[L_{1i}(\la),
 L_{2j}(\mu)]=0$~(see fig 2.1). 
 We shall see  that  this fact is actively used 
in constructing the QYBE.
On the other hand, the nonultralocal models,
which include also   famous models like
quantum KdV , supersymmetric models, nonlinear $\sigma$ model, WZWN model
etc. 
 are characterized by more general  
property with $~~[L_{1i}(\la), L_{2j}(\mu)] \not = 0 .$
This makes the formulation of their integrability theory difficult and the 
associated QYBE needs to be extended along with 
 specifying additional braiding
relations at different lattice points.
We   introduce  basic equations 
for quantum   integrability of both   ultralocal and 
 nonultralocal  models, which 
depend crucially  on the algebraic properties of the associated Lax 
operators.
We should mention however, that compared to ultralocal models the
integrability  of nonultralocal systems is more involved and  not yet 
well established.

 \subsection { Integrability of quantum ultralocal models: Yang-Baxter
equations 
 } 
For better understanding of  algebraic structures of  Lax operators
for ultralocal models,
let us consider first usual  matrices $A,~B,~C,~D$ satisfying  the obvious
relation \be~ ( A_1  B_2 )= ( B_2 A_1)
 ~~~\mbox{or}~~~  ~[A_1,B_2]=0,\ll{matrix1}\ee
 and  the standard  multiplication rule
   \begin {equation}
({A} \otimes  {B})(C \otimes D) = (AC \otimes BD)
\ll{matrix2}
\end {equation}
$ \mbox{ where}~~ A_1 \equiv  A \otimes 1,~~ B_2= 
1 \otimes  B$. 
 If we choose now
 $A=L_i(\la),
B=L_i(\mu)$
 as Lax operators at the same lattice points and 
check for  the relation (\re{matrix1}), we 
see  immediately   its nonvalidity
due to the  noncommuting  nature of the
matrix elements of the quantum Lax operators. The equation in  effect 
turns
into the famous QYBE (see fig. 2.2 for $i=3$) 
\begin {equation}
R_{12}(\lambda , \mu) {L}_{1j}(\lambda) {L}_{2j}(\mu)
~ = ~  {L}_{2j}(\mu)~{L}_{1j}(\lambda) R_{12}(\lambda , \mu),
\ll{qybel}
\end {equation}
   where a matrix  $R_{12}$ with $c$-number elements 
appears   to compensate
for the noncommutativity of quantum $L$.
  This is the basic reason
for the  nontrivial algebras underlying such integrable systems.
 There are  two main types of  $R(\la,\mu)$-matrices we will be interested
in, which are   
 given through  trigonometric or  rational functions of the 
spectral parameters $\la-\mu$. The explicit forms  of such matrices 
will be  
obtained below.

Note however that for the  choice     
 $A=L_{i+1}(\mu),
B=L_i(\la)$  at different lattice points, 
  the trivial  commutation  (\re{matrix1})
is indeed satisfied due the
ultralocal nature of the Lax operators. 
Let us  choose   
 the matrices  through Lax operators again at different lattice
points as 
 $~A=L_{i+1}(\la),
 B=L_{i+1}(\mu),
 C=L_{i}(\la),
 D=L_{i}(\mu) ~~ $  and turn to  
the multiplication rule (\re{matrix2}). We observe that 
  this   rule  holds whenever    
  $B_2$ and $C_1$ commute  and therefore, it  remains valid
also for our choice  due to the ultralocality 
condition on  Lax operators resulting (see fig. 2.3)
\be
 (L_{1 i+1}(\la)
 L_{2 i+1}(\mu))
 (L_{1 i}(\la)
 L_{2 i}(\mu))
= (L_{1 i+1}(\la)
 L_{1 i}(\la))
( L_{2 i+1}(\mu)
 L_{2 i}(\mu))
\ll{mull}\ee
We show now that 
  combining  two commutation relations  (\re{qybel})
and (\re{mull}) of the local 
Lax operators, one can derive the global  QYBE essential for
proving  integrability, which in fact  is a global property. 
Start from QYBE 
(\re{qybel}) at point $i+1$ ( fig. 2.2),  multiply both sides 
by   $ (L_{1 i}(\la)
 L_{2 i}(\mu))$ and   apply  in the $rhs$  
   the same QYBE 
 at $i$ to flip the $R$-matrix to the extreme right (see fig 2.4).

We    use next the relation  (\re{mull}) (fig. 2.3) and 
the one obtained from it by interchanging  $1 
\leftrightarrow 2 $ and $ \la \leftrightarrow \mu$ to put together   
 the Lax operators acting in the same space $1$ or $2$ 
 in both the  $lhs$ and $rhs$. As a result we arrive
 again at the same form of QYBE but for a  nonlocal  object $T_a^{[i+1 ,i]}
(\la)=L_{ai+1}(\la)L_{ai}(\la)  $ (see fig. 2.5).

 Repeating this globalization  step  $N$ times we   finally obtain the
global  QYBE (fig 2.6a)
   \begin {equation}
R_{12}(\lambda , \mu)~ {T}_{1}(\lambda)~ {T}_{2}(\mu )
~ = ~  {T}_{2}(\mu )~{T}_{1}(\lambda)~ R_{12}(\lambda , \mu),
\ll{qybet}\ee
for the monodromy matrix $T(\la)= \prod_i^N L_i(\la)$ (fig 2.6b).
We shall come back to this important transition from local to global QYBE
related to the coproduct property of the underlying Hopf algebra  
in sect. 3.

Taking  now the trace from both the sides of (\re{qybet})
and using cyclic rotation of matrices under trace, 
 we  get  
\[ tr_{12}\l( { T}_1(\la)~ { T}_2(\mu)\r)= tr_{12}
 \l(R^{-1} { T}_2(\mu)~{ T }_1(\la)~ R\r) =
tr_{12}\l( { T}_2(\mu)~ { T}_1(\la) \r) \]
 Further with the obvious property  $tr_{12}( { T}_1 { T}_2 )=
tr_{1}( { T}_1) tr_2 ({ T}_2 ) $ we obtain the trace factorization
$ tr( { T}(\la))~ tr({ T}(\mu))
              =   tr( { T}(\mu))~tr( { T}(\mu)) $
or for the transfer
 matrix $\tau(\la)= tr T(\la)$ the trivial commutation (see fig. 2.7)
\be
[\tau (\lambda )~, \tau(\mu)~] = 0.\ll{tr-id}\ee 
 As in the classical case, 
 $\tau(\lambda )$  generates the conserved quantities 
  usually as  $\ln \tau(\lambda )=\Sigma_{n=0}^\infty
C_n \lambda^n  $. Therefore 
 the commuting transfer matrices at different spectral
parameters (\re {tr-id}) lead to the involution  
$ [C_n,C_m]=0$, i.e. to  the independence of the infinite number 
(in general) of  conserved
 quantities, which thus proves  the exact  integrability of the quantum system  
in the   Liouville sense. Tracing back
 this logical chain, we see therefore that the
integrability is guaranteed by the relation (\re{tr-id}),
 which   is  originated from
the global QYBE
(\re{qybet}) and that  in turn  is obtained from the local QYBE
 (\re{qybel})
  under the ultralocality constraint (\re{mull}). Consequently, 
 QYBE (\re{qybel})
 may be considered as the key equation for  ultralocal integrable systems
  represented by the  Lax operator $L_i(\la)$   and the quantum
  $R(\la,\mu)$-matrix.
For deriving an independent equation for the $R(\la,\mu)$-matrix, one can
 start with  
 the triple product $L_{1i}(\la)L_{2i}(\mu)L_{3i}(\nu)$
and  revert the ordering  by flipping 
successively  two operators at a time using (\re {qybel}).
Clearly this can be done   in
 two different ways
:$(12) \rw (13) \rw (23)$ and $(23) \rw (13) \rw (12)$, which due to   
associativity should be equivalent and  yield the equation
\begin{equation}
   R_{12}(\la,\mu)~R_{1{ 3}}(\la,\nu) ~R_{2{ 3}}(\mu,\nu) ~~
=~~R_{2 { 3}}(\mu,\nu)
    ~R_{1 { 3}}(\la,\nu) ~R_{12} (\la,\mu)~.
,\ll{ybe}\end{equation}
known  as the Yang-Baxter equation (YBE). The equivalence of these 
two {\it scattering} paths can also be visualized as in fig. 2.8.

 It should be mentioned that 
the trigonometric  solutions of YBE, we are interested in, 
 in the limit $\la \rw \pm \infty$ goes to 
 $  R_{12}(\la,\mu) \rw  R_{q12}^\pm $ with $q=e^{i \eta}$, the 
 spectral parameter free  solutions $R_q^\pm$ of   
\begin{equation}
   R_{12}~R_{1{ 3}} ~R_{2{ 3}} ~~
=~~R_{2 { 3}}
    ~R_{1 { 3}} ~R_{12} ~
\ll{braid}\end{equation}
given in  the upper and lower triangular form 
as 
\be
R_q^+ = \left( \begin{array}{c}
q \ \quad \ \quad \ \qquad \qquad \\
    \quad \ 1 \ \ \ q-q^{-1} \ \quad  \\
     \quad \ \quad \
    1 \ \quad \\
        \quad \quad \qquad \ \qquad \ \ q
          \end{array}   \right), \qquad R_q^-=P~(R_q^+)^{-1}~P
 = \left( \begin{array}{c}
q^{-1} \ \quad \ \qquad \ \quad \qquad \\
    \quad \ 1 \ \quad  \ \quad  \\
    \ \ \ -(q-q^{-1}) \
     1 \ \quad \\
     \qquad \    \qquad \ \quad \ \quad \ q^{-1}
          \end{array}   \right)
\ll{R+}\ee
with  permutation operator $P$. These solutions are 
related to the braid group representations  as  $\check R =P R_q^\pm.$ 

Note that the spectral independent 
 equation (\re{braid}) is obtained as the limiting case
$\la \rw \infty, ~\mu \rw \infty  ~$ of YBE (\re{ybe}), when
 the QYBE (\re{qybel}) also becomes  spectral parameter free. Taking
therefore different sets of limits 
$ \la \rw \infty , ~\mu \rw \infty, ~~ 
~\la \rw -\infty, ~\mu \rw  -\infty, ~~ 
\la \rw  \infty,  \mu \rw  -\infty ~~$ etc. with $ \la - \mu \rw  \infty$
we obtain from (\re{qybel}) the relations 
 \begin {equation}
R_{q12}^+ {L}_{1j}^\pm~ {L}_{2j}^\pm
~ = ~  {L}_{2j}^\pm~{L}_{1j}^\pm R_{q12}^+, ~~
R_{q12}^+ {L}_{1j}^+~ {L}_{2j}^-
~ = ~  {L}_{2j}^-~{L}_{1j}^+ R_{q12}^+,
\ll{qybe+}
\end {equation}
where $L(\la)_{\mid \la \rw \pm \infty} \rw L^\pm$ are
 spectral parameter free upper (lower)-triangular matrices. We shall
return to such equations in sect. 3 for describing the underlying 
algebraic structures
and in sect. 5 for the  construction  of  Lax operator
 through Yang-Baxterization.
The spectral parameter dependent Lax operator $L(\la)$ and the
 $R(\la,\mu)$-matrix  are sufficient, in principle,  for constructing
  infinite number  of conserved quantities
and also for  solving  the corresponding eigenvalue
 problem exactly. We  demonstrate these features in  subsequent sections
and  show how     exploiting the 
 underlying algebraic structures
  the Lax operators  can be generated
systematically.  
%____________________-
\subsubsection { Relation with statistical system, $R$-matrix solutions} 
The $(1+1)$ dimensional quantum systems are  intimately connected with $2$
dimensional classical statistical systems and the notion of integrability is 
equivalent in both these cases. While in the quantum systems
 the QYBE (\re {qybel}) is the central, for  statistical models
 similar role 
is played by the  YBE (\re {ybe}). 
Let us consider a $2$-dimensional array of $N\times M$
lattice points connected
by the bonds assigned with +ve (-ve) signs or equivalently, with right, up
(left, down) arrows  in a random way (see fig. 2.9).
 
For this classical statistical system known as vertex model, the
partition function $Z$ should give the probability of appearance of all
possible arrangements on the whole structure. However, for analyzing this
global property one should know first the local ones by finding the probability
of occurrence of a particular arrangement at a fixed lattice point $i$.
 For 
 two   allowed signs on each bond,
 $2^2=4$ possibilities  would appear in
the horizontal ($H$) and  $4$   
 in the vertical (V) direction. This  accounts  for the
 $4\times 4=16$  possible
arrangements at each lattice point. Setting the corresponding Boltzmann
weights as the matrix elements of a
 $4\times4$-matrix, we get the $R^{(i)}_{12}
$-matrix acting
on the product space (product of $H$ and $V$ spaces) $V_1 \otimes V_2 $.
Parametrization of Boltzmann weights makes a crucial dependence of 
the $R$ matrix  on  spectral  parameter $\la$.
 For finding out
 the configuration 
 probability for a string of $N$-lattice points, the $R^{(i)}$-matrices at
different sites in a row should be matrix multiplied to give the transfer
matrix 
$ \tau(\vec {\alpha},\vec{\beta})
= tr (\prod_i^N R^{(i)})$.
It is important to note that in this expression the horizontal indices $
\mu_i$
enter only as summation or repeated indices (see fig.2.9). The $tr$
 appears also
for the same reason from the extreme ends due to the periodic boundary
condition.  As a result the transfer matrix  $
\tau(\vec {\alpha},\vec{\beta}) $ depends solely on the $V$-space (see fig.
2.9).
   $V$-space is usually called  the {\it quantum space} due to its  
significance for quantum models, while the $H$-space is named
as the {\it auxiliary space}. For finding  the partition function
involving $M$ such strings, one
has to repeat  the  procedure $M$ times to give 
$Z = tr (\prod^M T )=  tr ( T^M)
$.

The YBE (\re {ybe}) restricts the solution of the $R$-matrix 
 to  integrable models. However  
the $R$ matrix with $16$ different Boltzmann weights, 
 representing in general
a $16$-vertex model is difficult to solve. Therefore 
usually   some extra symmetry and conditions are
 imposed on the $R$-matrix to 
simplify its solution.
 For example, we shall require
 the  {\it charge} conserving symmetry $R^{ij}_{kl} \neq 0, ~$, only  when 
$k+l=i+j$,
 i.e.  the nontrivial elements to exist only when the sum of
  incoming  charges = sum of outgoing  charges,  along 
 with   a charge or arrow
  reversing  symmetry (see fig. 2.10).

 Using an overall normalization it leads to  a $6$-vertex model 
 expressed as
\begin {equation}
R(\lambda) = \left( \begin{array}{c}
a(\lambda) \ \qquad \ \qquad \ \qquad \\
    \quad \ b(\lambda) \  \ \ \ 1 \ \quad  \\
     \qquad \ 1 \ \ \ \ \ \
     b(\lambda) \ \quad \\
        \qquad \ \qquad \ \qquad \ a(\lambda)
          \end{array}   \right).
\ll{R-mat}\end {equation}
where  $a(\la),b(\la)$ are  unknown functions  of the 
difference  $\la-\mu,$
and the   spectral parameters $
\lambda$ and $\mu$   
 parametrizing the  Boltzmann weights
correspond to different pairs
of spaces.
 Inserting this
  form of $R$-matrix in the integrability condition YBE (\re {ybe}),
   one gets  simple equations like 
    $~~a(\la-\mu)b(\la)-a(\la)b(\la-\mu) -b(\mu)=0~~, $
  giving  a
      solution 
through trigonometric functions in spectral parameters
  as
  \be a(\lambda)= \frac{\sin (\lambda+\alpha)} {\sin \alpha},  \quad
     b(\lambda)= \frac{\sin \lambda} {\sin \alpha},  \quad \ll{abtrig}\ee
Here  $\al$ is the  parameter fixed for a certain model, while the spectral
parameter $\la$ is a variable one.  
Apart from (\re
 {abtrig})  the YBE (\re {ybe}) allows also 
another simpler solution of  (\re {R-mat}) given through rational
functions in spectral parameters:   
  \be a(\lambda)= \frac{\lambda+\alpha} { \alpha},  \quad
     b(\lambda)= \frac{\lambda} {\alpha},  \quad \ll{abrat}\ee
    We do not take up   more general {elliptic}
 $R$-matrix solution \c{xyz,baxter}, except indicating  only  
 its relation with the Sklyanin algebra \c{skalg} and the elliptic
quantum group \c{eqg} in sect. 3  and 9.
Apart from the  $XYZ$  spin chain and
the $8$-vertex  statistical model,
 there seems to be not many  important  models associated with
 the elliptic case and 
 the underlying   algebraic picture, except  
some recent developments \c{eqg},
  is also 
not much clear.  

\subsubsection {Quantum Lax operators of ultralocal models}
To understand the structure of the Lax operators better and to 
illuminate the relationship between the $R$ and $L$ matrices appearing 
in the QYBE (\re {qybel}), we consider 
  in the above   picture  (fig 2.9)  a situation, 
when the quantum ($V$) and the auxiliary ($H$) spaces  
  become  inequivalent. That is  keeping the $H$-space to be 
of  dimension $4,$  the 
  $V$-space is extended to  
 arbitrary dimensions.  The $R$-matrix  then turns 
into  a $2\times 2$-matrix:
$~~
  R_{1{\bf 2}}(\la) \rightarrow  L_1(\la)
  \in  V_1 \otimes {\bf V} 
 ,\ \ $
   with
  its  elements becoming  operators acting in a Hilbert
  space,
 where  to distinguish we  denote the
arbitrary dimensional  quantum spaces by bold letters.
This $L$-operator  corresponds to the familiar quantum Lax operator,
 which similar to  the 
 $R$-matrices of the statistical models,  
  constructs  monodromy and transfer matrices 
 for the global description of the associated quantum system.

If we choose for example, the space {\bf 3} to be quantum and nonisomorphic
 to  $1$ and $2$, then 
we have     $R_{1{\bf 3}}(\la)=L_1(\la)$ and $
    R_{2{\bf 3}}(\mu)=L_2 (\mu)  ,$ with $R_{1{ 2}}(\la-\mu)$ 
remaining the same.
As a result, 
the  integrability equation YBE (\re {ybe}) for the  $R$ matrix 
 of the statistical systems  
 turns into the  QYBE (\re {qybel}) involving  both $R$ and $L$ 
and  describing  the  integrability of  the  quantum systems (compare fig.
2.2 and 2.8).
Conversely, when the quantum and the auxiliary
spaces become isomorphic,    the $L$ , $R$ matrices
 become equivalent and the YBE  and 
QYBE coincide. This happens in the  case of  fundamental models like 
spin models. 

The $R$-matrix appearing in the statistical and the
quantum systems however are equivalent and given by  (\re {R-mat}), which   
 can also be  rewritten   as 
\begin {equation}
R_{12}(\lambda-\mu) = \Sigma _{i=0}^3 \ \omega_i(\lambda-\mu)\sigma^i\otimes
\sigma^i,    \quad\quad \sigma^0 =\frac{1}{2} { I}
\ll {R-opr}\end {equation}
%(3.2)
with $\omega_i(\la)$ being the Boltzmann weights given by $
\omega_0(\la)= a(\la)+b(\la),\ \omega_3(\la)= a(\la)-b(\la),\
 \omega_1=\omega_2=1$
and $\si^i$ are the Pauli matrices. For inequivalent spaces
 we  may
expect   the associated Lax operator
to be of  the form 
\begin {equation}
%-------------------------------------------
L(\lambda)
= \Sigma_{i=0}^3 \ \omega_i(\lambda)\sigma^i\otimes
 S^i,    \quad\quad \sigma^0 =\frac{1}{2} { I}
\ll{L-sklya}\end {equation}
%(3.3)
by  replacing  $\sigma^i$'s  with 
some abstract operators  $S^i$ acting in the quantum space.
We shall see  in sect. 3,  
 that such operators generate certain  quadratic
algebra  intimately connected to the
well known quantum  algebra \c{drinfeld}. Different 
 realizations of (\re{L-sklya}) yield different 
 Lax operators representing a class of integrable quantum  models, as will
be demonstrated in sect. 4. 
 Remarkably therefore, 
   the {\em same} { trigonometric}
  $R$-matrix (\re {abtrig})  of the $6$-vertex model  is  associated  
 with a   variety  of integrable quantum models
 like the   $XXZ$ spin chain,
    sine-Gordon model, Liouville model, massive Thirring model,
    derivative nonlinear Schr\"odinger  equation,
    Ablowitz--Ladik model etc. This
    intriguing fact
    will be  elaborated further in sect. 4 and  5, while  
a list of  such  systems with some necessary information   
 is given below.
%-----------------------------

Before  looking  into  relevant limits of the
parameters, let us
clarify certain points
in view of some  confusing remarks that are often  encountered  
in the literature in this regard.
 First we  note  that,
in  quantum   models 
 the discrete Lax operator $L_n(\xi,\eta,\De)$ like (\re{L-sg})
contains the spectral parameter $\xi=e^{i\eta \la}$, 
parameter $\eta$ and the lattice spacing $\Delta$,
but not  directly $\hbar.$ On the other hand, the $R( \xi,\al)
$-matrix (\re{R-mat}) 
is independent of $\Delta$, but  depends on the parameter
$\al=\hbar \eta$, i.e
on $\eta$ as well as  on the quantum parameter
$\hbar$. The parameter $\al$  induced by the $R$-matrix is also 
 present in  the quantum commutators 
of the  operators involved in $L_n$.
 The  parameter
 $\eta$
     represents the coupling constant of the quantum models 
(e.g. anisotropy
     parameter in the $XXZ$
     spin chain) and  defines at the same time the deforming
     parameter $q=e^{i\eta}$
     of the underlying quantum algebra.

Therefore the classical
limit $\hbar \rightarrow 0 $ affects the $R$-matrix as
\be
R(\ti \la,\al)= I+\hbar r(\ti \la,\eta)+ O(\hbar)
\ee
yielding the classical $r(\ti \la,\eta)$ matrix with $\ti \la =\eta \la$
, which determines the structure constants in    the Poisson bracket
relations among the corresponding Lax operators.
 However,  the
 classical form
of the Lax operator stays almost  the same, except only for the obvious   
 replacement of the field operators  by field functions.

On the other hand at  $\Delta \rw 0$, 
the discrete
Lax operators get affected by reducing to the continuum limit 
\be
L_n( \ti \la, \eta,\Delta)= I+\Delta {\cal L} (x,\ti \la, \eta)+ O(\Delta),
\ll{contl}\ee
 while the $R$-matrix remains the same for both discrete as well as   
 the field models. For example,  at this limit
(\re{L-sg}) turns into the well known Lax operator
 of the sine-Gordon
field  model (\re{sg}), but with the same trigonometric $R$-matrix
(\re {abtrig}). 

At  $\eta \rw 0,$  when the deformation parameter $q=e^{i 
\eta} \rw 1+i \eta$
one gets
 an altogether  different class of models. Since both $R(\xi,\hbar \eta)$
 and
$L(\xi, \eta)$-matrices depend on this parameter,
 both of them transform at this limit
with  the spectral parameter reducing to 
 $\xi\equiv e^{i\eta \la}\rw 1+i\eta\la$.
The trigonometric  $R$-matrix
 (\re{abtrig}) reduces  clearly to the   {\em rational} solution
(\re {abrat}), while the $L$ operator (\re{L-sklya}) reducing to
%-------------------------------------------

\be L(\lambda)
=  \lambda +  \Sigma_{a=1}^3   \left (\sigma^a\otimes
 s^a \right), \ll{L-rat}\ee
 with an irrelevant scaling of the spectral parameter.  ${ S^a }\rw { 
 s^a}$  are simply  the generators of the undeformed $sl(2).$ 
 Expressing ${ s^a}$  through   spin or bosonic operators,  
  (\re{L-rat}) can generate 
  Lax operators of the isotropic $XXX$ 
 spin chain and the  NLS model.
A list of such models, associated 
   with the same rational $R$-matrix is given
below.

\subsubsection {List  of quantum integrable ultralocal models}
 We present
 for ready references a list of  important ultralocal 
models  with defining Hamiltonian or the evolution equation,
along with  their  representative Lax operators.
The  $2\times 2$ discrete Lax operators $L_n(\la) $ and the continuum ones
${\cal L}(\la)$,  given here are 
expressed through matrices 
\[ 
e^1\equiv \ha(1+\si^3)  = \left( \begin{array}{c} 1 ~~~
  0 \\
   0  ~~~ 0
    \end{array} \right),~~
e^2\equiv \ha(1-\si^3)  = \left( \begin{array}{c} 0 ~~~
  0 \\
   0  ~~~ 1
    \end{array} \right),~~\si^+=
 \left( \begin{array}{c} 0 ~~~
  1 \\
   0  ~~~ 0
    \end{array} \right),~~\si^-=
 \left( \begin{array}{c} 0 ~~~
  0\\
   1  ~~~ 0
    \end{array} \right) \]
  and $\si^3=e^1-e^2$.  The 
matrix  elements of the discrete Lax operators are realized in
 Pauli matrices $\si_n^a$,  
bosonic operators   $u_n ,q_n, \psi_n$ with commutation
relations
$[q_n,p_m]=[u_n,p_m]=~[\psi_n,\psi^\da_m]
={\hbar  \ov \De}\de_{nm} $ or in $q$-oscillators of different types.
The corresponding continuum Lax operators ${\cal L}$   
 involve bosonic fields  
 $u(x,t), \psi(x,t)$  with
 $[u(x),p(y)]=\hbar \de(x-y)$ or $~~[\psi(x),\psi^\da(y)]=\hbar
\de(x-y)$.  

These ultralocal models  are quantum integrable and associated with 
either trigonometric or the rational $R(\la)$-matrix and 
 ,in principle, exactly solvable by the quantum inverse scattering 
 method (QISM) \c{qism,fadrev}, based on the $R$ and $L$ matrix formalism and 
the  Bethe ansatz method described in sect. 7.
The underlying algebraic structures, generation of the Lax operators and the
construction of  Hamiltonians for these models will be discussed in the
following sections.
 However it should be
mentioned, that for some models the explicit quantum Hamiltonian is difficult
to find and therefore for them 
we provide only the  corresponding classical analogs.  

\sk

\ni  {\bf I.  Models  with trigonometric $R_{trig}(\la)$-matrix }

\sk

Listed below are some field as well as lattice models, all 
associated with the trigonometric $R$-matrix (\re {abtrig}). 
 We have used the notations $p\equiv u_t,~~\xi=e^{-i
\la}$ and $\De$ is the lattice spacing.
\smallskip

\noindent
i) { Field models:}\\ \\
1. {\it Sine-Gordon  model} (SG) \c{sgFad} 
\bea 
&u(x,t)_{tt}& - u(x,t)_{xx} = \frac {m^2}{\eta} \sin\eta u(x,t), \nonumber \\
 & {\cal L}_{sg} & =  ip~\si^3+  
  m  \sin (\la-\eta u)  \si^+ +
   m  \sin (\la+\eta u) \si^-  
\ll{sg}\eea
                     \noindent
2. {\it Liouville model } (LM)  \c{kulskly}  
\be
u(x,t)_{tt}- u(x,t)_{xx} = \frac {1}{2} e^{2\eta u(x,t)}, \qquad
  {\cal L}_{lm}  = i\left( \begin{array}{c} p  \qquad
   \xi e^{\eta u} \\
  \frac {1}{\xi}e^{\eta u}  \qquad -p
    \end{array} \right).
\ll{lm}\end{equation}
                     \ni
4. {\it Derivative NLS  model} (DNLS) \c{kb-dnls} 
\[
i\psi(x,t)_{t}- \psi(x,t)_{xx} + 4i (\psi^\da(x.t)\psi(x,t))\psi(x,t)=0,
\]
\be
  {\cal L}_{dnls}  = i\left( ({
  -\frac{1}{4} \xi^2+k_-\psi^\da \psi}~e^1) + 
   {\xi \psi^\da} \si^+ + 
  {\xi\psi} \si^- + 
  ({\frac{1}{4} \xi^2-k_+\psi^\da \psi}) ~e^2
     \right).
\ll{dnls}\end{equation}
                     \ni
5. {\it Massive  Thirring model (bosonic)} (MTM) \c{kulskly}  
\[
\mathsf { L}=
\int dx \bar\psi(i\gamma^\mu \partial_\mu-m)\psi-\ha g j^\mu j_\mu
\quad j^\mu= \bar\psi\gamma^\mu \psi, \quad
 \ \  [\psi_a(x).\psi^\da_b(y)]=\hbar \delta_{ab} \delta(x-y) \]
\be
  {\cal L}_{mtm}  = i\left( \begin{array}{c}{
  \frac{1}{4}(\frac{1}{\xi^2}- \xi^2)+k_-\rho_1- k_+\rho_2}  \qquad
  \ {\xi \psi^\da_1+\frac{1}{\xi}\psi^\da_2} \\
  {\xi \psi_1+\frac{1}{\xi}\psi_2}  \  \quad
  {-\frac{1}{4}(\frac{1}{\xi^2}- \xi^2)-k_+\rho_1+ k_- \rho_2}
    \end{array} \right).
\ll{mtm}\end{equation}

\medskip

\noindent
 ii) { Lattice Models:}  \\ \\
1.  {\it Anisotropic $XXZ$ spin chain  (related to  $6$-vertex  model)}
 \c{xyz} 
\bea
{ H} &=& J\sum_n^N(\si_n^1 \si_{n+1}^1+\si_n^2 \si_{n+1}^2 +\cos \eta
\si_n^3 \si_{n+1}^3), \nonumber \\
{L^{(xxz)}_n }(\xi) &=& i[\sin(\la + 
\eta \si^3 \si_{n}^3)+\sin \eta ( \si^+ \si_{n}^-
+\si^- \si_{n}^+ )].
\ll{XXZ}\eea

\ni
2.  {\it Wu-McCoy  spin chain} (corresponds to asymmetric $6V(1)$ model)
  (WM)\c{WM,6v1} 
\bea
{ H} &=& \sum_n^N \cos \th (\si_n^1 \si_{n+1}^1+\si_n^2 \si_{n+1}^2)
+\cos \eta
\si_n^3 \si_{n+1}^3+{i \ov 2}\sin \th (\si_n^+ \si_{n+1}^--
\si_n^- \si_{n+1}^+) , \nonumber \\
{L_n^{(wm)} }(\xi) &=&
  \left( \begin{array}{c}
  e^{-i{\th \ov 2}}(\frac{1}{\xi} \phi_n^++ { \xi}
\phi_n^-)
 \qquad  2i\sin \al \si_n^- \\
 2i\sin \al \si_n^+    \qquad
  -e^{i{\th \ov 2}}(\frac{1}{\xi} \phi_n^- + { \xi} \phi_n^+) 
    \end{array} \right), \ \
    \mbox {where \ \ }
\phi_n^\pm=e^{{i \ov 2}({\th \pm \al})\si^3_n}
\ll{WM}\eea
3.  {\it Lattice sine-Gordon   model} (LSG) \c{sglsg} 
\bea
  L_{n}^{(lsg)}(\xi) & = &
 (  g(u_n)~ e^{ip_n \Delta }~e^1
+e^{-ip_n \Delta }~g(u_n)~ e^2) \nonumber \\ & + &\si^+ 
 m\Delta  \sin (\la-\eta u_n)\si^+ +  
   m\Delta  \sin (\la+\eta u_n)\si^- 
     ),  \ll{L-sg}\eea
 {where}  $ \ \ g(u_n)=~\left [ 1 + {1  \over  2}  m^2  \Delta^2
\cos  {2 \eta (u_n+{1 \over 2}) } \right ]^{1 \over 2} $.

\smallskip

\ni
4. {\it Lattice Liouville model} (LLM) \c{llm} 
\begin{equation}
  L_{n}^{(llm)}(\xi)  = \left( \begin{array}{c}   e^{p_n \Delta }~f(u_n)~
 \qquad  {\Delta}{\xi}e^{\eta u_n} \\
 \frac{\Delta}{\xi}e^{\eta u_n}   \qquad f(u_n)~  e^{-p_n \Delta }
    \end{array} \right) \quad \mbox {where} \ \ f(u_n)=[
     1 + {\Delta^2} e^{\eta(2u_n+i)}  ]^{{1 \over 2}}.
\ll{Llm}\end{equation}
     
\smallskip

\ni
5.{\it Lattice version of DNLS  model} (LDNLS)
 \noindent ({ $q$-oscillator model})\c{kb-dnls} \begin{equation}
  L^{(ldnls)}_n(\xi)  =
   (
  \frac{1}{\xi}q^{-N_n}- \frac{i \xi\Delta}{4}~q^{N_n+1} )~e^1 +(
  \frac{1}{\xi}q^{N_n}+ \frac{i \xi\Delta}{4}~q^{-(N_n+1)} )e^2,+
  {\kappa}A^\da_n \si^++ 
  {\kappa}A_n \si^-  
\ll{Ldnls}\end{equation}
     where \ \ $  A_n, A^\da_n \  $ are $q$-oscillators expressed through
  standard    bosonic operators as $\  A_n =
     (\frac{\Delta}{\hbar})^\ha~\psi_n \sqrt\frac{[N_n]_q}{N_n}, \ N_n=
   \frac{\Delta}{\hbar}  \psi^\da_n\psi_n$

\smallskip

\ni
6. {\it Ablowitz-Ladik model} (ALM) \c{ALM} 
\bea
{ H}= \sum_n b_n^\da( b_{n+1}-b_{n-1})+\frac{2\hbar}{\ln(1+\hbar)}
\ln(1+b_{n}^\da
b_n), \nonumber \\
L_n^{(alm)}(\xi) =\left(   { 
\xi^{-1} }~e^1 +
{ b_n^\dagger }\si^++ {b_n}\si^- + \xi ~e^2
            \right)
,\ll{ALm}\eea
 where $b_m , b_n^\dagger$ are the $q$-bosonic operators with commutators
$~~ [ b_m , b_n^\dagger ] =~ \hbar ( 1-b_n^\dagger b_n ) ~\delta_{m,n}.~~$

\sk

7.  {\it Relativistic quantum  Toda chain} (RTC)  (corresponds also 
to discrete-time 
Toda chain (DTTC)) \c{qrtoda}
\[
H=\sum_i\left(\cosh 2\eta p_i
+\eta^2 \cosh \eta (p_i+p_{i+1})e^{\eta (p_{i+1}-p_i)+(q_i-q_{i+1})}
\right),\]
\begin {equation}
L_n^{(rtc)}(\xi) = \left(( 
  \frac {1}{\xi}e^{2\eta p_n}
-\xi)~e^1 + \eta e^{q_n}\si^+
 -  \eta e^{2 \eta p_n-q_n
} \si^-  \right).
\ll{rtoda}\end {equation}

\ni 8. {\it  Discrete-time quantum  Toda chain like model} (DTTCL)
\c {qrtoda} 
\bea
L_{i}^{dttcl}(\lambda)& = &
\frac { i\gamma } { [i]_\eta} [\lambda + N_i 
+\omega_i]_\eta  ~e^1 
+ \frac {\alpha_i \beta_i}{\gamma}
       \cosh (\eta(N_i-\lambda))~e^2 \nonumber \\ &+&  \beta_i    f(N_i) p_i \si^+
 + \alpha_i \cos  \eta \ q_i  f(N_i)\si^-,
\ll{dttcl}\eea
%26
with $N_i=p_iq_i, ~~f^2(N)=
 \frac {i } {2\cos  \eta \ [i]_\eta}
\frac {1} {N}\left( [2N+\omega- i]_\eta
-[\omega- i\hbar]_\eta \right)
$

\ni {\bf II.  Models  with rational $R_{rat}(\la)$-matrix }

Here we list  ultralocal models associated with the rational $R$-matrix
(\re {abrat}).

\smallskip

\noindent
i){ Field models:}\\ \\
 1. {\it Nonlinear Schr\"odinger equation} (NLS) \cite{nls}
\begin{equation}
{\cal H}=\int dx \l( \psi(x,t)^\dagger_{x} \psi(x,t)_{x} + \eta
 (\psi^\da(x.t)\psi(x,t))^2 \r),
      \quad
{\cal L}_{nls}(\la)  = \left( \begin{array}{c} i\la 
\qquad \eta^\ha \psi^\da \\
\eta^\ha  \psi \qquad -i\la
    \end{array} \right).
\ll{nls}\end{equation}

\noindent
 ii) { Lattice Models:}  \\ \\
1. {\it Isotropic $XXX$ spin chain }
 \c{xyz}
\bea
{ H}&=& \sum_n^N(\si_n^1 \si_{n+1}^1+\si_n^2 \si_{n+1}^2 +
\si_n^3 \si_{n+1}^3), \nonumber \\
L_n^{(xxx)}(\la) &=& i( \la { I} + ( \si^3 \si_{n}^3+
\si^+ \si_n^-
+\si^- \si_n^+))
\ll{XXX}\eea

\ni
2. {\it Lattice NLS  model} (LNLS)  \c{sglsg} 
\be L^{(lnls)}(\la) = \left( \begin{array}{c}
\la +s- \De \psi^\da \psi   \qquad    \De^\ha(2 s- \De\psi^\da
\psi)^\ha\psi^\da
 \\
\De^\ha\psi (2 s- \De\psi^\da \psi)^\ha  \qquad \quad \la -s +\De \psi^\da \psi
          \end{array}   \right), \qquad s=-{2 \ov \kappa \De}.
\ll{lnls}\end {equation}
\sk

\ni
3. {\it Simple lattice NLS  type model} (SLNLS) \c{krjp94} 
\bea H
&=&{1\over\Delta}
  \sum_k (
  \phi(k+1) \psi(k -1) -  (N(k)+N(k+1))
  \phi(k+1) \psi(k)+ 
  (3\kappa\Delta^2)^{-1} N(k)^3 )
\nonumber \\
{ L}_{n}(\xi) & = & \xi \l( ( -{i \ov \xi} +{N(n)\over  \Delta })e^1
+  {{1\over\Delta}}~e^2
- i  \kappa^{1 \over 2}\phi(n) \si^++ 
i  \kappa^{1 \over 2}\psi(n)\si^-
\r)
\ll{slnls}\eea 
 with
 $N(k)=  1+\kappa \Delta^2 \phi(k) \psi(k).$ 

\sk

\ni
4.  {\it Toda chain  } (TC) (nonrelativistic, time-continuous) \c{toda}   
\be
H=\sum_i\left(\ha p^2_i
+ e^{(q_i-q_{i+1})}
\right),\qquad
L_n^{(tc)}(\la) = \left( \begin{array}{c}
  p_n
-\la \qquad    e^{q_n}
 \\- e^{-q_n
}
 \qquad \quad 0
          \end{array}   \right).
\ll{toda}\end {equation}
 \sk

\ni 5.
{\it  Quantum Toda chain like model} (TCL) \c{qrtoda,closetc}
\be
L_{i}(\lambda) = (\lambda +\gamma p_iq_i+\omega_i) ~e^1
+ \beta_ip_i \si^+   
 +  \alpha_i q_i  \si^- +\frac {\alpha_i \beta_i}{\gamma}~e^2
\ll{todalike}\end {equation}
%22
with constant parameters $\ga, \al_i, \bet_i, \omega_i$

\medskip

\ni
 6. {\it Tamm-Dancoff $q$-bosonic model (TDqB)} \c{krjp94}

An integrable $q$-bosonic model with    $q \equiv e^{i \th}$-deformed
rational $R$-matrix is given by 
 the Lax operator 
\begin{equation}
L^{q}(\la)= \left(  { (1 -i \la  \Delta +
  \kappa N) f(N)}~(e^1+ e^2)  
 {- i \sqrt { \kappa} c} \si^+ +
{ i \sqrt { \kappa} b   } \si^- 
     \right),  
\ll{slnlsq}\end{equation}
with $~~ f(N)=q^{\ha(N(n)-\ha)}~~$ 
involving Tamm-Dancoff type $q$-bosonic operators \c {TDqb}:

Historically  the above models and  their Lax operators 
were discovered  in  isolated  and seemingly unrelated way.
 However it will be shown    
  in sect. 5  using their underlying algebraic structure,
how  these models 
are interrelated and how they can be generated  systematically 
as {\em descendants} of some more general integrable {\em ancestor} models
(see fig. 5.1).

\subsection {
Integrability of quantum nonultralocal models: braided Yang-Baxter
equations }

As already mentioned, 
there exists  another  large  and important class  of models, called 
{\it nonultralocal}
models. Such models might  have inherent nonultralocality  in the
sense that the   basic fields may  satisfy  current like  algebras 
with derivatives of $\delta$-functions appearing in their commutators,
or they may be
  like   fermionic  or anyonic fields  noncommuting 
 at different lattice points. Alternatively,   
the  models with  standard canonical 
fields might also acquire   nonultralocality
due to 
   gauge transformation from  ultralocal models. Note that 
, the   gauge
transformations  in the quantum case unlike classical models,  might
   change the 
ultralocality property leading to a nonequivalent system.
 
 Since the ultralocality condition is
a key requirement in 
 deriving the representative   QYBE,  
the    above scheme  must be significantly modified   
 to make it  suitable for nonultralocal systems.
 We  introduce here  a scheme with extended QYBE and other
  relations for describing   nonultralocal quantum systems, serving 
 to some extend  
as universal  
equations  for a wide variety of such integrable models.
We provide     a list of  important nonultralocal 
models  at the end of this section, deferring  the details    
  to sect. 8.

\subsubsection {Braided structures  and  extensions of  QYBE}
For understanding the algebraic structures in nonultralocal  systems 
we return  to the simple matrix relations (\re {matrix1},\re {matrix2})
and see immediately, that both of them   
fail for nonultralocal quantum models, since now the Lax operators   
do not  commute at the same as well as at different points. Therefore 
the   algebra   with 
somewhat trivial multiplication property (\re{matrix2}) needs
to be generalized
as
   \begin {equation}
({A}\otimes  {B})(C \otimes D) =\psi_{BC} (A(C \otimes B)D)
\ll{multnm}
\end {equation}
where the noncommutativity of $B_2, C_1$ should be 
 taken into account. However the underlying coproduct structure, 
responsible for the  globalization of  QYBE must be preserved.
 Such an idea was realized  in the  braided algebra \c{majid,majidq,hlavqb}
 and  
was implemented in formulating the integrability theory of 
nonultralocal models \c{khijmp96}.
The basic point is to 
complement the commutation rule for the Lax operators at the same site
e.g. the local QYBE  with
the braiding properties  describing commutators  at different
lattice sites. 
The local QYBE  must also  be generalized to incorporate the  
 braiding relations, such that
 the transition from local to the global 
QYBE as in the  ultralocal case 
becomes possible.
 In general, since the braiding at different  sites  may  differ
widely  with   arbitrarily  varying  ranges,
 the  picture 
might  be too complicated for explicit description. Therefore  
we  limit
 here    to  {\it two} types of braiding only,  showing however
  that this limited choice is enough to describe most of the physical
models. Referring to
\c{maillet,khijmp96} for more details we present the braided QYBE (BQYBE)
as  (see fig. 2.11)   
\begin{equation}
{R}_{12}(\la-\mu)Z_{21}^{-1}L_{1j}(\la)\tilde Z_{21}L_{2j}(\mu)
= Z_{12}^{-1}L_{2j}(\mu) \tilde Z_{12}L_{1j}(\la){R}_{12}(\la-\mu).
\ll{bqybel}\end{equation}
where the 
 nearest neighbor braiding is described   by  matrix $Z$ , while  the 
braiding between other neighbors, taken to be the same for all pairs,  
are given  by a single  matrix $\tilde Z$. Note that these braiding matrices 
in general  can also have spectral parameter  dependence as
$Z_{21}=Z_{21}(\la,\mu),~~~Z_{12}=Z_{12}(\mu,\la)$ and similarly
$\ti Z_{21}= \ti Z_{21}(\la,\mu),~~~\ti Z_{12}=\ti Z_{12}(\mu,\la)$.
We  however suppress this explicit dependence in all the relevant equations
for brevity.  
To describe  nonultralocal systems, the BQYBE  must be  complemented 
 by the  braiding relations  
\begin{equation}
L_{2 j+1}(\mu) Z_{21}^{-1}L_{1 j}(\la)
=\tilde Z_{21}^{-1}L_{1 j}(\la)\tilde Z_{21}
 L_{2 j+1}(\mu)\tilde Z_{21}^{-1}
 \ll{zlzl1u}\end{equation}
  at   nearest neighbor
sites and by
\begin{equation}
 L_{2 k}(\mu)\tilde Z_{21}^{-1}L_{1 j}(\la)
=\tilde Z_{21}^{-1}L_{1 j}(\la)\tilde Z_{21}
 L_{2 k}(\mu)\tilde Z_{21}^{-1}
\ll{zlzl2u}
\end{equation}
with $k>j+1$ 
  answering for all nonnearest neighbors (see fig. 2.12).
Note that along with  the conventional  quantum $ R_{12}$-matrix 
as in ultralocal systems,  
additional  $ \  \tilde Z_{12} , \ Z_{12}$ matrices 
 appear here, which may or may not depend on  spectral parameters.
When $Z$ matrices are spectral independent, the above 
BQYBE and braiding
equations can be obtained through Yang-Baxterization \c{khijmp96}
from the more fundamental spectral parameterfree
algebraic relations like
\bea
{R}_{q12}^+Z_{21}^{-1}L_{1j}^\pm \tilde Z_{21}L_{2j}^\pm
&=& Z_{12}^{-1}L_{2j}^\pm \tilde Z_{12}L_{1j}^\pm {R}_{q12}^+
\nonumber \\ L_{2 j+1}^\pm Z_{21}^{-1}L_{1 j}^\pm
&=&\tilde Z_{21}^{-1}L_{1 j}^\pm \tilde Z_{21}
 L_{2 j+1}^\pm \tilde Z_{21}^{-1}
\ll{bqybe+}\eea
etc. Note that the relations like (\re{bqybe+}) are also obtained from the 
BYBE (\re{bqybel}) etc. at the limit $\la \rw \pm \infty$. 

The braiding relations (\re {zlzl1u}-\re {zlzl2u}) are the 
generalizations of the
ultralocality condition and  bear the true 
essence of the nonultralocal systems,
where nonnearest and nearest neighbor braiding can be treated differently.
In analogy with the derivation of YBE (\re {ybe})
  as the consistency condition of
the QYBE, one can derive also the corresponding
relations   for nonultralocal models consistent with (\re {bqybel}). 
Such YBE type equations involving
$R$ and $Z$ matrices
are 
\begin{equation}
   R_{12}(\la-\mu)~Z_{2{ 3}} ~Z_{1{ 3}} ~~
=~~Z_{1 { 3}}
    ~Z_{2 { 3}} ~R_{12} (\la-\mu)~,~~~
Z_{13}~Z_{1{ 2}} ~R_{2{ 3}}(\mu-\nu) ~~
=~~~R_{2{ 3}}(\mu-\nu)
    ~Z_{1 { 2}} ~Z_{13},~~~
\ll{bybe12}\end{equation}
with similar equations relating  $R,~$ and $~ \ti Z$.
 Moreover the $Z, \ti Z$ 
 and $R$ matrices  also  satisfy  individually  
 the  YBE (\re {ybe}).
Strictly speaking, the validity of YBE (\re{ybe}) for the $Z$-matrix
is a sufficient but not
necessary condition. One gets rather the YBE for  
 matrix $\ti R_{12}(\la)=Z_{1 { 2}}
R_{12}(\la) Z_{{ 2}1}^{-1}$. This  typical situation can be seen in the 
example of nonultralocal quantum mapping listed below and
discussed in sect. 8. 

  The $R$-matrices associated with the nonultralocal 
models   are the same trigonometric (\re {abtrig}) 
and rational (\re {abrat}) solutions 
 (elliptic solutions  are not considered here) found earlier 
for ultralocal models. 

The  set of relations (\re{bqybel}-\re{zlzl2u}) represent the universal 
equations for the integrable nonultralocal models
within a certain class of braiding. We shall show in sect. 8 that 
much in parallel to the established ultralocal case, 
particular choices for $R,L, Z, \tilde Z$  can  
derive systematically from these  relations the 
 basic equations for  concrete physical models.
 It is readily seen  that the trivial choice  $Z=\tilde
Z={ I} $ 
reduces the above set into the standard QYBE (\re{qybel}) together
with the ultralocality  condition, while  the nontrivial
 $Z$'s would lead to different types of nonultralocal models.
 For example, 
 particular  homogeneous 
braiding with
 $Z=\tilde Z \neq {I}$  correspond to    
 SUSY and   anyonic models, while   
the choice $\tilde Z={ I}$ with  $Z \not = { I}$ 
 describes  nonultralocal models like WZWN, quantum mKdV
 etc. with   $ \delta'$ 
function appearing in their fundamental commutation relations. The 
case like $Z={ I},~~\tilde Z \neq 0$  
 coincides  formally with the  reflection  
equation  \c{skly-r}.

For establishing the role of 
 the extended  QYBE along with the braiding relations
for describing   the  integrability of nonultralocal models,
 we have to show first that the important  
transition from local to the global QYBE is  
possible 
for the  monodromy matrix
\be T_a^{[k,j]}(\la) =L_{ak}(\la) L_{a,k-1}(\la)\ldots
   L_{aj}(\la)\ll{Tkju}\ee
acting in the space
${\cal H} \equiv V_k\otimes V_{k-1}\otimes\ldots\otimes V_j.$
Similar to our derivation for the ultralocal models 
we start from the local BQYBE (\re {bqybel}) at point $j+1$, multiply 
both sides by  
   $Z_{21}^{-1}L_{1j}(\la)\tilde Z_{21}L_{2j}(\mu)$ (see fig 2.13a, 2.14a) 
and   using 
   the same BQYBE 
 at $j$ in the $rhs$   transfer  $R_{12}(\la-\mu)$ to the extreme right
(fig 2.14a-b). 

We    use next the braiding 
for nearest neighbors  (\re{zlzl1u}) (fig. 2.13a-b)
 and its $1 \leftrightarrow 2$
interchanged form (fig. 2.14 b-c)  
to drag  $L_{j}$ through $L_{j+1}$ in both the sides. Some $ Z$ matrices
gets canceled with $Z^{-1}$ allowing us to  arrange together  
 the Lax operators acting in the same space $1$ or $2$ to  
give  again the same form of BQYBE  for  $T_a^{j+1 j}
(\la)=L_{aj+1}(\la)L_{aj}(\la)  $ (fig. 2.13c, 2.14d).
 We  repeat further the same steps
multiplying  by
  $Z_{21}^{-1}L_{1j-1}(\la)\tilde Z_{21}L_{2j-1}(\mu)$ again 
from both sides and   using 
 the similar arguments. We have to remember only that  together  with
(\re {zlzl1u}) for $L_j$ and $L_{j-1}$ we have to use also 
the braiding relation (\re {zlzl2u}) for commuting $L_{j+1}$ and $L_{j-1}$.
Continuing for $j-k+1$ steps we  get 
the global  BQYBE is the same form as the local one:
\begin{equation}
{R}_{12}(\la-\mu)Z_{21}^{-1} T_{1}^{[k,j]}(\la)\tilde Z_{21}
T^{[k,j]}_{2}(\mu)
= Z_{12}^{-1} T^{[k,j]}_{2}(\mu) \tilde Z_{12}
T^{[k,j]}_{1}(\la){R}_{12}(\la-\mu)
\ll{rztzt}\end{equation}
confirming the existence  of the coproduct structure
of the underlying braided algebra 
in  nonultralocal systems.

For showing the quantum integrability one has to find the commuting
transfer matrices through trace factorization, which becomes a
nontrivial task due to the appearance of $Z$ matrices in the BQYBE
(\re {rztzt}). We shall see in sect. 8 how to tackle
 this problem and  modify the BQYBE 
for periodic models.

Investigations of some nonultralocal systems from  other angles can be found
in \c{reshet,goddard,volkov95}
\subsubsection {List of quantum integrable nonultralocal  models}
 Nonultralocal
 models are mostly nonfundamental ones, defined  in the  Hilbert space.
They may correspond to 
 genuinely integrable systems with spectral parameter dependent
$R(\la)$-matrix and the Lax operator or one may be interested only in their
nonultralocal algebra (\re{bqybe+}) without invoking any spectral parameters.  
We  collect  such quantum integrable models, which  can be  
described fairly well  by  the present formalism 
described by the relations (\re{bqybel}-\re{zlzl2u})
and list them below mentioning briefly  the  
  choices for their  braiding 
 $\tilde Z, \ Z$  apart from the $R$-matrix and the Lax operators.
 Details of  these models 
together with the systematic derivation of  their basic equations 
will be given in sect. 8.

\sk

\ni {\bf I. Systems with spectral parameterless $R$-matrix} 
 
Since the algebraic structures of the models are of concern here,
we may take the spectral parameterfree equations 
like (\re{bqybe+}) involving  $R^+_q$
matrix (\re {R+}) for describing the nonultralocal features.

\sk

\ni  1. {\it Current algebra in WZWN model} \c{wzwn}

The model involves   the nonultralocal current algebra
\be \{L_1(x),L_2(y)\}= \frac {\gamma} {2} [C,L_1(x)-L_2(y)]\delta(x-y)
+ \gamma C \delta'(x-y) \ll{current}\ee
with $C_{12}= 2P_{12}-1$, where 
$L= \frac {1} {2} (J_0+J_1)$ with $J_\mu= \partial_\mu gg^{-1},$
being the  current. 
For  describing the discretized algebra
 one needs the spectral parameter independent 
choice $R_{12}=Z_{12}=R_{q21}^+$  with 
 $\tilde Z=1.$
\\ \\
 2. {\it Coulomb gas picture of  CFT}  (CG-CFT)\c{babelon}

The associated  nonultralocal 
Lax operator is  \be  {\cal L}_{cg}(x)= P(x)- {\cal E}_+,
 \quad  {\cal E}_+=\sum_{\alpha} E_\alpha,
~~ \{ P(x) \otimes, P(y) \}= \delta'(x-y) \sum_i H_i \otimes H_i 
\ll{pbcft}\ee
  Integrable  description is made by the 
 solution $R=R_{q12}^+ $    
with the choice $\tilde Z=1$ and
 $Z_{12}=
q^{-\sum_i H_i
\otimes H_i}
$.
\\ \\
  3. {\it Braided  algebra} (BA) \c{majid}

The braided algebras described in sect. 3 can be reproduced from the
spectral-free   BQYBE (\re{bqybe+}) as a particular case, by choosing
homogeneous braiding $~~~\ti Z_{12}= Z_{12}= R_{21}~~$. As $R_{12}$  one
may take various  solutions of (\re{braid}) like (\re{R+}), nonstandard
(\re{nstR}) , Jordanian solution etc. to reproduce results of \c{majid}. 
\\
\\
\ni {\bf II. Models with rational  $R_{rat}(\la)$-matrix}

 Nonultralocal  quantum models associated with the
spectral-dependent rational  $R(\la)$-matrix (\re{abrat}) are listed below.
\sk

\ni 4. {\it Nonabelian Toda chain } (NATC) \c{natoda}

 The Lax operator of the model is given by
\be
L_{k}^{(ntc)}(\lambda) = \left( \begin{array}{c}
\lambda - A_k
\qquad  \ \ -B_{k-1}
 \\ I
\qquad \ 0
         \end{array}   \right), \ \ A_k=\dot{g}_k g_k^{-1}, \ \  B_k= g_{k+1}
 g_k^{-1}
\ll{Lkor}\ee
along with  the rational  
$~R(\lambda)=ih P\ot \pi-\lambda$ and  the  braiding
  $\tilde Z=1$  and $Z_{12}=
 {\bf 1} + i { h }(
 e_{22}\otimes e_{12})\otimes \pi$
\\ \\
 5. {\it Nonultralocal quantum mapping }\c{Nijhof}

The  system is  
associated  with $\la$ dependent Lax operator
$L_n^{(qm)}=V_{2n}V_{2n-1}$, with 
\be V_{n}=\Lambda_n (\la) \left(1+\sum^N_{i>j=1}v_{i,j}(n) E_{i,j}\right).
\ll{Lqm}\ee
with $~\Lambda_n= \la_n  E_{N,1} +\sum_{i=1}^{N-1} E_{i,i+1},~ \la_{2n}=
 \la_{2n+2}=\la~$,  
involving nonultralocal algebra 
\be [v_{i,j}(n),v_{k,l}(m)]=
 h(\delta_{n,m+1}\delta_{k,j+1}\delta_{i,N}\delta_{l,1}-
 \delta_{m,n+1}\delta_{i,l+1}\delta_{k,N}\delta_{j,1})\ll{algqm}\ee
For integrable description one needs  the rational $R(\la)$-matrix
 with $\tilde Z=1$
and the  spectral parameter dependent choice
 $\  \ Z_{12}(u_2)= { I}+ \frac { h }{u_2}\sum_\alpha^{N-1}e_{N \alpha
}\otimes e_{\alpha N}. \ \ $
\\ \\
\ni {\bf III. Models with trigonometric $R_{trig}(\la)$-matrix}

Nonultralocal integrable quantum models 
associated with the spectral dependent trigonometric $R(\la)$-matrix
(\re{abtrig}) are listed below.

\sk

\ni 6. {\it Quantum mKdV model  } (QMKDV) \c{kmpl95}

The nonultralocal Lax operator
\be
{\cal L}_{mkdv}(x,\zeta)= \frac {i}{2} (-v^-(x) \sigma^3+\zeta \sigma^2)
, ~~~~~~~  [v(x), v(y)]= \delta'(x-y)\ll{Lcmkdv}\ee
 describes this integrable  model with classical equation
\be \pm v_t+v_{xxx}-6v^2 v_x=0 .\ll{mkdve}\ee
The  quantum model corresponds to  the trigonometric
 $R$-matrix   and 
 $\tilde Z=1$ , 
$ Z_{12}=  Z_{21}= q^{-\frac {1}{2} \sigma^3\otimes \sigma^3
},$
\\ \\
 {\bf IV. Models with homogeneous braiding}

 Nonultralocal models with 
homogeneous braiding have  no distinction between 
braiding at different sites.

\sk 

\ni 7. {\it Integrable model on moduli space } (IMMS) \c{alex}

 This  integrable model, motivated by the  Chern-Simons theory, 
 may be  introduced  by the 
 Lax operator  
\be
L_n(\la) = K^n + \la I , \ll{Lmoduli0}\ee 
where  $K^n$ matrices are related to  $L^\pm$  generating 
quantum algebra.
This abstract  nonultralocal model
may  be described by the  trigonometric
$R_{trig}(\la)$ matrix along with the 
  braiding   $\tilde Z_{12}=Z_{21}^{-1}=R^+_{12q}$ 
\\ \\ 
8. {\it Supersymmetric models } \c{SUSY} (SUSY)

 SUSY models can be defined from the nonultralocal  point
by choosing 
 $Z=\tilde Z=\sum \eta_{\al \bet} g_{\al \bet}$, where $ \eta_{\al \bet}= 
e_{\alpha \alpha}\otimes e_{\beta \beta}$ and $ g=
 (-1)^{\hat \alpha \hat
 \beta}$ with  supersymmetric grading  $\hat \alpha.$
 $R$-matrix is either  rational
or the nonstandard trigonometric  type (\re{nstR}).
\\ \\
9. {\it Integrable Anyonic  model} (IAM)

Generalizing the SUSY model  one may choose
$Z=\tilde Z=  
 \sum \eta_{\al \bet} \tilde g_{\al \bet}$, with  
   anyonic phase $\th$ included in $\tilde g_{\al \bet}=e^{i \theta 
\hat \alpha \hat
 \beta}$. Integrable   anyons correspond to  the rational  $R$ matrix
,  while  their $q$-deformations are associated with  
the trigonometric case.
\\ \\
\ni {\bf V. Systems with specific braiding}

Here we list integrable 
nonultralocal models models, which correspond to braiding a bit  different from
$Z, \ti Z$ used in the  present formulation. 
\\ \\
 10. {\it Kundu-Eckhaus equation} (KEE) \c{kun84}

Classically integrable equation   
\be i\ti \psi_t+ \ti \psi_{xx}+ \eta  (\ti \psi^\dagger \ti\psi)
\ti \psi
+ \theta^2   (\ti\psi^\dagger \ti \psi)^2 \ti \psi +2i \theta
 (\ti\psi^\dagger\ti \psi)_x \ti\psi =0, \ll{kuEk}\ee
corresponding  to the Lax operator 
\be {\cal L}_{ke}(\la)  = \left( \begin{array}{c}i ( \la -
 \th \ti \psi^\dagger \ti\psi) 
\qquad \eta^\ha  \ti \psi^\da \\
\eta^\ha \ti \psi \qquad - i(\la 
- \th \ti \psi^\dagger \ti\psi)
    \end{array} \right)
\ll{lkuEk}\end{equation}
 is nonultralocal as quantum model due to noncanonical field $\ti \psi$
 and can be described  
by the   rational $R(\la)$ matrix along with braiding expressed through 
 $~~~
Z= diag (e^{i\theta},1,1,e^{i\theta})~~$.
 The success and difficulties 
with the model are discussed in sect. 8.
\\ \\
11. {\it Reflection equation} (RE) \c{skly-r}

The reflection equation may also be viewed as nonultralocal model 
  given through a different type of braiding. However formally
it fits  with the  present formulation 
for  nonperiodic models 
with braiding  $Z=1, \tilde
Z=R(\la_1+\la_2)$.

 We shall elaborate on all the above  models in sect. 8 showing their
systematic classification in fig. 8.1.

\vspace* {1cm} \fbox{fig.2.1} 

\vspace* {1cm}
{\bf Figure 2.1} \ {\it Ultralocality condition on the Lax operators showing
trivial comuutation of operators defined  at points $i \neq j$ .}

\vspace* {1cm} \fbox{fig.2.2} 

\vspace* {1cm}
{\bf Figure 2.2} \ {\it Local quantum Yang-Baxter equation (\re{qybel}).
$R,L$ matrices act at the intersecting lines 
with $1,2$ representing auxiliary and ${\bf 3}$ quantum spaces}.

\vspace* {1cm} \fbox{fig.2.3} 

\vspace* {1cm}
  {\bf
Figure 2.3 } \ {\it Multiplication rule (\re{mull}) induced by the
ultralocality condition of the Lax operators.}

\vspace* {1cm} \fbox{fig.2.4} 

\vspace* {1cm}
  {\bf
Figure 2.4 } \ {\it Flipping the $R$-matrix using QYBE
 (\re{qybel}) at point $i$.}

\vspace* {1cm} \fbox{fig.2.5} 

\vspace* {1cm} 
 {\bf Figure 2.5 }\ {\it QYBE for the nonlocal objects $T^{[i+1 i]}(\la)~$ 
defined at two points $i+1, i$.}

\vspace* {1cm} \fbox{fig.2.6} 

\vspace* {1cm}
  {\bf
Figure 2.6 }\ a) {\it Global QYBE (\re{qybet}) for the monodromy matrix 
$T(\la)$.}\  b) {\it $T(\la)$  is a global object defined through the Lax
operators $L_j$ at points $N$ to $1$.}

\vspace* {1cm} \fbox{fig.2.7} 

\vspace* {1cm}
 {\bf
Figure 2.7 } \  {\it Trivial commutation of transfer matrices $~\tau(\la)=
tr T(\la)$ at different $\la$. This feature of integrability  ensures 
involutiveness of conserved quantities.}

\vspace* {1cm} \fbox{fig.2.8} 

\vspace* {1cm}

 {\bf
Figure 2.8 }\ {\it Yang-Baxter equation (\re{ybe}) showing equivalence of 
two scattering processes to reach from the arrangement of particles (1,2,3)
to (3,2,1)}.

\vspace* {1cm} \fbox{fig.2.9} 

\vspace* {1cm}
  {\bf
Figure 2.9 } \ {\it Vertex model with periodic boundary condition.
Due to  repeated summation indices  $\mu_i$ ( H-space indices), 
the transfer matrix $\tau$ 
depends only on the quantum indices $(\alpha, \beta)$ }.

\vspace* {1cm} \fbox{fig.2.10} 

\vspace* {1cm}

  {\bf
Figure 2.10 } \  {\it Boltzmann weights of the $6$-vertex model with charge
conservation and charge inversion symmetry, constituting the elements of the
$R$-matrix}

\vspace* {1cm} \fbox{fig.2.11} 

\vspace* {1cm}
 {\bf
Figure 2.11} \  {\it  Local braided QYBE
(\re{bqybel}). Intersecting (wavy) lines $1,2$ with crossing from below
denote $Z_{12} $ or $\ti Z_{12}$, while from above denote their 
inverses.}

\vspace* {1cm} \fbox{fig.2.12} 

\vspace* {1cm}
  {\bf
Figure 2.12 } \  {\it 
Braiding relation (\re{zlzl2u}) showing nontrivial commutation 
between  Lax operators $L_{2k}(\mu)$ and $L_{1j}(\la)$ at different
points $k>j+1$. For $k=j+1$, i.e. for  (\re{zlzl1u}),  in the $lhs$
of the figure $\ti Z$  should be replaced by $Z$.}

\vspace* {1cm} \fbox{fig.2.13} 

\vspace* {1cm}
 {\bf
Figure 2.13 } \  {\it  Successive steps  of the 
 BQYBE for transition from its local to the 
two point form for nonlocal} $T^{[j+1,j]}$
 (The figure shows the
{\it lhs} of the equation).
\\ a){\it 
 Local BQYBE for $L_{j+1}(\la) $ multiplied by Lax operators  at site} $j$.
\\
b){\it  Use of braiding relation (\re{zlzl1u}) to drag $L_j$ through
} $L_{j+1}$( use fig. 2.12 in the boxed portion of a)-b)).\\
c){\it  BQYBE for} $T^{[j+1j]}=L_{j+1}(\la) L_{j}(\la) $.

\vspace* {1cm} \fbox{fig.2.14} 

\vspace* {1cm}

 {\bf
Figure 2.14 } \  {\it 
 Successive steps  of the 
 BQYBE for transition from its local to the 
two point form for nonlocal} $T^{[j+1,j]}$
 (The figure shows the
{\it rhs} of the equation).
\\ a) {\it Local BQYBE at  $j+1$ multiplied by objects at}  $j$.~~
b) {\it Flipping of $R$-matrix using BQYBE at} $j$ (use fig. 2.11 in the
dashed boxes in a)-b)).\\
c) {\it Use of braiding  (\re{zlzl1u}) with $1 \rightleftharpoons 2 $
(apply fig 2.12 in line boxes in b)-c).
}~~ d){\it 
 Transition to  BQYBE for $T^{[j+1j]}$.( Use of $Z_{12} Z^{-1}_{12} =I$}
unknots the lines simplifying the picture of b)) .

%_________________________________________\input {preport3.tex}
%Preport3.tex (sec 3 : 19 pages in 1.8 skip
\setcounter {section} {2. }
\section {Algebraic aspects}
In this more or less self-contained section we present
some useful  and elegant algebraic structures
hidden behind the integrable theory and intimately connected with it.
The application of these formal objects for constructing basic relations in
quantum integrable systems, will be considered  in subsequent sections.
Though the  subject matter is    rather
abstract,
we  try to maintain our presentation at   simple physical level.
\subsection{  Sklyanin algebra}
We  focus on  the local QYBE (\re{qybel})
$~~~ \
R(\lambda , \mu)~ L_1(\lambda)~ L_2(\mu )
~ = ~  L_2(\mu )~ L_1(\lambda)~ R(\lambda , \mu),
~~$
 the key relation in  quantum integrable ultralocal systems and consider
first the  general $R$-matrix solution of (\re{ybe}) given through the elliptic
functions:
\bea
R(\lambda-\mu)& =& \Sigma _{i=0}^3 \ \omega_i(\lambda-\mu)\sigma^i\otimes
\sigma^i, \qquad  \sigma^0 = { I}
\nonumber \\
\om_0+\om_3&\equiv& a(\la)=\sn (\la+\al), \
\om_0-\om_3\equiv b(\la)=\sn \la, \nonumber \\
\om_1+\om_2&\equiv& c(\la)=\sn \al, \
\om_1-\om_2\equiv d(\la)= k ~ \sn \al ~ \sn \la ~ \sn (\la+\al)
\ll{xyz}\eea
The corresponding  solution for the $L(\la)$-operator
may be chosen  in analogy with (\re{xyz}) as (\re{L-sklya}): $~~
L(\lambda) = \Sigma_{i=0}^3 \ \omega_i(\lambda)\sigma^i\otimes
 S^i.\ ~~$
The algebraic properties of  $S^i$, which  are yet unspecified operators,
are to be determined from the QYBE.
Inserting the above $R(\la-\mu)$-matrix 
and the $L(\la)$-operator 
 in QYBE (\re{qybel}), we  notice that  except operators $S^i$
 all other elements entering in
this equation  are  known objects. Comparing  therefore 
 the terms with independent functions
 of $\la,\mu$,  we  arrive
at a set of algebraic  relations for  $S^i$,
free from  spectral parameters:
\be
[S^0,S^\al] = i J_{\bet \ga} [S^\bet,S^{\ga}]_+ ,\qquad
  [ S^{\al} , S^{\bet} ] = i[S^0,S^{\ga}]_+
\ll{sk-ealg}\ee
involving  commutator $[\quad,\quad]$ and anticommutator 
$[\quad,\quad]_+ $  along
 with the Casimir operators
\begin {equation}
\bar {C}_1=\Sigma _{i=0}^3 (S^i)^2 , \quad
\bar {C}_2=\Sigma _{\al=1}^3J_\al (S^a)^2 .
\ll{c12}\end {equation}
The triplet of indices $ \ (\al, \bet, \ga)\ $ represents any
cyclic permutation of $\ (1, 2, 3)\ $ and
the structure constants $J_{\al \bet }={\om_\al^2 -\om_\bet^2 \ov
\om_\ga^2 -\om_0^2}$ may be parametrized  by
$\quad  J_{\al \bet} ={J_\al-J_\bet \ov J_\ga}\ .\ $ 
and obey the constraint $~~ J_{12}+J_{23}+J_{31}+~ J_{12}J_{23}J_{31}=0$.
Note
 that in contrast to a Lie algebra,  (\re{sk-ealg}) is a
  quadratic  algebra, which depends
 on the parameters $\alpha, k$. This quadratic algebra,  discovered
in  
 \cite{skalg} is known as the {\it Sklyanin algebra}.

\subsection { Quantum algebra and its  origin   in 
integrable theory}
We will be interested mostly in the trigonometric
 limit: $k\rw 0$ of the Sklyanin algebra     (\re{sk-ealg},\re{c12}),
 since at this limit  it acquires a clear  Hopf algebra
property and gets linked with the quantum algebra.

Since at $k \rw 0,
\quad \sn x \rw \sin x, \ \cn x \rw \cos x \  $ and $\ \dn x \rw 1 \ ,$
the elements of the $R$-matrix (\re{xyz} ) 
turn into trigonometric functions as
 \[ a(\la)=\sin (\la+\al), \
 b(\la)=\sin \la,
\ c(\la)=\sin \al, \
 d(\la)= 0 .\]
The corresponding trigonometric limit of the Sklyanin algebra
 may be obtained from  
the limiting values of the parameters
 giving
 the structure constants  
 \be J_{12}=0, \ J_{23}=- J_{31}=\tan^2 \frac {\alpha} {2}.\ll{jtrig}\ee
 This   clearly  reduces (\re{sk-ealg}) to
\begin{eqnarray}
[S^0,S^3] = 0, \quad  [S^3,S^{\pm}] = \pm [S^3,S^{\pm}]_+ , \nonumber \\
   \ \quad [ S^{0} , S^{\pm} ] =
 \mp \tan^2 \frac {\alpha} {2} [ S^3 , S^{\pm}]_+,\
 \quad [ S^+, S^-]=4 S^0 S^3.
\ll{tsa}\end {eqnarray}
The  Casimir operators  (\re{c12}) reduce to
\begin {equation}
\bar {C}_1 \rw C_1=\Sigma _{i=0}^3 (S^i)^2 , \quad
\bar {C}_1- \bar {C}_2 \rw
C_2 =  \cos^2 \frac {\alpha}{2} \ (S^0)^2 + \sin^2\frac {\alpha}{2}
  \  (S^3)^2
.\ll{tc12}\end {equation}
This { trigonometric Sklyanin algebra} (TSA) 
is also  quadratic 
and depends  on
a single parameter  $\alpha.$ 

Instead of
obtaining the TSA as a limit of (\re{sk-ealg}),
  we may
get it directly by using the trigonometric solution of the 
 $R$-matrix and  constructing in analogy
the  corresponding
 $L$ operator.
To complete this  chain of relationships, we  should connect the 
 TSA with the well known quantum algebra (QA) and
 that would     link  the origin of the QA
  to the
 Yang--Baxter equations and thus to  the  quantum 
integrable theory.
Let us notice first that  the Casimir operator
  $  C_2 
  $ given   in the form of an 
   ellipse suggests  a
    simple  realization (putting $C_2=\frac {1}{4} { I}$)
\begin {equation}
  S^0=  \frac {\cos (\alpha s^3)}{2\cos \frac {\alpha}{2}},\quad
  S^3=  \frac {\sin( \alpha s^3)}{2\sin \frac {\alpha}{2}}
\ll{sk-real}\end {equation}
through a single  operator $s^3$.
  This realization together with  the same  $S^{\pm}, $
  reduces the quadratic algebra  (\re{tsa})
 to 
\begin {equation}
 [s^3,S^{\pm}] = \pm S^{\pm} ,\quad
  [ S^+, S^-]= [2 s^3]_q.
\ll{sl2qa}\end {equation}
 The remaining Casimir operator 
in (\re{c12}) reduces to  
\begin {equation}
 [j]_q [j+1]_q = [s^3]_q [s^3-1]_q + S^+S^-.\ll{cas-qa}\end {equation}
where the $q-bracket$ is defined as \[ [x]_q= \frac {q^x-q^{-x}}{q -q^{-1}}
=  \frac {\sin( \alpha x)}{\sin \alpha} , \ \ q=e^{i \alpha}. \]
This    
  in fact is  the well known
     quantum  algebra   $U_q(sl(2))$  \c{drinfeld,qa}, 
and that  
   fulfills our aim of finding the origin of QA in  integrable systems.

 At the limit $q\rightarrow 1$ (or
  equivalently at $\alpha \rightarrow 0 $ ) \
, when  
\[ [2 s^3]_q \approx 2 s^3+\frac {\alpha^3}{3}(2 s^3)^3+ \cdots ,\]
  the deformed algebra (\re{sl2qa}),  clearly goes to
    the familiar  $su(2)$ spin algebra
\begin {equation}
 [s^3,s^{\pm}] = \pm s^{\pm} ,\quad
  [ s^+, s^-]= 2 s^3.
\ll{sl2}\end {equation}
The related
       Casimir operator  reduces to 
$  ~~
 j(j + 1) = s^3(s^3-1)+ s^+s^-
~.~$
We should remember that
   the  quantum algebra,  
    sometimes called as the   {\em quantum group}, 
        is in fact
     neither { quantum} nor a  { group}.
     Nevertheless, it has
      got  some
     analogy with the quantization procedure (see for example
    \c{fadq} as well as with group properties.

Quantum or quantized algebras  may be introduced in different ways.
 However the way it has been obtained here linking it with  the integrable
  systems, 
gives perhaps more physical insight of this abstract mathematical structure.
 The 
 $q$-deformation   may be viewed as  $q=e^{i\hbar \eta}$ 
by relating   the deforming parameter $\eta$ 
to the coupling constant  of the physical models and $\hbar$ 
to the Plank constant like quantum characteristic.
We will see   that  the QA is not
     given by just {\it any} deformation, but   
       is constrained by 
  important  algebraic properties  like the Hopf algebra and that   
     makes such deformations  a unique one in the simplest case.
Note that  even in  our familiar example  (\re{sl2}), 
the sum of two commuting spins 
    yields again a spin solution, i.e.
\begin {equation}
 \Delta s^{\pm} = s^{\pm}\otimes { I}+{ I}\otimes s^{\pm} ,\quad
 \Delta s^3 = s^3\otimes { I}+{ I}\otimes s^3
\ll{co-su2}\end {equation}
also satisfies     (\re{sl2}).
In mathematical language this is the definition of the 
     coproduct $\Delta$,
 which takes an algebra ${\cal A}$
    to its  tensor product and represents the 
    main feature of a Hopf algebra. 
    The familiar  spin algebra 
     therefore is a Hopf algebra, though only a
    cocommuting Hopf algebra (characterized by the
    explicit symmetry of the first and the second spaces in relation
    (\re{co-su2})).
   However, the  {\em  superposition} rule (\re{co-su2}),
     as can be easily checked by direct insertion,
    is no longer true for the
    quantum algebra (\re{sl2qa}).
     Therefore
    the idea is to generalize  this 
     rule as
   \begin {equation}
 \Delta S^{\pm} = S^{\pm}\otimes q^{s^3} + q^{-s^3}\otimes S^{\pm} ,\quad
 \Delta s^3 = s^3\otimes  I+{ I}\otimes s^3
\ll{co-slq2}\end {equation}
to make it compatible with (\re{sl2qa}).
This generalization indeed is quite nontrivial, since it
   defines now a noncocommuting Hopf algebra (notice the asymmetry
   in the two spaces in the coproduct relation (\re{co-slq2}))
   and makes the
   quantum algebra $U_q(sl(2))$ unique.

     A simple but fascinating  effect of this nontriviality
      may perhaps be given by the
   following example \c{selzub}. In  anisotropic spin chain
%\begin {equation}
$~~~H_{xxz} = J \
\Sigma _{n}^N [ (\sigma_n^+\sigma _{n+1}^- +\sigma_n^-\sigma _{n+1}^+) +
\cosh \alpha \sigma_n^z \sigma _{n+1}^z]
,~~$
%\end {equation}
%(4.7)
the  rotational nonsymmetry is  immediate.
 However, if we {\it spoil}
the symmetry still further
by adding another  term to it  as
$~~~\ H=
H_{xxz} +
\sinh \alpha (\sigma_1^z -\sigma _{N}^z)
, \ ~~~$
curiously the rotational symmetry is restored again.
 The fact is that, the Hamiltonian 
$H$ 
is    related  to the Casimir operator of the corresponding quantum
    algebra and  consequently, it is 
      symmetric  under $U_q(SU(2))$ and hence 
    naturally under $SU(2)$. This type of Hamiltonian
     has been found useful in solving open chain problems \c{openc}.
\subsubsection{ Quantum algebra as Hopf algebra } 

In a more formal   approach 
    quantum algebra $U_q(g)$ can be   defined as a deformed (or quantum) 
universal enveloping
     algebra of some Lie algebra $g$.
 Example is  $U_q(sl(2))$ as given above.
     With the following properties it turns also into   a noncommuting
and     noncocommuting  Hopf algebra ${\cal A}$.
%\begin {enumerate}

%  \item
 \sk

\ni i.
Coproduct
 $\ \Delta : {\cal A} \rightarrow {\cal A} \otimes {\cal A},$
       example  (\re{co-slq2}).

 % \item
\sk

\ni ii.. Multiplication $\ m :
 {\cal A} \otimes {\cal A}\rightarrow {\cal A},$
example (\re{matrix2}).
%  \item
\sk

\ni   iii.
Counit $\epsilon: {\cal A} \rightarrow {\bf K}, $ example  $\epsilon (
  S^{\pm}) =0 ; \ \epsilon (q^{\pm s^3})=1.$

\sk

\ni v.
 Antipode ($ \approx $ {\it inverse}) $s: {\cal A} \rightarrow {\cal A},$
example $s ( S^{\pm}) = -q^{\pm 1} S^{\pm}, \ \ s( q^{\pm s^3})=  q^{\mp s^3}.$

% \end{enumerate}
%\eit
       \sk
\subsubsection{    Quantum algebra through noncommuting geometry} 
This is a {\em group} like approach  starting from the noncommutative
    coordinate space having  the relation $ xy=  q^{-1} yx$,
     which can also be expressed
    as $\hat {R}^+ (\vec {x} \otimes\vec {x})=(\vec {x} \otimes\vec {x})
      $ in the matrix form. Here $\vec {x}= \left(\begin{array}{c}
      x \\ y \end{array} \right) $ and  $\hat {R}^+ = P {R}^+
       $ is the braid group representation, while 
  $ {R}^+$ is the upper-triangular solution (\re{R+}).
If  we demand now the invariance of the above
defining  relation for $\vec x$ under the
    transformation $\vec x '=T \vec x$ (recall that  the
    vector transformation $x'=g x ,~~g \in$ (rotation group)
  in usual commuting space preserves  the length of
      vector $x$), we obtain the relation on
    $T$ as
\begin {equation}
\hat R^+\ T_1 T_2=  T_2 T_1 \ \hat R^+
,\ll{R+tt}\end {equation}
%(4.9)
  which is a {group like}  relation of quantum algebra ${\cal
A^*}$ 
  in the matrix form (recall 
     the property $ g^tg=1$ for a rotation group).
    The relation (\re{R+tt}) written in the elementwise form
     through noncommuting matrix elements of
      $
      T = \left( \begin{array}{c}
a \ b  \\
    c \ d \end{array} \right) $     with quantum determinant 
$det_q T= ad-qbc$ 
    usually taken as the definition of  the quantum group:
\begin {eqnarray}
ab=qba \ , \ ac=qca \ ,\ bc=cb \ ,\
bd = q^{-1} bd \ , \nonumber \\ dc=q^{-1} cd \ , \ ad-da= (q-q^{-1}) bc.
\ll{qg}\end {eqnarray}
%(4.10-11)
Let us stress again that,  though this is not a group, 
  the  property that the coproduct $\Delta T=T\cdot T',
$
    for $[T,T']=0$
    satisfies again the relation (\re{R+tt}) bears similarity with the
    standard group
  multiplication rule.

In recent years  success 
 in establishing  conventional exponential 
relation  between the  quantum algebra
 and group has been  achieved \c{qalg-gr}.

\subsubsection {  Faddeev--Reshetikhin--Takhtajan (FRT) formulation } 
      The FRT formulation \c{frt} , closely
 related to the theory of quantum integrable
    systems, bridges between  
above two approaches by noting that 
the algebras ${\cal A}$ and ${\cal A}^*$ are dual to each other. 
Coproduct $\De$ in ${\cal A}^*$ induces multiplication $m$ in ${\cal A}$:
$~~\l(l_1l_2,a\r)=l_1l_2(a)=(l_1\otimes l_2)(\De(a))~~$, where $
l_1, l_2  \in  {\cal A} ,~~  a\in {\cal A}^*
$.
Setting $L^\pm \in M_n({\cal A})$ 
 the duality condition links between them  as 
$~~ \l(L^{(\pm)}, T\r)=~R_q^{\pm}. ~~$  The main relations of
    the FRT algebra may be given by
  \begin {equation}
 R^{+}~L^{(\pm )}_1~ L^{(\pm)}_2 ~=~L^{(\pm)}_2 ~L^{(\pm)}_1~R^{+}
 ~,\\~~
 R^{\pm}~L^{(\pm)}_1~ L^{(\mp)}_2 ~=~L^{(\mp)}_2 ~L^{(\pm)}_1~R^{\pm}
,\ll{frt}\end {equation}
%(4.12)
and an equation  similar to the first of (\re{frt}) 
with  $R^+$ replaced by   $R^-$ .   
 The $R^\pm$ matrices are   same as $R_q^\pm$ in (\re{R+}),
     $L^{\pm}$ are upper/lower-triangular matrices,
     the operator elements of which are expressed through the generators
    $({ I}, s^3,S^{\pm})     $ of  quantum  algebra (\re{sl2qa}) as
\begin {equation}
L^{(+)} = \left( \begin{array}{c}
q^{-s^3}  \quad   -(q-q^{-1}) S^- \\
    \qquad \qquad  q^{s^3}
          \end{array}   \right), \qquad
  L^{(-)} = \left( \begin{array}{c}
q^{s^3}  \qquad  \qquad   \\
 (q-q^{-1}) S^+ \quad  q^{-s^3}
          \end{array}   \right)
.\ll{l+-}\end {equation}
%(4.14)
    Thus (\re{frt}) with (\re{l+-}, \re{R+})
    reproduces the  quantum  algebra (\re{sl2qa}),  though
     given in a matrix form  similar to the defining relation  for
    the quantum group  (\re{R+tt}).

Another important feature of the FRT algebra is that, it derives elegantly
 the
coproduct structures of the quantum algebra (\re{co-slq2}).
Defining operators $X^a_n={ I}\otimes \cdots \otimes X^a\otimes\cdots
{ I} $ we may  evaluate 
  \be
L^{(+)}_{1n_{i}}~ L^{(+)}_{1n_{i+1}} ~=~
 \left( \begin{array}{c}
q^{-(s_{n_{i}}^3+s_{n_{i+1}}^3)},  \qquad   -(q-q^{-1})\left( S_{n_{i}}^-
\otimes q^{s_{n_{i+1}}^3} +q^{-s_{n_{i}}^3}\otimes S_{n_{i+1}}^- \r) \\
    \qquad \qquad  \qquad q^{s_{n_{i}}^3+s_{n_{i+1}}^3}
          \end{array}   \right).\ee
Comparing with (\re{co-slq2})   it is immediate that
  \be
L^{(+)}_{1n_i}~ L^{(+)}_{1n_{i+1}} ~=~
= \left( \begin{array}{c}
\Delta q^{-s_{(n_in_{n+1})}^3}  \quad   -(q-q^{-1})\De S_{(n_{i}n_{i+1})}^- \\
    \qquad \qquad \De q^{s_{(n_{i}n_{i+1})}^3}
          \end{array}   \right)  =  \De  L^{(+)}_{1n_{i}n_{i+1}}.
\ll{coprod}\ee
 Similarly we get $  L^{(-)}_{1n_{i}}~ L^{(-)}_{1n_{i+1}}
 ~=~ \De  L^{(-)}_{1n_{i}n_{i+1}}$
 and also at the same time  
\be  L^{(\pm)}_{1n_{i+1}}~ L^{(\pm)}_{1n_{i}} ~=~\tilde
 \De  L^{(\pm)}_{1n_{i}n_{i+1}}, \ll{Ln1Ln} \ee
 where $\ \tilde \De =
 {\cP} \De {\cP} \ $ is
 another possible permuted
 form of  the coproduct with ${\cP} (x\otimes y) {\cP} = y\otimes x$.

Note
   that the same $R_q^+$-matrix (\re{R+})
 appears in some or other form in the variety of relations
    concerning quantum algebra described above.
    We will see  in sect. 4 that the  $R_q^\pm $ matrices  representing
     structure constants of the quantum algebra  
    enters also  as the building blocks for the $R(\la)$-matrix   constructions in 
integrable theory.
      The associativity of the FRT algebra
      gives  the spectral parameterless YBE (\re{braid}).  

\subsubsection {Multiparameter deformation of general reductive Lie algebras}
Apart from the
simplest one parameter deformation of $sl(2)$ discussed above,
multiparameter deformations of  general Lie algebras 
can be formulated \c{reductive}.
We give below a brief account of it.
\medskip

Let $  g  $ be a reductive Lie algebra of rank $N$. Namely $g$ is the
direct sum of, say, $M$ simple Lie algebras plus an abelian center.
$H_1 , H_2, \ldots , H_N $ is the basis
of the Cartan algebra  of which $ H_i (1 \leq i\leq N_1) $
span the semisimple part and the remaining $N-N_1$ number of
 $H_{\alpha} , (N_1 < \alpha \leq N)$ belong to the center of the algebra.
Let $ {\bf a}_k$ be vectors with $N$
 components $a_{\ell k}$, such that  $a_{\alpha k} = 0$ for all $k$ and
  $ (N_1 < \alpha \leq N)$  , while  $a_{ij}= 2(\alpha _i \cdot\alpha _j)/
(\alpha _i \cdot \alpha _i)$ with $ (1 \leq i, j \leq N_1)
  $ are the entries of
the Cartan matrix related to the semisimple part. Let $X_i^{\pm}$
be generators associated to the simple roots $\alpha_{i}$.
The universal deformation $\uqg$
 can be defined so that
each simple component remains the same as that of the standard
one parameter quantization with relations \c{dobrev}
\be
[X_i^+,X_j^-]= \delta_{ij} { q_{ii}^{H_i\over 2} -
 q_{ii}^{ -{H_i\over 2}} \over
   q_{ii}^{1\over 2} - q_{ii}^{ - {1\over 2}}  } \equiv  \delta_{ij}
   [ H_i ]_{q_{ii}}
\ll{qa-gen}\ee
and $ f({\bf H}) X_i^{\pm} =  X_i^{\pm}  f({\bf H}\pm {\bf a}_i) $
plus the Serre relations
\be
\sum_{0 \leq k \leq n}  (-1)^k
  \pmatrix{ n \cr k}_{q_{ii}} \left(  X_i^{\pm}  \right)^{n-k}
     X_j^{\pm} \left(   X_i^{\pm}  \right)^k = 0 , \qquad i\ne j
\ll{serre}\ee
with $\ n= 1-a_{ij}
 $ and the notation $q_{ii}=e^{h_{\rho(i)}(\alpha_i\cdot
\alpha_i)}  $, where $(1 \leq  \rho(i) \leq
M)$ counts the number of simple components. Since $\rho$ is constant
on each simple component, we may use also the notation
$q_{\rho} = e^{h_{\rho (i)}}$.
The remaining deformation
 parameters on the other hand can be relegated to the
coalgebra structure
 defining the corresponding coproducts as
\bea
 \Delta ( X_i^{\pm})& = & X_i^{\pm} \otimes   \Lambda _i^{\pm}
 +( \Lambda _i^{\pm})^{-1} \otimes    X_i^{\pm}
\nonumber \\
 \Delta ( H_i) &= & H_i \otimes { I} +   { I}  \otimes   H_i ,
\ll{co-gen}\eea
where $\Lambda _i^{\pm} $, containing the parameters $v_{i\alpha}, t_{ik}$
$(t_{ik} = - t_{ki})$, has the form
\[
\Lambda _i^{\pm} \equiv q_{ii}^{  H_i\over 4} e^{\pm  { 1\over 2}
  \left( \sum_{j, k } t_{ik}  a^{kj} H_j  +  \sum_{\alpha}  v_{i \alpha}
    H_{\alpha} \right)}.\]
The antipode ($S$) and co-unit ($\epsilon$) are given by
\[
 S( X_i^{\pm}) =-  q_{ii}^{\pm {1\over 2}}
   X_i^{\pm}, \ \
  S( H_i) =-  H_{i}
\ \   \hbox{and }  \ \ \epsilon (X_i^{\pm}) = \epsilon (H_i)=0.
\]
This defines the  deformation of general Lie algebra  as a Hopf algebra and we
will show below  the existence of universal ${\cR}$-matrix in this case.
\medskip

Sometimes it is convenient to use  {\it q-deformed} Cartan-Weyl basis
\c{dobrev}
for related constructions. Such  generators satisfy the commutation relations
\be
[e_{\gamma},e_{-\gamma}]= a_\gamma { q_{\gamma\gamma}^{h_\gamma /2} -
 q_{\gamma\gamma}^{ - {h_\gamma /2}} \over
   q_{\rho} - q_{\rho}^{ - 1}  }
\ll{cw-alg}\ee
In the above expression, if the index $\gamma$ corresponds to the
non-simple root $\alpha_\gamma = \sum \alpha_i$ for certain simple
roots $\alpha_i$, then $h_\gamma = \sum h_i$ and the relations between two
basises are
\be h_i= {(\alpha _i \cdot \alpha _i) \over 2} H_i,  \qquad
 e_{\pm i}= 
\left[ {(\alpha _i \cdot \alpha _i)\over 2}\right] _{q^2_\rho} ^{1/2}
\;   q_{ii}^{-1/4} X^\pm_i q_{ii}^{\mp H_i/4} \ll{e-x}\ee
\subsection{Yangian algebra}
The Yangian $Y(g)$ \c{drinfeld} associated to a Lie algebra $g$ is another deformation of
the universal enveloping algebra  of polynomial maps  $g[\la]$: $C \rw g$.
The  general  representation theory of the Yangian is discussed in
\c{cheri},
though we will be interested   only  in its simplest realizations  related
to the rational $R$-matrix solution of the  YBE given by (\re{abrat}):
$~~R(\la)=\la-\eta P~~,$ due to its direct consequence in the integrable theory.   

If we consider the global QYBE (\re{qybet}) with the rational
$R$-matrix, 
the monodromy matrix $T(\la)$ satisfies the quadratic relation \be~~
(\la-\mu)[T^{ab}(\la),T^{cd}(\mu)]=\eta(T^{cb}(\la)T^{ad}(\mu)-T^{cb}(\mu)
T^{ad}(\la))~~~\ll{YT}\ee
 and the quantum determinant defined as $~~
det_qT(\la)=T^{22}(\la-\eta)T^{11}(\la)-T^{21}(\la-\eta)T^{12}(\la)~~$
commutes with all matrix elements of $T(\la)~~
$. Normalizing $
~det_qT(\la)=1~~$ and expanding \be
~T^{ab}(\la)= \de^{ab}+ \eta \sum_{k=0}^\infty \la^{-(k+1)}  ~t_{(k)}^{ab}
~~ \ll{expyang}\ee
 we get a realization of Yangian $ Y(sl(2))$ generated by the elements
$ ~t_{(k)}^{ab}$. The algebraic relations given by  (\re{YT}) are equivalent
to 
\bea
 [~t_{(0)}^{ab}, ~t_{(k)}^{cd}] &=& \de^{cb} ~t_{(k)}^{ad}-
\delta^{ad} ~t_{(k)}^{cb}, \nonumber \\
{ [ t_{(k+1)}^{ab}, ~t_{(l)}^{cd}] } - [~t_{(k)}^{ab} ,~t_{(l+1)}^{cd}]&=&
 ~t_{(l)}^{cb} 
 t_{(k)}^{ad} -~t_{(k)}^{cb}t_{(l)}^{ad}.
\ll{yangian}\eea
Note that these are like recurrence relations and 
due to the constant quantum determinant  all  generators
of the algebra can be constructed starting only from its first two
components  $t_{(0)}^{ab}$ and $t_{(1)}^{ab}$. Clearly the Yangian algebra
(\re{yangian}) is a deformation of the loop algebra. We shall see in 
sect. 5 some useful 
finite dimensional representation of this algebra realized in
 integrable models.
\subsection {Universal ${\cR}$-matrix}
 Notice that for deriving the algebraic relations like
  $~~~
 R_{12}^{+}~L^{(\pm )}_{1n_{i}}~ L^{(\pm)}_{2n_{i}}
 ~=~L^{(\pm)}_{2n_{i}} ~
 L^{(\pm)}_{1n_{i}}~R_{12}^{+(-)} ~$
in the FRT formulation (\re{frt}) we use the  $ R_{12}$-matrix
acting  only in the  {\em auxiliary}
spaces,
while 
the operators $L_{a n_i}$ are defined in both 
 the {\em auxiliary} space 
 ( $a=1$ or $2$)
and the {\em quantum} space
$n_{i}.$ For obtaining
the coproducts we need to multiply 
$L$ operators in the same auxiliary space, but
in different quantum spaces  $n_{i}, n_{i+1}.$
However if we allow certain  abstraction and suppose the existence of
a  ${\cR}_{n_{i}n_{i+1}}$-matrix,  such that
  $~~~~
 {\cR}_{n_{i}n_{i+1}}~L^{(\pm )}_{1n_{i}}~ L^{(\pm)}_{1n_{i+1}}
 ~=~L^{(\pm)}_{1n_{i+1}} ~
 L^{(\pm)}_{1n_{i}}~R_{n_{i}n_{i+1}}, ~~ $
then we can have, as seen from (\re{Ln1Ln}), the relation
$~~~{\cR}_{n_{i}n_{i+1}}~\De \l(L^{(\pm )}_{1(n_{i}n_{i+1})}\r)~
= \tilde \De \l(L^{(\pm )}_{1(n_{i}n_{i+1})}\r)~{\cR}_{n_{i}n_{i+1}}. \ ~~~ $
In operator form such ${\cal R}$-matrices  satisfy therefore
the intertwining
relation 
\be {\cal R}~\De (a)= \tilde \De (a)      {\cal R}.\ll{ur1}\ee
between coproducts $\De$ and $ \tilde \De.$ 
This  abstraction of the quantum $R$-matrix satisfying (\re{ur1})
 is called the {\em universal }
 ${\cR}$-matrix. A Hopf algebra endowed with a universal ${\cR}$-matrix
 is called a {\em quasitriangular } Hopf algebra. As a consequence of the
 coproduct property we then have
 \be
 (\De \otimes I) {\cR}={\cR}_{{\bf 13}}{\cR}_{{\bf 23}},\qquad
 (I \otimes \De) {\cR}={\cR}_{{\bf 13}}{\cR}_{{\bf 12}},
\ll{ur2}\ee
which would yield
 \be
{\cR} (\De \otimes I) {\cR}={\cR}_{{\bf 12}}{\cR}_{{\bf 13}}
{\cR}_{{\bf 23}},\quad
,\ll{ur3}\ee
where we have denoted the quantum space indices by bold face numbers
as in sect. 2, for convenience. Notice the ordering of indices
in the two relations in (\re{ur2}),
using which together with 
 the intertwining property (\re{ur1})
we get
\be
{\cR} (\De \otimes I)({ \cR}) =   (\ti \De \otimes I)({ \cR })
{\cR}_{{\bf 12}} 
={\cR}_{{\bf 23}}{\cR}_{{\bf 13}}{\cR}_{{\bf 12}}\ll{ur4}\ee
 Equating (\re{ur3},\re{ur4}) we get  finally 
 \be
{\cR}_{{\bf 12}}{\cR}_{{\bf 13}}{\cR}_{{\bf 23}} =
{\cR}_{{\bf 23}}{\cR}_{{\bf 13}}{\cR}_{{\bf 12}}
\ll{ybeR}\ee
as abstract generalization of all YBE's.
Therefore 
 this {\em universal} equation
and  the universal ${\cR}$-matrix ,as we will see below,
 may be taken as the  basic objects  for
 constructing
  other important structures like FRT algebra, $L^{\pm}, R^{\pm}$
matrices etc. They in turn would generate the 
  key elements in integrable theory, like the spectral
parameter dependent
   Lax operators $L(\la)$ and 
 quantum $R(\la)$-matrices as well as  the QYBE and YBE.
%\subsection{Solution for $U_q(sl(N))$ }

Though in general it is difficult to find the 
   solutions for the universal ${\cal R}$-matrix is the explicit form,
  following the prescription of
Drinfeld \c{drinfeld}  such an object
may be obtained  exploiting basically the defining
relation (\re{ur1}) and using (\re{co-slq2}). The simplest of 
such  solutions for $U_q(sl(2))$
 is given by \c{urslq2}
\be
 {\cal R}
= q^{\ha (\si^3 \otimes \si^3) }
 \exp_{q^{-2}} ((q -q^{-1})
   S^- q^{\si^3 } \otimes
    q^{-{\si^3 }} S^+  )
    \ll{ur-slq2}\ee
 with the  notations
$\exp_{q} (x) = \sum_{n\ge 0} {x^n \over (n)_q !}$
where
$(n)_q := {q^n -1\over q-1}$. 
%--------------defintions_________________
%\quad  and$[n]_q = {q^{n\over 2} - q^{-{n\over 2}} \over
% q^{1\over 2} - q^{-{1\over 2}}}$ \quad using  the  identity
%$\ [n]_q ! = (n)_{q^{-1}}! \; q^{n(n-1)\over 4} \ .$
%__________________________________________________________
\subsubsection{Twisting transformation and  general  ${\cR}$-matrix
 solution}
For obtaining a general universal ${\cR}$-matrix solution corresponding
to all deformed reductive Lie algebras
 discussed above,  let us   first look into the {\em twisting}
 transformation introduced in \c{twist}.
 Under such  transformation one gets a twisted    ${\cR}$-matrix
 starting from the original one:
$~~~ {\cR}_0\rw {\cR}_{\bf 12}= {\cal F}^{-1}_{12} {\cR}_{\bf 12}
{\cal F}^{-1}_{12} ~~~~$
provided the  twisting operator ${\cF}$ satisfies the  conditions
 \be
{\cR}_{{\bf 12}}{\cF}_{{\bf 13}}{\cF}_{{\bf 23}} =
{\cF}_{{\bf 23}}{\cF}_{{\bf 13}}{\cR}_{{\bf 12}}, \qquad
%.\ll{twist1}\ee        and
{\cF}_{{\bf 12}}{\cF}_{{\bf 13}}{\cF}_{{\bf 23}} =
{\cF}_{{\bf 23}}{\cF}_{{\bf 13}}{\cF}_{{\bf 12}}
\ll{twist2}\ee
along with 
 the antisymmetry property $\ {\cF}_{{\bf ab}}= {\cF}^{-1}_{{\bf ba}}$.
The corresponding 
Hopf algebra  also gets twisted with a transformed coproduct 
$\De={\cF} \De_0 {\cF}^{-1}, \ $ 
such that the defining relations (\re{ur1}) and
(\re{ybeR}) are  valid again for the  twisted ${\cR}$ matrix.

We  find first the   ${\cR}_0$ related to
the semisimple Lie algebras. 
Implementing then  suitable   twisting $\ti { \cal G}$  
 we  get the general solution for the
  reductive case with multiparameter deformations as
\be
                  {\cal R }(q_{ii}, t_{ij}, v_{i\alpha})=
    \ti { \cal G}^{-1}(t_{ij}, v_{i\alpha})
           {\cal R }_0 (q_{ii})
\ti {\cal G}^{-1}(t_{ij}, v_{i\alpha}).\ll{ur-gen} \ee
We may  construct the universal   ${\cal R}_0$-matrix 
  for the  semisimple Lie algebras in the form \c{urslqn,tolstoy}
$~~~
{\cal R}_0 = {\check R} K,
~~~$
where $K$ is expressed only in terms of the Cartan generators as
\be
 K = \exp \left(\sum_{ij}
 h_{\rho(i)} {(\alpha_i \cdot \alpha_i)\over 2}
{(\alpha_j \cdot \alpha_j)\over 2}     d^{ij}    H_i {\otimes}H_j \right),
\ll{K0}\ee
with $    d^{ij}=  ( d^{-1})_{ij} $, $d_{ij}= (\alpha_i
 \cdot \alpha_j )$ representing  symmetrized Cartan matrix. On the other
hand 
$\check R$ is expressed  in the factorized form 
\be
     \check R  = \prod_{\rho=1}^M (  \check R^{(\rho)}), \quad \ \hbox {with}
      \ \     \check R^{(\rho)} =   \prod_{\gamma \in \Delta_+^\rho
      } (  \check R^{(\rho)}_\gamma), ~~~
~~\check R^{(\rho)}_\gamma = \exp_{q^{-1}_{\gamma \gamma}}
\left( a_\gamma^{-1}( q_\rho - q_\rho^{-1})( e_\gamma \otimes
e_{-\gamma}) \right)
\ll{cR0}\ee
where $\Delta_+^\rho$ is the set of {\it all} positive roots belonging
to the $\rho$-th simple component with the prescribed normal ordering
 \c{tolstoy}.
For finding now the general solution  (\re{ur-gen})
we may choose  the twisting operator  as
\bea         \tilde {\cal G}(t_{ij}, v_{i\alpha}) &\equiv &
        {\cal G}(v_{i\alpha}) {\cal F}(t_{ij}) \nonumber \\ =
\exp \{ {1\over 2} ( & \sum_{j \alpha} &
      v_{j\alpha}   a^{jk} (  H_k {\otimes}H_\alpha -
H_\alpha {\otimes}H_k )   +
     \sum_{ijkr}
      a^{ki}   t_{kr}  a^{rj}   H_i {\otimes}H_j )\}
\ll{cG}\eea
with $ a^{ij} = (a^{-1})_{ij}.$ It can be checked easily  that $\tilde {\cal G}$
exhibits the required antisymmetry  property   
 $\tilde {\cal G}_{21}^{-1}=
   \tilde {\cal G}_{12}$
   and satisfy the required conditions (\re{twist2}).
    It is  exciting to check, that the
   general solution ${\cR}$  (\re{ur-gen})   with the insertion of 
 (\re{K0}, \re{cR0},\re{cG}) is the 
   intertwining operator (\re{ur1})
   for the deformed general reductive Lie algebra with coproducts (\re{co-gen}).

Note that  specializing 
to a particular  Lie algebra, which would be reflected in the corresponding
choice of root systems,  Cartan matrix $a_{ij} $  and the normal
ordering in $ {\Delta}_+  $, one can derive
from the general  universal ${\cal R}$-matrix obtained here, 
the solutions for all the classical   algebras of $
A,B,C,D $ etc. types together with
their semisimple
 as well as reductive generalizations. For example 
in the case of  quantum algebra $U_q(gl(2)) $,
 by applying the  twisting transformation
$ \
    {\cal F}=e^{i\theta (s^3\ot Z-Z \ot s^3)}$ 
on  (\re{ur-slq2}),
one derives the universal ${\cR}$ matrix
\be
{\cR}=q^{i2(s^3 \ot s^3+\theta (s^3\ot Z-Z \ot s^3))}
\sum_{m=0}^{\infty } {(1-q^{-2})^m \ov (m)_{q^{-2}}!}(q^{s^3-Z} S^+)^m \ot
(q^{-(s^3+Z)} S^-)^m
,\ll{urglq2}\ee
where $Z$ is  central to the algebra.

 In the next section we  take up
 important applications to integrable systems using different representations
 in the case $U_q(gl(n+1)). $

\subsection {Hecke algebra, Birman-Wenzl-Murakami algebra}
Such algebraic structures inherent to  $R^\pm$ matrices play important
role in integrable systems.
Hecke  and Birman-Wenzl-Murakami
 algebras may be represented by the braid group
along with certain extra conditions. The braid group in  turn is given by
the relations
\be
 b_{i}~b_{i+1}~b_{i }
  ~~=~~b_{i+1}~b_{i}~b_{i+1}, \qquad b_i~b_j=b_j~b_i, \ \mid i-j \mid \geq
  2, \ll{braid0}\ee
 where $b_i, \quad i=1,2,\cdots , n-1 \ $  are braiding operators
 defining a group, the elements of which are $n$-braid.
    The braid group  can be used for obtaining  link
     polynomials, which have   importance in the knot theory
     for    classifying  links and knots \c{wadati}. 
We are interested however  in the integrable theory and  
 observe that, 
defining   $~{\hat R}~=~ P R_q^{+}~$
the spectral-free equation   (\re{braid}) rewrites as
            \be
 \hat  R_{12}~\hat R_{23}~\hat R_{12 }
  ~~=~~\hat R_{23}~\hat R_{12}~\hat R_{23},
  ~~~~~~ \quad [ \hat  R_{12}, \hat  R_{34}] =0. \ll{braidr}\ee
Representing $\hat  R_{i i+1}=b_i$ the equivalence of (\re{braidr}) and
(\re{braid0}) becomes  immediate.

The {\em Hecke algebra}
may be defined  by the braid group relations
(\re{braid0}) along with an extra Hecke condition
\be h_i^2+(1-t)h_i=t \ll{hecke}\ee
with $\ep =const.$
Notice  that, if we demand the 
$R_q^\pm$-matrices to satisfy in  addition to (\re{braid})
the condition
\be R^+-R^-=(q-q^{-1}) P, \qquad R^-=P (R^+)^{-1}P,\ll{hecke-r}\ee
then one finds  easily that 
$~~~~~\hat R- \hat R^{-1} = (q-q^{-1}){ I} ~~~$
or $~~~~(q \hat R)^2+(1-q^2)(q \hat R) = q^2 { I} ~~~~$
which is exactly the Hecke condition (\re{hecke}).
Though the   condition (\re{hecke-r})
may appear to be  rather restricted, it is interesting to
check, that not only the matrices (\re{R+}), but also 
   the whole class of
$R^\pm$-matrices related to $U_q(gl(n+1))$ satisfy 
this condition. In the next  section 
 we will be concerned  basically  with such $R^\pm$-matrices,
 since they through Yang-Baxterization allow an easy and
systematic construction of Lax operators
and quantum $R$-matrices  representing  integrable systems.

Birman-Wenzl-Murakami (BWM) algebra \c {bwm} 
is of more general structure  and may be defined  through
the braid group relations (\re{braid0}) along with the condition
\c{wadati}
\be
(R_i-c_1)(R_i-c_2) \cdots (R_i-c_n)=0
,\ll{bm-alg}\ee
where $c_r, r=1,2, \cdots, n$ are  constant eigenvalues.
 Note that in the restricted case of $n=2,$ condition (\re{bm-alg})
reduces to (\re{hecke}), though otherwise it is more     general
and applicable   to all other  $R^\pm$-matrices  not
belonging to the  Hecke class. Some prominent examples of such $R$-matrices
are those of the  higher spin models as well as of
the models related to algebras other than $gl(N)$. For the
Yang--Baxterization
and  other applications  of the  BWM algebra, 
the readers are referred to \c{jones,wadati,mlg}.

\subsection {Significance of Hopf algebra in integrable systems}
To  clarify  the  significance  of  the  Hopf  algebra  in  integrable 
 theory   
let us    return   to  the  coproduct  and multiplication
properties  of the algebra  and  the transition from the local to the 
global QYBE,
  explained
in sect. 2.

We will discuss first the ultralocal case.
Let $S^a_{n_1}=\underbrace{I\ot I\ot I\ot \cdots \ot S^a}_{n_1}
 \ot  I \cdots \ot I$ and
similarly $S^a_{n_2}$ be two copies of generators of the 
quantum algebra, which trivially commute. 
Due to the Hopf algebra structure  
the coproduct $\De S^a_{(n_1n_2)}$  acting nontrivially on both the spaces 
$n_1$ and $n_2$ satisfies 
 again  the same  algebra. Continuing such tensor products one obtains the 
coproduct acting globally on all the spaces
 (see fig 3.1 a,b,c).
 Note now, that 
   $L^{(\pm)}_{1n_1}(S^3_{(n_1)},S^\mp_{(n_1)})$  and  ~$L^{(\pm)}_{1n_2}
(S^3_{(n_2)},S^\mp_{(n_1)})$  satisfy  the  FRT
relations
$~ \
 R^{+}~L^{(\pm)}_{1n_i}~ L^{(\pm)}_{2n_i} ~=~L^{(\pm)}_{2n_i}
 ~L^{(\pm)}_{1n_i}~R^{+}
,$ for $ ~ i=1,2 ~$
due to the related $ S^a_{n_i}$ satisfying  the  algebra (\re{sl2qa})
or (\re{sl2}).
 The ultralocality condition 
$~~
 [L^{(\pm)}_{1n_1} ,      ~L^{(\pm)}_{2n_2}]=0, ~~~$ holds due to 
the validity of  $~~~      
[S^a_{n_1}S^b_{n_2}]=0~~$. Finally, 
   the     coproduct         $\De
L^{(\pm)}_{1(n_1n_2)}=L^{(\pm)}_{1n_1}L^{(\pm)}_{1n_2}$
defined on both the lattice sites $(n_1n_2) $ is  again in
upper (lower)-triangular  form  and  due to the Hopf structure
of the elements $\De S^a_{(n_1n_2)},$ becomes a solution
of the same FRT algebra
\be
 R^{+}~(\De L^{(\pm)}_{1(n_1n_2)})~(\De L^{(\pm)}_{2(n_1n_2)})
 ~=~(\De L^{(\pm)}_{2(n_1n_2)})
 ~(\De L^{(\pm)}_{1(n_1n_2)})~R^{+}.\ll{frt2}\ee
Thus the coproduct $\De: {\cal A}\rw {\cal A} \otimes {\cal A}$ 
of the Hopf algebra induces {nonlocal operators}, continuing  which 
as ${\cal A}\rw {\cal A} \otimes {\cal A}\otimes \cdots \otimes {\cal A},$
we get the  global object $
\De L^{(\pm)}_{1(1 \ldots N)}(\De S^a_{(1 \ldots N)}) $  defined  on
the
whole interval $[1,N]$ (see fig. 3.2 a-c).
Such a globalization of the FRT algebra   
caused by  the coproduct of the Hopf algebra (fig. 3.1 a-c),
 in fact represents in the
Yang-Baxterized form the 
construction of the global
QYBE from the local one, described in sect. 2 (fig. 2.2-2.6).
 Let us now focus  on another important feature of a Hopf
 algebra, namely the multiplication rule $m:{\cal A} \otimes
 {\cal  A}  \rw  {\cal  A}$ and look into 
 the  ultralocality condition we have
crucially
used. Though its action is easy to define for the quantum algebra:
$ m:(a \otimes b)
\rw a\cdot b  $,  this  operation  for  the higher  tensor  products,  is 
given in  a  bit
nontrivial way involving the the permutation operator 
 $ \tau: b \times c \rw  c \times b $
$~~
m^{\otimes} \equiv (m \times m) \ : (id \times \tau \times id ).
~~~~$
Thus we actually  arrive at our familiar relation (\re{matrix2}) 
 as 
\be
(a \otimes b) \cdot (c \otimes d)=
(m \times m) (a \otimes  \tau (b \otimes c) \ot d) 
= m( a\ot c) \times m(b \ot d) 
= a c \otimes b d \ll{abcd} \ee
where we have 
assumed  the {\it trivial} permutation property
$ \tau( b \ot c ) =  c \ot b, $  which is  also true for the quantum
algebra. Note that  transition from
local (\re{frt}) to global (\re{frt2}) 
 and similarly for the QYBE's related to the
ultralocal models 
   mimic 
the multiplication property (\re{abcd}).

It is remarkable that even for nonultralocal models, as we will see below,
the Hopf algebra structure of the underlying algebra induces again the
transition from the  local
to the global  braided QYBE (fig. 2.13, 2.14 ) 
crucial for the quantum
integrability.

\subsection {Braided  and quantized braided algebra}
	In   general  the permutation
property of $\tau$ in (\re{abcd}) 
 may become nontrivial :$ \tau_{23} \rw \psi_{23}$ and 
unlike  quantum algebras can modify 
the multiplication rule as
\be
m_{B \otimes B }=(m \times m) \cdot 
 (I \otimes  \psi_{23} \otimes I ) 
 \ll{mbraid} \ee
 which  is the main essence of the  braided algebra  
 \cite {majid}.
Generalizing the notion of superalgebra, the braided algebra
$B(R)$ may be defined through the 
 commutation relation
\be~~{R}_{21}T_{1} R_{12}T_{2}
= T_{2}R_{21}T_{1}{R}_{12}~~ \ll{braided1}\ee
for braided matrix $T$ 
with constant solution $R_{12}$ of YBE (\re{braid}). If $T'$ is another
independent braided matrix having the braided statistics 
\be~~T_{2 }R_{12}^{-1}T'_{1}R_{12}
=R_{12}^{-1}T'_{1}R_{12}T_{2}~~,\ll{braided2}\ee then 
$\De(T)=T T'$ is also a braided matrix satisfying  the same 
algebra $B(R)$. This shows that the braided algebra, though have complicated 
multiplication rule, has the
coproduct structure, an  important ingredient   of the  Hopf
algebra. The matrix elements $a,b,c,d$ of the $2 \times 2$-matrix $T,T'$
generates different braided algebras for different constant 
 $R$-matrix solutions  like standard, nonstandard, Jordanian forms etc.
\c{majid}.  

Generalizing further to the quantized braided group $QB(R,Z)$
 \c{majidq,hlavqb},
along with  $R$-matrix another $Z$-matrix 
were introduced with $R,Z$ being the 
solutions of  constant YBE's  $[RRR],[ZZZ],[RZZ]$ and $[ZZR]$.
Here we have used the obvious notations \c{hlav94}
 for writing constant YBE's like
(\re{braid}). The defining 
algebraic relations for $QB(R,Z)$ may be obtained
from equations like (\re{bqybe+}) as a particular case when $\ti Z=Z$.
This type of relations are responsible for the homogeneous braidings 
for all copies of the algebra. 

For  incorporating different braidings for nearest and 
further neighbors, $Z$ and $\ti Z$ matrices were included 
 along with the  $R$ matrix  
in the  generalization of the quantized braided algebra in \c{khijmp96}.
Yang-Baxterizing the braided FRT  relations given by (\re{bqybe+}),
one may arrive at the spectral parameter dependent BQYBE's
(\re{bqybel}-\re{zlzl2u}) etc. suitable for describing integrable
nonultralocal models. 
Therefore, in one hand such equations serve for describing physical models
including 
their conserved  quantities and   on the other hand at the  spectralfree limit
(\re{bqybe+}) reproduce the abstract mathematical structures like quantized
braided algebra \c{hlavqb} for $\ti Z=Z$ and the braided algebra \c{majid}
for $\ti Z_{12}=Z_{12}=R_{21}$. 
%--------------------

The Hopf algebra structure of the braided algebra plays the key role through
its coproduct in constructing the BQYBE for the monodromy matrix leading to
the quantum integrability of the nonultralocal models similar to the
ultralocal case.

\vspace* {1cm} \fbox{fig.3.1} 

\vspace* {1cm}
   {\bf
Figure 3.1 }\  {\it Coproduct property of quantum algebra.}\\
a){\it Generators $S^a_{n_i}$ act nontrivially of the $n_i$-th space.}~~
b){\it  The coproducts  $\De S^a_{(n_1n_2)}$ act nontrivially on both 
 $n_1$ and $n_2$  spaces}.~~ 
 c) {\it  The coproducts  $\De S^a_{(1 \cdots N)}$ act globally
 on all the   spaces}.

\vspace* {1cm} \fbox{fig.3.2} 

\vspace* {1cm}
  {\bf
Figure 3.2 a-c)} \  {\it FRT algebra and its tensor product induced by the
coproduct structure.}

%_________________________________________________\input {preport4.tex}
\setcounter
{section} {3}
\section {Construction of models from universal ${\cR}$-matrix }
%%%%%%%%%%%%%%%%%%%%%%%%%%%%%%%%%%%%%%%%%%%%%%%%%%%%%%%%%%%%%%%%%%
\setcounter{equation}{0}

In sect. 2 we have given a list  of ultralocal models along with their
 Lax operator $L(\la)$ and the  associated quantum $R(\la)$-matrix.
Here we  focus on their origin and construct them 
for a large class of models. This scheme 
\\ i)     reveals  the
    algebraic aspects of the ultralocal quantum integrable theory and
\\ ii) using  underlying  algebraic structures
    constructs
  spectral parameter dependent representative
Lax operators  in a  systematic way. 

It also   highlights   the following   basic  questions pertinent to 
the  integrable systems 
\\ iii)  explaining { why} 
 a wide range of          integrable models like
  sine-Gordon , derivative nonlinear Schr\"odinger  equation,
 massive Thirring model  etc.
share the {\em same} trigonometric 
$R(\la)$-matrix, while 
another large class including NLS model, Toda chain , isotropic spin-$\ha$
chain etc. are all associated with the same rational $R(\la)$ matrix   
 (see the list in sect.  2).
\\
iv) Thus it  focuses on 
  the {\em commonness}  between such     diverse  models and
   tries  to find    the  guiding principle
 for classifying different classes
of  integrable models.
\\ v) It identifies an algebraic  
   criterion for defining the   {\em integrable nonlinearity}, i.e.
 the  nonlinearity that 
 makes a  system
 integrable.

\subsection {Useful representations of universal ${\cR}$-matrix }

The aim is to start from  abstract    algebraic structures
 and   build systematically the objects of
  direct relevance in   
integrable ultralocal  models. 
We look first into the
 universal ${ \cal R}$-matrix solutions
 like 
 (\re{ur-slq2}, \re{ur-gen})
of the { universal} Yang--Baxter type equation (\re{ybeR})
and  denote them
  by $ {\cal R}^+$ and the related equation
by $(+++)$, since
 their 
 finite-dimensional representations
as derived for example from (\re{ur-slq2}), are  of upper-triangular  form.  
It is  easy to  check that, if we define ${\cal R}_{{\bf 12}}^- = 
 ({\cal R}^+)_{{\bf 21}}^{-1}$,\ it would correspond 
to the representation in   lower-triangular   form
 and be a solution of
  the universal $(---)$ equation. \
 More generally one would get the
universal equations
\be
 {\cR}^{\ep 1}_{{\bf 1 2}}~{\cR}^{\ep 2}_{{ \bf 1 3}} 
~{\cR}^{\ep 3}_{{ \bf 2 3}}
 ~~=~~{\cR}^{\ep 3}_{{ \bf 2 3}}
    ~{\cR}^{\ep 2}_{{ \bf 1 3}}{\cR}^{\ep 1}_{{ \bf 1 2}} ~,
\ll{ur+-}\ee
of {\em six} different forms with $\ep 1, \ep 2, \ep 3$ occuring
in the sequence  $ \ +++,  \ +--, \ ++-, \ ---, \ --+, \ -++ \ $. The
 validity of these equations   can be proved starting 
 from the original  one $+++$, taking the inverse of ${\cR}$ and
renaming the quantum spaces ${\bf 1},{\bf 2},{\bf 3}$ in a suitable order.
  At the same time one concludes also about
   the nonvalidity of $(\pm \mp \pm)$ type equations, which can
  hold only for triangular
 Hopf algebras with  $  R_{{ \bf 2 1}}=R_{{\bf 1 2}}^{-1}$.

It is  important now to show that,  algebraic objects related to the
quantum algebras can be generated from the above universal equations
and the  ${\cR}$-matrices through  direct  reductions.
 For finite-dimensional fundamental
representations $\pi$ acting
 in the space ${\bf 1} $ and  ${\bf 2}$, while     ${\bf 3}$ kept 
  as the quantum space, one gets
\be (\pi \otimes \pi) {\cal R}_{{\bf 12}}
=   R^+_{ 12} , \qquad 
   (\pi \otimes I\otimes I) {\cR}^\pm_{{\bf 1 3}} =R^\pm_{1{\bf 3}} \equiv
 L^{(\pm)}_{ 1{\bf 3}}\ll{r-l} \ee
which  reduce (\re{ur+-}) clearly to  
the FRT algebra relations (\re{frt}). 
 Consequently, taking  the universal    solution (\re{ur-slq2}) and 
using the map   $\pi$   one may get the  
$ R^{\pm}$ and $ L^{(\pm)}$ matrices in  explicit form and hence derive 
 the  quantum  algebra (\re{sl2qa})
 from  the quadratic relations among the elements of
 $L^{(\pm})_{a{\bf 3}}$.
At the same time why   FRT relation like $(+ - +)$
does not  exist,
  is also   answered  promptly from  the nonvalidity of such
universal equation.

Note  that  the complementary action  of $\pi$  on 
             $ {\bf 3}\rightarrow
   { 3} $ with  ${\bf 1}, {\bf 2}$ kept as quantum spaces,
    derives  from the same universal relations (\re{ur+-})
   the intertwining relation of the universal
  $ {\cal R}$-matrix (\re{ur1})
and from the solution (\re{ur-slq2})
 the  coproduct structures
of the quantized algebra  (\re{coprod}).
The action of $\pi$ on all the spaces reduces (\re{ur+-}) to the spectral
parameterfree YBE (\re{braid}).
\subsection {Yang--Baxterization and construction of 
 $R$-matrix and Lax operator}
It is   crucial to observe   that
 the FRT algebra relations (\re{frt}) are quite similar to
the QYBE (\re{qybel}) and the algebra related objects
$L^\pm$ and $R^\pm$  resemble  closely to
 the Lax operator $L(\la)$ and the quantum $R(\la, \mu)$-matrix
associated with  the integrable systems. However
the main distinguishing feature  of the former algebraic entries
is their spectral independence. The spectral parameter on the other hand 
is  an essential
item in  the integrable theory.
Therefore, for  application  to
integrable systems
 we need to introduce spectral parameters  in the formal algebraic
structures  or
in other words {\em Yang--Baxterize} the FRT algebra.

To frame the scheme for   generating integrable models we adopt 
sequentially  the following steps.

\b{description}
\item i)  Start from the  universal $ {\cal R}$-matrix and
 reduce it to  $ R^{\pm}$ and $ L^{(\pm)}$
  involved in the FRT algebra.
\item ii) From the reduced universal YBE
determine the
 underlying quantum algebra  and the related
coproducts.
                             \item iii)
Yang-Baxterize
 the FRT algebra to construct the
{ Lax operator} $L(\lambda)$ and the trigonometric
$R_{trig}(\lambda-\mu)$-matrix
 of a generalized and exact  integrable model on a lattice,
 which may serve as an
 ancestor model.
 \item  iv)
Find different
 realizations of the  related    quantum algebra  and 
 reduce consequently the ancestor
model to  generate
 explicit Lax operators  for variety   of  integrable lattice models.
Since the Yang-Baxterized ancestor Lax operator is a solution of the 
QYBE, all the models  generated from it  must also 
 be quantum integrable and share naturally the same quantum
$R$-matrix.

\item  v)
Continuum models are obtained from  their lattice versions at the continuum
limit  $\De \rw 0.$ 
\item vi)
For generating the class of    models associated with the  rational
$R_{rat}(\la-\mu)$-matrix,  
we may take    $ q \rightarrow 1$ limit in the above scheme.
As a result 
the underlying quantum algebra  becomes a
realization of the Yangian algebra
and one recovers consistently the relevant objects
for   the rational class of models.
\e{description}

This gives  a systematic
 scheme  for constructing the Lax operators and the
quantum $R$-matrices  of different families of
 exactly integrable quantum models
 of ultralocal class, both at the lattice and the continuum  level.
However, as we will see   below,
 one  encounters  certain practical   shortcomings of this
scheme  and has to prescribe some  working remedies.

For demonstration we take the simplest  $U_q(sl(2)) $ case 
and  as itemized above, look first 
for the
universal ${\cal R}$-matrix
solution  (\re{ur-slq2}).
 Noting that, since  $\pi (S^\pm)= \si^\pm, \pi(s^3)=\si^3$
maps into  Pauli matrices, we  obtain 
using  (\re{r-l})  explicit forms of  $L^{(\pm)}$ and $R^\pm$
 reproducing  the FRT algebra with
 (\re{R+}) and (\re{l+-}), which  yield naturally 
  
   the well
known quantum algebra (\re{sl2qa}) as the underlying algebra.

%---------for YB
The
Yang--Baxterization of the FRT algebra
, i.e.  introduction of   spectral parameters
$\xi=e^{i\eta \lambda},\  \zeta=e^{i\eta \mu}$,
 is in general  a difficult task.
However in the present  case
 it takes the  simple form
\c{jones}
\be   R (\lambda-\mu ) ~=~ \frac {\zeta}{ \xi } R^+ -
     \frac{ \xi}{\zeta} R^-  ~,\ll{YB-r}  \ee
 for  the quantum $R$-matrix
     and similarly  \c{kbjp92,fad95}
     \be
 L (\lambda)  ~=~ \frac {1}{ \xi  }  L^{(+)} + \xi L^{(-)} ~
, \ \ \  \  \ll{YB-l} \ee
for constructing the Lax operator.
 For showing  that  $R(\la-\mu)$ and $ L(\la)$
 thus obtained  indeed  represent   integrable systems,
 one has to check  them as  solutions of the   QYBE.
Therefore   inserting  the
 above Yang--Baxterized forms in (\re{qybel})
 and comparing    the  coefficients
of various powers of spectral parameters,  we    arrive
at a set of seven algebraic relations involving $R^{\pm}$ and $~L^{(\pm )}$~.
Six of them are  same as the  FRT relations (\re{frt}) and  hold
automatically, while the remaining  one is
\be   R^{+}L^{(-)}_1 L^{(+)}_2 ~-~L^{(+)}_2 L^{(-)}_1 R^+ ~=~
   R^{-}L^{(+)}_1 L^{(-)}_2 ~-~L^{(-)}_2 L^{(+)}_1R^- ~\ll{extrafrt} \ee
As we see, it is a combination of   $(\pm \mp \pm)$ type relations   
and clearly does not belong
to the FRT algebra
as explained above.

However the  key  observation is  that, if we
demand 
the $R^\pm$-matrix to satisfy the Hecke algebra
 (\re{hecke-r}),  we may
 replace in  (\re{extrafrt}) $R^\pm$
through each other, 
  using the Hecke condition
$~   R^+  =~R^-  +c ~{ P }$, which results
\[   R^{-}L^{(-)}_1 L^{(+)}_2 ~-~L^{(+)}_2 L^{(-)}_1 R^-
(c { P})L^{(-)}_1 L^{(+)}_2 ~-~L^{(+)}_2 L^{(-)}_1 (c P) \] \[~=~
   R^{+}L^{(+)}_1 L^{(-)}_2 ~-~L^{(-)}_2 L^{(+)}_1R^+ +
  (-c P) L^{(+)}_1 L^{(-)}_2 ~-~L^{(-)}_2 L^{(+)}_1(-c P). \]
The above relations hold  obviously    due to the permutation property of
$P$ and the  FRT relations (\re{frt}).
We see thus that  the    FRT relations can be
Yang--Baxterized  as
(\re{YB-r}), (\re{YB-l})   to give the  QYBE
for the class of $R$-matrices belonging to the Hecke algebra. At the
same time  one generates also the spectral parameter dependent genuine 
$R(\la)$-matrices and the Lax operators of quantum integrable systems.
As mentioned, 
 all   $R$-matrices related to the general    $U_q(gl(n+1))$ 
exhibit  the
 Hecke property.

It is illustrative to check 
  that, the
$R(\la)$-matrix  built through  Yang--Baxterization (\re{YB-r})
 from the basic solutions (\re{R+}) has the form
\be
R_{11}^{11}=R_{22}^{22}= 
{\sin ( \lambda+\al)},    \quad R_{11}^{22}=R_{22}^{11}=
{\sin ( \lambda)},  \  \ \ R_{12}^{21} ={\sin \alpha}e^{i 
    \la},  \quad  R_{21}^{12}=
      {\sin \alpha}e^{-i \la} \ \ \ \
\ll{rtrig}\ee
where  $~~\al =\hbar \eta~~$ and $ \la $ is scaled by $ \eta .$ 
 Notice that this 
is  the same trigonometric solution  (\re{R-mat},\re{abtrig})
 obtained   as a direct  solution of 
 YBE (\re{ybe}). Similarly  the
spectral  dependent
 Lax operator, Baxterized  as (\re{YB-l}) from $L^{(\pm)}$ (\re{l+-}): 
\be L(\la)= \left( \b{array}{c}\sin (\la+ \eta s^3),
\qquad \sin \eta S^- e^{i \la} \ \\
\sin \eta S^+ e^{-i \la}, \qquad \sin  (\la -\eta s^3) \e{array}
\right) \ll{lslq2}\ee
 coincides  with the expression  (\re{L-sklya}), when   the
  Sklyanin algebra is realized  through the generators of the  
 quantum algebra as (\re{sk-real}).
It should be remarked,  that   the  above Yang--Baxterized
  $R(\la-\mu)$ and $L(\la)$ matrices are in fact 
slightly different from their earlier forms, while 
  the exact coincidence  is achieved  by a simple and allowed   similarity
transformation \[R_{12}(\la)\rw A_{12}(\la)R_{12}(\la)A^{-1}_{12}(\la),
\quad \mbox{
and} \quad
L_{1{\bf 2}}(\la)\rw A_{1{\bf 2}}(\la)L_{1{\bf 2}}(\la)
A^{-1}_{1{\bf 2}}(\la)\]
with $A_{1{\bf  2}}(\la) =
e^{i {\la \ov 2}( \si^3 \otimes {\bf I}- { I}\otimes s^3 )}.$

Now we come to the points v) and vi), i.e. to the 
 practical applications  of the scheme for
constructing
 explicit Lax operators   of  integrable lattice as well as 
field models. For this we 
  find different
 realizations of the   underlying   quantum  algebra  $U_q(sl(2))$ (\re{sl2qa})
 generated by  $S^\pm, s^3$. Inserting them subsequently 
in the Baxterized $L(\la)$ (\re{lslq2}), which serves as an ancestor model,
 we generate easily 
the representative Lax operators of other  models. Since  the ancestor model 
is  quantum integrable with trigonometric $R_{trig}(\la)$-matrix
(\re{rtrig}),
 all its different descendants obtained through 
particular realizations would evidently  be quantum integrable
and related to the same $R_{trig}$.

\subsection{Models with trigonometric $R_{trig}(\la)$-matrix}

\subsubsection {$XXZ$ spin-$\ha$ chain} 
Simplest realization of the generators $S^\pm, s^3$  
through Pauli matrices gives 
$S^\pm =\si^\pm, ~~s^3= \si^3$ and  since
$e^{\pm i\eta \si^3}=\cos \eta \pm i \si^3 \sin \eta$ one gets
  $ [\si^3]_q=\si^3$, which  reduces     
the quantum algebra (\re{sl2qa}) evidently to the spin algebra
 (\re{sl2}). The Baxterized $L(\la)$ operator reduces correspondingly 
 to the form  
\be L_{xxz}(\la)= \left( \b{array}{c}\sin (\la+ \eta \si^3),
\qquad \sin \eta \si^- \ \\
\sin \eta \si^+ , \qquad \sin  (\la -\eta \si^3) \e{array}
\right) \ll{XXZ1}\ee
 Note that this is the Lax
operator of the quantum integrable 
anisotropic $XXZ$ spin-$\ha$ chain  listed in (\re{XXZ}).
The Hamiltonian of the model is  derived in sect. 6 and the
eigenvalue problem is solved in sect. 7.   

\subsubsection {Sine-Gordon: lattice and field models} 
An important realization of the $U_q(sl(2))$  quantum algebra
in bosonic operators $u_n,p_n$  is 
\be e^{i \eta s^3}= \sqrt {2 s}e^{i \eta u}, S^-=g(u) e^{i\De p},
 S^+=(S^-)^\da,~~ g(u)=[1+2s^2\cos 2\eta (u+\ha)]^{\ha}, \ll{qa-up} \ee
where $s={m \De \ov 2}$.
Using the commutation relation  $[u_n,p_m]={i \ov \De}\de_{nm},$ 
it is easy to check that the mapping (\re{qa-up}) satisfy the 
$q$-deformed  algebra (\re{sl2qa}). Therefore we may insert 
(\re{qa-up}) in the general  $L(\la)$ (\re{lslq2}) to  obtain
\begin{equation}
  L_{n}(\la)  =
  \left( \begin{array}{c} m\Delta  \sin (\la+\eta u_n)  
 \qquad    g(u_n)~ e^{ip_n \Delta } \\
       e^{-ip_n \Delta }~ g(u_n) \qquad
    m\Delta  \sin (\la-\eta u_n)\end{array} \right), 
\ll{L-sg1}\end{equation}
with $g(u_n)$ given as in (\re{qa-up}). The lattice Lax operator generated 
by $\si^1 L_n(\la)\equiv L_n^{(lsg)}$ represents exactly  integrable
 lattice version of the sine-Gordon model listed in (\re{L-sg}).

At the continuum limit $\De \rw 0$, as explained in sect. 2,  
we obtain    
$ L_n^{(lsg)}(\la)= I-i\De {\cal L}_{sg} $. Using the limits
up to the order $\De$ giving $u_n \rw u(x), p_n \rw p(x), ~~  g(u_n) \rw 1, 
 ~~e^{\pm ip_n \Delta }~ \rw I \pm i \De p(x) ~~$  one    
derives   finally  from
(\re{L-sg1}) the well known
 Lax operator of the  sine-Gordon field model as 
$~~{\cal L}_{sg}(\la)  =  p(x) \si^3 + k_0 \sin \eta u(x)\si^2 + 
k_1 \cos \eta u(x) \si^2, ~~$  
with the momentum vector $k_0= m \cos \la, ~ k_1=i m \sin \la$ 
satisfying the  relativistic condition $~~  k_0^2-k_1^2=m^2 ~~$. The
 energy spectrum  and the  physical significance 
 of the SG field model are discussed in sect. 7.

Note that both the lattice and the field SG models are associated with the
same trigonometric $R(\la-\mu)$ (\re{abtrig}) obtained through
Yang-Baxterization. It should however be remembered that, while the lattice 
version (\re{L-sg1}) is an exact solution of the QYBE, the field Lax
operator ${\cal L}_{sg}$  satisfies QYBE only up to first order in $\De$,
 which 
is enough for the field limit. Similarly, the  quantum algebra
 realized exactly in the  lattice SG  trivializes also at the continuum
 limit, though
with some mysterious affine quantum group 
symmetry appearing in the conserved quantities 
\c{lecber}, which goes beyond the present scope.  
\subsubsection {$q$-oscillator model}
$q$-oscillators or $q$-bosons
 may be defined through the commutation relations \c{qoscl}
 \be  [A,N] = A,~[A^{\dagger }, N ] = \ -
A^{\dagger},~~
~AA^{\dagger } -  q^{\mp} A^{\dagger }A = q^{\pm N} \ll{qoscl}\ee
An interesting realization  of the quantum algebra (\re{sl2qa}) in
$q$-oscillators may be given  through 
the $q$-Holstein-Primakov transformation \c{qhpt}
\be S^+=([2s-n]_q)^{\ha} A,~~ S^-=A^\da ([2s-n]_q)^{\ha} ,~~
 s^3=s-n, ~~A^\da A=[n]_q \ll{qhpt}\ee
while  another simpler one through  
\be ~~ s^3 = N+\ha ,~ S^+= \kap A  q^{{N \ov 2}} ,~S^-= \kap  q^{-{N \ov 2}}
 A^{\dagger},
~~~\kap =   (q-q^{-1})^{\ha}. \ll{S-qoscl}\ee
Using such realizations it is possible to construct 
  from (\re{lslq2})
  quantum integrable  lattice models involving
$q$-oscillators.
\subsection {Multicomponent  models}
For  constructing  
  $\mathsf { N\ot N}$ matrix 
Lax operators and $\mathsf {N^2\ot N^2}$ $~~R(\la) $-matrices associated with
 multicomponent integrable  models, we may start with  
  the
 universal  ${\cR}_0$-matrix 
$\in U_q(sl(n+1))\otimes U_q(sl(n+1)) $  given by  
(\re{K0},\re{cR0}), where $\mathsf {N}=n+1.$
For finding  finite-dimensional fundamental representations  
(\re{r-l})
of ${\cR}_0,$ we may use the mapping 
\bea \pi(H_i)  &=& E_{ii} -E_{i+1 i+1} ,~~
\pi(q_{ii}^{\pm H_i})  =1+ E_{ii} (q_{ii}^{\pm 1}- 1) +E_{i+1 i+1}
(q_{ii}^{\mp 1}- 1)  \nonumber \\
 \pi(X_i^+) & =& E_{i i+1} , \quad   \pi(X_i^-) =E_{i+1 i}, \quad
  \pi(X_{ij}^+)  =E_{i j+1} ,\quad  \pi(X_{ij}^-) =E_{j+1 i}, \ll{pix}\eea
where
 $(E_{ij})_{kl}=\delta_{ik}\delta_{jl} ,$ 
along with the relations (\re{e-x}).
Since the $R^\pm$-matrices obtained are again of the Hecke type,
 we may follow the steps
(\re{YB-r},\re{YB-l}) to  construct
 the spectral parameter dependent operators
 \be R(\la- \mu)
= \sin(\la-\mu+\al) 
\sum_{k=1}^{n+1} 
  E_{kk} \otimes  E_{kk}  +
   \sum_{ i\neq j}^{n+1}( \sin (\la-\mu) 
  E_{jj} \otimes  E_{ii}   +\sin \al
  E_{ij} \otimes E_{ji})  \ll{rtrign}\ee
and 
 \be L^{(N\times N)}(\xi)=
\sum_{k=1}^{n+1} ( \xi\omega_k+\xi^{-1}\omega_k^{-1}) E_{kk}
  + \sin \al  \sum_{i<j}
\xi E_{ij+1} \omega_i { e}_{ji}
-\xi^{-1}E_{j+1i}   { e}_{ij}\omega_{i}^{-1}
\ll{ltrign}\ee
where
\be~~ \omega_k=q^{
\left(\sum_{s=k}^n H_s-
 \sum_{i=1}^n {iH_i\over n+1} \right)},
 \   \
\omega_{n+1}=q^{-
 \sum_{i=1}^n {iH_i\over n+1}  }\ , ~~
 e_{\pm i}= 
   q^{-\ha} X^\pm_i q^{\mp {H_i\ov 2}} \ll{exgln}\ee
   for $k=1,\cdots,n \ $ 
and the operators $e_{ij}$ correspond to the nonsimple roots $\sum_{s=i}^j \al_s$.
For generating  multicomponent integrable lattice 
 models from (\re{ltrign})
associated with the quantum $R$-matrix  (\re{rtrign}), we have  
 to consider different  realizations of the underlying quantum
algebra in more
  physical variables like bosonic, $q$-bosonic
 or spin variables.
\subsubsection{Generalized spin  model}
For fundamental models  the Lax operator
(\re{ltrign}) coincides with (\re{rtrign}) 
due to the mapping (\re{pix})  and represents quantum integrable
model with interacting generalized spin operator $E_{ij}$ \c{sun}.
 Due to the
regularity condition $R(\la=0)=P$ the Hamiltonian of the model can be
constructed following sect. 6 and the eigenvalue problem can be solved using nested
Bethe ansatz described in sect. 7. 
\subsubsection {Multicomponent $q$-bosonic model}
 Consider  
 $q$-bosonic
realization  of the higher rank quantum algebra $U_q(sl(n+1))$:
\bea
H_{i}&=& N_{i}-N_{i+1},\ \ \  N_{1}=(n+1)s-\sum_{k=2}^{n+1}N_k
\ \hbox{for } \ i=1,2\ldots,n  \nonumber \\
 X^+_k&=&
 A^\dagger_k A_{k+1}, \quad X^+_1=( [N_1]_q)^{{1\over 2}}A_2, \ \
 X^-_k=( X^+_k)^\dagger
\ll{qa-qo} \eea
%4.12
with  $k=2,3\ldots,n.$
Here $s$ is an arbitrary parameter and $A^\dagger_k,A_k$  are $n$
number of  $q$-bosonic
operators with deformed commutation relations
\be
[A_k,A^\dagger_l]=\delta_{kl} {\cos(\alpha(N_k+ {1\over 2}))\over \cos
{\alpha\over 2}},\quad [N_k,A^\dagger_l]=\delta_{kl}A^\dagger_l
,\  [N_k,A_l]=-\delta_{kl} A_l.
\ll{qoa} \ee
%4.13
With such realization  and  relation (\re{exgln}),  (\re{ltrign})
 represents the Lax operator $L_n(\la)^{(multi-q)}$ of
 a  quantum integrable multimode $q$-bosonic model \c{ktjp95}.

\subsubsection {Toda field model}
Let us consider a  realization of the above $q$-bosons in canonical
variables $ [u_n^a,p_m^b]=i \delta_{nm}\delta_{ab}$  as
\be
A_i
=q^{(\vec \rho_i \ \vec p)} \ [N_i]_q, \ \
 A_i^\dagger
= [N_i]_q \ q^{-(\vec \rho_i \ \vec p)} \
\quad N_k
=\vec {\alpha}_k \  \vec {u} +s
\ll{qo-cal} \ee
%4.20
with $i=2,3,\ldots,n+1$ and  $k=1,2,\ldots,n+1$, where $\vec \alpha_i$ and
$\vec \rho_i$ are the simple  roots and fundamental weights of $sl(n+1)$,
respectively with the relation
$\ (\vec \alpha_i\cdot\vec\rho_j)=\delta_{ij}, ~~ i,j=2,3,\ldots,n+1. \ $
Here we have assumed $\vec \alpha_1=-\sum_{i=2}^{n+1}\vec \alpha_i $
and $\vec \rho_1=0$.
It is curious to note that this realization of $q$-boson  
  reduces the  Lax operator $L_n(\la)^{(multi-q)}$
or equivalently, the original $L$-operator
(\re{ltrign}), with the  use of   (\re{qa-qo})   to 
 the well known lattice regularized
version of the Toda field model $L_n(\la)^{(todafield)}$ \c{babelontoda}.
Obviously for   $n=1$ we  recover the 
 lattice SG model.

By rescaling $\vec p \rightarrow\Delta \vec p$ and adjusting the
parameters one can introduce the lattice constant $\Delta$ and go to the
continuum limit at $\De \rw 0$, to generate the Lax operator of the Toda field
models.

\subsection{Models with twisted $R_{trig}^\th(\la)$-matrix}

For constructing    more
general  integrable models we may consider      
${\cR} \in U_q(gl(2)\otimes U_q(gl(2))$ constructed in (\re{urglq2}).
 However instead of repeating
the Yang-Baxterization \c{ktjp95}, we can 
obtain the spectral parameter dependent 
 $\ti R$-matrix and the Lax operator   by  twisting 
  the previous Yang-Baxterized forms ({\re{rtrig},\re{lslq2})
with suitable  representations of  the operator  $ \
    {\cal F}=e^{i\theta (s^3\ot Z-Z \ot s^3)}$. This gives through $
 \ti  R^{\th}_{trig}(\la-\mu)=F(\th) R_{trig}(\la-\mu) F(\th)
~~$ with $F_{12}(\th)=e^{i\theta (\si^3\ot I-I \ot \si^3)}$ 
the twisted quantum $R_{trig}^{\th}(\la)$-matrix 
\be
\ti R_{11}^{11}=\ti R_{22}^{22}= 
{\sin ( \lambda+\al)},    \quad \ti R_{11}^{22}=
{\sin ( \lambda)}e^{-i \th}, ~~~\ti R_{22}^{11}=
{\sin ( \lambda)}e^{i \th}
,  \  \ \ \ R_{12}^{21} =
  R_{21}^{12}=
      {\sin \alpha} \ \ \ \
\ll{th-r}\ee
and similarly  the twisted Lax operator  as
\be  \ti L_{a{\bf n}}(\la,\th,Z_n)=
F_{a{ n}}(\th,Z_n) L_{a{ n}}(\la)F_{a{n}}(\th,Z_n), \quad
F_{a{ n}}(\th,Z_n)= diag \l(e^{i\th(Z_n-s^3_n)},~~e^{-i\th(Z_n+s^3_n)}  \r)
\ll{lglq2}\ee
where  $L_{a{ n}}(\la)$ is given by (\re{lslq2}). We have chosen here $\pi
(Z_a)=1, ~ a=1,2$ to  exclude the color parameters, which will be considered below.
\subsubsection{Generalized Wu-McCoy model}
As examples of concrete  models  we  may consider simply the fundamental
representation of the Lax operator    (\re{lglq2}) coinciding with 
(\re{th-r}), which would 
 correspond to the spin Hamiltonian 
\be    H=\sum_{m}^{}{1\over 2} \cos \alpha
\sigma_{m}^{3}   \sigma_{m+1}^{3}   +
e^{ {1\over 2}(Z_{m}+Z_{m+1})\th}
\sigma^{+}_{m}
 \sigma^{-}_{m+1}   +
e^{- {1\over 2}(Z_{m}+Z_{m+1})\th}
\sigma^{-}_{m}
 \sigma^{+}_{m+1}  .\ll{Wu-m}\ee
Note that for $Z_{m}=1$ it  reduces to the Wu-McCoy model
(\re{WM}) of ferroelectric in a constant external electric
 field $\th$, the Hamiltonian construction of which is dealt in 
(\re{WM1}). Therefore, (\re{Wu-m})
 may be considered as   the Wu-McCoy  or the asymmetric $6$-vertex model
\c{asym6}
placed in an
 inhomogeneous electric field $Z^{(m)}\th$ and  solvable through
 the Bethe ansatz. Similarly, one may choose more general
 twisting 
with  color
representations  $\pi (Z_a)=\bar \la_a,$ relating 
  to the spectral parameters
as $\bar \la_1=c \la,~ \bar \la_2=c \mu $ and  construct the corresponding 
Lax operator from 
(\re{lglq2}). In the simplest case (see (\re{twicol})),
 such  color parameters induce
 interaction with an external magnetic field.    

\subsubsection{Generalized lattice sine-Gordon model}
The Lattice sine-Gordon model obtained above can be generalized ($\th$LSG)  
to include  nontrivial  parameters $\th$ and $ Z_n$,
 retaining its quantum integrability.
 The corresponding Lax operator can be constructed 
by  a $\th$ twist of 
    $
L_{n}^{(lsg)}(\lambda) $   (\re{L-sg1}) related to  the lattice SG model 
as \be
L_{n}^{(\th lsg)}(\lambda,Z_n)= 
F_{{ n}}(\th,Z_n) L_{{ n}}^{(lsg)}(\la)F_{{n}}(\th,Z_n), \quad
F_{{ n}}(\th,Z_n)= diag \l(e^{i\th (Z_n-u_n)},~~e^{-i\th(Z_n+u_n)}  \r)
\ll{tlsg}\ee
\subsection{Multiparameter and color generalizations}
 We can construct integrable multiparameter models 
by introducing parameters in the multicomponent 
Lax operator (\re{ltrign}) through twisting transformation.
The concerned universal ${\cR}$-matrix belongs to  
$\in U_q(gl(n+1))\otimes U_q(gl(n+1)) $  and  may be obtained  
 using the  twisting operator  ({\re{cG}) as
\be          {\cal G}(\phi_{ij}, v_{j}) =
\exp \l(- {1\over 2} ( \sum_{j k}
      v_{j}   a^{jk} (  H_k {\otimes} Z -
Z {\otimes}H_k ) -
     2\sum_{ij}
        \phi_{ij}     H_i {\otimes}H_j \right)
\ll{Ggln}\ee
by specializing $M=1, q_{ii}=q^2, H_\al=Z, a_{ij}=2 \de_{ij}-\de_{ij+1}-
\de_{ij-1}$.

 Here the  $ \phi_{ij}, v_{j}$ are external parameters   
along with a central operator $Z$ responsible for the color parameters.
Note that,  nontrivial   $\phi_{ij}$ can be introduced  starting from 
$n\ge 2$ only, i.e. at least  from $gl(3)$, while the parameter 
$v_1\equiv \th$
 can appear even 
in  $gl(2)$, as considered above. 
Remarkably for 
 reductive type algebras
  some additional   nontrivial
 parameters  might  arise
 due to the  presence of central operators $Z$, 
since  one  can consider  ${Z} $
 to have different eigenvalues in
different spaces, i.e. $(\pi_{\bar \la}\otimes { I}){ Z}= \bar\lambda
 I\otimes{ I} ,$
 while $ ( { I}\otimes \pi_{\bar \mu}){ Z}= \bar \mu{ I}\otimes I .$
  Parameters $\bar \lambda, \bar \mu $ may be taken as the {\em color
 } parameters.

Therefore,   we may  construct the multiparameter Lax operators  by twisting
(\re{ltrign}) through the operator \be
( \pi_{\bar \la} \otimes { I}\otimes I)
{\cal G}_{{\bf 13}}={ G}_{1 {\bf 3}}(\bar \la,\phi_{ij},v_j)=
\sum_{k=1}^{n+1}T_k(\phi_{ij}) W_k(\bar \la,v_j) 
  E_{kk} 
\ll{gcol}\ee
where
\be
T_k(\phi_{ij})=e^{-  \sum_{j=1}^{n}( \phi_{kj}- \phi_{k-1j} ) H_j }
,\quad
\
W_p(\bar \lambda, v_j)=e^{  {1\over 2}\left(\bar
\lambda \sum_{jl}^n v_ja^{jl} H_l+{\bf Z}(
 \sum_{j=1}^n {jv_j\over n+1} - \sum_{s=p}^n v_s)\right)
}, \ll{TW}\ee
for $p=1,\cdots,n \ $ with $\phi_{0j}=\phi_{n+1 j}=0$, while $
W_{n+1}(\bar \lambda)=e^{
 {1\over 2}\left( \bar \lambda \sum_{jl}^n v_ja^{jl} H_l+{\bf Z}
 \sum_{j=1}^n
 {jv_j\over n+1}\right)    }.$    
The Lax operator of  multicomponent multiparameter quantum integrable
models thus can be generated 
from  the $N \times N$ Lax operator 
$ { L}^{(N\times N)}_{1 {\bf 3}}( \xi) $   
 (\re{ltrign}) using  (\re{gcol}) as
\be \ti L_{1{\bf 3}}(\xi,\bar \la,v_j,\phi_{ij})
={ G}_{1 {\bf 3}}(\bar \la,\phi_{ij},v_j)
{ L}_{1 {\bf 3}}^{(N\times N)}( \xi)
{ G}_{1 {\bf 3}}(\bar \la,\phi_{ij},v_j)
\ll{lgnl}\ee
 where   
$\phi_{ij},v_j$ are additional  
parameters apart from  the color parameter $\bar \la.$ As mentioned, the
contribution from $T_k(\phi_{ij})$ introducing parameters $\phi_{ij}$ starts
 for the higher rank algebras with $n \geq 2$, while parameters $v_j$
through $W_p$ can enter even for $U_q(gl(2))$.

\subsubsection{Multiparameter multimode $q$-bosonic model}
Multiparameter generalization of the multimode $q$-bosonic Lax operator
$L_n^{(multi-q)}$
 constructed above, can be obtained by transformation like 
(\re{lgnl}) with  twisting operators $ { G}^{qoscl}_{ {n}} $ obtained from
(\re{gcol},\re{TW}) by using the $q$-bosonic
realization  (\re{qa-qo}).
\subsubsection{Multiparameter Toda field  model}
Performing  twisting  (\re{lgnl}) on the discretized  Toda field Lax
operator $L_n^{(todafield)}(\la)$ \c{babelontoda} obtained above, it can be
generalized to include parameters $~~ \bar \la,\phi_{ij},v_j,~~
 i, j = 1, 2, \ldots,n~~$
 along with the constant inhomogeneous field $Z_n$. For this one has to
make  only  the replacement
 through bosonic operators $H_i=u_i$ in the expressions (\re{TW}).
Note that $n=1$ case recovers the generalized lattice SG  model.

It should  however be remarked that, though such multiparameter quantum models 
are exactly  integrable at the discrete level, whether they 
lead to any sensible   field models 
at the continuum limit  is not yet clear.

\subsubsection
{Multiparameter  trigonometric $R(\la)$-matrix} 

%---------------------
The multiparameter models  are associated with the multiparameter 
generalization of the trigonometric $R(\la-\mu)$-matrix, which can be
obtained by twisting of  (\re{rtrign}):
$\ti  R_{12}=G_{12}
 R_{12}(\la-\mu)G_{12}
~~$
where $G_{12}  \equiv G_{12}(\bar \la, \bar \mu, \phi_{ij},,v_j) )
$  is obtained as $ (I \otimes \pi) { G}_{1 {\bf 2}}(\bar \la,\phi_{ij},v_j)
={ G}_{1 {\bf 2}}(\bar \la,\bar \mu, \phi_{ij},v_j)$ from ${ G}_{1 {\bf 2}}$
(\re{gcol}).  In the explicit form the  
$\ti R$-matrix is given by 
  \bea &\ti R&({\la-\mu},\bar\lambda, \bar \mu,\phi_{ij},v_j)
=  \sin(\la-\mu+\al)
\sum_{k=1}^{n+1}   g_{kk}^2
  E_{kk} \otimes  E_{kk}  +\sin(\la-\mu)
   \sum_{ k>j}^{n+1} ( g_{kj}^2
   \Phi_{kj}  \nonumber \\
 & & E_{jj} \otimes  E_{kk}    +
    g_{jk}^2\Phi_{kj}^{-1}
  E_{kk} \otimes  E_{jj} )+ \sin \al
  \sum_{i\neq j}  g_{ij}g_{ji}
  E_{ij} \otimes E_{ji} 
  \ll{col-r} \eea
  with the notation \be
  g_{ij}
 = {f_{i}^2(\bar \mu, \{v_n \})\over  f_{j}^2(\bar \lambda, \{v_n \})}, \
       \qquad f_l(\bar\lambda, \{v_n \})=
e^{  {\bar \lambda\over 2}\left(
 \sum_{j=1}^n {jv_j\over n+1}-
  \sum_{t=l}^n v_t
 \right).          }
\ll{gr}\ee
Note that the $\th$-deformed $R$ (\re{th-r}) can be recovered  from the
general  (\re{col-r}) as a particular case.

   Interpreting the color parameters  as
  spectral parameters: $\bar \lambda=c \la ,\bar \mu= c \mu,$ 
 one can  get 
  from (\re{col-r}) a   $R$-matrix with
  two sets of spectral parameters
   with nonadditive dependence.
  Recall that in standard  integrable models  the
   known quantum
  $R(\lambda-\mu)$-matrices  (see (\re{R-mat})) depend
additively on spectral parameters.

For the special
colorfree  choice
 $f_k(\lambda)=1$ and $f_k(\mu)=X_k,$  (\re{col-r}) reduces to
\bea   R( \la)&= & \sin(\la +\alpha)
\sum_{k=1}^{n+1} X_{k}^2
  E_{kk} \otimes  E_{kk}    +
\sin \la   \sum_{k> j}^{n+1}
   (   \Phi_{kj}X_{k}^2
  E_{jj} \otimes  E_{kk}  \nonumber \\ & +&  \Phi_{kj}^{-1} X_{j}^2
  E_{kk} \otimes  E_{jj}  ) +
  \sin \alpha \sum_{i<j}  X_{i}X_{j}(
  E_{ij} \otimes E_{ji}+   E_{ji} \otimes E_{ij} ),
\ll{perks}\eea
 recovering
the well known Perk-Schultz model \c{perks}. Thus it 
shows a remarkable   connection
between an integrable statistical model constructed in a traditional way
 and the representation of  the universal
$\cal R$-matrix  constructed through twisting
transformation.
The additional parameters $\epsilon_\alpha=\pm 1$ appearing in the
diagonal elements of the Perk-Schultz model can be generated if one starts
from a universal $\cal R$-matrix related to noncompact groups (nonstandard
$R$-matrix \c{nstR}).
%%%%%%%%%%%%%%%5
\subsection {Models with rational $R_{rat}(\la)$-matrix}
As explained in sect. 2, at the limit $q \rw 1$ both $R,L$ operators 
are reduced to their rational limits. The rational  $R$-matrices
can be obtained from the trigonometric  (\re{rtrign}) or (\re{col-r})
by taking the limit $\eta \rw 0$ in $\la\equiv \eta \la, \al =\eta \hbar$,
which amounts to the formal replacement of $\sin (\la-\mu) \rw
\la-\mu, ~~ \sin (\la-\mu+\al ) \rw \la-\mu+\hbar $ etc. 
As a result  (\re{rtrign})   gives
\be R_{rat}(\la-\mu)= (\la-\mu) I + \hbar P, \qquad P= \sum^N_{ij}
   E_{ij} \otimes E_{ji}    
\ll{rationalr}\ee
with $P$ being the permutation operator.

On the other hand at the rational limit  the quantum algebra
 generators reduce to   spin operators, which in turn transforms 
the $L$  operator (\re{lslq2})    to its rational form 
\be L(\la)= \left( \b{array}{c} \la+s^3 \qquad  s^-   \\
 s^+  \qquad  \la-s^3 \e{array}
\right), \ll{lsl2}\ee
with $  s^\pm,s^3 $  being the  generators of the standard  $sl(2)$ algebra
(\re{sl2}). 
Different realizations of the spin operators generate different integrable
models from  (\re{lsl2}) and all of them are associated with the same rational
$R$-matrix (\re{rationalr}) for $N=2$, which also  coincide
 with (\re{abrat}) obtained
earlier as a direct  solution of YBE.

\subsubsection {$XXX$ spin-$\ha$ chain} 
Spin-$\ha$ representation 
through Pauli matrices, 
 rewrites (\re{lsl2}) into
\be ~~~
L_n^{(xxx)}(\la) = i( \la { I} + ( \si^3 \si_{n}^3+
\si^+ \si_n^-
+\si^- \si_n^+) ~~ \ll{lxxx}\ee
 generating  the Lax operator of the $XXX$ spin
chain listed in (\re{XXX}) and analyzed  in sect. 6 and 7. Note that
$L^{(xxx)}$
can also be obtained from (\re{XXZ1}) at the rational limit.

Similarly, the  multiparameter isotropic generalized   spin models
 can be obtained at the
rational limit of (\re{rtrign})   or (\re{col-r}).
\subsubsection {Nonlinear Schr\"odinger equation: lattice  and field models} 
Through bosonic realization of spin operators 
called {\it Holstein-Primakov} transformation 
\be s^+=(\De(2s-\De \psi^\da \psi))^{\ha} \psi,~
~ s^-= \psi^\da (\De(2s-\De \psi^\da \psi))^{\ha} ,~~
 s^3=s-\De \psi^\da \psi, \ll{hpt}\ee
with $~~[\psi_n, \psi^\da_m]= \hbar 
{\de_{nm} \ov \De}, ~~$ one can construct from (\re{lsl2})
the Lax operator  
of the  exact lattice version of the NLS model as  
\be L_n^{(lnls)}(\la) = 
\la +(s- \De \psi^\da_n \psi_n )\si^3+ ~~  \sqrt { \De}(2 s- \De\psi_n^\da
\psi_n)^\ha\psi_n^\da \si^+
 +\sqrt
{\De}\psi_n (2 s- \De\psi^\da_n \psi_n)^\ha \si^-
\ll{lnls1}\end {equation}
as listed in (\re{lnls}). We return to this model in sect. 6 
for  the  Hamiltonian  constructing 
 and in sect. 7 for its exact Bethe ansatz solution.
 Considering the spin parameter $s ={ \kap \De}^{-1}$ and
taking the continuum limit 
$\De \rw 0,$ we  obtain $ ~~
  s^+ \rw \sqrt \kap \psi^\da ,~
~  s^-=\sqrt \kap  \psi^\da, ~~ 
 \De s^3 \rw 1 $ with $\psi_n \rw i \psi (x)$,
 leading to ${\De  \ov \kap} \si^3 L^{(lnls)}_n  \rw I+
-{i \ov \kap } \De {\cal L}_{nls}(\la)$ and thus generating the Lax operator of the NLS
field model
\begin{equation}
{\cal L}_{nls}(\la)  =  i\la \si^3+ 
 \kap^\ha (\psi^\da(x) \si^+ +
 \psi (x) \si^-.
\ll{nls1}\end{equation}

The vector NLS model can be constructed from the rational limit of
(\re{ltrign}) with proper bosonization.   
Another possibility of the vector NLS will be considered in the next section.

\subsection{Models with twisted rational $R_{rat}(\la)$-matrix}
Note that the twisting operator $F(\th,Z_n)$ involved in transforming the
$R$-matrix  and the Lax operator remains the same at the rational limit,
since it  does not contain  the deformation parameter
$\eta$.
The rational $R$-matrix (\re{rationalr}) therefore 
twisted into 
\be \ti R^{\th}_{rat}(\la-\mu,\th)= (\la-\mu) T + \hbar P,~~~
 T= diag (1,e^{-i\th}, 
e^{i\th},1) 
\ll{thratr}\ee
while the Lax operator (\re{lsl2}) turns into 
\be L(\la)= \left( \b{array}{c} (\la+s^3)e^{i\th s^3}, \qquad 
e^{i\th (s^3+\ha)} s^-   \\
 s^+e^{i\th (s^3+\ha)} , \qquad  (\la-s^3)e^{i\th s^3} \end{array}
\right) \ll{thlsl2}\ee
It is clear that the spin-$\ha$ representation will lead to the Lax operator
of a $\th$ deformed $XXX$ chain given by (\re{thratr})  and 
 corresponds to the Hamiltonian
\be H_{\th xxx}=\sum_{m}^{}{1\over 2} 
\sigma_{m}^{3}   \sigma_{m+1}^{3}   +
 e^{i \th}
\sigma^{+}_{m}
 \sigma^{-}_{m+1}   +
e^{- i\th}
\sigma^{-}_{m}
 \sigma^{+}_{m+1}  \ll{Wu-mxxx}\ee
obtained also from (\re{Wu-m}) at the isotropic limit.

On the other hand the Holstein-Primakov transformation (\re{hpt})
would lead (\re{thlsl2}) to a $\th$-deformed lattice NLS model
 ($\th$-LNLS). Such  $\th$-LNLS 
 would represent an exact quantum integrable discrete bosonic
model associated with the twisted $R^{\th}_{rat}$-matrix 
(\re{thratr}).
Twisting transformations on 
multicomponent systems  likewise would generate multiparameter
models. However it is not clear,  whether  proper  
 continuum limit exists for such discrete  systems  leading to 
  any new field
models.

Note that the systematic  constructions described here
 have 
 partially answered the points  raised in the beginning of this section.
 We  shall   elaborate
 on this issue
  after  the constructions in the next section.

%__________________________________________\input {preport5.tex}
%Preport5.tex (sec 5 : 19 pages in 1.8 skip
\setcounter
{section} {4}
\section{Construction of models  from extended trigonometric Sklyanin algebra}
\setcounter{equation}{0}

 A systematic 
scheme for constructing integrable ultralocal models
starting from  the universal ${\cR}$-matrix 
is  formulated in the previous section.
We  detect however its  shortcomings by  noticing  that, 
 following this approach  we could
 discover not all the ultralocal   models listed in sect. 2 , but
only a few of them, which  constitute certain symmetric class.
Such models    include spin models, sine-Gordon, Toda field models 
and the nonlinear Schr\"odinger equation.
The reason  is simply the fact that, since we are confined  to the  universal
${\cR}$-matrix belonging  to
$ U_q(gl(n+1)),$  we could generate   only
those models which can be  obtained as  realizations of the   
quantum algebra, i.e. the models with Lax operators having  certain
 symmetric form.
 On the other hand, as seen clearly from the list in sect. 2,
there are  models with
 diverse nonsymmetric structures. This demands 
therefore  that  the present  
scheme should be modified   suitably to incorporate also the generation of
these models.
It seems that one can do so by proposing a more general type of
deformed  algebra \cite{kbmpl92}, but
at the cost of loosing some
 generality.
%--------------

It is found that 
 that at the representation (\re{r-l}), when the universal  YBE
 (\re{ur+-}) reduces to the relations (\re{frt}), it
 acquires some extra freedom allowing  more general choice
    for $ L^{(\pm)}_{1{\bf  3}}$ than (\re{l+-}) obtained
  as a  direct realization of   ${\cal R} $-matrix 
 related to the known quantum algebra.
 To identify such general algebraic structures 
let us  choose $L^\pm$  in  { upper/lower-triangular}
 form  
 \be
L^{(+)} = \left( \begin{array}{c}
  {\tau_1^+} \quad    {\tau_{21}} \\
    \quad \quad   {\tau_2^+}
          \end{array}   \right), \quad
\quad L^{(-)} = \left( \begin{array}{c}
 {\tau_1^-} \quad  \quad   \\
  {\tau_{12}} \quad {\tau_2^-}
          \end{array}   \right), \ll{etsa-l+-} \ee
 with yet unspecified operators $\vec {\bf \tau}$.
 Note however that
  $R^{\pm}_{12}=R^{\pm}_{q} $  are
 unchanged from (\re{R+}) and keeps track of the
     ${\cal R}$ from which it has been originated through representation
(\re{r-l}). Interestingly,
  Using   $L^{(\pm)}$  operators as (\re{etsa-l+-}) along with (\re{R+}) 
  one extracts from the    FRT relations (\re{frt}) 
  a new type of quadratic
  algebra     
\begin {eqnarray}
\left[ \tau_{12}, \tau_{21} \right] ~=~
- (q-q^{-1}) \left (  \tau_1^+ \tau_2^-  -  \tau_1^- \tau_2^+ \right )
 ~,  \nonumber \\
\tau_i^{\pm }\tau_{ij} ~=~q^{\pm 1 } \tau_{ij} \tau_i^{\pm }~,~\ \
\tau_i^{\pm }\tau_{ji} ~=~q^{\mp 1 } \tau_{ji} \tau_i^{\pm }~,~
\ll{etsa}\end {eqnarray}
%(3-4 )
for $i,j=(1,2) ; \ q=e^{i\alpha}$. The related  Casimir operators
may be given by
\begin{equation}
D_1= \tau_1^+ \tau_1^- \ , \  D_2= \tau_2^+ \tau_2^-  , \
  D_3= \tau_1^+ \tau_2^+ \ , \
  D_4=
  2 \cos {\alpha} \left (  \tau_1^+ \tau_2^-  +  \tau_1^- \tau_2^+ \right )
- \left[ \tau_{12}, \tau_{21} \right]_+
\ll{etsa-cas} \end{equation}
%(5)
To study  the properties of this algebra one should  find
the  coproduct  $\Delta$, antipode $S$ and the
  counit $\epsilon$  related to it.
 Using the  formulas like (\re{coprod}) 
it results
\be
\Delta(  \tau_i^{\pm })=\tau_i^{\pm } \otimes \tau_i^{\pm } ,\
\Delta(  \tau_{21})=\tau_1^{+ } \otimes \tau_{21 }
                 +   \tau_{21} \otimes \tau^+_{2 }
, \ \Delta(  \tau_{12})= \tau_2^{- } \otimes \tau_{12 } +
  \tau_{12} \otimes \tau^+_{1 }
,\ll{Detsa}\end{equation}
%(6)
  \begin{equation}
  S (  \tau_{21})= - (\tau_1^{+ })^{-1} \tau_{21 } (\tau_2^{+ })^{-1}, \
  S (  \tau_{12})= - (\tau_2^{- })^{-1} \tau_{12 } (\tau_1^{- })^{-1}, \
S (\tau_i^{\pm })  = (\tau_i^{\pm })^{-1}
  \end{equation}
  %(7)
  and $ \epsilon  (\tau_i^{\pm }) = 1, \epsilon (\tau_{ij}) = 0 $.
This  establishes the essential
 Hopf algebra properties of (\re{etsa}), though its quasi-triangularity
  i.e.  the existence of  universal ${\cal R}$-matrix
  is still an open problem.

For relating   (\re{etsa}) with known  algebras,
 one may check that it maps into 
  (\re{tsa}) for the   symmetric reduction
 \bea  \tau^+_1 = - \tau^-_2 &=&   {S^3 \ov 2 \cos { \alpha  \over 2 } }
 + {S^0 \ov 2i \sin { \alpha \over 2 } }  , ~~ 
~\tau_{12} = 2i \sin \al S_{- }, \ \ ~\tau_{21} =  2i \sin \al S_{+ }  
\nonumber \\
  \tau^-_1 &=&  - \tau^+_2 =  {S^3\ov 2 \cos { \alpha  \over 2 } }  -
  { S^0 \ov 2i \sin { \alpha \over 2 } },
 \ll{etsa-tsa}\eea
along with its Casimir operators.
Therefore the quadratic algebra (\re{etsa}) is 
 an extension of the
 trigonometric Sklyanin algebra (ETSA) \c{kbmpl92}. It is  related  also through 
symmetric reduction (\re{etsa-tsa}) and the mapping (\re{sk-real})
to the standard quantum algebra (\re{sl2qa}) with 
 generators $(I, s^3, S^\pm)$ 
as
\be
  ~\tau_1^\pm ~=~- \tau_2^\pm = \pm \Omega^{-1} q^{\pm s^3 }, \ \
, \ ~\tau_{12} = -\Omega S^+, \ \tau_{21} = \Omega S^- ~,~ \Omega
\equiv (q-q^{-1})
 \ll{etsa-qa} \ee
with $q=e^{i \al}$

\subsection{Ancestor model with trigonometric $R_{trig}(\la)$-matrix }
Note that   compared to the previous procedure,
 the $L^{(\pm)}$ operators   have  been changed here,
 while  the $R^\pm$
matrices remain untouched. Therefore the above Yang--Baxterization
procedure, valid  for  $R$-matrices with Hecke algebra must also go through
 in the present scheme.   As a result
 one  arrives  at the same 
 $R_{trig}(\la-\mu)$ matrix  (\re{rtrig}), while  the
related $L(\la)$ operator takes  the form
\begin {equation}
L_{trig}(\xi) = \left( \begin{array}{c}
\xi {\tau_1^-} + \frac {1}{\xi}{\tau_1^+} \qquad    {\tau_{21}} \\
   {\tau_{12}} \qquad \xi {\tau_2^-} +\frac {1}{\xi} {\tau_2^+}
          \end{array}   \right),
\ll{l-anc}\end {equation}
 with the upper and lower triangular
  parts coming from $L^+$ and $L^-$
 are combined through the  spectral parameter  
$\xi=e^{\eta \la}$. Since  by construction 
this $L(\la)$ operator along with the
 $R$-matrix solution
 (\re{rtrig}) satisfy
 the QYBE (\re{qybel}), one can take (\re{l-anc})
 as  the Lax operator
 of some generalized ultralocal 
 integrable model. Such a quantum model would 
involve  abstract operators 
$\{ {\vec \tau } \} $ satisfying the $q$-deformed algebra  (\re{etsa})
and would be associated with the trigonometric 
 $R_{trig}(\la-\mu)$  matrix.
For physical applications,
 (\re{l-anc}) can be looked as  
an {\em ancestor} model,
which under different reductions and
suitable realizations through  bosonic,
$q$-bosonic  or spin operators 
 can  lead to various ultralocal  {\em descendant} models.
All such
descendant models would  naturally  be  quantum integrable,
 satisfy the  QYBE's (\re{qybel},\re{qybet}) and 
share the same quantum $R_{trig}(\la)$-matrix inherited from the ancestor model.

Among different reductions  of the ETSA (\re{etsa}), there
is  a special { symmetric} one  
 \be  ~\tau_1^+ ~=~ - \tau_2^- =~-(\tau_1^-)^{-1} ~=~ (\tau_2^+)^{-1},
~~~ \tau_{12} =(\tau_{12})^\da \ll{red1s}\ee
which corresponds  to  (\re{etsa-qa})
and links  it to the  standard $ \ U_q(sl(2)) \ $.
Such  realization in  quantum algebra generators  reproduces clearly from 
(\re{l-anc}) the symmetric  ancestor model  (\re{lslq2}) and
 generates the important class of
 integrable models with the underlying $U_q(sl(2))$ symmetry.
Recall that such models constructed in the previous section 
from  (\re{lslq2}) and its rational 
limit  (\re{lsl2}) include $XXZ$ and $ XXX$
 spin-$\ha$ chains, SG and NLS models etc.
\subsection{Integrable reductions of ancestor model} 
Note that though  ETSA  can be mapped into the  quantum  algebra as
 (\re{etsa-qa}) for  invertible nonsingular generators,
    it allows in general
     richer
   possible  realizations   through nonsymmetric reductions. Consequently, 
   much     wider class
 of  models can be generated, which are 
 not reachable  from the
  quantum algebra without undergoing limiting procedures.
Some of such interesting cases are listed below, though in no way
they exhaust all possibilities.

\ni Let us consider a particular nonsymmetric    reduction
\be
  \tau_2^+ ~= 0, \  \
 ~\tau_1^+ ~=~ \tau_2^- ~= c (~\tau_1^-)^{-1} , \  \
~\tau_{12} = (\tau_{21})^\dagger
 ,~~c=const. \ll{red2}\ee
 which  corresponds to the quadratic algebra
\be
\left[ \tau_{12}, \tau_{21} \right] ~=~
- 2i \sin \al  (\tau_1^+)^2
 ~, \quad
\tau_1^{+ }\tau_{12} ~=~e^{i \al} \tau_{12} \tau_1^{+}~,~\ \
\tau_1^{+ }\tau_{21} ~=~e^{-i \al} \tau_{21} \tau_1^{+}~,~
\ll{etsa2}\ee
reducible from (\re{etsa}).
Through   bosonic and $q$-bosonic realizations
of this algebra we construct below   nontrivial  models.

\subsubsection { Liouville lattice and field models}
Consider   realization of (\re{etsa2}) with $~\tau_1^- =0$ 
in canonical operators
having  $ ~ [u,p] = {i \hbar
  \over  \Delta  } ~ $ in the form
\be \tau_1^{+ } = \Delta ~ e^{i \eta u },~~ \tau_{21 } = e^{-i \Delta p} f(u)  ,  ~~
\tau_{12 } = f(u) e^{ i \Delta p},~~~~~~~ 
 f(u) = {  \left( 1 + {\Delta}^2 ~ e^{i  \eta (2 u  +  i\hbar )  }
\right)  }^{1\over 2}~ \ll{map-llm}\ee
This   results  from (\re{l-anc})
\begin{equation}
  L_{n}^{(llm)}(\xi)  = \left( \begin{array}{c}  
{\Delta}{\xi}e^{\eta u_n}
 \qquad f(u_n)~  e^{p_n \Delta }  \\e^{-p_n \Delta }~f(u_n)~
    \qquad {\Delta}{\xi}e^{\eta u_n} 
    \end{array} \right), \quad \mbox {where} \ \ f(u_n)=[
     1 + {\Delta^2} e^{\eta(2u_n+i)}  ]^{{1 \over 2}}
\ll{Llm1}\end{equation}
giving   Lax
operator of the exact   lattice version of the quantum Liouville  
model (LLM) listed in (\re{Llm}).
At the continuum limit:
$ \Delta \rightarrow 0 $, when $(u_n,p_n) \rw (u(x), p(x))$ with 
$[u(x),p(y)]
=\hbar \de (x-y) $ and 
 $e^{ \pm i \Delta p_n} \rw I + \pm i\De p(x),~~ f(u_n) \rw I, ~~$
the lattice version (\re{Llm1}) 
 would lead to
$ L_{n}^{(llm)}(\xi) \si^1 \rw I +\De {\cal L}_{lm}(\xi)~~$
yielding the Lax operator
\be {\cal L}_{lm}(\xi)
= -p\si^3 +({1 \ov \xi} \si^+ +\xi \si^-) e^{\eta u_n} \ll{lm1}
\ee
of the well known Liouville field model (LM) listed in (\re{lm}).
Both  the lattice as well as the field Liouville models have 
 the same $R_{trig}(\la)$ as the ancestor model (\re{l-anc}).

Note that for the choice $c=-\De^2$ in 
 reduction (\re{red2}) one  gets yet
another simple and similar realization
\be
 \tau_1^+ =
 \De  e^{-i \eta u }~,~~\tau_{21} ~=~e^{i \De p }~,~~
\tau_{12} ~=~f^2(u) ~e^{-i \De p } , \ll{llm2} \ee
with $f(u)$ being as defined in (\re{map-llm}).
This  leads to (after multiplying by  $\si^1$ from the left)
 \be
L^{llm1}(\la) = \left( \begin{array}{c}
  {e^{i \De p }} \quad  \qquad \De {e^{i\eta( \la- u)}} \\
  -2i\De\sin\eta(\la+u) \quad   {f^2(u) ~e^{-i \De p }}
          \end{array}   \right), \quad
           \ll{l-llm1} \ee
representing  the Lax operator of a nontrivial spectral parameter dependent
lattice  Liouville model, solved recently using
 algebraic Bethe ansatz in \c{liuFad}, where 
this model has been   derived  directly from the 
sine-Gordon model invoking   some nontrivial limiting procedures.

We consider next  partially symmetric
  reduction of (\re{etsa})
  \be \tau_1^+  = - c~ (\tau_1^-)^{-1} = (\tau_2^+)^{-1}
  = c^{-1}~ \tau_2^-, \ \tau_{21}= \tau_{12}^\da  \ll{red3}\ee
giving the algebra
 \be [ \tau_{21}, \tau_{12} ] = -2ic~ \sin \alpha ~ ( \tau_1^{+2} +
 (\tau_1^+)^{-2} ) ~ . \ll{etsa3}\ee
\subsubsection { $q$-oscillator model}
Note that though for  $c\neq 0$ (\re{red3}) can be mapped into the  quantum
algebra (\re{sl2qa}), 
 in general it  gives more freedom of choice and for $c={i\De q \ov 4}$
through   realization
\begin{equation}
 ~\tau_1^+ ~=~\frac{4 i q}{\Delta} \tau_2^- ~= q^{-N}, \  \
 ~\tau_2^+ ~=~-\frac{4 i }{\Delta q} \tau_1^- ~= q^{N}, \  \
~\tau_{12} = - \kappa A , \ \ \tau_{21}=  \kappa A^\dagger
\ll{red3-qosc} \end{equation}
where  $\kappa=(\ha \De \sin 2\al)^\ha$ and  $A, A^\da$ are $q$-oscillators
or $q$-bosons with the commutation relations (\re{qoscl}), we can construct 
an integrable  lattice $q$-oscillator model 
\begin{equation}
  L^{(qoscl)}_n(\xi)  =
  \left( \begin{array}{c}
  \frac{1}{\xi}q^{-N_n }- 
\frac{i \xi\Delta}{4}~q^{N_n +1} 
 \qquad  {\kappa}A^\da_n \\
  -{\kappa}{A_n}   \qquad
  \frac{1}{\xi}q^{N_n }+
 \frac{i \xi\Delta}{4}~q^{-(N_n +1)} 
    \end{array} \right), \qquad
\ll{Ldnls1}\end{equation}
 listed in  (\re{Ldnls}).

Another such  
realization  may be given  as reduction    (\re{red2}) by
\be \tau_1^+ = q^{N+\ha },~ \tau_{21} = A g(N) ,~\tau_{12}
= g(N) A^{\dagger},
\ll{red2-qosc}\ee
where  $ ~ g^2(N)  =   (  q-q^{-1}  ) ~ q^N ~ $.
The Lax operators $L_n^{(qboson)}$ of the corresponding
 quantum integrable  lattice models  involving
$q$-bosons can be obtained by inserting (\re{red2-qosc}) in the
ancestor operator (\re{l-anc}).
\subsubsection { Derivative NLS: lattice and field models}
Curiously  the $q$-oscillator model (\re{Ldnls1}) 
can be interpreted as the lattice DNLS model. To see this we 
express $q$-oscillators through bosonic operators 
with 
$ [\psi_n , \psi_m^\dagger ] = {\hbar \over \Delta }\de_{nm}$
as 
  \be  A_n =
     (\frac{\Delta}{\hbar})^\ha~\psi_n \sqrt\frac{[N_n]_q}{N_n}, \ N_n=
   \frac{\Delta}{\hbar}  \psi^\da_n\psi_n \ll{qo-b}\ee
and inserting  in (\re{Ldnls1})   obtain the Lax operator $
  L^{(ldnls)}_n(\De,\xi)$ of an integrable bosonic model, which satisfies the
QYBE in all orders of $\De$ with  the trigonometric $R$-matrix (\re{rtrig}).
To identify $  L^{(ldnls)}_n(\De,\xi)$ as the exact lattice 
regularized version of the
DNLS field model (LDNLS), we take the limit 
$~~\lim_{\De \rw 0 }~ \xi L^{(ldnls)}_n(\De,\xi) ~= 
I+ \De {\cal L}^{dnls}_n(\la)~$
by using $q^{\pm N_n}\rw I \pm \eta \psi_n^\da \psi_n, ~~\kap A_n \rw \De
\psi_n$ etc. with $q=e^{i \eta}$
and demand it to satisfy the QYBE up to order $\De$.
However we face an initial    difficulty  due to the fact,  that
such naive limit does not satisfy   the QYBE
up to the required order  
and the coupling constants needs to  be regularized 
as
\[ 1+i\kappa_- h= e^{-i\al} ,\quad 1-i\kappa_+h= e^{i\al},
 \quad h=-\sin \al. \]
This gives
\begin{equation}
  {\cal L}^{dnls}_n(\xi)  = i
  \left( \begin{array}{c}
  -\frac{1}{4}\xi^2 +\kappa_-(\psi^\da_n\psi_n) ,
 \qquad  {\xi}\psi^\da_n \\
  {\xi}\psi_n    ,\qquad
  \frac{1}{4}\xi^2 -\kappa_+(\psi^\da_n\psi_n)
    \end{array} \right)
\ll{Ldnl}\end{equation}
and defining the field as
 $\De \psi_n=\int^{x_n+\De}_{x_n} \psi(x) dx$ with
the commutator $[\psi(x),\psi^\da(y)]= -\sin \al \delta (x-y), $ we 
finally arrive from (\re{Ldnl}) at the Lax operator 
\begin{equation}
  {\cal L}^{dnls}(\xi)  = i
  \left( 
  -\frac{1}{4}\xi^2 \si^3  +\psi^\da(x)\psi(x) {\bf \kap }
 + {\xi}(\psi^\da(x) \si^+ + \psi(x) \si^-) 
     \right),~~ {\bf \kap}= diag (\kappa_-, -\kappa_+)
\ll{dnls1}\end{equation}
of  the DNLS field model 
listed in 
 (\re{dnls}). The model is   solved exactly through Bethe ansatz in sect. 7.
The explicit forms of the conserved
 quantities including the Hamiltonian  of the model $H^{dnls}$
are 
\bea C_0 &=&  \int_{- \infty }^{ +\infty } \psi^\da \psi ~dx ~,
~C_1 = 4i \int_{- \infty }^{ +\infty } \psi^\da  \psi_x ~dx ~, ~ \nonumber \\
~H^{dnls}&=& {1 \ov 8} C_2 =
  \int_{- \infty }^{ +\infty } (~ \psi^\da \psi_{xx} ~-~
2ic (\psi^\da \psi ) \psi^\da \psi_x ~) ~dx ~ ,\ll{cdnls}
\eea
 etc.

It is  remarkable that, while the standard derivative NLS \c{dnls1}
is   nonultralocal  \cite{kb-dnls} and could not be solved through 
QIST,
the present   DNLS model generated 
from the   underlying
quantized algebra (\re{etsa}) is ultralocal and Bethe ansatz solvable.  

\subsubsection {Massive Thirring model}
 Notice an important    symmetry of the   algebra (\re{etsa}), that  
 the relations remain unchanged under the replacement \be 
(\tau_1^+ , \tau_1^-)\rightleftharpoons (\tau_2^-,\tau_2^+) .
\ll{etsasym}\ee
Using this property 
we can have  another realization  like (\re{red3-qosc}) in
an independent $q$-oscillator  mode $(A_2,A_2^\da)$ giving  
another  Lax operator
$L^{qoscl(2)}$
similar to $L^{qoscl(1)}$ (\re{Ldnls}).  Since the product of mutually
commuting  $L$-operator solutions of QYBE is itself a  solution,
 $L^{2-qoscl}_n=L^{qoscl(1)}_nL^{qoscl(2)}_n$ must be a solution of QYBE and 
consequently would represent the  Lax operator of   an  
integrable model with
two independent $q$-oscillator modes (\c{kbplb92})
\begin{equation}
  { L}^{2-qoscl}_n(\xi)  = 
  \left( \begin{array}{c}
  L_{11} ,
 \qquad L_{12} \\
  -L^\da_{12}  ,\qquad
  L^\da_{11}
    \end{array} \right)
\ll{Lmtm0}\end{equation}
with the elements
\bea
  L_{11} & =&
  q^{-N_1+N_2 }- 
\frac{\Delta}{4i \xi^2}~q^{-(N_1 +N_2+1)} -
\frac{\Delta \xi^2}{4i}~q^{N_1 +N_2+1} +
\frac{\Delta^2}{16}~q^{N_1 -N_2} -\kap^2 A^\da_1 A_2 \nonumber \\ 
  L_{12} & =&
\frac{\kap}{ \xi} \l(~q^{-N_1} A_2^\da  -
\frac{\Delta}{4i}~A_1^\da q^{N_2+1} \r) +
{\kap}{ \xi} \l(A_1^\da ~q^{-N_2}  -
\frac{\Delta}{4i}~ q^{N_1+1}A_2^\da \r) 
\ll{L1112}\eea
and their complex conjugates.
It is exciting to note that, when mapped through bosons  as (\re{qo-b}),
the Lax operator (\re{Lmtm0})  represents
 an exact lattice version of the 
 massive  Thirring model (LMTM) (bosonic),
 which satisfies QYBE in all orders of 
$\De$ with  trigonometric $R(\la)$-matrix.

 At the
continuum limit  the Lax operator  of the corresponding
field model reads
\be
  {\cal L}_{mtm}  = i\left( 
  \frac{1}{4}(\frac{1}{\xi^2}- \xi^2)\si^3 +{\bf \kap }
\rho_1- \ti {\bf \kap} \rho_2  +
  \ ({\xi \psi^\da_1+\frac{1}{\xi}\psi^\da_2}) \si^+ + 
  ({\xi \psi_1+\frac{1}{\xi}\psi_2})\si^- 
     \right).
\ll{mtm1}\end{equation}
with $ \rho_a=(\psi^\da_a\psi_a), ~~
{\bf \kap}= diag (\kappa_-,- \kappa_+),$ and $ ~~
\ti {\bf \kap}= diag (\kappa_+,- \kappa_-).$
 (\re{mtm1})   coincides with the two component relativistic field model 
\c{rel2} and also  generates the well known  massive Thirring model
(bosonic) (MTM)
 with the Lagrangian
$$~~
\mathsf { L}=\int dx \bar\psi(i\gamma^\mu \partial_\mu-m)\psi-\ha g j^\mu j_\mu
\quad j^\mu= \bar\psi\gamma^\mu \psi,~~$$ where $\psi= (\psi_1, \psi_2),
 \quad
   [\psi_a(x).\psi^\da_b(y)]=\hbar \delta_{ab} \delta(x-y). $

Let us consider one more nonsymmetric 
 reduction 
\be \tau_1^- =\tau_2^-=0, \tau_1^+ =(\tau_2^+)^{-1}, \ [\tau_{21},
\tau_{12}]=0 \ll{red4}\ee
and its  complementary one, obtained through the symmetry (\re{etsasym}),  
both of which  answer to  the  simple algebra
\be
\left[ \tau_{12}, \tau_{21} \right] ~=~ 0
 ~, \quad
\tau_1^{+ }\tau_{12} ~=~e^{i \al} \tau_{12} \tau_1^{+}~,~\ \
\tau_1^{+ }\tau_{21} ~=~e^{-i \al} \tau_{21} \tau_1^{+}~.
\ll{red45}\ee
This 
  allows generation of  
various  other models through different   realizations.
  For example,
a lattice version of the  light-cone sine-Gordon model (LCSG)
studied in  \c{sgKS}
is obtained for   \be \tau_1^+= e^{-i\De p}, \ \ \tau_{12}=\tau_{21}^\da=
i\De e^{i\eta u}.  \ll{lcsg}\ee
 On the other hand,  the realization
\be
 \tau_1^+ =-(\tau_1^-)^{-1}=-e^{\eta p}, \ \tau_{21}=
-(\tau_{12})^{-1}\eta^2= \eta e^u \ll{dtl1} \ee
yields
a discrete-time  or relativistic Toda chain (RTC) with
    the Lax operator
\begin {equation}
L_{0}(\lambda) = \left( \begin{array}{c}
  \frac {1}{\xi}e^{\eta p}
-\xi e^{-\eta p}\qquad   \eta e^{q}
 \\-\eta e^{-q
}
 \qquad \ 0
          \end{array}   \right).
\ll{dtc-sym}\end {equation}
%16
The Hamiltonian and momentum of the model are given by
\begin {eqnarray}
H=\sum_i\left(\cosh 2\eta p_i
+\eta^2 \cosh \eta (p_i+p_{i+1})e^{q_i-q_{i+1}}
\right), \nonumber \\
P=\sum_i\left(\sinh 2\eta p_i
+\eta^2 \sinh \eta (p_i+p_{i+1})e^{q_i-q_{i+1}}
\right).
\ll{hdtc-sym}\end {eqnarray}
%17,18
 derivation of which  from the Lax operator is shown in sect. 6. 
A family of
such  relativistic quantum Toda chains
are   constructed  below.

\subsubsection{Discrete-time quantum Toda like model}  
We consider  now another nonsymmetric  
realization
of (\re{etsa})  through $q$-oscillators at lattice site  $i$ as 
\be
\tau_{1i}^{\pm }=\mp \frac {\hbar\gamma}{2\sin \al}
 e^{\mp \eta(N_i+\omega_i)} ,
\  
\tau_{2i}^{\pm }= \frac {\alpha_i \beta_i}{2\gamma}
            e^{\pm \eta N_i}, \ \  
\tau_{21i}= \beta_iA^+_i,\  \tau_{12i} = \alpha_i \cos \al \
A^-_i,
\ll{rtodalike}\ee 
%23
where $\al\equiv \hbar \eta, \hbar, \ga, \eta$ are homogeneous and
$\al_i, \bet_i$  inhomogeneous constant parameters,  
$A^{\pm}$ are  $q$-oscillators, a bit different from \c{qoscl} and 
  given by the commutation 
relations
\begin {equation}
[N_k,A^{\pm}_l]=\pm i \delta_{kl} \hbar \  A^{\pm}_k , \ \
\ A^{-}_k
 A^{+}_l -  e^{\mp 2i \hbar \eta }  A^{+}_l A^{-}_k= i \delta_{kl} \hbar
 \ e^{\pm\eta (2 N+ {\omega- i\hbar})}
\ll{qoscl2}\end {equation}  
%24
Through canonical variables these operators may be expressed as 
\be
N=p q,  
\  A^{-}= q f(N), \ \ 
 A^{+}= f(N) p,\ll{qoscl2-b}\ee
 {with} $  \qquad
 f^2(N)= \frac {i \hbar} {2\cos \hbar \eta \ [i\hbar]_\eta}
\frac {1} {N}( [2N+\omega- i\hbar]_\eta
-[\omega- i\hbar]_\eta )~~~~$
 and   $[ x ]_\eta=\frac {\sinh \eta x} {\sinh \eta}$.
 Inserting 
(\re{rtodalike}) in (\re{l-anc})
and assuming $\xi=e^{\eta \lambda}$
 one derives the Lax operator 
\begin {equation}
L_{i}^{dttcl}(\lambda) = \left( \begin{array}{c}
\frac { i\gamma \hbar} { [i\hbar]_\eta} [\lambda + N_i +\omega_i]_\eta 
\qquad  \ \ \beta_i  A^+_i
 \\ \alpha_i \cos \hbar \eta \ A^-_i
\qquad \ \frac {\alpha_i \beta_i}{\gamma}
       \cosh (\eta(N-\lambda))   \end{array}   \right),
\ll{dttcl1}\end {equation}
%26
of a $q$-oscillator model 
associated with  $R_{trig}(\la)$.
Remarkably, this  quantum integrable model with mapping 
(\re{qoscl2-b}) may be interpreted as the 
  discrete-time as well as the  quantum
generalization of a  Toda chain like model (DTTCL) 
proposed in \c{closetc} and listed in (\re{dttcl}).
At the time-continuous
limit: $\eta \rightarrow 0$, as  seen from  (\re{qoscl2-b}),
$~~N=p q,  
\  A^{-}\rw q, \ \ 
 A^{+}= \rw p~,~ $ one recovers the Lax operator (\re{todalike1}), which is a
 quantum 
version of
\c{closetc}.

We stress again that all the above  models
obtained as { descendants}  from the same ancestor model ({\re{l-anc})
  are quantum integrable by construction, since
  they
 satisfy the QYBE 
as different  exact realizations of ETSA (\re{etsa}). Moreover they
are associated  with the same quantum $R_{trig}(\la)$-matrix (\re{rtrig}) of
trigonometric type, inherited from the ancestor model. 
The  family of integrable models
is specified thus by the associated  $R$-matrix  and 
the classification of integrable systems could therefore be based on the
classification of the quantum $R$-matrix   assigned to  them.
Search for any new family consequently should  start
with a new  solution of
YBE (\re{ybe}).
For this reason we   switch next to
the twisted $R$-matrix solution related to $gl(2)$
for constructing other kinds of   models.

\subsection {Twisted quadratic algebra and related models}
Twisted $R^\th_{trig}(\la)$-matrices as  (\re{th-r})    can
 be obtained  from the
 original trigonometric $R_{trig}(\la)$-matrix (\re{rtrig})
 by a simple twisting  transformation
$
R^\theta_{trig}(\la)= F(\theta)
R_{trig}(\la) F(\theta), ~~~  F(\theta)=
 e^{i\theta(\sigma_3 \otimes 1-1\otimes \sigma_3)}$, which
takes a solution of the  YBE to another solution and  introduces
an extra  parameter $\theta$.

For constructing the associated  Lax operator $L_{trig}^{\th}(\xi,
\ti {\vec {\tau}})$ of the  
ancestor model, it may be taken  in the same
form (\re{l-anc}),  while the $\ti {\vec {\tau}}$ operators
involved  satisfy now a $\theta$-deformation of the quadratic algebra ETSA
given as
\bea
t~\ti \tau_{12}\ti \tau_{21} 
 -t^{-1}\ti \tau_{21} \ti\tau_{12} ~&=&~
- 2i \sin \al \left ( \ti \tau_1^+\ti \tau_2^-  - \ti \tau_1^-
\ti \tau_2^+ \right )
 ~, \nonumber \\
\ti \tau_i^{\pm } \ti \tau_{ij} ~=~t~q^{\pm}
\ti \tau_{ij} \ti \tau_i^{\pm }~&,&~\ \
\ti \tau_i^{\pm }\ti \tau_{ji} ~=
~q^{\mp}t^{-1}\ti \tau_{ji}\ti \tau_i^{\pm }~
\ll{thetsa}\end {eqnarray}
where $q=e^{i\al}, t=e^{i\th}.$
Note that the coproduct structure  for this algebra remains the same as
(\re{Detsa}).
Redefining generators it is however possible  to shift the
 $\theta$-dependence
from the algebra to its coalgebra as done in the general treatment of sect.
3. It is evident
 that at $\th=\pm \al$, i.e.  when the additional $t$-deformation 
coincides
 with  the original $q$-deformation,  the algebra (\re{thetsa}) is much
 simplified. Interestingly, most of the  physically significant  models,
 as we see below, are generated at 
 this particular {\it resonating } situation.

We may   consider  as before different   
  reductions of the $\th$-deformed ETSA for finding
relevant realizations.
Notice  that the symmetric reduction  
\bea  
\ti\tau_1^+ ~&=&~- \ti\tau_2^- = -ie^{i(\al+\th)s^3}, \ \
~\ti\tau_1^- ~=~-\ti \tau_2^+=ie^{-i(\al-\th)s^3} \nonumber \\,
\ti\tau_{21}&=& e^{i\th (2s^3+1)} S^-, \ \ti \tau_{12}= S^+e^{i\th (2s^3+1)}
\ll{thetsa-qa} \eea
is related to the 
  multiparameter quantum  algebra known  as  
 $U_{q,p}(gl(2))$) we have already  considered. 

Through  bosonic realization of (\re{thetsa-qa})
we  construct the   $\th$-deformed
 lattice sine-Gordon model, derived previously in sect. 4  starting from  
the  universal ${\cal R}$-matrix related to deformed $gl(2).$ 
Similarly  realizing (\re{thetsa-qa})
in Pauli matrices we get  the Wu-McCoy model (\re{WM}).

On the other hand, the nonsymmetric   reductions  (\re{red2}-\re{red4})
may lead to  the  $\th$-deformation of all other 
 integrable   lattice  models
like  LDNLS, LMTM,
 LCSG, LLM, $q$-oscillator models  etc. obtained earlier 
having the same   
quantum $R_{trig}^\th(\la)$-matrix (\re{th-r}).

\subsubsection { Ablowitz-Ladik Model}
Such a $\th$-deformation of     
the $q$-oscillator realization  (\re{red2-qosc}) gives
\be \ti \tau_1^+ = ~\ti \tau_2^-  = 
e^{-i( \theta + \alpha ){N+\ha }},~ \ti \tau_{21} = A \ti g(N) ,
~\ti \tau_{12}
= \ti g(N) A^{\dagger},
\ll{red2-thqosc}\ee
where  $ ~ \ti g^2(N)  = \Omega    e^{i(\al- \theta) N }~,
\Omega \equiv (  q-q^{-1}  ) ~ $.
Observe that for the special value $ \theta = - \alpha,
$
(\re{red2-thqosc}) is simplified to $ \ti \tau_1^+ = ~\ti \tau_2^-
= 1 , \ti \tau_{21} =  b ,~~\ti \tau_{12} = b^{\dagger } $ , where
$ b \equiv A q^{{N \over 2}} 
\Omega^{1\over 2} $
and $b^{\dagger} \equiv
q^{{N \over 2}} 
\Omega^{1\over 2} A^{\dagger } ~~~ $
is  another  form  of    $q$-oscillator  with  the  algebraic relations 
\c{qoscl}    $ ~  q^2  b
b^{\dagger} - b^{\dagger }b = q^2 - 1 ~ $.
Remarkably, this deformed oscillator model represented by 
\begin{equation}
  { L}_n(\xi)  = i
  \left( \begin{array}{c}
  \xi^{-1} ,
 \qquad  b^\da_n \\
  b_n    ,\qquad
  \xi
    \end{array} \right)
~ \ \ [ b_m , b_n^\dagger ] =~ \hbar ( 1-b_n^\dagger b_n ) ~\delta_{m,n}
\ll{Ab-lad}\end{equation}
with $~ \hbar = 1 - q^{-2}~ $ 
is the
Lax operator of the well known    Ablowitz-Ladik model (ALM)
\c{ALM}, proposed as a discretization of the NLS equation years before the
discovery of $q$-oscillators.

\subsubsection { Family of  relativistic quantum Toda chains}
Exploiting  $\th$-deformed quadratic algebra (\re{thetsa})
one can construct   a family of   quantum integrable  
  discrete-time or relativistic 
 generalizations
of the  periodic   Toda chain, apart from (\re{dtc-sym}) obtained above.
These models  are 
canonically related \c{suris}
 to the  relativistic Toda chain  of Ruijsenaars \c{ruij}
 and 
can  be generated  from a single ancestor model
at different realizations of the underlying  algebra \c{qrtoda}.

Considering reduction
 complementary to (\re{red4}) with  $\ti  \tau_2^{\pm}=0 $,
one   simplifies (\re{thetsa})  to
\be
 \ti   \tau_{12} \ti  \tau_{21}= e^{-2i\theta} \ti  \tau_{21} \ti  \tau_{12}
~, ~~
\ti    \tau_1^{\pm }\ti  \tau_{12} ~=~e^{i(\pm \alpha+\theta)}
 \ti  \tau_{12}\ti   \tau_1^{\pm }~,~~
\ti  \tau_1^{\pm }\ti  \tau_{21} ~
=~e^{i(\mp \alpha-\theta)} \ti  \tau_{21} \ti   \tau_1^{\pm }~
\ee
involving only Weyl type relations.
This observation facilitates to find the
  realization 
\begin {eqnarray}\ti  
\tau_1^{+ }=e^{(\eta+\epsilon) p} , \ 
\ti  \tau_1^{- }=-e^{-(\eta-\epsilon) p },
\ \
\ti    \tau_{21}=\eta e^{-c \epsilon p+q}, \  
\ti  \tau_{12} = -\eta e^{(2+c)
 \epsilon p-q
}\ll{DTC-alg}\end {eqnarray}
%6
  where
$\alpha=\eta \hbar, \theta= \epsilon \hbar$.
We remark that  certain operator ordering should
be maintained in  all such  expressions
 and  consider  the  periodic boundary condition.
 (\re {DTC-alg}) gives from  (\re{l-anc}) the
explicit form  for the Lax operator
\begin {equation}
L_{(\epsilon,c)}(\xi) = \left( \begin{array}{c}
  \frac {1}{\xi}e^{(\eta+\epsilon) p}
-\xi e^{-(\eta-\epsilon) p}\qquad   \eta e^{-c \epsilon p +q}
 \\-\eta e^{(2+c) \epsilon p-q
}
 \qquad \quad 0
          \end{array}   \right).
\ll{genrtoda}\end {equation}
%7
  depending  on
the additional parameters $\epsilon$
and $c$ and associated with the twisted  quantum $\ti R(\theta, \la)$-matrix
 (\re{th-r} ).
 The Hamiltonian $H$
 and the  momentum
$P$ of the generalized Toda  system may be given by 
\begin {eqnarray}
H&=&\sum_i(\cosh 2\eta p_i
+\eta^2 \cosh \eta (p_i+p_{i+1})e^{\epsilon ( 1+c)(p_{i+1}-p_i)+(q_i-q_{i+1})}
), \nonumber  \\
P&=&\sum_i\left(\sinh 2\eta p_i
+\eta^2 \sinh \eta (p_i+p_{i+1})e^{-\epsilon ( 1+c)(p_i-p_{i+1})+(q_i-q_{i+1})}
\right).
\ll{dtc-cq}\end {eqnarray}
%9,10
 This  integrable quantum model 
   is able to generate the whole
family of relativistic Toda chains for different choices of  parameters
 $c, c_0$ with $\epsilon=c_0 \eta$.
  Moreover, since (\re{th-r}) 
is independent of   $c$, models with different
values of
$c$ would share the same  quantum
$\ti R_{trig}(\th, \la)$-matrix (\re{th-r}).

Some members of this  quantum integrable 
relativistic Toda chain  family 
deserve special attention.
The symmetric reduction with $\th=0$, 
corresponding to 
$c=c_0=0$ has  already been found with 
the most simplification as 
 (\re{dtc-sym}).
The related Hamiltonian and momentum  are  expressed  in  symmetric form
(\re{hdtc-sym}).
Interestingly, 
 even for  nontrivial  $\th$ or $\epsilon,$
 one can achieve  the same symmetric  conserved quantities
(\re {hdtc-sym})
 by choosing
   $c=-1$, as seen clearly from  (\re{dtc-cq}).
 However 
 the corresponding $\ti R$-matrix  will
be  different
 with a more involved  Lax operator.

For   $c_0=1$ and $c=0$,
(\re{DTC-alg}) reduces to
\begin {eqnarray}
\ti \tau_1^{+ }=e^{2\eta p} , \ \ti \tau_1^{- }=- 1
\ \
 \ti \tau_{21}=\eta e^{q}, \ \ti \tau_{12} = -\eta e^{2\eta
  p-q
}\end {eqnarray}
%14
with  the Lax operator easily obtainable  from 
(\re{genrtoda}).
The conserved quantities are similarly reduced from (\re{dtc-cq})
   simply as 
\begin {eqnarray}
I^{+}=\sum_i\left(e^{ 2\eta p_i}\left(
1+\eta^2 e^{
(q_{i-1}-q_i)}
\right)\right),\quad
I^{-}=\sum_i\left(e^{ -2\eta p_i}\left(
1+\eta^2 e^{
(q_i-q_{i+1})}
\right)\right).
\ll{qsuris}\end {eqnarray}
%15
with $H= \frac {1}{2}(I^++I^-)$ and $P= \frac {1}{2}(I^+-I^-)$. 
This  model 
 corresponds to the    quantum version 
of the  discrete-time Toda chain  \c{suris} and  
the associated quantum $\ti R_{trig} (\alpha, \la )$-matrix coincides
with  (\re{th-r})  for $\theta =\alpha$.

Another  quantum generalization of  \c{suris}
is  achieved  for  $c_0=1$ but $c \not = 0$, for which 
the  Lax operator and the  conserved quantities 
are obtained from the general forms (\re{genrtoda},\re{dtc-cq}).
This model though  different from the above due to different $c$,
 shares  the same quantum $\ti R(\al, \la)$-matrix.

Finally, one  observes
 that the Hamiltonian and momentum (\re{dtc-cq}) of
the generating model is transformed exactly to (\re{qsuris})
 under canonical transformation
$(p,q)\rightarrow (P,Q)$ as
\begin {equation}
p =P, \qquad \ q +(\eta -\epsilon(1+c))p =Q
\end {equation}
%21
Therefore, since   the other models are obtained from the generating
model  at different
 parameter choices, all of them are
canonically equivalent  to  (\re{qsuris}), which in turn
has been shown \c{suris} to be  equivalent 
to the  Ruijsenaars model. Therefore 
 the whole family of
  discrete-time or relativistic
 Toda chains discussed here
  are  canonically equivalent  to the relativistic generalizations
of the Toda chain of \c{ruij}.

In sect. 7 
it will be shown, how  to achieve the separation of variables 
in the eigenvalue problem for such  quantum Toda chains by  applying 
  functional Bethe ansatz.
Thus from the ancestor models 
through  suitable realizations, one can generate systematically 
the Lax operators of various quantum integrable 
ultralocal models of 
 the trigonometric class.
\subsection{ Models with rational $R_{rat}(\la)$-matrix}
 We have seen already  
that at $q \rw 1$ or $ \eta \rightarrow 0 ,$ 
   the trigonometric $R_{trig}(\la)$-matrix  reduces
to its  rational form (\re{rationalr}). We are interested to 
examine here how at this rational limit 
  the underlying algebraic structures
(\re{etsa}) are changed  
and how (\re{l-anc})  is reduced 
   at this limit  to yield   the  Lax operator for the
 ancestor model representing  the
rational class.

It is not difficult to find the  consistent  
 limit of the operators,
 when the deforming parameter $\eta \rw 0$ as 
\[(  \tau_i^{+ }+  \tau_i^{- }) \rw K^0_i, ~~
(  \tau_i^{+ }-  \tau_i^{- }) \rw   {1 \ov \eta}    K^1_i ,~~ 
   \tau_{ij}\rw K_{ij}. \]
Using such operators  along with the reduction of the spectral parameter 
 $\xi\rw 1+i \eta \la, $ 
 the $L$ operator (\re{l-anc}) 
undergoes a smooth transition and  simplifies to
\begin {equation}
L(\lambda)_{rat} = \left( \begin{array}{c}
{K_1^0 + i  { \lambda  } K_1^1 }
\qquad   \ \  {K_{21}}
  \\     {K_{12}}
 \ \ \qquad  {K_2^0 + i  { \lambda } K_2^1 }
          \end{array}   \right).
\ll{l-anc0}\end {equation}
The $ {\bf K }$ operators  satisfy  the 
algebra,  obtained at 
 the $q \rightarrow 1$ limit of (\re{etsa}):
\[  [ K_{12} , K_{21} ]
=  ( K_1^0 K_2^1 - K_1^1 K_2^0 ) , ~~ [ K_1^0, K_2^0 ] = 0
              \]  
\be       [K_i^0, K_{12}] = 
 \ep_i K_{12} K_i^1 , ~ [K_i^0, K_{21}]  = - \ep_i K_{21} K_i^1,
  \ll{k-alg} \ee
with $\ep_1=1, \ep_2=-1$ and  $ K_1^1, K_2^1 $ serving as Casimir operators.
The algebra (\re{k-alg}) corresponds to a representation  
of the Yangian, more general  than $Y(sl(2))$ 
 presented  in sect. 3.

Since    $L(\la)$ operator given in the general form  (\re{l-anc0})
satisfies QYBE and associated with the rational $R_{rat}(\la-\mu)=
\la-\mu+\hbar P$
 matrix,
 it may be taken as   the  ancestor model
for generating  
  integrable  models belonging to the rational class.

 Realization through spin operators: 
\be  K_1^1 =K_2^1   =  1,
~~ K_1^0  =  -K_2^0= s^3,  ~~ \  K_{12}=  \ s^+, \    K_{21}=  \ s^-
\ll{K-s}\ee
 recovers   the    $sl(2)$  algebra (\re{sl2}) corresponding to 
    (\re{yangian}).
 The ancestor Lax operator (\re{l-anc0}) for 
 this   symmetric  reduction  goes  to (\re {lsl2}) considered 
above. This 
produces the $XXX$  chain (\re{lxxx}) 
for the  spin-$\ha$ representation 
in Pauli matrices and the NLS lattice and field models 
for the Holstein-Primakov transformation (\re{hpt}).

 The full algebra (\re{k-alg}) on the other hand allows 
more   freedom for constructing models with nonsymmetric reductions.
For example if we consider   $ K_1^1 =  i,~ K_2^0  =  K_2^1
=0 , $ (\re{k-alg}) simplifies to 
 \be [K_1^0 , K_{12}] =  i K_{12},~~[K_1^0 , K_{21}] = - i K_{21}
~ [K_{12} , K_{21} ] = 0 ~ \ll{altoda}\ee
and  admits the   bosonic realization  \be  K_1^0
=  p , K_{12} = e^{-u},~~K_{21} = e^{u}. \ll{k-b}\ee This   yields from 
(\re{l-anc0})
the Lax operator of the well known Toda chain (TC)
\be L_n^{(tc)}(\la) = \left( \begin{array}{c}
  p_n
-\la \qquad    e^{q_n}
 \\- e^{-q_n}
 \qquad \quad 0
          \end{array}   \right).
\ll{toda1}\end {equation}
 listed in (\re{toda}).
 We return to this model
in sect. 6 and 7 for highlighting other aspects.

\subsubsection{ Quantum Toda chain like model} 
Another nonsymmetric  realization 
through canonical variables $p_i,q_i,~~[q_i,p_j]=\de_{ij}~$:
\be K_2^1=0, ~~ K_1^1=-i, ~~ K_1^0= \gamma pq+\omega, ~~ 
K_2^0=\frac {\alpha \beta}{\gamma},~~
 K_{12} =  \alpha q,~~K_{21} = \bet a p \ll{ktoda-b}\ee
derives from (\re{l-anc0})
 the Lax operator 
\begin {equation}
L_{i}(\lambda) = \left( \begin{array}{c}
  \lambda +\gamma p_iq_i+\omega_i 
\qquad  \ \ \beta_ip_i  
 \\ \alpha_i q_i
\qquad\qquad \ \frac {\alpha_i \beta_i}{\gamma}
          \end{array}   \right),
\ll{todalike1}\end {equation}
%22
with constant parameters $\ga, \al_i, \bet_i, \omega_i$. 
(\re{todalike1}) thus constructed represents the 
 quantum generalization of a Toda like classical model (TCL)   proposed  
in \c{closetc}.  The model   can also be obtained directly 
from (\re{rtodalike}) at $\eta \rw
0$. 
\subsubsection{ Simple lattice NLS model }
The lattice NLS generated in (\re{lnls1}) through symmetric reduction from
(\re{l-anc0}) has rather involved structure and the corresponding 
Hamiltonian of the model, as  will be shown  in the next section, is also  
much complicated. Notably, the algebra (\re{k-alg}) allows  a 
 nonsymmetric reduction realized as 
\be K_0^1=\kappa \Delta^2 \phi \psi+1, ~~ K_1^1=-\De, ~~ K_2^0= 1, ~~ 
K_2^0=0,~~
 K_{12} = i \De \sqrt{\kap} \psi,~~K_{21} = - i \De \sqrt{\kap} \phi
 \ll{kslnls-b}\ee
where the operators $~\psi, \phi~$ obey the canonical commutation relation
$~[\psi(n),\phi(m)]= {\kap \ov \De} \de_{nm}$. Therefore, 
 (\re{l-anc0})  generates   a AKNS type integrable model  (SLNLS) given by the
Lax operator
\begin{equation}
{ L}_{n}(\la)^{(slnls)}  = \left( \begin{array}{c} { 1 -i \la  \Delta +
  \kappa \Delta^2 \phi (n) \psi (n)} 
\qquad {- i \sqrt { \kappa} \De \phi(n)} \\
{ i \sqrt { \kappa} \De \psi(n)   } \qquad {1}
    \end{array} \right).
\ll{slnls1}\end{equation}
Through  an irrelevant scaling $L \rw L={1 \ov \la
\De} L$ (see (\re{slnls}),  it
 is convenient to construct the explicit conserved quantities 
of  the model in a much simpler way as will be shown in the next 
section. 
Though at the discrete level the operators $\psi $ and $\phi$ are
independent from each other,  at the continuum limit : $
{ L}_{n}(\la)^{(slnls)} \rw 1 -i \De {\cal L}^{nls}(x, \la) +O(\De^2) ~~$,
we get 
\be~~{\cal L}^{nls}(x, \la)= {1 \ov 2} (1+\si^3) \la +  
\sqrt { \kappa}  \phi \si^+ -
\sqrt { \kappa}  \psi \si^- \ll{snls}\ee  
 Due to the restoration of the
symmetry,  one can put $\phi=\psi^\da$ 
 in the corresponding conserved quantities, which 
 recovers the standard NLS field
model \c{krjp94}.

All the above models constructed as descendants of (\re{l-anc0}), 
 share the same rational $R_{rat}(\la)$-matrix inherited from the ancestor
model.

\subsubsection {Models with twisted rational $R^{\th}_{rat}(\la)$-matrix}
 We consider  $\theta$-deformed  
rational $R_{rat}^{\th}(\la)$-matrix given by (\re{thratr}), which induces 
 through QYBE a deformation 
  of   algebra  (\re{k-alg})  to 
\bea  t K_{12}  K_{21}- t^{-1} K_{21}  K_{12} 
&=& ( K_1^0 K_2^1 - K_1^1 K_2^0 ) , 
 K_i^1 K_{12}  = t K_{12}K_i^1, ~~
 K_i^1 K_{21}  = t^{-1} K_{21}K_i^1, \nonumber \\ 
  K_i^0 K_{12}  = t K_{12}(K_i^0  \ep_i  K_i^1)
 , &  & K_i^0 K_{21}  = t^{-1}K_{21}(K_i^0  - \ep_i K_i^1),
  \ll{thk-alg} \eea
with $t= e^{i \theta }$ and $  
~~i=1,2~~$. Here 
   operators $ K_i^0, K_i^1 $ form a commuting set, though 
noticeably,   $K^1_i$'s are  no longer  central elements
 of the algebra. (\re{l-anc0})
 with such $K$ operators would represent the Lax
operator
$L^{\th}_{rat}(\la)$ of another ancestor model generating  rational  
 class of quantum integrable $\th$ deformed  models, associated with the
twisted $R_{rat}^{\th}(\la)$. 
For example, symmetric reduction with realization in spin operators  
\be  K_1^1 =K_2^1   =   t^{ s_3 },
~~ K_1^0  =  -K_2^0= s^3 t^{ s_3 },  ~~ \  K_{12}=  \ s^+t^{ s_3 +\ha }, \ 
   K_{21}=t^{ s_3 +\ha }  \ s^-
\ll{thK-s}\ee
reproduces  the Lax operator
(\re{thlsl2})  and recovers the $\th$-deformed $
XXX$ and the lattice NLS   
 models constructed before.  Through  nonsymmetric reductions of (\re{l-anc0}) however 
we can generate wider class of twisted  models.
 
 \subsubsection {$\th$-deformed Toda chain }

 The bosonic realization
\be  K_2^0  =  K_2^1= 0 ,~ K_1^1 = i e^{\theta p} ,~ K_1^0 
= p e^{\theta p},
~K_{12 } = e^{  - \left( u + { i\over 2 } \theta \right) + \theta p
 } ,~~K_{21 } = e^{   \left( u + { i\over 2 } \theta \right) + \theta p
 }
\ll{thtoda}\ee
 satisfying  (\re{thk-alg}) rewrites
  (\re{l-anc0}) as the Lax operator $L^{\th}_{tc}
$ of a quantum integrable 
 deformed Toda chain ($\th$-TC).

 \subsubsection {Tamm-Dancoff $q$-bosonic model}
By deforming the lattice NLS model ${ L}_{n}(\la)^{(slnls)}$  found above,
 through the  
 parameter $q \equiv e^{i \th}$
one  gets the Lax operator 
\begin{equation}
L^{q}(\la)= \left( \begin{array}{c} { (1 -i \la  \Delta +
  \kappa N) f(N)} 
\qquad {- i \sqrt { \kappa} c} \\
{ i \sqrt { \kappa} b   }  ~~~~~ \qquad { f(N)}
    \end{array} \right), \qquad  f(N)=q^{\ha(N(n)-\ha)} 
\ll{slnlsq1}\end{equation}
of a quantum integrable system (TDqB}
involving Tamm-Dancoff type $q$-bosonic operator  \c {TDqb}
satisfying
$~~
[b,N]=b,~~
[c,N]=-c,~~ bc-qcb=q^N ~~$. The
$q$-bosonic models, as we have seen, are usually 
 related
to the  quantum algebras and associated with 
trigonometric $R_{trig}(\la)$-matrix. However the
present  TDqB  model is  
 connected with the rational matrix (\re{th-r}) with $\th=\al$.
 Such $q$-bosons
can be mapped easily to the NLS fields as 
$~~b=f(N) \psi,~~c= f(N)\phi $ with $~[\psi,\phi]=1 ~$ leading to the
relation 
$L^{q}(b,c,N)= f(N) { L}_{n}(\psi,\phi)^{(slnls)}$. 
Such $\th$-deformed 
models though correspond to the
rational class 
 with $q=1$ algebra,
they  exhibit properties resembling   quantum  algebra.

The ultralocal quantum integrable models generated systematically
following the above procedure, from the 
ancestor models with trigonometric and rational $R(\la)$-matrices as well as 
with their twisted forms are presented  
  schematically  in fig.5.1. The abbreviations used to denote the models are 
 as  in their list  in sect. 2. 

Thus  the following results  are achieved through the
present scheme matching to some extend the original aims and answering the 
questions raised at the beginning of   sect. 4. 
                       \\  \\
i) Ancestor
 Lax operators of exactly integrable
quantum lattice models are constructed in a rather systematic way. Through
consistent realizations of these   models  one could generate different 
 families
 of      quantum integrable ultralocal 
systems representing  known as well as new lattice
 and field models.
\\   
ii) The descendant models generated from 
 the same ancestor model share the same inherited $R$-matrix.
And this classifies the models into two broad classes: trigonometric and
rational,   explaining thus  the  {\it  commonness} of diverse models 
pointed above.
\\ 
iii)  The notion of {\em integrable nonlinearity} seems to be determined
by  concrete realizations of the
 quantized algebras (\re{etsa},\re{thetsa}) (or
their $q\rw 1$   limit (\re{k-alg},{thk-alg}), as given for example
 by  (\re{qa-up},\re{map-llm},\re{k-b}) etc. Transition to  continuum
limit usually distorts such {\em exact} nonlinearity 
leaving only some remnants.

  Extension of the scheme presented here
  to higher dimensions  and its application 
for constructing  ( $(N\times N)$) Lax operators  will be taken up 
in sect. 9.

\vspace* {1cm} \fbox{fig.5.1} 

\vspace* {1cm}
  {\bf
Fig. 5.1 } \ {\it Generation of quantum integrable models from  respective
ancestor models with Lax operators 
$L_t^{\theta},~~ L_t $ related to the trigonometric 
$ R_t^{\theta},~~R_t$ matrices  and with Lax operators 
 $L_r^{\theta}, L_r $
 related to the rational  
$ R_r^{\theta},~~R_r$ matrices  obtained at particular limits. The
subscript $t$ stands for the trigonometric  and $r$ for the rational
class.}

%_____________________________________________\input {preport6.tex}
\setcounter{section}{5}
\section{  Hamiltonian and other conserved quantities from Lax operator}
\setcounter{equation}{0}
We have seen how the representative Lax operators belonging to an 
integrable class, determined by the  associated 
  $R$-matrix can be constructed
in a systematic way. However, since  the Hamiltonian of
the model   is  an important physical object,
 we focus here on  its
construction along with other conserved quantities, 
starting from  a more abstract element, namely the Lax operator.
Recall that (see sect. 2) 
the conserved quantities $\{C_n \}, n=1,2, \ldots$ are generated 
from the transfer matrix $\tau(\la)$ as the expansion coefficients
of $\ln \tau(\la)=\sum_{n=0}^\infty C_n \la^n$ and therefore 
$C_n= const.~{d^n \ov d \la^n}\ln \tau(\la)_{\mid \la=\la^0}$, where 
the expansion point $\la^0$ may be $0$.
 The global operator $\tau(\la)$ is constructed
from the local Lax operators  $L_{aj}(\la),~j=1,\ldots N$ as  
\be
\tau (\la)=tr_a\l(L_{a1}(\la)\ldots  L_{aN}(\la)\r),
\ll{tau}\ee
where $\tau (\la)$ acts on the total quantum space ${\cal H}=
\otimes_{n=1}^N
h_n. $ 
Therefore, from the knowledge of Lax operators it is possible to derive
expressions for conserved quantities including the Hamiltonian. In
general, they will be nonlocal objects and physically uninteresting. However
some
special properties of the Lax operators allow to  overcome this difficulty
and obtain Hamiltonians with localized interactions. 
The Lax operator $L_{aj}(\la)$
 acts on the product space $V_a \otimes h_j$, of  the common auxiliary
space $V_a$  and   the quantum space $h_j$ at site $j$. The models 
with   $V_a$ and all 
$h_j , j=1,\dots N$  isomorphic are called {\it fundamental}, while
the rest with inequivalent quantum and auxiliary spaces are known as 
 the {\it nonfundamental} models.   
For fundamental models since $V_a \approx h_j$, 
the Lax operators $ L_{aj}(\la)$ 
formally coincide with their
corresponding $ R_{aj}(\la)$-matrices.
\subsection {  Fundamental models }
Lax operator  $ L_{aj}(\la)$  of   such models acts on 
$ V\otimes V$ and therefore the
auxiliary and quantum spaces may be interchanged. To  
  utilize this feature the {\it regularity } property 
  of the Lax operator $L_{aj}(\la=0)= P_{aj}$, i.e.
$L_{\al \bet}^{ab}(0)=\de_\bet^a\de_\al^b$  is demanded, where 
the permutation operator  $P_{aj}$ can interchange the spaces as 
$P_{aj}L_{ak}
=L_{jk}P_{aj}$ with the property $P^2=1$ and $tr_a(P_{aj})=1$.
Therefore, for regular models the transfer matrix  (\re {tau}) yields
\be
\tau (0)=tr_a\l(P_{aj} P_{aj+1}\ldots  P_{aj-1}\r) 
\ =(P_{jj+1}\ldots  P_{jj-1})tr_a(P_{aj}) 
\ll{tau0}\ee
for any $j$, using the freedom of cyclic rotation of matrices under the 
trace.
Taking derivative with respect to $\la$ in (\re {tau}) we similarly get
\be
\tau' (0)=tr_a \sum_{j=1}^N 
\l(P_{aj} L'_{aj+1}(0)\ldots  P_{aj-1}\r) \ \ = \
\ \sum_{j=1}^N(L'_{jj+1}(0)\ldots  P_{jj-1})tr_a(P_{aj}) ,
\ll{tau1}\ee
where we have assumed the periodic boundary condition: $
 L_{aN+j}= L_{aj} $. Defining  now 
 $H=c~ C_1= c{d \ov d \la} \ln \tau(\la)_{\mid \la=0}
 =c~ \tau' (0)\tau^{-1}(0),~c  =const.$

 we construct the Hamiltonian  from   (\re {tau0}) and 
(\re {tau1}) as 
\be
H=c \sum_{j=1}^N(L'_{jj+1}(0) P_{jj+1})
\ll{htau01}\ee
with only $2$-neighbor interactions, 
due to cancelation of all other nonlocal factors and the property  $P^{-1}=P$.

Similarly taking higher derivatives of $\ln \tau(\la)$  higher conserved
quantities $C_n, n=2,3,\ldots$ may be generated. 
Note that the conserved operator $C_n$
involves the interaction of $n+1$ neighbors. 

\subsubsection { Hamiltonian construction}
We take  the Lax operator coinciding with the   $R(\la)$-matrix 
  in $4  \times 4 $-matrix form  
\be L(\la)= \left( \begin{array}{c}
  a_+(\la)
 \qquad \qquad \qquad    
 \\  \qquad b_+(\la)
\ \  ~
   c_+(\la)  \qquad \\ \qquad c_-(\la)\ \ ~    b_-(\la) \qquad \\
     \qquad \qquad\qquad a_-(\la) 
          \end{array}   \right), ~~
P= \left( \begin{array}{c}
  1
 \qquad \qquad \qquad    
 \\  \qquad 0
\ \  ~
   1 \qquad \\ \qquad 1 \ \ ~   0 \qquad \\
     \qquad \qquad\qquad 1 
          \end{array}   \right), ~~
\ll{LP}\end {equation}
 with more  general entries and   demand  the regularity
condition $ L_{jj+1}(0) = c P_{jj+1}$.
Multiplying $L'_{jj+1}(0)$ with $P_{jj+1}$ and inserting in (\re{htau01})
we get the  explicit  expression for the Hamiltonian  
\bea
{ H}& =& const. \sum_{n=1}^N
b'_+(0) \si_n^+ \si_{n+1}^- +b'_-(0)\si_n^- \si_{n+1}^+ 
 +
 ( I^+ -J^+)\si_n^3 \si_{n+1}^3  \nonumber \\ &+& (I^+ +J^+) {\bf I}
+2 I^- \si_n^3 
 + J^- (\si_1^3 -\si_{N+1}^3) 
\ll{hgeneral}\eea
which is general enough to  cover all  spin models considered below. 
 Here
the parameters are   $I^\pm =
\qr( a_+'(0) \pm a_-'(0))$ and $J^\pm =
\qr( c_+'(0) \pm c_-'(0))$. 
It should be noted       that, for the models with 
 periodic boundary condition: $\vec \si_{N+1}=\vec \si_{1}$
, in  which  we are  mostly  interested,
  the last term in  (\re{hgeneral}) 
representing  a boundary term simply drops out.
On the other hand, the Hamiltonian with open boundary condition 
    takes the form  
\be
{ H} =- \sum_{n=1}^N
\si_n^+ \si_{n+1}^- +\si_n^- \si_{n+1}^+ 
 +
 \qr (q+q^{-1})(\si_n^3 \si_{n+1}^3 -{\bf I})   
 - \qr (q-q^{-1}) (\si_j^3 -\si_{j+1}^3), 
\ll{hqgroup}\ee
which  acquires an interesting $U_q(su(2))$  quantum group symmetry.
Such models  were investigated 
in connection with  open chain problems
 \c{hqg,openc}

\subsubsection {  $XXZ$ spin-$\ha$ chain}  

The Lax operator for this model is given by the trigonometric $R$-matrix
(\re {rtrig}) with regularity condition and corresponds to 
  $ a_+'(0)= a_-'(0)=\cos \eta,~~
 b_+'(0)= b_-'(0)= 1$ and  $   c_+'(0) = c_-'(0)=0.$
This clearly reduces (\re{htau01}) or similarly (\re{hgeneral})  to 

\be
{ H_{xxz}} =c \sum_n^N\l(\ha \cos \eta ({ I}+ \si_n^3 \si_{n+1}^3)
+ (\si_n^+ \si_{n+1}^- +\si_n^- \si_{n+1}^+ \r),
\ll{hxxz1}\ee
which   after neglecting the constant term
reproduces exactly the anisotropic $XXZ$ Hamiltonian (\re{XXZ}).

\subsubsection { $XXX$  spin-$\ha$  chain}  

Since in this case the  Lax operator may be given by the rational $R$-matrix 
(\re{rationalr}) having  the regularity condition and with the values $
 a_+'(0)= a_-'(0)=
 b_+'(0)= b_-'(0)= 1$ and  $   c_+'(0) = c_-'(0)=0$,
 one gets from  (\re{LP})
$L'_{jj+1}(0) = {\bf I}$ reducing 
 (\re{htau01})  simply to
\be H=
c \sum_{j=1}^N P_{jj+1}.\ll{xxx0}\ee
This is clearly equivalent to
 the known Hamiltonian of the isotropic spin-$\ha$ chain model listed in
  (\re{XXX}). It can also be  obtained from (\re{hxxz1}) as the degenerate 
case  $\eta=0$:    
\be
{ H_{xxx}} =c \sum_n^N\ha  \si_n^3 \si_{n+1}^3
+ \si_n^+ \si_{n+1}^- +\si_n^- \si_{n+1}^+ 
\ll{xxx1}\ee

\subsubsection {   spin-$1$ chain models}

A simple way of getting an Integrable spin-$1$
 model  is to construct   
the Lax operator
as  a  $SU(3)$ related rational $R$-matrix.
This would  lead to 
  (\re{xxx0}) with the permutation operator
$P_{jj+1}=\sum_{a,b=1}^3 E^j_{ab}E^{j+1}_{ba}$, which when expressed
in
spin-$1$ operators ${\bf S}$ takes the form 
\be
{ H_{spin 1}} =const. \sum_n^N\l(({\bf S}_n  {\bf S}_{n+1})
+\ep ({\bf S}_n  {\bf S}_{n+1})^2 \r)
\ll{spin1}\ee
with $\ep=1$. 
Another more interesting possibility is to use the  Lax operator obtained 
 through fusion of spin-$\ha$ Lax operators $R^0$ \c{smodel} as
\be
R_{{\bf a b}}(\la)= (P^+_{\bf a}\otimes P^+_{\bf b}) \pr _{j=1}^{s}
\pr _{k=1}^{s} R_{a_j b_k}^0(\la+i\eta (2s-k-j)) 
 (P^+_{\bf a}\otimes P^+_{\bf b})  g_{2s}^{-1}(\la)
\ll{R-s}\ee
with spin $s=1$ \c{spin1}, where $
 P^+_{\bf a}$ is the symmetrizer in the fused spin-$s$ space 
 ${\bf a}$ and $g_{2s}(\la)$ is some
multiplicative factor.

The corresponding Hamiltonian is given in the form (\re{spin1}) with
$\ep=-1$.

\subsubsection {  Impurity model}

Simple integrable impurity models may be
 constructed by replacing the Lax operator $L_{aj}(\la)$ of the regular
model by an inhomogeneous  
$L_{am}(\la-\nu)$ 
at a particular lattice site $m$,
where $\nu$ corresponds to the impurity at  $m$.
 We see immediately, that 
the regularity condition  no longer holds at $m$ and consequently, 
the terms involving $m-1,m,m+1$ in (\re{htau01}) become nonregular.    
 As a result, instead of
 (\re{htau01}) one gets the Hamiltonian
 of the impurity model as
\bea
H_{imp}&=& \sum_{j\neq m-1,m}^N(L'_{jj+1}(0) P_{jj+1})
+( L'_{m-1m}(\nu) L^{-1}_{m-1m}(\nu)\nonumber \\ &+&
 L_{m-1m}(\nu) L'_{m-1m+1}(0) P_{m-1m+1} L^{-1}_{m-1m}(\nu)) 
\ll{impure}\eea
with next-to-nearest neighbor interaction.
Using therefore 
the $L$-operator of the $XXX$ or the  $XXZ$  chain one can derive
from (\re{impure}) the  Hamiltonian of the respective impurity  model. 

\subsubsection {  Inhomogeneous  model}
Inserting different impurities $\nu_j$ at sites $j$ we can get an
inhomogeneous spin model, where instead of (\re{tau}) one obtains
\be
\tau (\la)=tr_a\l(L_{a1}(\la-\nu_{1})\ldots  L_{aN}(\la-\nu_{N})\r).
\ll{tauIH}\ee
Note that at $\la=0$ 
due to the loss of regularity at all points:
 $L_{aj}(-\nu_{j})\neq P_{aj}$,  
 the corresponding 
Hamiltonian becomes nonlocal and   as such of no physical interest.
However interestingly at $\la=\nu_{j}$ the regularity  is restored and 
one can use the  dual 
conditions on  the 
$L$-operator at $\eta=0$ as
 \be 
L_{aj}(\la-\nu_{j}, \eta)\mid_{\la=\nu_j}= P_{aj}  ~~~~  
  L_{ak}(\nu_{j}-\nu_{k},\eta)\mid_{\eta=0}= {\bf I}
\ll{PIIH}\ee
This allows to  construct a new set of  Hamiltonians 
$H_j= {d \ov d \eta} \tau(\nu_j)\mid_{\eta=0}= \sum_{k \neq j}
{d \ov d \eta}L(\nu_{j}-\nu_{k},\eta)\mid_{\eta=0}.$
Notice that though this is a bit different definition of the conserved
quantities, their commutativity also follows from that of $\tau(\la)$ 
at different $\la$'s.

We can check easily the property (\re{PIIH}) for the $L$-operators 
of the  $XXZ$ and $XXX$ models. The long range Hamiltonian for such
inhomogeneous $XXX$ model therefore can be constructed as 
%\c{wadatiIH} 
$H_j=\sum_{k \neq j}
{1 \ov (\nu_{j}-\nu_{k})} \vec \si_{j}\cdot \vec \si_{k}$
exhibiting similarity with the Haldane-Shastry model (see sect. 9).
\subsubsection { Alternating  spin chain}
The idea is to replace the spin-$\ha$ Lax operators $L_{\si \si'}$
 at every alternate
lattice sites by a Lax operator $
L_{\si s}$ with the same auxiliary space $\si$
 but  different   quantum space $s,$ corresponding to the 
spin-$s$ operators.  Since at alternating sites the regularity condition 
does
not hold, in analogy with the impurity model one gets  a Hamiltonian
with interaction between different spins ranging to 
 next to nearest neighbors \c{alters}.

\subsubsection { Twisted and colored  models}

Twisted as well as color deformation of Lax operators have been considered in
sect. 4. It is interesting to see their effects in the Hamiltonian of the
models. Note  that, under twisting  
 transformation   $\ti R_{ab}(\la,\th)=
 F_{ab} R_{ab}(\la) F_{ba}^{-1},$  the regularity condition is
preserved as  $$\ti R_{ab}(0)=
 F_{ab} P_{ab} F_{ba}^{-1}= P_{ab} F_{ba} F_{ba}^{-1}= P_{ab}$$
while the term 
 $L'(0)$ in (\re{tau1}) is replaced by 
\be F_{jj+1} L'_{jj+1}(0) F_{j+1j}^{-1}
+ 2 F'_{jj+1}(0) F_{jj+1}^{-1}(0)  P_{jj+1}.\ll{twistc}\ee
  We consider 
 the realization of the twisting operator as   
\be
 F_{aj+1}(\th, \la) =e^{\qr((i\th
+c \la)\si^3_{j+1} -i \th \si^3_a )}\ll{twicol}\ee
with the choice of the color parameter through the spectral parameter  $\la$
as  
 $  \bar \la_a=c\la$ and all  other  parameters $  \bar \la_j=0$. 

If we consider first the  twisting   involving
 $\th$ parameter only  and no $\la$, then due to  $
F'_{jj+1}(0)=0$ 
we get from (\re{htau01}) and (\re{twistc})  the expression
$$ 
H=\sum_{j=1}^NF_{jj+1}(\th)(L'_{jj+1}(0) P_{jj+1})F_{jj+1}^{-1}(\th).$$
This leads to a   twisted form of  the $XXZ$ chain 
given by  the Wu-McCoy model (\re{WM})
\be
H^{wm} = const. \sum_{n=1}^N
e^{i\th} \si_n^+ \si_{n+1}^- +e^{-i\th}\si_n^- \si_{n+1}^+ 
 +
 \ha \cos \eta \si_n^3 \si_{n+1}^3.
\ll{WM1}\ee
 One can derive also the same model
directly from (\re{hgeneral})
 by considering   twisting of  the  parameters 
\be 
 b_\pm(\la)=e^{\pm i\th} b(\la),~~  a_\pm (\la)=
a(\la),~~c_\pm=c. \ll{twiab}\ee  
Note that by transforming $\si_{n}^+= e^{in\th} \ti \si_{n}^+$
the Hamiltonian (\re{WM1}) can be reduced to the untwisted form
(\re{hxxz1}), however with the quasiperiodic  boundary condition 
$\ti \si_{N+1}^+= e^{-iN\th}\ti \si_{1}^+
$. More general type of twisted boundary conditions
in integrable spin chains have  been studied in \c{twistbc}. 

On the other hand, 
for twisting  transformation  (\re{twicol}) involving
spectral parameter  $c\la$ induced  through the  color parameter as 
$F_{jj+1}(\la) =e^{\qr c \la \si^3_{j+1} }$,
 we get  
$~~~ 2 F'_{jj+1}(0)F_{jj+1}^{-1}(0)=\ha c\si_{j+1}^3 .~~$ This
  yields an extra term $\ha c  \sum_j^N \si_{j}^3$ in the
 Hamiltonian, signifying the interaction with    an external 
 magnetic field $\ha c$
 along
the $z$ direction. Note that this is consistent  with the choice 
\be 
 b_\pm(\la)=e^{\pm {c \ov 2}\la } b(\la),~~  a_\pm (\la)=
e^{\mp {c \ov 2}\la } a(\la),~~c_\pm=c\ll{colab}\ee  
which gives   $I_-= {c \ov 4}$ and can derive directly  from  
(\re{hgeneral}) the Hamiltonian of the model as 
$H=H_{xxz}+{c \ov 2}  \sum_j^N \si_{j}^3$.
\subsubsection { Free fermion models }
It is easy to check, that 
if we choose the parameters of the $L$ operator in (\re{LP}) 
 as \be
a_+(\la )=a_-(\la)=\cos \la, ~~b_+(\la)=b_-(\la)=\sin \la
,~~c_+=c_-= 1
\ll{ffR}\ee
 satisfying the free fermion condition $a^2+b^2=c^2$,
the regularity condition of the corresponding 
 Lax operator  $l^{xy}(\la)$ is valid and the 
 Hamiltonian (\re{hgeneral}) reduces simply to the $XY$
chain 
  \be
{ H}_{xy} =c \sum_j^N
 \si_j^+ \si_{j+1}^- +\si_j^- \si_{j+1}^+. 
\ll{hxy}\ee
The explicit equivalence of the $XY$ chain  with the free fermion model 
can be shown 
by   realizing the Pauli matrices though the anicommuting 
fermionic operators $c_j, c_j^\da$ using the  
 {\it Jordan-Wigner transformation}
\be
c_j= \l(\si_1^3 \cdots \si_{j-1}^3\r) \si_j^-  
\ll{JWT}\ee
This clearly leads to 
\be
c^\da_jc_j\equiv n_j= \ha(\si_j^3 +1), ~~
c^\da_jc_{j+1}=- \si_{j}^+ \si_{j+1}^-,  ~~
c_jc^\da_{j+1}= \si_{j}^- \si_{j+1}^+,  ~~
\ll{JWT1}\ee
etc. with $c_jc^\da_{j+1}=-c^\da_{j+1} c_{j}$
and rewrites (\re{hxy}) as the free fermion model
  \be
{ H}_{ff} =-c \sum_j^N
 c_j^\da c_{j+1} +c^\da_{j+1} c_{j}
\ll{freexy}\ee

Another free fermion model can be constructed remarkably from the 
 supersymmetry related nonstandard $R$-matrix \c{nstR} consistent 
 with  the parametrization   \be a_+ (\la)=
\sin(\la+\eta), ~a_-(\la)=
\sin(-\la+\eta), 
\ll{nstR}\ee
 with   other parameters as in  the    $XXZ$ model. 
 The regularity  condition for  this model follows  due to $a_+(0)=a_-(0)$ 
and we   
 notice that, since  $a_+'(0)=-a_-'(0)$ and
$c_+'(0)=c_-'(0)=0,$ the interacting 
term $\sum \si^3_j\si^3_{j+1}$ in (\re{hgeneral}) drops out from the
Hamiltonian. This  leads  again to a $XY$ chain placed 
in a magnetic field given  by the term $\cos \eta \sum_j^N \si_j^3$.
The transformation  (\re{JWT})  reduces  the Hamiltonian clearly 
 to  a free fermionic model.
\subsubsection { Hubbard model}
A nontrivial example of the regular models is given by the Lax operator of
the  one-dimensional integrable Hubbard model, where
the Lax operator does not coincide with the $R$-matrix and moreover 
 the spectral dependence of 
 the $R$-matrix is both   of the difference as well as  the sum.
The Lax operator corresponding to the model  may  be given  as 
\be 
L^{Hub}_{aj}(\la)= \l(l^{xy}_{a_1j}(\si\mid \la) \otimes 
l^{xy}_{a_2j}((\tau \mid  \la)\r)e^{h(\la) \si_j^3 \tau_j^3},
\ll{lhub}\ee
where 
$l^{xy}(\si\mid \la)$ and $l^{xy}(\tau \mid \la)$ are the Lax operators of
two $XY$ spin chains involving mutually commuting spin-$\ha$ operators 
${\bf \si}$ and ${\bf \tau}$ and 
 ${\sinh 2h(\la) \ov \sin \la}={U \ov 4}
$.
 
We check first the regularity condition using the property 
 of the 
$XY$ Lax operator $l^{xy}(\la)_{\la=0}=P$ as $L^{Hub}(0)=
l^{xy}(\si \mid 0)\otimes
l^{xy}(\tau \mid 0) =P(\si) \otimes P(\tau)=P_{Hub}$, where we have assumed
$h(0)=0$ for the function $h(\la)$. Therefore following the above procedure
, apart from the separate contributions of $l^{xy}$
related to  $\si$ and $\tau$ operators, we also get an interacting 
 term $h'(0)\si_j^3 \tau_j^3$ in the Hamiltonian coming from $\tau'(0)$.
This finally gives 
the Hubbard Hamiltonian in the form
  \bea
H_{Hub}&=& { H}_{xy}(\si)+{ H}_{xy}(\tau)+{ H}_{int}(\si\tau)
\nonumber \\
& =&  \sum_j^N \l(
 (\si_j^+ \si_{j+1}^- +\si_j^- \si_{j+1}^+)
+(\tau_j^+ \tau_{j+1}^- +\tau_j^- \tau_{j+1}^+)+{U \ov 4}\si_j^3
\tau_j^3\r). 
\ll{hhub0}\eea
 We can transform  Hamiltonian (\re{hhub0})
 into its conventional form (\re{hhub})
by applying  the Jordan-Wigner transformation (\re{JWT}-\re{JWT1}).
\subsection {Nonfundamental models}

When the auxiliary space $V$ and the quantum spaces $h_j\equiv h $ are not
isomorphic, the regularity of the Lax operator $L_{a \a j}(\la)$
does not hold  
and the construction of the Hamiltonian becomes difficult. However there
are few methods, that can in principle  derive the conserved quantities
from the Lax operators of such models, which are mostly 
 bosonic  models.
\subsubsection { Generalized spin model method}
The first method is based on the  idea of considering bosonic  models 
as the  generalized spin models  with arbitrary spin \c{TTF}.
One has to take the  fused $R_{{\bf a b}}(\la)$ matrix (\re{R-s}) with spin
$s$ along with  possible gauge transformation  and
replace  the auxiliary spaces by the quantum ones. We thus  
  recover the regularity in the quantum space : $ 
R_{{\bf a b}}(0)=P_{{\bf a b}}$ and
 may construct  the Hamiltonian of such generalized 
 spin  models
with nearest neighbor interactions.
The isotropic  and the anisotropic  cases with arbitrary spins
 are determined by fusing $R$-matrices  corresponding to $XXX$  or  
 $XXZ$ spin-$\ha$ chains and belong to two different
integrability classes: rational and trigonometric.  
 In the first case realizing the spin $s$,  $su(2)$  generators
    through bosons  we construct  the 
Hamiltonian of the corresponding   bosonic model.
Similarly in the second case 
 the bosonic realization of the  $U_q(su(2))$  generators 
is required, which would lead to  the
  nonfundamental model of  the trigonometric class.

\subsubsection { Projector method}
In a more straightforward  approach, known as the  projector
method \c{sglsg,Coker},  
 one starts directly with the 
Lax operator $L_{a  j}(\la)$ of the discrete bosonic model  without demanding any
regularity condition. However from the vanishing condition
$det_qL_j(\la)=0$ at points $\la=\nu $, $\nu^*$
 of the
quantum determinant  
     \be 
det_qL \equiv {1\over2}
\l((L_{11}\tilde L_{22} +L_{22}\tilde L_{11})-(L_{21}\tilde L_{12}+ L_{12}
\tilde L_{21})\r) \ll{detq}\ee with $\ti L=L(\la+i \kappa)$,
 we
represent the Lax operator at these points both  as a direct and an 
inverse 
projector. Therefore at  $\la= \nu,$   $L^{\al \bet}_j(\nu)=r_\al(j) s_\bet(j)
=v_\bet (j) u_\al(j)$ with $r,s,u,v $
being two-dimensional vectors and similarly for $ \la=\nu^*$.
Therefore,  the transfer matrix  at these points becomes 
\bea & & \tau(\nu) = tr \l(K(j-2,j+2 \mid \nu) L(j+1 \mid \nu) L(j \mid \nu)
L(j-1 \mid \nu)\r)
\nonumber \\  &=&K_{\al \bet}(j-2,j+2 \mid \nu) r_\bet(j+1)  s_\ga(j+1) 
 r_\ga(j)
 s_\de(j)
 r_\de(j-1) s_\al(j-1)
\ll{proj1}\eea
where $K(j-2,j+2 \mid \nu)=L(j-2 \mid \nu) \ldots L(1 \mid \nu)L(N \mid \nu)
\ldots L(j+2 \mid \nu)$. We face with the problem, that in the quantum case
due to the noncommutativity of $r_i,s_i$ at the same site, the last operator
in (\re{proj1}) can not be dragged to the first place. However, using  
the inverse operator representation $ r_\de(j-1) s_\al(j-1)= v_\al(j-1)
u_\de(j-1)$ at the same site, we overcome this difficulty to get
$
\tau(\nu)=  Q_{j-1}^{j+1}(\nu)  M(j+1, j ,j-1)
$
where
\bea Q_{j-1}^{j+1}(\nu)&=& v(j-1)K(j-2,j+2 \mid \nu)r(j+1)\nonumber \\
 M(j+1, j ,j-1)&=&s(j+1)\cdot r(j)~s(j)\cdot u(j-1)
\ll{projQM}\eea
 with
$s(j+1)\cdot r(j)=s_1(j+1) r_1(j)+s_2(j+1) r_2(j)$ etc. 
 Similarly, we obtain
$
\tau'(\nu)= \sum_{j=1}^{N}
 Q_{j-1}^{j+1}(\nu)  \l( s(j+1)L'(j\mid \nu)u(j-1) \r)
$
to yield finally the conserved quantity
\bea C_1&=& c~ {d \ov d \la} \ln \tau(\la)_{\mid \la=\nu}
= c~ \tau^{-1}(\nu) \tau'(\nu)
\nonumber \\
&=&c~ \sum_{j=1}^{N}
 M^{-1}(j+1, j ,j-1)  \l( s(j+1)L'(j\mid \nu)u(j-1) \r)
 \ll{projc1}\eea
involving interactions of neighboring $3$ sites, 
where $M(j+1, j ,j-1)$ is defined in (\re{projQM}). Likewise one may derive
also the conserved quantity from the expansion at $\nu^*$.

Compared to the   generalized spin method  applicable  
to the symmetric models, the scope of the 
  projector method seems to be more wide.
\subsubsection { Lattice  nonlinear Schr\"odinger  model}
The LNLS  is a bosonic model with nonfundamental 
  Lax operator (\re{lnls1}), the Hamiltonian of  which can be constructed
with both the above methods. 

To apply the {\it generalized spin}  method to the lattice NLS,
 specialize to the  spin $s$  isotropic  model
  and
 consider the Holstein-Primakov transformation (\re{hpt})
with spin parameter $s=-{2 \ov \kappa \De}.$ 
  The   Hamiltonian derived from
$R_{{\bf a b}}(\la)$ by fusing $XXX$ spin-$\ha$   models takes the form 
 $H^s=\sum_j^{N} f_s({\bf S}_j  {\bf S}_{j+1})$, where $f_s$ 
 is a definite polynomial \c{TTF}.
Therefore from  the equivalence 
of the spin model Lax operator with that of the lattice NLS: 
$L^{lnls}(\la)= {1 \ov i s} \si^3 L^{xxx}\l(-{\la \ov \kappa}\r)$,
one  can obtain the corresponding Hamiltonian 
involving only the nearest neighbor interactions, which at the
continuum limit $ \De \rw 0 $ reproduces
 the well known  NLS field model \c{TTF}.

For constructing the  lattice NLS Hamiltonian through the 
{\it projector method},  the Lax operator $L_{n}(\la)$ is taken
in the form (\re{lnls}) with $s\equiv \ti \nu= -{2 \ov \kappa \De}- {(-1)^n
\ov 2}$ with 
the inclusion of a parity
factor $(-1)^n.$  Due to  $det_qL(\la)=\qr
\De^2(\la-\nu_1)(\la-\nu_2)=0,$ with $\nu_1={i \ti \nu  \kappa},~ \nu_2=
-i2 \kappa -\nu_1$   the Lax
operator  $L_n(\nu_1)$ at $\la=\nu_1$ reduces 
to the direct  projector at  even $n$ sites, while to the inverse projector
at the odd $n$. Similarly, the reduction of $L_n(\nu_2)$
is in the reverse order.   Thus the necessary requirement of 
representing the Lax operator at the same lattice point
 as both direct and inverse projectors for the
same spectral parameter value fails here. However, this difficulty is 
overcome by defining a new Lax operator as product of two Lax operators at 
 even and odd sites: ${\cal L}(j\mid \la)=
 L(2j\mid \la) L(2j-1\mid \la)$. At new points $\nu= -{2i  \ov \De}+ 
{i  \kappa}$ and $\nu^*$ the necessary condition is fulfilled and 
 the Hamiltonian for the lattice NLS 
model is obtained finally 
by adding the conserved quantities $C_1$ coming 
 from the expansions at $\la=\nu$ and $\la=\nu^*$  along with 
a normalization factor involving $\De$ and an additional 
 term proportional to the number operator $N=\sum_{j=1}^N
\psi_j^\dagger \psi_j $\c{sglsg,Coker}.
This discrete Hamiltonian $H(\De)$ is localized with interaction 
ranging over $five$ neighboring lattice sites  and  has a complicated form,
though 
 at the continuum limit it  reduces to 
the known field Hamiltonian  (\re{nls}).

\subsubsection { A simple lattice NLS type model}
To demonstrate that the projector method is  applicable also  to models with
lesser symmetry, we look into the Lax operator  
\begin{equation}
{ L}_{n}(\xi)  = \left( \begin{array}{c} { 1 +{N(n)\over  \Delta }\xi
  } 
\qquad {- i  \kappa^{1 \over 2}\phi(n)\xi} \\
{i  \kappa^{1 \over 2}\psi(n)\xi  } \qquad {{1\over\Delta}\xi}
    \end{array} \right).
\ll{nl3l}\end{equation}
 with
 $N(k)=  1+\kappa \Delta^2 \phi(k) \psi(k)$,  listed in (\re{slnls})
and constructed   
as the AKNS type integrable model (\re{slnls1}) from the
rational ancestor model  in sect. 5. 
At the  degenerate point $\xi=\nu_2=0$ when $det_qL=0, the $ 
 Lax operator (\re{nl3l}) becomes remarkably simple 
and turns  into a field independent projector:
$ L(0)=P^+$.
As a consequence, the  construction  becomes much easier
resulting a  local
and  simple  Hamiltonian $H$ along with 
 other conserved quantities like the number operator $N$ and the momentum $P$
\c {krjp94}:
 \bea 
H=  C_3
&=&{1\over\Delta}
  \sum_k (
  \phi(k+1) \psi(k -1) -  (N(k)+N(k+1))
  \phi(k+1) \psi(k)+ 
  (3\kappa\Delta^2)^{-1} N(k)^3 )
 \nonumber \\
  P=C_2&=& \sum_k(
  \phi(k+1) \psi(k) - {1\over 2\kappa\Delta^2} N(k)^2), ~~N=C_1=
  {1\over\Delta\kappa}  \sum_k N(k).
  \ll{nl3cn}\eea
 It may be noted that the above conserved quantities of the discrete AKNS
type model are not
symmetric in $\phi$ and $\psi$. 
The Symmetry however is restored at the continuum limit allowing the reduction
$\phi = \psi^\dagger$ and recovering thus  
 the   known conserved quantities of the   NLS field model.
Note that the  expressions for the  $C_n$  (\re{nl3cn})
 including the discrete Hamiltonian   $C_3$ 
 are 
 much similar to  the corresponding 
 NLS field model.
\subsubsection {  Lattice  sine-Gordon  model}
For applying the {\it generalized spin } method to LSG,
we have to construct  first  the anisotropic spin  $s$ 
Hamiltonian by  fusing $XXZ$ chains. Subsequently, 
through  bosonic realization of the quantum $U_q(sl(2))$ 
(\re{qa-up}) one can, in principle, construct 
the Hamiltonian for the LSG model with nearest 
neighbor interaction \c{TTF}. However the construction  
 in its explicit form  is difficult to achieve.

The {\it projector} method can  be applied also to LSG model taking the
discrete Lax operator  as (\re{L-sg1}). However,    
 the problem of representing the Lax operators both as the direct and the
inverse projectors at the  same degenerate point in the spectral parameter
is difficult to solve and as a result  
 the corresponding Hamiltonian $H=\ha (C_1+C_{-1})$ and the momentum
 $P=\ha (C_1-C_{-1})$ obtained in this case 
turn out to be only   quasilocal functions 
\c{sglsg,Coker}.

\subsubsection { Relativistic and nonrelativistic Toda chains}

In some models the construction of conserved quantities become much simpler 
due to specific structure of their Lax operators. Both  relativistic and
nonrelativistic Toda chains with the Lax operators  constructed in sect. 5  are
such examples. 
Though these are nonfundamental models, due to the  triangular form 
of their Lax operators (see (\re{rtoda},\re{toda})) the local 
conserved quantities
 can be obtained directly as the expansion coefficients of the transfer
matrix   
$$\tau(\xi)=tr T(\xi)=tr (\prod^N L_n)  =
 \sum_n^N C_n\xi^n+C_{-n}\xi^{-n},
$$
by multiplying $N$ number of Lax operators given in the explicit form
 (\re{rtoda})
 and then taking the trace. 
Thus  we get for the  
{\it relativistic Toda chain} (RTC)
conserved quantities like  
\begin {eqnarray}
I^{+}=\sum_i\left(e^{ 2\eta p_i}\left(
1+\eta^2 e^{
(q_{i-1}-q_i)}
\right)\right),\quad
I^{-}=\sum_i\left(e^{ -2\eta p_i}\left(
1+\eta^2 e^{
(q_i-q_{i+1})}
\right)\right)
\ll{I+-}\end {eqnarray}
%15
where we have denoted $I^+ \equiv -C_{n-2}(C_n)^{-1}$ and 
 $I^+ \equiv -C_{-(n-2)}(C_{-n})^{-1}$.
The Hamiltonian and the momentum of this relativistic or discrete time 
 model is constructed as 
  $H= \frac {1}{2}(I^++I^-)
$ and 
$P= \frac {1}{2}(I^+-I^-).$
In the analogical way the conserved currents of all other models from the
family of RTC found in sect. 5 can also be derived.

Since $\eta \rw 0$ corresponds to 
 the {\it nonrelativistic} limit,
we obtain the corresponding conserved quantities  
 from the  limiting values  of $I^\pm$ in  (\re{I+-}), resulting 
 clearly the following   expressions for the momentum and Hamiltonian   
of the  standard Toda chain \c{toda0}
(\re{toda}) 
\bea
H&=& {\frac {1}{2}(I^++I^-)-1 \ov 2\eta^2}=
\sum_i \l(\ha   p^2_i+
\qr e^{
(q_{i-1}-q_i)}
+ e^{
(q_i-q_{i+1})}
\right),\nonumber \\
P&=& {\frac {1}{2}(I^+-I^-) \ov 2\eta}=
\sum_i    p_i
\ll{todaHP}\eea

%________________________________________\input {preport7.tex}

\setcounter{section} {6}
%{chapter} {6}
\section { Exact solutions of eigenvalue problem, Bethe ansatz}
\setcounter{equation}{0}

One of the most significant achievements 
of quantum integrable systems is that, 
  these are some of the few interacting  models, for which 
the  eigenvalue  problem   in principle can be solved
exactly. Though   the energy spectrum  of the Hamiltonian 
is the most important physical object, the integrable models allow to solve the
   eigenvalue  problem   for   all its
 conserved quantities. The most efficient method for
achieving this is the algebraic Bethe ansatz (ABA), though its coordinate
formulation (called coordinate Bethe ansatz (CBA)) is usually more 
effective for finding the energy spectrum of concrete models. 
We describe  both these useful methods and  compare 
them to  clarify their 
interrelations.

  \subsection{ Algebraic Bethe ansatz}
ABA  is an universal and to a great extent a model-independent approach 
for solving the eigenvalue problem of quantum integrable systems.
This method exploits significantly  the integrable structures
like the QYBE and  is designed to find   the eigenvalues of all conserved 
quantities including the Hamiltonian.

We describe  basic features of this method \c{qism} in a 
general way to emphasize on its universal applicability, though for
clarity we  orient mostly to the simplest examples of $XXZ$ and $XXX$
spin-$\ha$ chains. 
As we know,  the monodromy matrix  in ultralocal integrable systems
: $T(\la)_a=\prod_i^NL_{ai}(\la)$ satisfies a representative equation
given by the global QYBE (\re{qybet}), which in  
component form reads
\be
\sum_{ \{j\}} R^{i_1i_2}_{j_1j_2} \l( 
T^{j_1}_{k_1} \ti T^{j_2}_{k_2} \r)
= 
\sum_{ \{j\}}\l( \ti T^{i_2}_{j_2}
  T^{i_1}_{j_1} \r)  R^{j_1j_2}_{k_1k_2}. \ll{rtt}
\ee
Note that though  $T(\la)$ given through  explicit Lax operators
$L_i(\la)$  carries detailed information about the concrete model, at the
global QYBE level the  information needed is only about the associated 
$R$ matrix. Consequently,  the  
$R$ matrix (\re{rtrig}) with  
a multiplicative factor $b(\la)^{-1}$ for convenience:
\be
R^{11}_{11}=
R^{22}_{22} \equiv f=
{ a \ov b}={ \sin(\la+\eta) \ov   \sin \la }, ~~R^{12}_{12}=
R^{21}_{21} =1
, ~~R^{12}_{21}=
R^{21}_{12} \equiv  
f_1= {c \ov b}  ={ \sin \eta \ov   \sin \la}  
\ll{trm}\ee
describes the {\it trigonometric} class of models, while 
its $q \rightarrow 1$ limit 
\be
f = { \la+\eta \ov  \la},~~ 
 f_1 = {\eta \ov \la}  
\ll{rrm}\ee
answers for the models with {\it rational}  $R$   ( see sect. 2).
 We have witnessed  in  sect. 5,  that the  models may have 
 common algebraic origin and the same  ancestor models. 
We notice further in the ABA scheme, that the difference between
the descendant models of the same class  almost
 disappears,    showing an  elegant  universality of the scheme. 
Therefore for the choice (\re{trm})
the same ABA formulation with minor modifications  can describe  
the whole  
trigonometric class of models   
, e.g. 
$XXZ$ spin chain, sine-Gordon, DNLS 
 model, Liouville model, relativistic Toda chain 
 etc. Similarly, for (\re{rrm}) almost  the same scheme is applicable 
for the rational class,  e.g.  $XXX$  chain, NLS model, Toda chain   etc.

 ABA scheme aims to  solve some general  eigenvalue problem 
\be
  \tau(\la)\mid \phi>_m= 
  \Lambda_m(\la)\mid \phi>_m
\ll{evp}\ee
for the spectral parameter dependent 
transfer matrix $~~ \tau (\la )=tr_a T_a(\la)= \sum_{\al } T_\al (\la)=
A(\la)+D(\la),~~$
where we have introduced the short hand notation
 $$T(\la)_{11}=
A(\la),~~
 T(\la)_{22}=
D(\la),~~
 T(\la)_{12}=
B(\la),~~
 T(\la)_{21}=
C(\la).$$
The operators $ T(\la)_{ij}$ act on the  whole 
 Hilbert space ${\cal H}=\otimes_j^N
h_j,$ which is a product of $N$ local Hilbert spaces $h_j$ at site $j$.
Recalling that the expansion coefficients of $\ln \tau(\la)=\sum_n C_n
 \la^n
$,  generate the 
conserved quantities $C_n, n=1,2   \ldots,$ one can solve 
the eigenvalue problem for all  $C_n$ simultaneously  simply by expanding
$\Lambda(\la)$ as
\be
  C_1 \mid \phi>_m= 
  \Lambda_m'(0)\Lambda_m^{-1}(0)\mid \phi>_m
,~~C_2 \mid \phi>_m= 
  (\Lambda_m'(0)\Lambda_m^{-1}(0))'\mid \phi>_m
\ll{evpcn}\ee
etc. with  commuting set of operators $C_n$
 sharing the same eigenvectors. 
The model dependence comes  when a particular $C_j$ 
is chosen as  the appropriate Hamiltonian of the model.

The basic trick  in ABA  is in  making  an ansatz for the $m$-particle state
$ \mid \phi>_m $ as
\be
 \mid \la_1 \la_2 \ldots \la_m>=  B(\la_1)B(\la_2) \ldots B(\la_m)
\mid 0>,\ll{evm}\ee
 where $\mid 0>= \otimes_i^N \mid 0>_i,$ 
lying in the whole Hilbert space  is a {\it pseudovacuum} representing a
reference state. For the 
$XXZ$ and $XXX$ spin-$\ha$ chains
  for example, this is a state with all  spins up defined by 
$ \mid 0>_i= \left(\ba  1\\ 0 \ea \right),$ which coincides with the 
ferromagnetic ground
state of the models.  To see the justification 
for the ansatz (\re{evm}), we have to convince ourselves 
 that $B(\la_i)$ acts like 
a {\it creation} operator on the pseudovacuum 
$\mid 0>$, while the operator $C(\la_i)$
is its conjugate  representing  an {\it annihilation} operator.
For this we look into the example of $XXZ$  chain, though one may 
 consider other examples as well. 
Starting from its local Lax operator $L_j(\la)$ (\re{XXZ1})
 and  acting by the  local creation and annihilation 
operators on local pseudovacuum as  
$\si_i^-\mid 0>_i= \mid 1>_i,~~~
 \si^+\mid 0>_i=0$ along with 
 $ \si^3\mid 0>_i= \ha \mid 0>_i,$ one obtains
the upper-triangular form
\be 
L_i(\la)\mid 0>_i = \l( \ba \al(\la)\mid 0>_i \qquad \ga
 \mid 1>_i \nonumber \\
0 \ \ \qquad \bet (\la) \mid 0>_i \ea \r),
\ll{lvac}\ee
where $\al,\bet,\ga$ depend on the models.
Taking now the product of such local states we obtain  the  action 
of the monodromy operator on the global pseudovacuum
$ T(\la)\mid 0>=\prod_i^N L_i (\la)\mid 0>_i$, which   would again  be 
in the upper-triangular form. We conclude 
 therefore, that $C(\la)\mid 0>=0$ and 
$B(\la)\mid 0>=\mid \la>, $ where $\mid \la>= \sum_i^N g_i(\la)
((\otimes_{k<i}
\mid 0>_k ) \otimes\mid 1>_i  ( \otimes_{l>i}
\mid 0>_l),~~$ with $g_i=\ga \al^{N-i}\bet^i$
is the  superposition of single particle
excitations at different lattice points and
 represents
  a 1-particle state with rapidity $\la$. For diagonal entries we  
 obtain in general
\be A(\la)\mid 0>= \al^N(\la)\mid 0>, ~~
D(\la)\mid 0>= \bet^N(\la)\mid 0>.\ll{adv}\ee
The model dependence  comes at this stage through $\al^N,\bet^N$,
 since they
are determined by
 the structure of local $L_j(\la)$ operators as in  (\re{lvac}).

We observe  that, it is possible   to
 extract now the eigenvalues of $\tau(\la)=A+D,$ 
if it  can  be dragged  through the strings of $B's$ in (\re{evm}), without
spoiling their structure. This can be done  by 
using  commutation relations  like 
$A\ti B=\ti f\ti B A, ~~D\ti B=f \ti B D $ and hitting the vacuum $\mid 0>$
by $A$ and $D$.
The  property (\re{adv})  then  can   rebuild clearly the same  eigenvector
$  \mid \phi>_m $ and solves the eigenvalue problem. With this strategy 
in mind we write explicitly the QYBE (\re{rtt}) in  component form 
\bea
A\ti B&=& \ti f \ti B A - \ti f_1  B \ti A, \ll{ab} \\
D\ti B&=&  f \ti B D -  f_1  B \ti D, \ll{db} 
\eea
and $ [B,\ti B]  = 0$, 
where $A=A(\la),~B=B(\la),~D=D(\la)$ and 
 those with replaced arguments are 
 denoted by $\ti f=f(\mu-\la),~\ti f_1=f_1(\mu-\la), \ti A=A(\mu), 
~\ti B=B(\mu),$ etc. Here the functions 
 $ f={a \ov b} , f_1={c \ov b}$ and similarly $ \ti f, \ti f_1 $ are
contributions from the $R$-matrix (\re{rrm}) or (\re{trm})
, while the operators $A, B$ and $ \ti A, \ti B
$ etc. are
the components of $T(\la)$ and $ T(\mu)$ appearing in the QYBE.
We notice that (\re{ab}-\re {db}) 
are indeed the required type of commutation relations, but for the second
terms in both the $rhs$, where the argument of $B$
  has changed: $\ti B \rw B,$
spoiling the structure of the eigenvector. However, if we put  the
sum of all such {\it unwanted terms}= 0, we should be able to achieve our goal. 
Therefore  ignoring   the second terms for the time being and making  
use  of the first terms only, as argued  above we   obtain
\bea
A(\la) \mid \la_1 \la_2 \ldots \la_m>&=& \pr_{j=1}^m f(\la_j-\la) A(\la)\mid
0>\ll{a}\\
D(\la) \mid \la_1 \la_2 \ldots \la_m>&=& \pr_{j=1}^m f(\la-\la_j) D(\la)\mid
0>.
\ll{d}\eea
Considering  further  the action (\re {adv})  on the pseudovacuum,
  we  finally solve
 the eigenvalue 
problem to get  
\be
\tau (\la) \mid \la_1 \la_2 \ldots \la_m>=
\Lambda (\la, \{\la_j\}_1^m )\mid \la_1 \la_2 \ldots \la_m>
\ll{tev}\ee
where 
\be \Lambda (\la, \{\la_j\}_1^m ) =
 \pr_{j=1}^m f(\la_j-\la)
\al^N(\la)+
\pr_{j=1}^m f(\la-\la_j)
\bet^N(\la).
\ll{lambda}\ee
Note that equations (\re {a}-\re{d}) 
are universal within a class of models
specified by the form  of $f(\la-\mu)$ and hence determined by
the choice of the   $R$-matrix. In the  explicit eigenvalue expression 
(\re {lambda})
the   contribution from  $\al^N,\bet^N $ is the only  model-dependent part 
and in case of  $XXZ$ 
spin  chain  for example, as seen from (\re{lvac}), is  given by 
\be \al^N=\sin^N(\la+{\eta \ov 2})
, ~~\bet^N= \sin^N (\la- {\eta \ov 2}).
\ll{albet}\ee
Thus we have solved the eigenvalue problem for the whole set of conserved
currents.

 However we have left open  on the way two questions. Firstly, what
constraint one gets when the {\it 
unwanted} terms are equated to zero, as we have
done above. Secondly, what are the equations
 for determining the unknown parameters $\{
\la_j \} $. Curiously, the second  question  is answered by the
 first one itself
, i.e. the equations for $\la_j$'s can be   
obtained by  collecting meticulously all the {\it unwanted}
 terms following the
prescription of \c{xyz} and equating them to zero. It is however 
encouraging  that there is an alternative and easy way to derive 
the same equation.   
Clearly, the general eigenvalue  $\Lambda(\la, \{ \la_j \})
$ must be analytic in $\la$,
since it should generate the eigenvalues  
  of all $C_n$ through the expansion (\re {evpcn}).
Consequently, its residue at all singular 
points $\la=\la_j, j=1,\ldots, m$,   should
vanish, since $b(0)=0$ for both the rational as well as trigonometric
$R$-matrix. Therefore 
 using the residue rule $ res _{\la=\la_k} \prod_j f(\la_j-\la)
= \prod_{j \not =k}  f(\la_j-\la_k)(-a(0)) $ along with  the symmetry  $
b(-\la)=-b(\la)$, one gets 
  from (\re {lambda}) the equation  
\be 
\l({ \al(\la_k) \ov \bet(\la_k)}\r)^N=
 \pr_{l \not =k}{ f(\la_k-\la_l) \ov
 f(\la_l-\la_k)}
= ~ \pr_{l \not =k}{ -a(\la_k-\la_l) \ov
 a(\la_l-\la_k)}, ~~ k=1,2, \ldots ,m
\ll{be}\ee
 The set of relations (\re {be})
 called the {\it Bethe equation},  are  the determining
equations for the parameters  $\la_j, j=1,\ldots, m$ and  
 represents at the same time 
 the condition for vanishing of the {\it unwanted} terms.
 Thus we get  the basic equations of
	ABA  as  (\re {tev}-\re{lambda})
 giving the eigenvalues through parameters 
$\la_j$ and the equations 
(\re {be}) for determining such parameters.

\ni {\bf I.~ Lattice models}

\medskip

The eigenvalue problem 
 of various  lattice models   can be  solved readily from 
 the equations (\re {tev}-\re{lambda}) and (\re {be})
by knowing 
the functions $f(\la-\mu)$ and 
 calculating only $ \al (\la), \bet (\la)$ for concrete models.
Note that the function $f$ given by (\re {trm}) or (\re {rrm})
comes from the corresponding    $R$-matrix of the model
and therefore is the
 same for all models belonging to one integrable class.  Functions 
  $\al, \bet$    on the other hand
are related to the  Lax operator of  individual models and 
usually should be calculated separately in each case.
Let us demonstrate  this point again  on the examples  of spin models.

 \subsubsection {  $XXZ$ spin-$\ha$ chain}

The  anisotropic  $XXZ$ spin-$\ha$ chain belongs to the
trigonometric class with Hamiltonian and the Lax operators 
given by (\re {XXZ}).
We have seen in sect. 6 how the Hamiltonian of the model can be constructed
from the transfer matrix at an operator level as 
$H=C_1=\tau'(\la)\tau^{-1}(\la)_{\la=0}$. Here we aim to find its energy
spectrum. 
 For this we use the $R$-matrix input  (\re {trm}) along with 
the vacuum eigenvalue (\re {albet})   
in the Bethe ansatz equations 
 (\re {tev} ,\re{lambda},\re {be}) to obtain 
the eigenvalue 
\be \Lambda_{XXZ} (\la) =\sin ^N (\la+ \eta)
 \pr_{j=1}^m {\sin (\la_j-\la+{\eta \ov 2}) \ov \sin (\la_j-\la -{\eta \ov 2})}
 +\sin ^N \la
\pr_{j=1}^m {\sin (\la -\la_j+3{\eta \ov 2}) \ov \sin (\la-\la_j+{\eta \ov 2})}
\ll{lamxxz}\ee
by making a shift $\la \rw \la+{\eta \ov 2} $ 
and  the Bethe equation
\be 
\left(  {\sin (\la_k+{\eta \ov 2}) \ov \sin (\la_k -{\eta \ov 2})} \r)^N=
 \pr_{j \neq k}^m {\sin (\la_k-\la_j+\eta) \ov \sin (\la_k-\la_j - \eta)}
\ll{bexxz}\ee
for $j=1,2, \ldots , m.
$
Remembering that  the Hamiltonian for the $XXZ$ chain is
 $H_{xxz}=C_1,$ we get  the energy
spectrum
$~~
  H_{xxz}\mid \phi>_m= 
  E_{xxz}^{(m)}\mid \phi>_m
~~$ as 
\be
  E_{xxz}^{(m)}= \Lambda (\la)'\Lambda^{-1} (\la)\mid_{\la=0}=
\sin \eta\sum_{j=1}^m {1 \ov \sin (\la_j -{\eta \ov 2}) 
\sin (\la_j + {\eta \ov 2})}+ N\cot \eta.  
\ll{exxz}\ee

Depending on the values of the  anisotropy parameter $\De= \cos \eta,$ 
the model exhibits qualitatively different behavior. For $\De\geq 1$
 the ground
state of the chain is ferromagnetic, while $\De <1$ corresponds to the
antiferromagnetic ground state with total spin $=0$, attained at
half filling  $m={N \ov
2}.$ 
$\De=\pm 1$ is related to the isotropic case discussed below with $+1$
describing the ferromagnetic, while  $-1$ the antiferromagnetic chain.
\smallskip

\subsubsection { $XXX$ spin-$\ha$ chain}

 Isotropic $XXX$ Heisenberg spin-$\ha$ chain
 is  given by  the Hamiltonian and 
the Lax operator as in (\re{XXX})
and belongs to the integrable class with the rational $R$-matrix
 (\re{rrm}). However,
since  $q \rw 1$ limit of the
trigonometric model corresponds to the rational case,
 the  eigenvalue problem  (\re{evp})
 for the $XXX$ model can be solved cheaply 
 by  taking   this limit from  the anisotropic  result
  (\re{lamxxz}-\re{bexxz}). 
This in effect  amounts simply to the  formal replacement of 
  $~~\sin \la \rw \la,~~ \cos \la \rw 1,$
resulting 
\be \Lambda_{xxx} (\la) =(\la+ \eta)^N
 \pr_{j=1}^m { \la_j-\la+{\eta \ov 2} \ov  \la_j-\la-{\eta \ov 2}}
  +\la^N
\pr_{j=1}^m {\la -\la_j+3{\eta \ov 2} \ov \la-\la_j+{\eta \ov 2}}
\ll{lamxxx}\ee
for the eigenvalue expression along with  the Bethe equation
\be 
\left(  {\la_k+{\eta \ov 2} \ov \la_k-{\eta \ov 2}} \r)^N=
 \pr_{j \neq k}^m {\la_k-\la_j+\eta \ov \la_k-\la_j - \eta}
\ll{bexxx}\ee
for $j=1,2, \ldots , m.$
The energy spectrum for the $m$-spin   excited state
can also be    similarly obtained from (\re{exxz}) as 
\be
  E_{xxx}^{(m)}=
 \eta\sum_{j=1}^m {1 \ov \la_j^2-\eta^2}  
\ll{exxx}\ee
Note that  for the  
$XXX$ spin chain model (\re {XXX})
 the parameter is  to be set as $\eta=\pm {i \ov 2}$ in all the above formulas,
which correspond to the isotropic case $\De=\pm 1$ and is linked to 
 the   ferromagnetic
 or the  antiferromagnetic ground states, respectively.
Physical significance of the ABA 
results will become clear after we solve 
the  same problem   through the CBA and show their equivalence.

\subsubsection{
Spin chains with impurity }
For describing  integrability of the 
 impurity models considered in sect. 6, the Lax
operator $L(\la)$ at site  
$s$ in the monodromy matrix 
may be   replaced by $L(\la-\nu)$, keeping all other operators  the same.
 As a result,
for the eigenvalue solution of   spin chains with impurity,
 the model dependent part
 in the ABA relations (\re{lambda}) and (\re{be}) is changed
as $\al^N(\la) \rw\al^{N-1}(\la) \al(\la-\nu)$ and $
\bet^N(\la) \rw\bet^{N-1}(\la) \bet(\la-\nu)$.
 For $XXZ$ spin-$\ha$ chain with impurities , for example, 
 the expressions $\sin^N(\la\pm {\eta \ov 2}) $
in    (\re{albet})
 should be replaced by
$\sin^{N-1}(\la\pm {\eta \ov 2}) 
\sin(\la -\nu\pm {\eta \ov 2})$ influencing the result 
 (\re{lambda}-\re{be})  accordingly.  Therefore we may conclude, that 
the impurity parameter $\nu$
acts like an extra (but known)  rapidity parameter $\la_{m+1}$
in addition to the   set of such unknown parameters.   
The corresponding energy of the impurity model is
\be
E^{(m)}_{imp}=E^{(m)}_{hom}
+\cot ({\eta \ov 2}-\nu)-\cot {\eta \ov 2}
\ll{imxxz}\ee
where $E^{(m)}_{hom}$ is given by the homogeneous model (\re{exxz}) without
impurity.
\subsubsection{
	Inhomogeneous models}
In inhomogeneous models, extending the concept of 
impurity to every  site the arguments of all Lax operators are shifted
by $\nu_j$ as (\re{tauIH}). Therefore in the ABA equations one should replace
now 
\be
\al^N(\la) \rw \pr_{j=1}^N  \al(\la-\nu_j),~~
\bet^N(\la) \rw \pr_{j=1}^N \bet(\la-\nu_j).
\ll{vacIH}\ee
The energy spectrum for the inhomogeneous $XXZ$ model therefore  results to
\be
E^{(m)}_{inhom}=E^{(m)}_{hom}
+\pr_{j=1}^N \cot ({\eta \ov 2}-\nu_j)- N \cot ({\eta \ov 2})
\ll{ihxxz}\ee

 \subsubsection{
 Other examples of regular models}
Let us start with the more general regular
$L(\la)$ operator of the form (\re{LP}) and extend eigenvalue relation 
(\re{lambda}) for the spin models to  
\be \Lambda (\la, \{\la_j\}_1^m ) =
 \pr_{j=1}^m {a_+(\la_j-\la) \ov b_-(\la_j-\la)}
a_+^N(\la)+
\pr_{j=1}^m {a_-(\la-\la_j) \ov b_-(\la-\la_j)}
b_-^N(\la).
\ll{glambda}\ee
and the Bethe equation  (\re{be}) to
\be 
\l({ a_+(\la_k) \ov b_-(\la_k)}\r)^N=
 ~ \pr_{l \not =k}{ -a_-(\la_k-\la_l) \ov
 a_+(\la_l-\la_k)}, ~~ k=1,2, \ldots ,m
\ll{gbe}\ee
 
\ni i) {\it Wu-McCoy model}

This is a twisted $XXZ$ model constructed in (\re{WM1}), where the
parameters are given by (\re{twiab}) as
 $b_\pm(\la)= e^{\pm i \th} \sin \la, ~~a_\pm(\la)= \sin (\la+\eta)$.
Therefore  the Bethe equation (\re{bexxz}) of the original $XXZ$ chain  
,as seen from (\re{gbe}) 
is changed by the appearance of  an extra factor $e^{i \th N}$
 in its {\it lhs}, while   the energy spectrum 
remains the same as (\re{exxz}) due to cancelation of $\th$-factors.
 
\sk \ni ii) {\it $XXZ$ in a magnetic field}

We have seen in sect. 6 that the color
 twisting of the $XXZ$ chain amounts to putting it in a 
constant magnetic field.
Therefore,   using  the related  parameters
(\re{colab})  one finds that due to ${ a_+ (\la) \ov b_- (\la)}= 
{ a (\la) \ov b (\la)}$ and ${ a_- (-\la) \ov a_+ (\la)}= 
{ a (-\la) \ov a(\la)}$ the Bethe equation remains the same, while 
an extra constant term $\ha c N $ appears in the energy spectrum 
coming from $a^N_+(\la)$, which answers for  the effect of the  
constant magnetic field ${c \ov 2}$.

\sk \ni iii) {\it Free fermion models}

It is interesting to check how the {\it  freedom} of the free
 fermionic models are reflected in the Bethe ansatz equations.
From the parametrization (\re {ffR}) of such a simple  model 
we find that ${a_-(-\la) \ov a_+(\la)}={\cos (-\la)  \ov \cos \la}=1$,
which readily gives the {\it rhs} of the Bethe equation (\re{gbe})$=-1$,
signifying trivialization of the   $m$-particle scattering matrix and
 confirming thus the
noninteracting nature of the model.

Similarly for the free fermionic model generated from 
the nonstandard $R$-matrix with (\re{nstR}), we have
${a_(-\la) \ov a_+(\la)}={\sin (\la+\eta)  \ov \sin (\la+\eta)}=1$
yielding  again the trivial scattering matrix.
\subsubsection{ String solution and thermodynamic limit}
\ni i) {\it String Solution}

Though the general solution of the Bethe equations (\re {bexxz},
\re{bexxx}) is difficult to find,  a particular form of complex solution, 
called {the  string solution} of length $m=2M+1$ can be given as
\be 
\la_n =
 x+ i n, ~n=-M,-M+1, \ldots, M-1,M.
\ll{string}\ee
The corresponding excitation energy can be calculated  by inserting the 
 string solution in the expressions like (\re {exxz}, \re{exxx}).
For example,  the $XXX$ spin model gives 
\be
  E_{xxx}^{(str)}= 
 \sum_{n=-M}^M {1 \ov (x+ i n)^2 +{1 \ov 4}}= {2(M+ {1 \ov 2})   \ov x^2
 +(M+ {1 \ov 2})^2}   
\ll{strxxx}\ee

\sk

\ni ii) {\it Thermodynamic limit}

The thermodynamic limit or the large system limit at $N \rw
\infty$  with finite lattice spacing $\De$ 
is  important from the physical point, when the 
sum in the logarithm of the Bethe equations turn into an integral: $
 {1 \ov N }\sum_{j=1}^{m} f_i \rw 
\int_{x_0}^{x_1} dx f(x)$ and the equations become
 easier to
handle in many cases.  For example, in  antiferromagnetic 
$XXX$ chain with total spin zero for the ground state, we choose 
the half filling  $m= {N \ov 2}$
and for calculating the momentum $p_j$ take the 
 logarithm of the Bethe equations (\re {bexxx}) to get 
\be 
\tan^{-1} 2 \la_k=
 {\pi Q_k \ov N} + {1 \ov N }
\sum_{j=1}^{N \ov 2}\tan^{-1} (\la_k-\la_j),
\ll{belxxx}\ee
where the integers (or halfintegers) $Q_k$ 
take all  successive  values between
  $-{m-1 \ov 2 } \le Q_k \le  {m-1 \ov 2 }$ 
for the ground state.
  At the thermodynamic limit 
${ Q_k \ov N} \rw x, ~~  \la_k
\rw  \la(x),$  and since for the ground state 
${m-1 \ov 2 } \rw {N \ov 2}$, we get the ranges
 $-{1 \ov 4 } \le x \le  {1 \ov 4 },~~ -\infty \le \la  \le  \infty
$, which   reduce the Bethe equation  (\re {belxxx})
 to an integral equation
\be
~\tan^{-1}  \la (x) =
 {\pi x} + \int_{-{1 \ov 4}}^{{1 \ov 4}}d y~\tan^{-1} (\la(x)-\la (y)). ~~
\ll{tdeq1}\ee  Differentiating (\re{tdeq1}) 
and introducing the density 
$\rho (\la) = \l( {d \la(x) \ov dx }\mid_ {x=x( \la)} \r)^{-1}$ 
of $\la$ in the interval $d \la$  
one gets
\be
\pi \rho(\la) 
  + \int_{-{\infty}}^{{\infty}}d\mu~ {\rho (\mu) \ov 1+(\la-\mu)^2 }
= {2 \ov 1+ 4 \la^2}. 
\ll{tdeq2}\ee
The equation  can be  solved using the Fourier transform
\be~ \rho(\la)={1 \ov 2 \pi} 
\int_{-{\infty}}^{{\infty}}d\xi~ {e^ {-\ha \mid \xi \mid}
 \ov 1+e^ {- \mid \xi \mid} }e^ {-i\la  \xi }
  = {1 \ov 2 \cosh \pi \la }.\ll{rho}\ee
Total energy of the antiferromagnetic ground state therefore is
\be
E_{0}-{N \ov 2}=N \int h(\la) \rho (\la) d \la= -N \int_{-\infty}^\infty
d \la ({1 \ov  \la^2 + \qr}) \rho (\la)=-2 N \ln 2. \ll{tde}\ee 
The energy $E_0$  of an anisotropic chain in the antiferromagnetic regime
$-1<\De<1$ is calculated in a similar way \c{izumov}.
%-----------------
\subsubsection{
  functional Bethe ansatz }
For some integrable models though the $R$ matrix and $L$ operators are
known,
due the specific structures of the Lax operator or the boundary conditions,
sometimes it is difficult 
to construct the reference state $\mid 0>  for such models.$ 
As a consequence the standard  ABA method  
 is not applicable to them. Nevertheless, by the application of 
 another method called 
 { functional Bethe ansatz (FBA)} \c{toda,fba,baxter},  it is often possible to achieve
factorization  to reduce the many body problem  to a single particle one.
We briefly touch upon this method on the examples of Toda chains.

\sk \ni i) {\it
Nonrelativistic and  Relativistic  Toda chains }

 We have seen  the systematic  construction  of  the 
$R$-matrix and the Lax operators of the relativistic (or time discrete) 
 and nonrelativistic (time continuous) Toda chains as well as 
 the explicit derivation of the  respective Hamiltonians. However 
due to the presence of exponential functions of fields 
 in the Lax operators, the  reference state with $C(\mu)\mid 0>=0$
 is difficult to find for them.
We demonstrate in brief 
the application of FBA to the Toda chains following the  
argument of \c{toda}.

 Using QYBE (\re{rtt})  we find first
 $[C(\xi),C(\zeta)]=0$, a crucial relation for the application of  FBA, 
along with functional relations between $C(\xi), A(\zeta)$ analogous  to
(\re{ab}):
\be
A(\la) C(\mu)=  f(\la-\mu)  C(\mu) A(\la) -  f_1(\la-\mu)  C(\la)  A(\mu),
 \ll{ac} \ee
and similar relation between
$C(\xi),D(\zeta)$.
Defining now commuting operators $\theta_j, j=1,\ldots ,N-1$
 as $C(\lambda=\theta_j)=0$
and conjugate operators $\Lambda^-_j=A(\lambda\rightarrow \theta_j), \ \
\Lambda^+_j=D(\lambda\rightarrow \theta_j)
$
with the proper operator ordering prescription \c{toda}, we get from 
(\re{ac}) and from the relation between $C, D$ the simple equations
\be
\Lambda^-_j C(\mu)=  f(\th_j-\mu)  C(\mu) \Lambda^-_j ,~~ 
\Lambda^+_j C(\mu)=  f(\mu -\th_j)  C(\mu) \Lambda^+_j
\ll{acdc}
\ee
Let us consider the case of  the
 nonrelativistic Toda chain related to the rational $R$-matrix, for which  
we have $ f(\th_j-\mu)={ \mu-(\th_j-\eta)
\ov  \mu-\th_j}$ and interpolating  the zeros at $\th_j$ we construct 
$C(\mu)= e_N^- \pr_{j=1}^{N-1} (\mu-\th_j ),$ using the input of the 
trivial commutation of  $C(\mu)$ at different $\mu$.
  From (\re{acdc}) we conclude that  
 the operators $\Lambda^{\pm}_j$
shift the operators  $\theta_j$ by $\mp \eta$ and  since
it is true for the  symmetric polynomials in $\th_j$, one gets for the 
symmetric function $\phi (\th_1, \ldots ,\th_N)$ the important
relation \c{toda}
\bea
& & t(\th_j) \phi (\th_1, \ldots ,\th_{N-1})= (\kappa^{\ha}
\Lambda^-_j +\kappa^{-\ha}\Lambda^+_j) \phi (\th_1, \ldots ,\th_{N-1})
\nonumber \\
 &=&\kappa^{\ha}i^{-N}\phi (\th_1, \ldots ,\th_j-\eta,\ldots ,\th_{N-1})
+\kappa^{-\ha}i^{N} \phi (\th_1, \ldots ,\th_j+\eta,\ldots ,\th_{N-1})
\ll{fba}\eea
where $i^{\mp N}$ is the eigenvalues of $\Lambda^\mp_j$.
To  achieve separation of variables 
for the $N-1$-particle eigenfunction we represent 
$\phi (\th_1, \ldots ,\th_{N-1})= \prod_j^{N-1} \phi_j(\th_j)$ to get 
 \begin {equation}
t(  \th_j) \phi_j(\th_j)=
i^{-N} \kappa^{\ha}\phi_j(\th_j-\eta)+
i^{N}\kappa^{-\ha} \phi_j(\th_j+\eta)
\ll{fba1}\end {equation}
%27

The FBA method  described above  is applicable
 also to its  relativistic (discrete time) generalization \c{qrtoda} 
related to the trigonometric $R$-matrix by expressing    the
spectral parameter
as  $\xi=e^{i\lambda}$ and considering the zeros of $C(\mu)$ at $\th_n$ with 
$, \ \ \xi_n=e^{i\theta_n}$.
The   separation of variable equation takes in this case the form
 \begin {equation}
t(\xi_j) \phi_j(\xi_j)=
i^{-N} \kappa^{\ha}\phi_j(\xi_j e^{-i\eta})+
i^{N}\kappa^{-\ha} \phi_j(\xi_je^{i\eta})
\ll{fba2}\end {equation}

Possibility of such solutions is also related to the { pair propagation 
through a vertex} for the $R$-matrix \c{baxter}, i.e. the existence of  vectors
$h^{in(out)}, v^{in(out)} $ such that 
$R(\la)^{\nu \bet}_{\mu \al} v^{in}_\bet h^{in}_\nu
 = v^{out}_\al h^{out}_\mu$.
A recent application of FBA to solve $XXZ$ spin chain with antiperiodic
or more general boundary conditions  can be found in \c{fba,fba1}.

\ni {\bf  Field models}

\medskip

We focus here  on the specializations of the ABA, that one has to make
  for application to 
the field theoretic models,
 over that
   for the lattice
 models described above.
We have seen in sect. 4 
that  the integrable field models like NLS, SG , DNLS 
etc. can be lattice regularized with exact     Lax operators 
 exploiting their underlying algebraic structures. Therefore  such
 lattice versions of the field models defined in a finite interval $[-l,l],
~l=N \De$ with lattice spacing $\De$ 
satisfy the  global QYBE (\re {rtt})
and can be solved through the same ABA scheme described above.

 Our aim however is to explore the
continuum models with $\De \rw 0$ and at  infinite interval $l \rw \infty.$
Involvement of these two  limiting procedures makes the related ABA  a bit
different from and slightly difficult than the ABA  for the discrete  models.
However,  we achieve in turn some simplifications, e.g. we completely
get rid of the problem with 
{\it unwanted} terms and consequently of the Bethe equations.
We may consider first the $\De \rw 0$  limit by defining
\be T_l(\la)= \lim_{\De \rw 0} \pr^{2N}_{n=1} L_n(\la)
\ll{TL} \ee
which still satisfies  relation (\re{rtt}).
Another advantage is that, since we are interested in field models at 
$\De \rw 0$, in most of the cases 
 it is not necessary to deal with much complicated forms of the
exact lattice versions of the Lax operators (see the list in sect. 2), but
is sufficient to use $L_n$ correct up to order $\De,$ i.e.
$L_n=1+ \De U_n(\la),$ where $ U_n(\la)$ is the naive discretization of the Lax
operator $U(x,\la)$ 
 of the continuum model. 

Let us find now the asymptotic at $n \rw \infty,~~$ when 
 the field operator, say $\psi_n $ vanishes in the 
weak sense  $\mid \psi_n \mid \rw 0$, resulting 
\be L_n(\la)  \rw V(\la).~~\mbox {and}~~
~L_n(\la) \otimes L_n(\mu)  \rw W(\la, \mu),
\ll{vw}\ee
which may differ from $V(\la)\otimes V(\mu)$
due to normal ordering of the operators.
At the infinite interval limit the QYBE (\re{rtt}) is 
 needed to be regularized, for
which we define the scattering matrix as 
\be T(\la)= \lim_{N \rw \infty} V(\la)^{-N} T_l(\la) V^{-N}(\la)
=  \l( \begin{array}{c}  A(\la) \qquad  B(\la)\\
\ep B^\dagger(\la) \qquad A^\dagger (\la) \end{array} \r),~~ \ep= \mp.
 \ll{Tal} \ee
Observe that at $\infty$ interval limit  $SU(2) (SU(1,1))$ 
symmetry is usually restored for
the monodromy matrix. The regularized QYBE using the equation
$  R(\la-\mu) W(\la, \mu)= W( \mu, \la) R(\la-\mu) $ now looks like 
\be
R_+(\la-\mu)T(\la)\otimes T(\mu)= (I \otimes T(\mu))( I   \otimes T(\la))
R_-(\la-\mu)
\ll{qybe+-}\ee
with 
$R_\pm=\ti S_\pm^{-1} R S_\pm, ~$
where  $ S_\pm= S_\pm (\la, \mu) = \lim_{N \rw \infty} W^{\mp N}
(\la, \mu) V^{\pm N}(\la) 
\otimes V^{\pm N}(\mu)$ 
and $ \ti S_\pm=S_\pm(\mu,\la)$ carry information about the asymptotic behavior.
Unlike the  lattice models involving   the $R$ matrix only, 
 the explicit form of $S_\pm$ and consequently $R_\pm $ matrices  
become model
dependent. We consider below some examples which will clarify this point.
 $R_\pm $-matrix generally  takes the form
\be
 R_\pm =\l( \begin{array}{c}  f\qquad \ \ \ \ \ \ \ \\  \ \ 1  \ \   
f_{1\pm}\ \ 
 \\  \ \   \hat f_{1\mp} \ \ 1 \ \   \\
 \ \ \ \ \ \ \ \qquad  f  \end{array} \r),
 ~~\quad  f={ a \ov b} ,~~~ f_1={ c \ov a} \ll{r+-} \ee
with $\hat f_{1\pm}(\la,\mu)= \lim_{l \rw \infty} 
f_1 (\la,\mu)e^{ \mp i \rho (\la,\mu) {l \ov 2} },$ where  
 the function $\rho$  depends on the concrete model.
The limits of  such singular terms must be taken carefully in the
principal value sense
$$  \lim_{l \rw \infty}{\cal P} 
{ e^{ \mp i \rho  {l } } \ov \rho }= \pm i \pi
\de (\rho)  $$ and after properly normal ordering the operators.
\subsubsection{NLS model}
 Let us elaborate the subsequent  steps on the example  of NLS field
\c{nls,fadrev}
model involving  the fields
$ \psi, \psi^\da$ with $ [\psi (x), \psi^\da(y)]=\de(x-y)$, where 
\be
 \psi_n= {1 \ov \De} \int^{x_n}_{x_n -\De} \psi (x) dx
\ll{psin} \ee and  
described by the Hamiltonian  and the Lax operator as in  (\re {nls}).
The discrete Lax operator for our purpose may be chosen simply 
by discretizing (\re{nls}) as 
\begin{equation}
L_n(\la)= {\bf I}-i \De {\cal L}_{n}(\la)  = {\bf I}-i \De
\left( \begin{array}{c} \la \
\quad \eta^\ha \psi^\da \\
\eta^\ha  \psi \ \quad -\la
    \end{array} \right).
\ll{d- nls}\end{equation}
Therefore the asymptotic  expressions  at $\psi \rw 0$ give
$V(\la)={\bf I} {-i \De \la \si^3}, ~~W(\la.\mu)=
{\bf I} -i \De( \la \si^3\otimes I+ \mu I \otimes \si^3 -i\eta \si^- 
\otimes \si^+) $ which in turn construct $S_\pm$ and $R_\pm$ matrices.
Using  the 
 associated  rational $R$-matrix  (\re {rrm}) we get  $R_{\pm}$
in the form   (\re {r+-})   with 
$f_{1\pm}(\la,\mu)= \lim_{l \rw \infty} 
{\eta \ov \la-\mu }e^{i \mp  (\la-\mu) {l \ov 2} }=
  2 \pi i \eta \delta (\la -\mu),$  which yields from 
(\re {qybe+-})
the relations
\be
A \ti B= \ti f \ti B A ~~ -~~ 2 \pi i \eta \delta (\la -\mu) B \ti A 
\ll{abnls1}\ee
with $[A, \ti A] = 0 $  
and the notations same  as in (\re{ab}-\re{db}). 
In handling such singularities it is always better to extract the limits
only at the final stage after performing all other operations. 

Compared to   (\re{ab}-\re{db})
in place of {\it unwanted} terms
 there appears  in (\re{abnls1}) a term with $
\delta (\la -\mu),$ which  disappears for $\la \not = \mu. $
Therefore we can repeat here the  arguments of  
 the above ABA scheme
for constructing the reference state $\mid 0>$ as well as  the $m$-particle 
eigenstates as
$  \mid \phi>_m= 
\mid \la_1 \la_2 \ldots \la_m>
= B(\la_1)B(\la_2) \ldots B(\la_m)\mid 0> .
$ The eigenvalue expression consequently is given again by   
(\re {lambda}) 
 considering  the regularization (\re
{Tal}) and yields $\lim_{N \rw \infty}\al^N \mid 0>=\lim_{N \rw \infty} 
e^{i \la \De \si^3 N}T_l(\la)e^{i\la \De\si^3 N}=\mid 0>$
However,  due to the absence of  any constraint like Bethe equation,
parameters $\{ \la_j \}$ are   arbitrary in continuum models. This is
 an
important and major difference between  the ABA for the field 
 and 
the lattice models.    

 Note that  the NLS model is  intimately connected
with the spin models through Holstein-Primakov transformation (\re {hpt})
and belongs to the same rational class as the $XXX$ chain. We can formally
derive  the 
 eigenvalue of $A(\la)$ for the NLS  from (\re {lamxxx}) as
$ \Lambda_{nls} (\la) =
 \pr_{j=1}^m { \la-\la_j +i{\eta \ov 2}  \ov \la- \la_j -i{\eta \ov 2} }
$ considering $\al^N=1$ and a shift $\la \rw \la-i{\eta \ov
2}.$
 
Consequently  the eigenvalues for  the conserved quantities 
 for the NLS system are generated by the expansion  $ln 
\Lambda_{nls}(\la)=\sum_{n}
 C_{-n} \la^{-n}$
 at $\la \rw \infty $
and 
$~~
=  \sum_{j=0}^N C_{n} \la^{n}
 ~~$ at $\la \rw 0$ in correspondence to the respective 
 conserved opeartors.
Since $H_{nls}= C_{-3},$  one finds therefore the energy spectrum 
for the NLS model  as $
E_{nls}=  \sum_{j=1}^N \la^2_j.
 $
The case $\eta=- \mid\eta \mid <0$
 corresponds to the NLS model with {\it attractive} interaction
and allows also bound or {\it soliton} 
states given by the string like solutions
$   \la_j =P_0 -i \mid\eta \mid  (j-{m+1 \ov 2}), ~~j=1,\ldots, m$.
 Due to such choice of 
$\la_j$ the poles of the individual factors in $\Lambda_{nls}(\la)$ are
canceled by the zeros of the neighboring factors.  As a result  
we get $\Lambda_{soliton}(\la)= 
 { \la-P_0 -i{\mid \eta \mid m\ov 2}  \ov \la- P_0 +i{\mid \eta \mid 
m \ov 2} }
$, which though constructed from $m$-number of particles as a bound state, 
 mimics   clearly the structure of  a single particle and hence is called {\it
soliton}. Note the curious fact that the same structure  with 
 zeros and corresponding poles at discrete points in the complex $\la$
plane also  appear in the classical soliton solution
 for the scattering 
matrix element $a(\la),$  as mentioned in the introduction. 
  Thus   for such 
$m$-particle bound  state, the particle number $N=C_{-1}$, the 
momentum $P=C_{-2}$  and the energy spectrum $E_m=C_{-3}$ can be obtained as 
\be
 N=m,~ P=P_0, ~~ E_m= {P_0^2 \ov m}-{\eta^2 \ov 12}(m^3-m) \ll{solnls}\ee
The other set of conserved operators  $C_n,~n \geq 0$  leads usually  to
nonlocal quantities.

 Interestingly, the Hamiltonian of the NLS field model 
is equivalent to the 
$\de$-function Bose gas  as the   
  interacting $N$ body system  with the Hamiltonian 
\be H= - \sum_{j=1}^N {\pa \ov \pa x^2_j} + 2 \eta \sum_{i <j}\de
(x_i-x_j)  \ll{dbose}\ee
Therefore  
the energy eigenvalue problem $H\mid n>=E_n\mid n>$ for the NLS model \c{nls}
also coincides formally with that for the $\de$-function Bose gas solved 
in \c{dbose,korbook}. 
\subsubsection{ DNLS  model}
Unlike the well known  nonultralocal  derivative NLS \c{dnls1}, the DNLS 
model given 
by (\re{dnls}) is ultralocal    with the field commutation relation 
$[\psi(x),\psi^\da (y)]= -\sin h \de(x-y) $ and, as shown in sect. 5, has 
a  remarkable connection with the quantum algebras and $q$-oscillators. 
We show here how one can  solve  the eigenvalue problem 
of this quantum field model exactly  through the 
ABA scheme  \c{kb-dnls}.
Though the exact lattice version for  the Lax operator of the model 
is available as (\re{Ldnls1}), it  is sufficient to consider
  such  operator  correct up to order  $\De$ : $L_n(\la)= { I}+i \De {\cal L}_{n}(\la)$
by using (\re{psin}), 
where ${\cal L}_n$ is a simple disctretization of the field model
 ${\cal L}_{dnls}$ given in (\re{dnls}). The coupling constants 
$\kap_\pm$ are nontrivially renormalized in the quantum case as 
 $\kap_\pm= {e^{{i \mp \eta \ov 2}} \ov \cos {\eta \ov 2}}.$

 Since the model belongs to the
trigonometric class, the ABA  should  incorporate  the case
(\re{trm})  and  follow the
  formalism  described above for  the field models  involving the limits
$N \rw \infty, \De \rw 0$.
The structure of the Lax operator and the commutation
relations derive the asymptotic matrices (\re{vw}) as 
\bea V(\la)&=& \l({\bf I}-{i\De \ov 2} \xi^2 \si_3 \r) I,
\nonumber \\ W(\la,\mu)&=&
{\bf I}\otimes {\bf I}-{i\De \ov 4}\l( \xi^2 (\si_3\otimes I) +\zeta^2
 (I\otimes\si_3 )\r)
 +\De \xi\zeta ~ \sin \eta ~ \si_- \otimes \si_+  
\ll{vwdnls}\eea
The $R_\pm$ matrix has the structure of (\re{r+-})
with $$ f_{1\pm}(\xi,\zeta)= \lim_{l \rw \infty} 
f_1 (\la,\mu)e^{ \mp i (\xi^2-\zeta^2) {l \ov 2} }=2 \pi\xi^2 \sin \eta 
\de (\xi^2-\zeta^2) $$  
 and the QYBE relations have the  form (\re{qybe+-}) 
with  specializations for $R^\pm$ as  mentioned yielding 
\be
A \ti B= \ti f \ti B A ~~ -~~ 2 \pi \la^2 \sin \eta \delta (\la^2 -\mu^2)
 B \ti A .
\ll{abdnls1}\ee

Repeating the   steps described above and using further  
the conditions like $A(\la) \mid 0>= \mid 0>,$
 one gets  the eigenvalue solution  
$ \Lambda_{dnls} (\la) =
 \pr_{j=1}^m {\sinh ( \la-\la_j +i{\al \ov 2})  \ov 
\sinh (\la- \la_j -i{\al \ov 2}) }$. This 
  allows also the string-like solution
(\re{string}) as in the NLS model, 
 resulting the $m$-particle bound state with the spectrum
\be C_0= \eta m,
,~C_1= 2 p \sin (\eta m) ,~C_2= 8\eta H_m=  p^2 \sin (2 \eta m)
\ll{ednls} \ee
and the binding energy $$\De E= H_m-m H_1= {  p^2  \ov 8 \al}
(\sin 2 \al m -m \sin 2 \al)<0.$$
\subsubsection{ Sine-Gordon model}
This  relativistic field model  has been studied 
intensively highlighting its various aspects \c{sgFad,sgothers}.
 We have  generated its Lax operator in sect. 4 
 as a direct realization of the well
known
quantum algebra  and discussed   the problem of 
constructing its  quantum Hamiltonian in sect. 6.
Here we  apply  the ABA scheme to the  model 
for 
determining the  exact  energy spectra. 

\sk 

 \ni i) {\it Lattice regularization} 

For solving the SG field model by lattice regularization \c{sgFad},
 one finds that 
the naive disctretization does not work here and one  has to start 
  from the exact Lattice Lax operator given in 
 (\re{L-sg}). Since the SG model is associated with the trigonometric
$R$-matrix (\re{trm}) same  as the $XXZ$ spin chain, the discretized model 
follows the same ABA scheme giving  relations like (\re{ab},\re{db}),
 (\re{lamxxz}), (\re{bexxz}) etc.  with specialization 
$D_l(\la)=A_l(\la)^\da$ and $A_l\mid 0>=e^{ \al(\la) l} \mid 0>,
~~D_l\mid 0>=e^{ \al^*(\la) l} \mid 0>$. 
 However there are also some major
differences.
 Firstly,  the action of the discrete Lax operator 
(\re{L-sg}) on the pseudovacuum: $L_n\mid0>$ does not reduce to the
triangular form, which should be tackled by defining a new Lax operator
\be \hat L_n(\la)= L_n (\la) L_{n+1}(\la)\ll{l2}\ee
 as the product of two neighboring operators. 

Unlike the above discussed models,  the physical vacuum $\Omega$ is  
obtained here by filling the state $\mid 0>$ by particles as $\Omega=
\pr_\rho
B(\la_\rho+i {\pi \ov 2}) \mid 0>$ \c{sgFad}. Therefore $1$-particle state is
described as the excitation over the physical vacuum as
$ \mid 1>= B(\la_0)\pr_\rho
B(\bar \la_\rho+i {\pi \ov 2}) \mid 0>.$
 At the limit $l \rw \infty$, the second
term in (\re{lambda}) becomes negligible and the corresponding eigenvalue of
$\mid 1>$ normalized by that of the vacuum state gives
\be A_1(\la,\la_0)={\La(\la,\la_0) \ov \La(\la)}=
f(\la_0-\la){ \pr_{\rho \neq h}f(\bar \la_\rho+i {\pi \ov 2}-\la)
\ov  \pr_{\rho }f(\la_\rho+i {\pi \ov 2}-\la)
}\ll{A1}\ee
Now making a shift $\la \rw \la-{i\eta \ov 2}$ with a scaling of $\la, \la_0$
 , using the $R$-matrix input (\re{trm}) and finally
taking the length  $l \rw \infty$ in (\re{A1}),
we get for the $m$-particle excitation \c{sgFad}
\be A(\la) \mid m>= \pr_{k=1}^m {\tanh( {\la-\la_k -i\ga}) \ov 
\tanh( {\la-\la_k +i\ga}) }\mid m>
\ll{evsg}\ee
where $\ga={\pi \eta \ov 4 (\pi -\eta)} $ and $ \mid m>
=  B(\la_1) \ldots B(\la_m)\pr_\rho
B(\bar \la_\rho+i {\pi \ov 2}) \mid 0>$. Note that each factor
in (\re{evsg}) corresponding to a single particle excitation has  
 {\it two} zeros and corresponding poles in the complex $\la$ plane,
which 
describe   a {\it breather} like quantum  state representing bosonic
particle of the theory. One can consider also a $M$-particle bound state 
by considering a string-like solution for the parameters $\la_k$.
Curiously, the more fundamental {\it kink}
 state with nontrivial topological charge
is difficult to obtain in this model.  Expanding $\ln A(\la)$ from 
(\re {evsg}) as $\sum _n C_n
\la^n $  at $~\la \rw 0~~~
$ and as $~~\sum _n C_{-n}
\la^{-n}$ at  $ ~~\la \rw \infty,$
 we can get the energy and the momentum eigenspectra 
as
\[
H={i m \ov 8 \mid \ga' \mid } (C_{-1}-C_{1}),~~~~
P={i m \ov 8 \mid \ga' \mid } (C_{-1}+C_{1}).\]

\sk

\ni ii) {\it Continuum approach}
 
 Alternatively, 
   one can  solve the eigenvalue problem of the SG field model 
starting from    the continuum Lax
operator (\re{sg}) itself 
by  considering carefully the short distance behaviors   
\c{sgGS}. This more direct approach leads also to
the same particle spectra given through the breather state (\re{evsg}),
but draws an important conclusion that such solutions are possible only 
for the parameter value $\mid \eta\mid  <\pi$.
For getting the asymptotic solution $\lim_{n \rw \infty}L_n \rw V(\la)$
in the diagonal form, one has to make a gauge transformation over the
Lax operator  (\re{sg}) amounting  to a cyclic replacement of
$\si^a,a=1,2,3$. As a result the $R$-matrix changes to a form
\bea R(\la,\mu)&=&P\l({ I}\ot{ I}+X (\si^3 \ot \si^3 + \si^1 \ot \si^1)+
Y \si^2 \ot \si^2 \r)\nonumber \\
X&=& {1+\eta^2 \ov 1-\eta^2} g(\la,\mu),~~Y= \l( 1-
{2i\eta \ov 1-\eta^2} \tanh {\la-\mu \ov 2}\r) g(\la,\mu) \ll{sgr}\eea
where $ g(\la,\mu)= f(\la,\mu)\l( 1-
{2i\eta \ov 1-\eta^2} \coth {\la-\mu \ov 2}\r)^{-1}.$
Therefore in place of (\re{trm})  one gets  
the function $f= {a \ov b}$ as $f(\la,\mu)=
\l( {1-i\eta  \coth {\la-\mu \ov 2} \ov 
1-i\eta  \tanh {\la-\mu \ov 2}}\r)$, which after defining $\ga=2 \tan^{-1}
\eta $ and making a shift $\la \rw \la +i{\ga \ov 2}$ coincides with the
factors appearing in (\re{evsg}). 
At the continuum we construct  (\re{r+-}) using 
this modified $R$-matrix and the QYBE (\re{qybe+-}) to get   
\be
A (k)  B(k')=  f(k,k')  B(k') A(k) ~~ +~~ 2 \pi \ga k_0
 \delta (k_1-k_1') B(k)  A(k) 
\ll{absg}\ee
with $k_1=2\sinh \la,~k_0=2\cosh \la,$ which  consequently derives the 
same breather-like spectrum  (\re{evsg}) for the eigenvalue expression.
Note that the only difference appearing here  is in the value of the 
 renormalized coupling parameter
$\ga.$ 

The SG model in the light-cone coordinates using ABA has also been 
 solved  in \c{sgKS}.
\subsubsection{Liouville model}
This  relativistic field model is given as in (\re{lm}).
Lattice regularized version  is  usually represented by the exact discrete  Lax
operator (\re{Llm}). However in (\re{l-llm1})
 another such Lax operator is derived, 
  which appears to be more suitable for the
Bethe ansatz solution \c{liuFad}.
Like the  SG model one has to  use the 
product of  Lax operators  (\re{l2}) in constructing   the pseudovacuum.
Since the model belongs to the trigonometric class, 
 the ABA scheme goes parallel to the  spin-$\ha$ $XXZ$ chain described
above with eigenvalue (\re{lambda}), Bethe equation (\re{be}) etc. by
incorporating (\re{trm}).
 The  difference appears only  in the model-dependent part (\re{albet}) 
 defining the
 action of $A(\la)$ and $D(\la)$  on the pseudovacuum. This yields  the
$lhs$ of the Bethe equation  as
\be \l({\al \ov \bet}\r)^N= e^{i\eta N}\l({\sinh(\la_k-i{\eta \ov 2}) \ov
 \sinh(\la_k+i{\eta \ov 2})}\r)^N. \ll{beliu}\ee
Comparing with the Bethe equation for the spin-$S$ case 
 one may conclude, that  the lattice Liouville model corresponds formally 
to a
spin $(-\ha)~~XXZ$ chain
with an extra phase  factor $ e^{i\eta N}$ \c{liuFad}.
 Recall that, similar phase
factor arises under twisting  through (\re{gbe}).
The momentum and energy eigenvalues of the discrete Liouville  model
is the same (with opposite sign) to those of the spin $+\ha,~~XXZ$ chain.
Since discretized  model is defined in a finite volume $2 \pi= \De N$, the
$N \rw \infty$ limit would lead to the Liouville field model yielding
the corresponding eigenvalues at this limit.
Bound state solutions, conformal properties and other details of the model
can be found in  \c {liuFad}. 
\subsection  { Nested Bethe ansatz}
The ABA described above 
 is designed for the $SU(2)$ related models with $2
\times 2$ Lax operators. We mention in brief how to  generalize
this technique for models  with $SU \mathsf{(N),~ N} > 2$ having $
\mathsf{N \times N}
$ Lax
operators.   Examples of such models are  the 
Perk-Schultz \c{perks} type model 
associated with the trigonometric
$R$-matrix (\re{perks}) and  the vector NLS model with 
rational $R$.
The method,  called  
nested Bethe ansatz (NBA), consists of a recurrent  
application of the ABA or their {\it nesting} for subsequent 
solutions  of the higher 
rank problems using  those  of the lower  ones.
We demonstrate it for the case  $
\mathsf{N}=3, $
since the generalization to $
\mathsf{N}>3$ then becomes straightforward \c{nba}.

  We start
as in the standard ABA with the QYBE (\re {rtt}),
  where  the $R$ matrix is  taken now as a more general 
solution ( obtained also in (\re{rtrign}) through Yang-Baxterization) 
\be
R(\la)= \sum_{i,k}^3 f(\la) E_{ii}\otimes E_{ii} +
 E_{ii}\otimes E_{kk}
+f_1 (\la)(E_{ik}\otimes E_{ki}
  \ll{hrm}\ee
with $f,f_1$  given by the  same expressions as in  (\re{trm},\re{rrm}) and 
the monodromy matrix is  in the  $3 \times 3$ form
\bea
T(\la)= \l( \ba T_{00} \ \ \ \quad B_1 \ \quad B_2 \nonumber \\
C_1 \ \ \ \quad T_{11} \ \quad T_{12} \nonumber \\
C_2 \ \ \ \quad T_{21} \ \quad T_{22} \ea \r). \ll{T3} \eea
This gives the relations 
\bea
T_{00} \ti B_\al &=& \ti f \ti B_\al T_{00}  - \ti f_1  B_\al \ti T_{00},
 \ll{hab} \\
T_{\al \al } \ti B_\bet &=&  - f_1  B_\al \ti T_{\al \bet } D +  \ti 
B_\ga T_{\al \tau } R^{(1) \tau \ga}_{\al \bet },  \ll{hdb}\\
\mbox {and}\qquad  f B_\al \ti B_\bet & = &\ti B_\ga  B_\de 
R^{(1)  \ga \de }_{\al \bet }  
 \ll{hbb}\eea
with all Greek indices taking values   1 and  2 and with the  
notation $T_{00}=T_{00}(\la),~ \ti T_{00}=\ti T_{00}(\mu) $ etc.
 Note that the relation  (\re{hab})
is similar to   (\re{ab}), while  (\re{hdb},\re{hbb})
are complicated and unlike the usual  ABA case 
do not preserve the same set of 
 operators we started with. Nevertheless,  there
are similarities in  
the  structure and  the $2 \times 2$   matrix  
$~R^{(1)  \ga \de }_{\al \bet }~$  
appearing in the above relation   is again  a 
  $R$-matrix  solution  with one step lower rank.

Observing in (\re {hbb}) that a particular 
 string of $B$'s
generates its   other combinations, we guess the Bethe ansatz for the
eigenvector in the form of superposition of all such  strings of $B$'s as     
\be
  \mid \phi>= 
 \mid \la_1 \la_2 \ldots \la_n> = \sum_{ \{a \}} 
B_{a_1}(\la_1)B_{a_2}(\la_2) \ldots B_{a_n}(\la_n)\mid 0> F^
{{a_1} \cdots {a_n}}, \ll{hphi}\ee
 where the operator coefficients $F^{{{a_1} \cdots {a_n}}} $ would be determined later.
Using the similar pseudovacuum  reference state   (\re{adv}) one obtains
 \be 
T_{00}(\la)\mid 0> =\al^N(\la)\mid 0>, 
~~T_{11}(\la)\mid 0> =T_{22}(\la)\mid 0> =\bet^N(\la)\mid 0>.
\ll{hadv}\ee
For this model we  have $\al=f(\la)$ and $\bet=1.$
We assume  for the time being,  that the {\it unwanted} terms in the 
commutation relations  (\re {hab}-\re{hdb})
(those with changed arguments like $ B_\al,~ \ti T_{\al \bet })$
 are simply absent.  
We will find later that as in the ABA, the Bethe equations
 for this system  represent also
the condition  necessary for  vanishing of  such  
 {\it unwanted} terms.

Repeating the standard
 arguments described above we obtain the eigenvalue of 
 (\re {hab}) as
\bea
T_{00}(\la)\mid n> &=& \Lambda_{00}(\la \mid \la_1, \ldots , \la_n
 \mid n>,\nonumber \\ &=& \al^N(\la) \pr_{l=1}^n  f(\la_l-\la) \mid n>~~~ , 
\ll{hav}\eea
However,  the relation
 (\re {hdb}) can  not solve the eigenvalue problem
  directly and instead yields
\be
\sum_\al T_{\al \al}(\la)\mid n>=
  (\pr_{l=1}^n B_{b_l}(\la_l)) ~~ \tau^{(1)} (\la\mid \{ \la_l\}) ^{
b_l \cdots b_n}_{a_l \cdots a_n}~F^{a_l \cdots a_n}\mid 0> . 
\ll{hdv}\ee
 It is curious to note that the $\tau^{(1)} $ operator appearing here 
with  $\tau^{(1)} (\la\mid \{ \la_l\})
=\sum_\al T^{(1)}_{\al \al} (\la\mid \{ \la_l\})=
  tr ( T^{(1)} (\la\mid \{ \la_l\})), $
where
 \bea
T^{(1)}_{\al \al} (\la\mid \{ \la_l\})&=& 
R^{(1)  \ga_1 b_1 }_{\al a_1 }(\la-\la_1)  
\cdots R^{(1)  \ga_n b_n }_{\ga_{n-1} a_n }(\la-\la_n)  
T_{\al  \ga_n  }(\la)  \nonumber \\
\qquad &=& \l(L_{\al_1 }(\la-\la_1)  
\cdots L_{ \al_n }(\la-\la_n)  
T(\la) \r)_{\al \al} 
\ll{T1}\eea
 represents the transfer matrix of  
$N+n$ site chain with  $n$  inhomogeneous sites (coming from $R^{(1)}$
matrices).
 One can continue
the arguments further by attempting to solve the eigenvalue problem of 
$ \tau^{(1)} (\la\mid \{ \la_l\}) ^{
b_l \cdots b_n}_{a_l \cdots a_n}~F^{a_l \cdots a_n}\mid 0> ,$ since  
 we see immediately that  (\re {T1}) involves only  $2 \times 2$ monodromy
matrices and  hence  is the standard problem we have  already solved.
This {\it nesting} is the basic idea in NBA and   we choose therefore 
\bea
 F^{a_l \cdots a_n}&=& \mid \la^{(1)}_1, \ldots , \la^{(1)}_{n_1}
 >^{a_l \cdots a_n}  \nonumber \\
 \qquad  &=& (B^{(1)}(\la^{(1)}_1)
 \cdots B^{(1)}(\la^{(1)}_{n_1}) \mid 0^{(1)}>)^{a_l \cdots a_n}
\ll{F}\eea
with $~B^{(1)}(\la^{(1)}_j)~$ and  $ ~\mid 0^{(1)}>~ $ living
 on a lattice of $n$
sites and  thus is  an element of direct product of $n$ Hilbert spaces.
The part of $ \tau^{(1)}$ involving  $T_{\al \ga_n}$ however acts on
nonoverlapping  $N$ product Hilbert space  
 and hence commutes with $B^{(1)}(\la^{(1)}_j)$'s  and can  hit
directly
the vacuum $\mid 0>.$  

Using therefore the eigenvalue result (\re {tev}),  with trivial modification 
due to 
inhomogeneity
in $L_j (\la-\la_j) $ and its form (\re{hrm}) 
 affect the action on the vacuum as 
 $$ A^{(1)}(\la\mid   \{ \la_j \})  \mid 0^{(1)}> =
  \pr_{j=1}^n  f(\la-\la_j )\mid 0^{(1)}>, ~~
D^{(1)}(\la\mid   \{ \la_j \})  \mid 0^{(1)}> =
  \mid 0^{(1)}>,
$$
  along with   $T_{\al \al}\mid 0>$ as in  (\re {hadv}). Thus
we get the part of eigenvalues as
\be 
 (\tau^{(1)} (\la\mid \{ \la_l\})F )\mid 0>=  \bet^N(\la) \mid 0> \l(
\pr_{j=1}^{n_1}  f(\la^{(1)}_j-\la)\pr_{l=1}^{n}  f(\la-\la_l ) + 
\pr_{j=1}^{n_1}  f(\la^{(1)}_j-\la) \r) F.
\ll{tau11}\ee
The final expression for the eigenvalues of the original transfer matrix
$\tau(\la)= tr T(\la)= \sum_{a=0}^2 T_{aa}(\la)$ is   
\[\tau (\la) \mid n>=
\sum_{a=0}^2 \Lambda_{aa}\mid n>=
 \Lambda(\la \mid \{\la_l\} , \{\la^{(1)}_l\}) 
\mid n>
 \]
can be obtained in the closed form by adding (\re {hav}) and 
(\re{hdv}) incorporating (\re {tau11}).

We observe that  in the NBA   additional sets of 
parameters appear, for which  the determining equations called Bethe
equations 
are obtained  from  the
vanishing of {\it unwanted} terms (already  assumed). 
One can 
derive such equations by carefully keeping track of all such terms
throughout the nested steps \c{fabian} or an easy way out as before 
is to  obtain the same  Bethe equations from the vanishing conditions of the 
residues of  $\Lambda(\la \mid \{\la_l\} , \{\la^{(1)}_l\})$ at $
\la =\{\la_l\} ,j=1,\ldots, n, ~~$ and $~~\la =  \{\la^{(1)}_j\}
,l=1,\ldots, n_1.$ 
Thus from the condition
 $~~res \Lambda(\la )_{\la=\la_m}=
res (\Lambda_{00}(\la )+\Lambda_{11}(\la ))_{\la=\la_m}=0,$ (since 
$\Lambda_{22}(\la )$ gives no contribution) we obtain the $1st$
 set of Bethe equations
\be 
 f^N (\la_m)=
\pr_{l \not = m}^{n}{  f(\la_m-\la_l) \ov  f(\la_l-\la_m ) }~~ 
\pr_{j=1}^{n_1}  f(\la^{(1)}_j-\la_m) 
\ll{hbe1}\ee
where $ f(\la_m)
={a(\la_m) \ov b(\la_m)}, m=1, \ldots, n.$
The $2nd$ set of Bethe equations
 similarly can be obtained from
the condition  $res \Lambda(\la )_{\la=\la^{(1)}_k}=
res (\Lambda_{11}(\la )+\Lambda_{22}(\la ))_{\la=\la^{(1)}_k}=0$ as
\be 
\pr_{l=1}^{n}  f(\la^{(1)}_k-\la_l)= 
\pr_{j \not = k}^{n_1}{  f(\la^{(1)}_k-\la^{(1)}_j) \ov 
 f(\la^{(1)}_j-\la^{(1)}_k ) }~~ 
\ll{hbe2}\ee
with  $k =1, \ldots, n_1 .$

Choosing   the function $f(\la)$ as (\re{trm}) 
or as (\re{rrm}) we can  apply  the above result for solving 
concrete quantum models with internal  degrees of freedom 
and belonging to the 
 trigonometric or the rational class, respectively.   
Examples of such rational  models  are
the {\it vector} NLS model \c{vnls}, {\it matrix}  NLS model,
quantum $\mathsf 
{N}-wave$ system \c{nba,nwave} etc., while  the {\it Perk-Schultz} model
\c{schultz}
or the generalized six-vertex model \c{lopez}, the anisotropic 
$n$-vertex model \c{sun} etc. are    the examples of higher rank
models belonging to the trigonometric class.  
\subsection {Coordinate Bethe ansatz} 
The coordinate formulation   of the  Bethe ansatz (CBA) \c{Bethe,cba,BaxterHC}  
aims to solve the  Hamiltonian eigenvalue  problem 
    directly, 
 without assuming explicitly any integrability condition. However for
integrable models allowing ABA treatment, the CBA is also  found to work
successfully.
 Let us demonstrate it on the example of  anisotropic
$XXZ$ spin-${1 \ov 2}$ chain we have solved already through ABA. 
\subsubsection {$XXZ$ spin-$\ha$ chain}
The Hamiltonian of the model  (\re{hxxz1}) after addition of an irrelevant
constant term is written as  
\be
H_{xxz} = -\sum_n^N \si_n^+ \si_{n+1}^-+\si_n^- \si_{n+1}^+ +
 {\De  \ov 2} (\si_n^3 \si_{n+1}^3 -1),
\ll{HXXZ}\ee
where   values of the parameter $\De=\cos \eta$ defining the anisotropy of
the model
determines also its  anti-ferromagnetic or the ferromagnetic 
 behavior.
 We work now in the coordinate formulation, i.e. explicitly
on the lattice, where the spin states can be either {\it up} or {\it down}
 with 
basic vectors $~~ e^+=\l( \ba 1  \\  0 \ea \r)  ~~\mbox { or}~~ 
e^-=\l( \ba 0 \\  1 \ea  \r)= \si^-e^+ ,~~$ and  we consider the 
spin excited states $\mid x_1,\ldots,x_m>=\si^-_{x_1} \ldots \si^-_{x_m}
\mid 0>, $
with $m$ down spins $e^-$ 
at lattice sites $ x_1,\ldots,x_m$ in the background of the rest $N-m$ 
up spins. The set of such basis vectors are orthonormal satisfying 
\\$< x'_1,\ldots,x'_m \mid  x_1,\ldots,x_m>=0,$ for 
 different arrangements of  
$\{x'\},\{ x \}$ and \\ $~=~1$ for $\{x'\}\equiv \{ x \}.$ 
  The $m=0$  or the vacuum state  $\mid 0>$ state is   the ferromagnetic
 ground state with all spins up.

 The ansatz for the eigenvector is given as the superposition of
all such states: 
\be \Psi_m>=\sum_{ x_1<x_2 < \ldots <x_m }~\Psi (x_1,\ldots,x_m) \mid
 x_1,\ldots,x_m> \ll{psi}\ee
the explicit  form of  which will be specified later.
Equations for determining the functions $  \Psi (x_1,\ldots,x_m) $ can be
obtained from the eigenvalue equation $H \mid \Psi_m> =E_m 
\mid\Psi_m> $ by the
  Hamiltonian (\re {HXXZ}) acting  on the vector (\re {psi}).
We start from the
elementary  knowledge $\si_n^\pm e_n^\pm=0,~~\si_n^\pm e_n^\mp= e_n^\pm,~~
\si_n^3 e_n^\pm= \pm e_n^\pm,$
to obtain  \bea
(\si_n^\pm \si_{n+1}^\mp)e_n^\pm \otimes e_{n+1}^\mp = 0   
&,&     (\si_n^\mp \si_{n+1}^\pm )e_n^\pm \otimes e_{n+1}^\mp
=e_n^\mp \otimes e_{n+1}^\pm, \nonumber \\
(\si_n^3 \si_{n+1}^3 -1)e_n^\pm \otimes e_{n+1}^\mp
&=&-2 e_n^\pm \otimes e_{n+1}^\mp, 
 \nonumber \eea  along with the trivial actions of all the above operators 
on $e_n^\pm \otimes e_{n+1}^\pm $.
   An interesting conclusion  follows now, that  
on  local excitations involving the term $~~~
\ldots e^+ \otimes  e_n^- \otimes  e^+ \ldots \equiv 
\mid \ldots,0,x_{n},0,\ldots> $
the $XY$ part of the Hamiltonian
$~~h_{xy}=\sum_n^N h_{xy}^{n,n+1},~$ where $~ h_{xy}^{n,n+1} 
=-(\si_n^+ \si_{n+1}^-+\si_n^- \si_{n+1}^+)$
acts  as   
\be - h_{xy}^{l-1,l}\mid \ldots,0,
\underbrace{x_{l-1},0},\ldots>=\mid \ldots,0,\underbrace{0,x_{l}},\ldots>
\ll{x-1}\ee
 shifting it 
towards right by one site, provided that site is not already occupied.
In the analogous way  
\be -h_{xy}^{l,l+1}\mid \ldots,
\underbrace{0,x_{l+1}},0,\dots>=\mid \ldots,\underbrace{x_{l},0},0,\dots>
\ll{x+1}\ee
 acts with a 
similar left shift. On the other hand, 
the  remaining  part of the Hamiltonian
$~~ h_{z}=\sum_n^N h_{z}^{n,n+1},~~ h_{z}^{n,n+1}
=-{\De  \ov 2} (\si_n^3 \si_{n+1}^3 -1)
$
 preserves its position yielding
\bea - h_{z}^{l-1,l}\mid \ldots,\underbrace{0,x_{l}},0,\ldots>
&=&-\De \mid \ldots,\underbrace{0,x_{l}},0,\ldots>,\ll{xl}\\
 - h_{z}^{l,l+1}\mid \ldots,0,\underbrace{x_{l},0},\ldots>
&=&-\De \mid \ldots,0,\underbrace{x_{l},0},\ldots>.\ll{xr}\eea
For better understanding of the effect of spin interactions 
on the eigenvectors, let us look  into  the $m=2$ excitations  with
$
\mid \Psi^+_2>=\sum_{ x_1<x_2 }~\Psi (x_1,x_2) \mid x_1,x_2>, $
 before going to the general case, where  $+$ sign signifies the particular
ordering of $x_1<x_2$.
  Since the Hamiltonian involves only nearest neighbor 
interactions, we  consider  first the situation $    x_1+1<x_2 ,$ when
$  x_1, x_2 $ are {\it non-nearest} neighbors, resulting  the action of the  
Hamiltonian 
on  eigenvector $
\mid \Psi^+_2> $ as the  noninteracting 
one. Therefore implementing  the above  mentioned 
 properties  of $h_{xy}, h_{z} $  we
  obtain the action of the total Hamiltonian in the
eigenvalue equation. Considering further the 
orthonormality of the set of basic vectors
$\mid x_{1},x_2>$ we get an equation for the functions 
$\Psi ( x_{1},x_2)$ as
\be  
- (\Psi (x_1-1,x_2) +\Psi (x_1+1,x_2)+ \Psi(x_1,x_2-1) +\Psi(x_1,x_2+1)) 
+
4 \De  \Psi(x_1,x_2)=E_2 \Psi(x_1,x_2). \ll{a1} \ee
Note that since $x_1,x_2$ are not  nearest neighbors a particular state   
$ \mid x_1,x_2> $ can be reached under the action of $h_{xy},$ 
from both right and left sides as in 
(\re {x-1},\re {x+1}). This  results 
symmetric expressions  of
 $\Psi (\cdot ,\cdot)$  in (\re{a1}) and due to the same reason 
 the contributions of  $h_z$  in  (\re{a1})
are from both (\re {xl}) and (\re {xr}) summed to  
 $2 \De  \Psi(x_1,x_2),$ coming 
from each  of  the excitations at  $x_1$ and $x_2$.

For incorporating   interactions  we consider now $x_1,x_2$ as  
nearest neighboring points with $x_2=x_1+1$.  
It is crucial to note that any
state $ \mid x_1,x_2> \equiv \mid x,x+1>$ can  only be reached in this case 
  by a right shift of   $  x_1-1$ as in (\re {x-1})  
 and a left shift of $x_2+1$  as in (\re {x+1}), resulting the presence 
of only such functions   $\Psi (\cdot ,\cdot)$ in its 
eigenvalue equation.
  Furthermore, 
since only the left neighboring site of $ x_1$ is empty and correspondingly 
 the right site for $x_2$,   
 the contributions under  $h_z$  for the  excitation at  $x_1$ come
from (\re{xl}) only, while  that for     $x_2$
 is  only  due to (\re{xr}).  As a result compared  to (\re{a1}) less terms
appear now in  
the eigenvalue
equation giving  
\be  
- (\Psi (x-1,x+1)  +\Psi (x,x+2)) 
+
2 \De  \Psi (x,x+1)=E_2 \Psi (x,x+1). \ll{a2} \ee
However we demand that 
our eigenvector $~
\mid\Psi^+_2>$ should be the same for all values 
of $x_1<x_2$, i.e. (\re{a1}) by  setting formally $x_2=x_1+1$ 
should coincide with 
   (\re {a2}). Consequently  the 
condition for vanishing of  their differences
  gives the  constraint  equation
\be  
\Psi (x+1,x+1)  +\Psi (x,x) 
=
2 \De  \Psi (x,x+1). \ll{a3} \ee
The difference eqns. (\re{a1}) and  (\re{a3})
being at our disposal, an  ansatz for the solution is made in the form of
superposition of plane waves 
\be  
\Psi (x_1,x_2)=A_{12}e^{i(p_1x_1  +p_2x_2)}+A_{21}e^{i(p_2x_1  +p_1x_2)}, 
 \ll{ansatz} \ee
 where the coefficients $A_{12},A_{21}$ or rather their ratio
 and the momenta $ p_{1},p_{1}$
are unknown parameters. They will be 
 determined  as  shown below, by the equation
(\re{a3}) and the  boundary condition. Generally
speaking one should denote  $A_{12}\equiv A^+_{p_1p_2},~A_{21}\equiv A^+_{p_2p_1},$ to stress 
the ordering of $x_1,x_2$ and $p_1,p_2$. 
 Since the eigenvalue $E_2$ must not depend on the positions of 
the excitations, we may find it from (\re{a1}) inserting ansatz      
(\re{ansatz}) to yield 
\be  
E_2= 2\sum_{j=1}^2(\De-\cos p_j).
 \ll{e2} \ee
For determining the  coefficients we similarly obtain from (\re{a3})
\be  
{A_{12}   \ov  A_{21}}= -{g(p_2,p_1) \ov g(p_1,p_2) } \equiv
 -e^{i\theta (p_1,p_2)}, 
 \ll{a12} \ee
where
\be  
g(p_1,p_2)= 2 \De e^{i p_1} - e^{i (p_1+p_2)}-1, 
 \ll{g} \ee
 $\theta(p_1,p_2)=-\theta(p_2,p_1)$
 may be interpreted as  the phase change suffered in the scattering
of two pseudoparticles (magnons) due to the exchange of  momenta.

For deriving now the  equations for  the parameters $p_1,p_2,$
we invoke the boundary conditions considering them to be periodic. That is 
we assume the chain to be closed by identifying the states 
at sites $x$ and $x+N$ given by
$ ~
\mid \Psi^+_{x_1<x_2}>=~
\mid\Psi^-_{x_1+N=x'_1>x_2}>$, where $~\mid \Psi^->~$ denotes the
state with  reverse ordering of $x_i'$s. However it is easy to
 see that 
$~
\mid\Psi^->$ does not give any new solution and can be  obtained from 
$\Psi^+>$ by
just  interchanging $x_1 \leftrightarrow x_2.$ Therefore  
the periodic boundary condition  is compatible with the relation
\be 
 \Psi(x_1,x_2)= \Psi(x_2,x_1+N),
 \ll{pbc2} \ee
which with the   substitution (\re{ansatz}) yields the system of equations
\be 
  e^{ip_1N} =- e^{i\th (p_1,p_2)},~~e^{i p_2N} =- e^{i\th (p_2,p_1)}
 \ll{p12} \ee
defining $p_1,p_2.$

Thus we have completely solved the eigenvalue problem for the $m=2$ case by
finding explicitly the energy spectrum and the eigenvector. Repeating  
similar arguments one can solve the general case for arbitrary $m<N,$ 
yielding the  energy value 
\be  
E^{xxz}_m= 2\sum_{j=1}^m(\De-\cos p_j)
 \ll{em} \ee
 and the eigenvector (\re{psi}) with
\be  
\Psi(x_1,\ldots,x_m)=\sum_{P}A_{P(1,\ldots, m)}~e^{i\sum_{j=1}^m ~P (p_j) x_j }
 \ll{ansatzm} \ee
where $P(1,\ldots, m)$ stands for the 
permutation of all $m$ indices $1,\ldots, m$
involving $m!$ number of terms in the expression and $P (p_j)$  corresponds to the
the same set of  permuted indices $j$ in $p_j$. The amplitudes are connected 
through  the phase factors    $\th$ as 
 \be
{ A_{1',\ldots, m'}
\ov A_{1,\ldots, m}}= \pm e^{i\sum_{j,k} ~\th (p_j,p_k)} 
\ll{apa0}\ee
where the summation of the $\th$-factors are extended over those pairs of 
$j,k$ which are necessary to interchange to arrive from the arrangement 
of indices $1',\ldots, m'$ to the initial arrangement $1,\ldots, m
$. More  physically, for each interchange of particles at $x_j$ and $x_k$ 
the
scattering amplitude picks 
up a phase factor $e^{i\th (p_j,p_k)}$  and the total
 change in amplitude is given by the multiplication of such phase factors 
due to two-body scattering. This fascinating fact of the factorization of 
{\it many-body} scattering into the {\it two-body} one is the essence of the
integrability \c{zamfac} and has been  witnessed in all integrable models. 
The $\pm$ sign appearing in (\re{apa0}) is determined by the parity 
of the permutation. In particular case of $m=2$ we have derived 
this formula with $-$ sign in (\re{a12}). 
The periodic boundary condition, as explained already, gives the equations
for determining the parameters $p_j,~ j=1,\ldots,m$ generalizing (\re{p12}) 
as
\be 
  e^{ip_aN} =(-1)^{m-1} e^{i\sum_{k=1}^m \th (p_j,p_k)},
 \ll{pj} \ee
accounting for the total phase change incurred in scattering of particle at
$x_j$ through all the rest round the chain and returning to its original
position.
 For considering the  thermodynamic limit
it  is convenient to take first  the logarithm
of (\re{pj}) to get 
\be 
  p_jN =2 \pi Q_j+~i\sum_{k=1}^m \th (p_j,p_k).
 \ll{pj1} \ee
where $ Q_j=$ {\it half integers} for $m$ {\it even} and {\it  integers}
for  $m$ {\it odd}. The expression for $\th (p_j,p_k)$ is given as 
 (\re{a12},\re{g}) 
by replacing $1,2$ by $j,k$ or equivalently as

\be 
\th (p_j,p_k)=2 \tan^{-1}\l({ -\De\sin \ha (p_j-p_k) \ov 
\cos \ha (p_j+p_k)- \De \cos \ha (p_j-p_k)}\r)
  \ll{thjk}\ee
\subsubsection { $XXX$ spin-$\ha$  chain}
All the formulas for solving the isotropic spin model  can be derived
 easily by just putting  the anisotropy
parameter $\De=1$  in the above equations. This simplifies considerably
the corresponding expressions giving for example    
\be  
E^{xxx}_m= 2\sum_{j=1}^m(1-\cos p_j),~~
\th^{xxx} (p_j,p_k)=2 \tan^{-1}\ha \l( \cot \ha p_k -\cot \ha p_j \r)
 \ll{ethxxx} \ee
etc. with the same ansatz (\re{ansatzm}) together with (\re{apa0}).
\subsection  { Interrelation between algebraic and coordinate Bethe ansatzes} 
Though the ABA and the CBA differ considerably 
in their approaches, with ABA dealing with 
global objects having more universal and model-independent formulations, 
while CBA being concerned with concrete Hamiltonians and their  local actions,
 there exist also intimate connections between them. Though 
the above results  obtained for the same models 
using both  the methods  look different, we intend to 
show  here their
equivalence basically  through a  change in  variables.
 Indeed the spectral parameter related  
{\it rapidity} variables $\la_j$ 
involved in the ABA may be  linked to the momentum parameters $p_j$ of
the CBA by the relation
\be 
  e^{ip_j}= f(\la_j)\equiv { a(\la_j) \ov
 b(\la_j)}= ~ {\sin (\la_j + {\eta \ov 2}) \ov \sin (\la_j - {\eta \ov 2})},
\ll{plambda}\ee
for the trigonometric models like $XXZ$ spin$-\ha$ chain. 
 Similarly 
for the rational models like 
spin$-\ha$ $XXX$ chain,  one gets the relation $
e^{ip_j}= f(\la_j)= ~ { \la_j + {i \ov 2} \ov \la_j - {i \ov 2}}$
or  more simply as $\la_j=\ha \cot {p_j \ov 2}.$  With  mapping (\re {plambda}) between the variables 
we derive from (\re {thjk})  interesting relations for the scattering
amplitudes as 
\be 
  e^{i\th (p_j,p_k)}= ~- {\sin (\la_j -\la_k + {\eta \ov 2}) \ov 
\sin (\la_j -\la_k- {\eta \ov 2})},~~\mbox {and}~~ 
 e^{i\th(p_j,p_k)}= 
 -{\la_j -\la_k + {\eta \ov 2} \ov 
\la_j -\la_k- {\eta \ov 2}}
\ll{thf}\ee
 for $XXZ$ and  $XXX$ models, respectively. 
 concluding in general $ e^{i\th (p_j,p_k)}=    
-{ f(\la_j-\la_k) \ov
 f(\la_k-\la_j)}$. Note the
 remarkable fact, that 
the scattering amplitude becomes a function  
 only  of the differences    
 in the rapidity  variables $\la_j$ and is linked  
 with the function 
  $f(\la -\mu)$ of the $R$-matrix. Therefore function $f$,
 which is also    relevant for the eigenvalue problem, 
  is  the most important element
in the  $R$-matrix. 

Using the basic  relationship (\re {plambda}) between the variables, we can
 easily verify  the equivalence between the eigenvalues
 (\re {exxz})  and (\re {em}) obtained for the same $XXZ$ model through 
ABA and CBA methods, respectively. Similarly  one checks the equivalence 
between (\re {exxx})
and (\re {ethxxx}) for the $XXX$ model, which also  gives a closed relation
$E^{xxx}_m=- \sum_{j=1}^m {d p_j \ov d \la_j}$. 
 Using the  relationship  (\re {plambda}),  (\re {thf}) 
we   establish the 
intriguing fact  that  the 
Bethe equation  (\re {be}) imposed in the ABA approach
 for eliminating {\it  unwanted  terms} 
by using an involved argument 
  of vanishing residues for the  eigenvalue $\Lambda(\la)$ of the transfer
matrix, 
coincides exactly  with condition (\re {pj}), obtained
in CBA  as a consequence of the periodic
boundary condition.

The Nested Bethe ansatz applicable for systems with internal  degrees
has been described above in the ABA formulation. CBA approach can also be 
formulated similarly  for such systems. We consider below such an  example, where 
the  Hubbard model is solved through CBA. The ABA treatment of
this model remained unclear until very recently \c{abaHub}.

\subsection{Hubbard model}
One dimensional Hubbard model is an integrable system, for which $R$-matrix
, Yang-Baxter equation and even ABA formulation is now available
\c{shastryH,
wadatiH,abaHub}. We have constructed the Hamiltonian (\re{hhub0})  
of the Hubbard model starting from its Lax operator  in sect. 6 .  
Our aim  here is to present in brief its 
eigenvalue solution through CBA \c{LWHub,izumov}  mainly 
to demonstrate the applicability of this 
method also  to the systems with internal  degrees of freedom.  
 
The model is  described conventionally by the Hamiltonian  
\be
H_{hub} = \sum_i^N \l( t~\sum_{\si= \pm}
c^\da_{ \si i} c_{ \si i+1}+c^\da_{ \si i} c_{ \si i-1} + 
U~ n_{ + i}n_{ - i}     
\r)
\ll{hhub}\ee
with fermionic creation and annihilation  operators 
$c^\da_{ \si i}, c_{ \si i} $ and the number operators 
$n_{ \si i}=c^\da_{ \si i}, c_{ \si i}
$ with spin projections $\si=\pm .$  The equivalence of  
(\re{hhub})
(scale it
 by $t$ and put $t=-1$) 
 and the Hamiltonian (\re{hhub0}) constructed before can be shown
through the Jordan-Wigner transformation (\re{JWT}).

Eigenfunctions for the Hubbard model 
may be chosen similar to 
 the above   CBA ansatz. However we notice, that  in constructing 
 the    $m=2$ case
 due to the internal 
indices $\si$, 
 the states    $\mid \Psi_{12}(x_1<x_2)>$ and $ \mid \Psi_{21}(x_1>x_2)>$
unlike the  above description
 correspond now to two independent  solutions. This is  because 
 together with coordinates the corresponding 
internal  indices  also get permuted. Therefore we have to keep track of both
the corresponding  functions in the eigenvalue equation. 
Considering  first 
 the nonnearest neighbor  situation  $x_1+1<x_2,$  when the
interacting
$U$ term does not play any role, we obtain
\be  
- (\Psi_{12} (x_1-1,x_2) +\Psi_{12} (x_1+1,x_2)+ \Psi_{12}(x_1,x_2-1) 
+\Psi_{12}(x_1,x_2+1)) 
=E_2 \Psi_{12}(x_1,x_2) \ll{hub1} \ee
and a similar equation  for $\Psi_{21}.$
Note that the equation coincides with   (\re {a1}) for  $\De=0$. 
This may be explained by  the fact that 
 the action of 
$h_{xy}$ from the spin Hamiltonian  is  the 
same as the  $H_{hub}$ with $U=0$ (their operator equivalence can  be shown
through the Jordan-Wigner transformation \c{shastryH}.

The case  $x_2=x_1=x$ invokes the interacting term and mixes the 
$\Psi$-functions to yield 
\be  
- (\Psi_{12} (x-1,x) +\Psi_{21} (x+1,x)+ \Psi_{21}(x,x-1) 
+\Psi_{12}(x,x+1))+ U~\Psi_{12}(x,x) 
=E_2 \Psi_{12}(x,x). \ll{hub2} \ee

Both the functions $ $  and $ \Psi_{21} (x,y)$ arise here depending obn the
situation $x \leq y$ or $x \geq y $  and the consistency demands
 \be~ \Psi_{12}(x,x)=\Psi_{21}(x,x).
\ll{hub3} \ee
Repeating the CBA argument described before, we put coinciding $x'$s in
 (\re {hub1})
and take its difference from  (\re {hub2})
to get 
\be  
- (\Psi_{12} (x+1,x) +\Psi_{12}(x,x-1))+\Psi_{21} (x+1,x)+ \Psi_{21}(x,x-1) 
- U~\Psi_{12}(x,x) 
=0. \ll{hub4} \ee
In analogy with (\re{ansatz}) we make 
 the ansatz 
\be  
\Psi_{12} (x_1,x_2)=A_{12}(p_1p_2)e^{i(p_1x_1  +p_2x_2)}
+A_{12}(p_2p_1)e^{i(p_2x_1  +p_1x_2)} 
 \ll{hubans2} \ee
 for $\Psi_{12}$
and similarly for $\Psi_{21},$ with the corresponding coefficients taken as   
 $A_{21}(p_2p_1)~$ and $A_{21}(p_1p_2) $. 
The energy spectrum can be derived  from  (\re {hub1})
  as 
$~~
E^{hub}_2=- 2\sum_{j=1}^2(\cos p_j),~~ $
with obvious  coincidence with  (\re {e2}) for $\De=0$. The ansatz also 
extracts from (\re {hub3}) the relation 
$A_{12}(p_1p_2)-A_{12}(p_2p_1)=A_{21}(p_2p_1)-A_{21}(p_1p_2)
$ and using it derives from 
(\re {hub4}) the important interrelation between the coefficients
\be
A_{\si_1\si_2}(p_2p_1)=
\sum_{\si'_1\si'_2} ~ S^{\si_2\si'_1}_{\si_1\si'_2}(p_1p_2)
A_{\si'_1\si'_2}(p_1p_2)
\ll{hubasa}\ee
where the scattering matrix may be written in the operator form 
\be
S_{12}(\la)_{\mid \la=\la^0_1-\la^0_2}=
 g(\la)(\la+i{U\ov 2}P_{12})_{\mid \la=\la^0_1-\la^0_2}
\ll{hubs}\ee
with the notation $\la^0_j=\sin p_j$ and $g(\la)=(\la+i{U\ov 2})^{-1}$ 

The periodic boundary condition $\Psi_{12}(x_1,x_2)=\Psi_{21}(x_1+N,x_2)$
 induces  additionally the  relations like $
e^{ip_2N}A_{21}(p_1p_2)=A_{12}(p_2p_1)=
\sum_{\si'_1\si'_2} ~ S^{2\si'_1}_{1\si'_2}(p_1p_2)
A_{\si'_1\si'_2}(p_1p_2),$ which should be diagonalized to 
determine the equations for  $p_j$'s. Note that the $S$-matrix is exactly
like  the  rational $R$-matrix (\re{rrm}) or the 
Lax operator $L_{jk}(\la)\mid_{\la= (\la^0_j-\la^0_k}$  
associated with the $XXX$ spin-$\ha$ chain. 

For the general $m$-particle excitations, since 
 the scattering matrix is factorized into the
two-particle matrices as described above,  
the periodic boundary condition gives  the operator form
$e^{ip_jN}A(I)=T_j A(I)
$ where $T_j=\pr_{k\neq j}^n L_{jk}(\la^0_j-\la^0_k)$
posing the  problem of diagonalizing the transfer matrix 
$T_j=T(\la=\la^0_j)$ of a
 inhomogeneous $XXX$ spin-$\ha$ chain,
  which however has already been  solved. Therefore without repeatation we 
present the final result for determining $\{p_j \}, j=1,\ldots,n$ as 
\be 
e^{p_j N} =\pr_{s=1}^{n_1}  f(\la_s-\la^0_j),~~ 
\pr_{l=1}^{n}  f(\la_t-\la^0_l)= 
\pr_{s \not = t}^{n_1}{  f(\la_t-\la_s) \ov 
 f(\la_s-\la_t ) } 
\ll{hubbe}\ee
where $\la^0_j\equiv  \sin p_j$ and the function $f$ is in our standard
notation and given here by 
$ f(\la)= {\la+i{U \ov 2} \ov \la}$.

%______________________________________\input {preport8.tex}
\setcounter {section}{7}
\section{ Nonultralocal models and  
quantum integrability}
   \setcounter{equation}0
As we have mentioned,   ultralocal  systems  received more attention  
and the integrability theory 
 is well established for them \c{fadrev}, while   
 parallel progress has not been 
achieved   for nonultralocal models and   the related   theory is   
  still in the stage of development. However  
 based on recent results \c{maillet, khijmp96},
some progress  in formulating such a scheme  
 is reported  in sect. 2 
along with a list of nonultralocal  models  proposed 
 in different contexts, for which the scheme can be successfully applied.

Recall that  the basic equation representing integrability for 
nonultralocal models is the braided QYBE (\re{bqybel}) together with
braiding relations (\re{zlzl1u}-\re{zlzl2u}), which lead to the global
equation (\re{rztzt}).  
 The underlying
algebra  is the braided or the quantized braided algebra 
  \c{majid,majidq,hlavqb}.  
 Our aim here is to  
focus on  the  concrete  models listed before and  
show how 
 their basic  equations  can be derived from the general formulation 
in a systematic way. 
%---------------------------------------
\subsection {Braided  Yang-Baxter equation for periodic models }
For closer  contact with the 
 physical models
we incorporate  the  periodic boundary condition
$ L_{aj+P}(\la)= L_{aj}(\la)$   and define 
 the global monodromy matrix $ T_a(\la) \equiv
  T^{[N,1]}_{a}(\la)$ for the closed chain with $N$ sites.
 Notice  now, that for deriving
equation for the monodromy matrix $T_a(\la)$ 
 along with  (\re{zlzl1u},\re{zlzl2u})
  one requires also the braiding relation like
\begin{equation}
 L_{b j}(\mu)\tilde Z_{ba}^{-1}L_{a j+1}(\la)
=\tilde Z_{ba}^{-1}L_{a j+1}(\la) Z_{ab}^{-1} L_{b j}(\mu)\tilde Z_{ba}^{-1}
,\ll{zlzl3}\end{equation}
since $L$ at the extreme ends again become  nearest neighbors 
due to the periodicity $ L_{1}(\la)= L_{N+1}(\la).$
 The compatibility of (\re {zlzl3}) with (\re {zlzl1u})
demands the constraint
$ \tilde Z_{ab} \tilde Z_{ba}= {I}
$ putting  some  restriction on $\ti Z$. Therefore
 one may continue the  globalization beyond equation (\re{rztzt})  
 repeating  similar steps and  
using  (\re{zlzl1u},\re{zlzl2u}).
 Finally at  $j=N$
 applying   (\re{zlzl3}),
 the braided QYBE for the monodromy matrix for periodic models   can be
expressed as  
\begin{equation}
{R}_{12}(\la-\mu)Z_{21}^{-1} T_{1}(\la)Z_{12}^{-1} T_{2}(\mu)
= Z_{12}^{-1} T_{2}(\mu) Z_{21}^{-1} T_{1}(\la){R}_{12}(\la-\mu).
\ll{bqybet}\end{equation}
Note that  the $\ti Z$ matrix  has disappeared from the equation.
%----------------------------------------------------------
\subsection{Trace factorization  and quantum integrability}
For ensuring quantum integrability we need a
commuting set of conserved quantities.
However  by taking the trace of (\re{bqybet}) we arrive at
\begin{equation}
tr_{12}\left(Z_{21}^{-1} T_{1}(u)Z_{12}^{-1} T_{2}(v)\right)
=tr_{12}\left( Z_{12}^{-1} T_{2}(v) Z_{21}^{-1} T_{1}(u)\right),
\ll{trace}\end{equation}
which due to the appearance
of $Z_{ab}$  
does not  readily allow the trace  factorization . It is obvious that  for
  $Z\equiv { I}$ corresponding to the  ultralocal
 models, the above trace identity is trivially factorized proving the
integrability.
For tackling   this problem for  general $Z$ we consider 
the following  factorization procedures.
\subsubsection {Factorization by k--trace}
The procedure is
 based on
the generalization \c{hlavopsc,maillet} of 
  reflection--type algebras \c{skly-r}.
Let  $T(\la)$  satisfy  the quadratic relation
\be A_{12}(\la,\mu)T_1(\la)B_{12}(\la,\mu)T_2(\mu)=
T_2(\mu)C_{12}(\la,\mu)T_1(\la)D_{12}(\la,\mu)),
\ll{atbt} \ee
and commute with  $[T_1(\la),K_2(\mu)] = 0$, 
where  $A, B, C, D \  $ are numerical matrices.
 Let   
 operator $K$ be a realization of a similar algebra 
\be A_{12}^{-t}(\la,\mu)K_1^{t_1}(\la)\tilde B_{12}(\la,\mu)
K_2^{t_2}(\mu)=K_2^{t_2}(\mu)
\tilde C_{12}(\la,\mu)K_1^{t_1}(\la)D_{12}^{-t}(\la,\mu)
\ll{akbk} \ee
where
\be \tilde B_{12}(\la,\mu)={({(B_{12}^{t_1}(\la,\mu))}^{-1})}^{t_2},\ 
A_{12}^{-t}(\la,\mu)
=A_{12}^{-t_1 t_2}(\la,\mu), \
 \tilde C_{12}(\la,\mu)={({(C_{12}^{t_2}(\la,\mu))}^{-1})}^{t_1}.
\ll{abct} \ee
Then
 it is not difficult to show by direct calculation,  that 
 the {\it k--trace} $ \ti \tau(\la)=tr (\ti T(\la))=tr \l(K(\la)T(\la)\r)$
 forms a commutative
subalgebra
\[ [\ti \tau(\la),\ti \tau(\mu)] = 0. \]

We like to mention that the  spectral parameter dependent
braided extensions of QYBE's for nonultralocal models can  be  derived
through Yang-Baxterization of algebraic relations like braided FRT algebra
etc. for $R^\pm_q$ matrices satisfying Hecke condition \c{khijmp96}.
The  procedure is  similar to that
  of the ultralocal case  presented in  sect. 4,
only  with the drawback that the  spectral parameter dependence of the 
$Z$-matrices can not be achieved in this way. 
The $R(\la-\mu)$-matrices associated with 
 the nonultralocal models is   the same   
(\re{rtrign}) as for the ultralocal models, which with the inclusion of the parameters
  $\epsilon_\alpha=\pm 1$ 
is given by
 \be R_{trig}(\la)=\sum_{\alpha,\beta}\left(\sin \la
E_{\alpha \alpha}\otimes E_{\beta \beta} +\epsilon_\alpha
\sin( \la+\epsilon_\alpha\eta)
E_{\alpha \alpha}\otimes E_{\alpha \alpha}
+\sin \eta
E_{\alpha \beta}\otimes E_{\beta \alpha}\right).\ll{R(u)ep}\ee
 All the models discussed below are associated 
with this trigonometric $R(\la)$-matrix or  its rational form at at
 $q \rightarrow 1.$  
 The 
{\it standard} form (with all $\ep=+1$) is the  most 
used one, while  the  {\it nonstandard}  solutions (with some $\ep =-1$)
are needed usually for the SUSY models.
Such $R$-matrices exhibit the symmetry
  \be [R_{12}(u),s_1s_2]
=0, \quad s=
\sum_{\alpha}
e_{\alpha \alpha}\phi_\alpha.
\ll{strace}\ee
Therefore whenever  $s$ commutes with $Z$, along with the commuting  traces
$tr \ti T(\la)$ 
 the {\it s-traces}  $tr(s \ti T(\la))$  also naturally represent
a commuting family. This
property is  used    for example,
in  SUSY models  for 
constructing conserved quantities, due to physical reasons.

%----------------------------------------------+++

\subsubsection {Simplifying ansatzes for $Z$}
For 
  deriving  commuting transfer matrices from 
 (\re{bqybet}),
we notice  by comparing with  (\re{atbt}) that 
\[A_{12}(\la,\mu)= Z_{12}R_{12}(\la-\mu)Z_{21}^{-1},\ \ B=Z^{-1},\ \
 C=Z_{21}^{-1},\ \ D(\la,\mu)=R(\la-\mu).\]
For finding the operator $K(\la),$ which 
 commutes with $T(\la)$ and satisfies (\re{akbk}),
the simplest possibility is $K(\la)=k(\la){\bf I},$
where $k(\la)$ is a numerical matrix  with constraint
\be k_1(\la)\hat Z_{21} k_2(\mu)Z_{12}R_{12}(\la-\mu)=
R_{12}(\la-\mu)k_2(\mu)\hat Z_{12}k_1(\la)Z_{21}, \ll{rzkzk} \ee
where
$~~ \hat Z={({((Z^{-1})^{t_1})}^{-1})}^{t_1}.~~$
It is true that equation (\re{rzkzk})  determining the
condition for factorizability of  traces
is rather difficult to solve  for
general $Z$.
Nevertheless,
 for some special forms
 of the $Z$-matrix listed below, 
one gets  $\hat Z= Z$  simplifying  the  equation.
 These forms for $Z$ are
\\  a){  { the diagonal form }},
\\   b){ { \ factorized form}: \ \ $ Z=
A\otimes B$},
 \\c)   { specifically chosen form as}
 $   \ \  Z_{12}={\bf I}+\sum_iA_i\otimes B_i, \
 \ (B_i)^2=0,$ 
% \ll{case2}
\\ 
\noindent d) { $Z$ chosen as  rhe $R$-matrix}:
 $  \ \  Z_{12}=  R^+_{q21}
,$ \\
% \ll{ex4} 
\noindent  e) { $Z$ expressed as}
 $~~ \quad \ \  Z_{12}=
e^{h\sum_i A_i
\otimes B_i}~~$
% \ll{ex5} 
with arbitrary invertible matrices $A_i, B_i$.

It may be  checked easily  that for all the above ansatzes
 the condition   $\hat Z= Z$ holds 
and (\re{rzkzk}) reduces to 
\be R_{12}(\la-\mu)k_2(\mu) Z_{12}k_1(\la)Z_{21}=
k_1(\la) Z_{21} k_2(\mu)Z_{12}R_{12}(\la-\mu).
\ll{rzkzk1} \ee
Consequently any solution of $k(\la)$ obtained from (\re{rzkzk1}) 
for concrete $R,Z$ related to 
nonultralocal quantum models would lead to 
transfer matrices 
 $\tau(\la)=Tr (k(\la) T(\la))$  commuting for different $\la$ and that 
confirms the integrability of the system.
%--------------------------------------------------------
\subsubsection {Factorization by gauge transformation}
{ As stated, the nonultrlocality 
may be acquired through gauge transformation,
which can be used  for trace
factorization.    Let the transforming operator 
  $\hat l$ be a matrix of operators
 acting in the Hilbert space  and satisfy 
the relations
\be [{R}_{12}(\la),\hat l_{1 }\hat l_{2}]=0 ,\  \
[\hat l_{1 },\hat l_{2}]=0, \ \
 Z_{21}^{-1}T_1=\hat l_2T_1\hat l_2^{-1}.
\ll{sym3}\ee
%.\ll{case3}\ee
 Starting from
(\re{bqybet})   and  using  (\re{sym3})  the factors
$Z_{ba}^{-1}T_a, ~a,b =1,2$ appearing in the equation can be  replaced by  
$\hat l_bT_a\hat l_b^{-1}$. Multiplying from the right by $\hat l_1\hat l_2$
and from the left by its inverse and using further from (\re{sym3}) 
the commutation
relations
between $\hat l_a, {R}(\la) $ and between $\hat l_a$'s themselves
we  arrive at
\begin{equation}
R_{12}(\la-\mu)\hat l_{1}^{-1} T_{1}(\la)\hat l_1
\hat l_{2}^{-1} T_{2}(\mu)\hat l_2
=\hat l_{2}^{-1} T_{2}(\mu)\hat l_2
\hat l_{1}^{-1} T_{1}(\mu)\hat l_1
R_{12}(\la-\mu).\ll{trace3}\end{equation}
One detects the appearance of  gauge transformed monodromy matrices  
$\tilde T(\la)=\hat l^{-1} T(\la)\hat l$
and 
taking trace of (\re{trace3}) finds 
 the required commutation relation $[tr \tilde T(\la), tr \tilde T(\mu)]
=0$.
\subsection {Examples  of nonultralocal models
 and their basic equations}
We focus on  the examples of nonultralocal models listed in sect. 2
and show that the basic equations representing their quantum integrability 
can be obtained systematically 
from the set of relations (\re{bqybel}-\re{zlzl2u}), (\re{bqybet})  etc.
  derived before  from  general arguments.
These  examples 
 involve either  spectral parameter independent $R^\pm$ 
or the spectral dependent
 trigonometric matrix (\re{R(u)ep}) or its rational limit. At the same
time they
 correspond also to different braiding, i.e.  
different choices for  $\ Z, \ti Z $ matrices.
 Curiously   most of these choices for the existing models  
 agree  with the above forms of $Z$, proposed on the argument of  
 simplicity.

\subsubsection
 { Supersymmetric models }
  Quantum integrable supersymmetric
(SUSY) theory  \c{SUSY,fabian,SUSYq}
 is possible to formulate in a convenient alternative way
 expressing the gradings in a matrix form  \c{Liao}. We
  observe
that such a formulation is in agreement with  the  framework
 of nonultralocal models for the choice of homogeneous  braiding
 $Z=\tilde Z=\eta$ with 
$\eta$ as the diagonal matrix 
  \be \eta_{12}=\sum_{\alpha,\beta}
g_{\alpha \beta}
e_{\alpha \alpha}\otimes e_{\beta \beta}
~~~\mbox {with}~~~ g_{\alpha\beta}=(-1)^{\hat \alpha \hat
 \beta}.\ll{susy}\ee
 Here $\hat \alpha=0(1)$   is the supersymmetric grading
 depending on the even (odd)-ness of the
indices.
Due to the obvious property $\eta_{12}\eta_{21}= 1$,
 we get from (\re{rzkzk1}) $k=1$ and 
  integrability follows from the commuting traces.
 However supertraces,   obtained from the s-traces
 (\re{strace}) by choosing $\phi_\alpha=(-1)^{\hat \alpha}$
are usually used for defining conserved quantities for such models.
The  super QYBE's for the
SUSY models can be derived from the general relations 
 (\re{bqybel} ,\re{bqybet}) as
\begin{equation}
{R}_{12}(\la-\mu)\eta_{12}L_{1j}(\la) \eta_{12}L_{2j}(\mu)
= \eta_{12}L_{2j}(\mu) \eta_{12}L_{1j}(\la){R}_{12}(\la-\mu)
\ll{bqybels}\end{equation}
and
\begin{equation}
{R}_{12}(\la-\mu)\eta_{12} T_{1}(\la)\eta_{12} T_{2}(\mu)
= \eta_{12} T_{2}(\mu) \eta_{12} T_{1}(\la){R}_{12}(\la-\mu).
\ll{bqybets}\end{equation}
On the other hand, 
 the nonultralocal braiding
 (\re{zlzl1u},\re{zlzl2u})
correspond to  the supercommutation relations 

\begin{equation}
 L_{2 k}(\mu)\eta_{12}L_{1 j}(\la)
=\eta_{12}L_{1 j}(\la)\eta_{12} L_{2 k}(\mu)\eta_{12}
\ll{zlzl24}\end{equation}
for $k\neq j$.
For  comparing  with the conventional results  on SUSY models,
  we express the QYBE  (\re{bqybels})
and the braiding relation   (\re{zlzl24}) 
 in   matrix elements to give 

\begin{equation}
{R}_{{a_1}a_2}^{{b_1}b_2}(\la-\mu)(L_{b_1}^{c_1}(\la))_j 
 (L_{b_2}^{c_2}(\mu))_j
(-1)^{\hat b_2 (\hat b_1+\hat c_1)}
=(-1)^{\hat a_1 (\hat a_2+\hat b_2)}(L_{a_2}^{b_2}(\mu))_j 
 (L_{a_1}^{b_1}(\la))_j
{R}_{{b_1}b_2}^{{c_1}c_2}(\la-\mu)
\ll{bqybelse}\end{equation}
and \be
(L_{a_2}^{b_2}(\la))_k  (L_{a_1}^{b_1}(\mu))_j
=(-1)^{(\hat a_1 +\hat b_1)(\hat a_2 +\hat b_2)}
(L_{a_1}^{b_1}(\la))_j  (L_{a_2}^{b_2}(\mu))_k
,\ee
respectively. Note that this  reproduces 
 exactly the results of \c{SUSY,fabian} and
\c{SUSYq} with 
  $R$-matrix as rational (\re{Rany})
or the  trigonometric solutions (\re{R(u)ep}) of  nonstandard type 
with proper choice of  $\ep_\al=\pm 1$ depending on the specific gradings.
%-----------------------------------
\subsubsection {Integrable anyonic  models}
Quantum integrable anyonic models  (IAM)
can be introduced by generalizing (\re{susy})
 in the nonultralocal description of the SUSY models
 by choosing 
  $~~ g_{\al \bet}=
e^{i\theta \ \hat \alpha \hat \beta},
~~$
%\ll{Zany}\ee
where $\theta$ is the arbitrary anyonic phase and $\hat \alpha=1$ is the
anyonic grading, while  $\hat \alpha=0$ gives bosonic commutations.
For such a $\tilde Z=Z$-matrix the braiding
relation describing  commuting relations between
Lax operator at different
lattice sites $k>j$ is given by
\begin{equation}
L_{b_2(k)}^{a_2}(\mu) L_{b_1( j)}^{a_1}(\lambda)
= e^{-i \theta (\hat a_1- \hat b_1)(\hat a_2- \hat b_2)}
  L_{b_1( j)}^{a_1}(\lambda) L_{b_2(k)}^{a_2}(\mu)
,\ll{brany}\end{equation}
with all matrix indices running from $1$ to $\mathsf {N}=m+n$.
The algebraic relations at the same point $l$ on the other hand
are given by the
corresponding braided QYBE
\begin{equation}
\sum_{\{k\}}R^{k_1k_2}_{a_1a_2}(\lambda-\mu)
L_{k_1(l)}^{b_1}(\lambda) L_{k_2(l)}^{b_2}(\mu)
e^{i \theta \hat k_2(\hat b_1- \hat k_1)}
=\sum_{\{k'\}}
e^{i \theta \hat a_1(\hat k'_2- \hat a_2)}
  L_{a_2(l)}^{k'_2}(\mu)L_{a_1(l)}^{k'_1}(\lambda)
R^{b_1b_2}_{k'_1k'_2}(\lambda-\mu)
,\ll{bqybelany}\end{equation}

  The quantum $R$-matrix may be chosen in the trigonometric form
(\re{R(u)ep})
  or as its rational limit
  \be
  R_{12}(\lambda)=t_{12} \lambda+\frac {\hbar}{2}P_{12}
  ,\ll{Rany}\ee
where $P$ is the permutation matrix and $t={ I}$ for the {\it standard}
and $t_{12}= \eta_{12}$ for the {\it nonstandard} solutions.
The consistency equations for $Z$ and $R$ matrices obviously hold for
all these solutions.
A possible  Lax operator of the anyonic 
model may be proposed as
\be
L_{a(l)}^{b}(\lambda)
=\lambda \delta_{ab} \ p^{0(l)}_{b}+ \frac {\hbar}{2}
e^{i\theta (\hat a \hat b)} p^{(l)}_{ba}.
\ll{Lany}\ee
Using  rational $R$-matrix
(\re{Rany}) one may derive from (\re{bqybelany}) and (\re {brany})
the anyonic  super algebras
by matching the coefficients of different powers of the spectral
parameter   
\c{khijmp96}. For standard and nonstandard solutions of $R$-matrix
these algebras also differ.
They are  nontrivial generalization of the 
 well known graded super algebra \cite{SUSY},
where together with an  anyonic parameter $\theta$ a set of
 additional $N$ operators $  p^{0(l)}_{a}    $  appear.
  Nonstandard $R$-matrix and the  choice  $ \theta=\pi$ recovers
the known super algebra.

For constructing quantum integrable models involving $m$ number of
bosons $b^{(k)}_a$
and $n$ number of
anyons $f^{(k)}_\alpha $,
one may consider a realization of the generators as
 \be
p^{(k)}_{ab} = b^{\dagger(k)}_a \ b^{(k)}_b, \quad
p^{(k)}_{\alpha a} = f^{+(k)}_\alpha \ b^{(k)}_a,\\  \quad
p^{(k)}_{a \alpha } = b^{\dagger(k)}_a \ f^{(k)}_\alpha,\quad
p^{(k)}_{\alpha \beta} = f^{+(k)}_\alpha \ f^{(k)}_\beta.
\ll{realany}\ee
The bosons satisfy the standard  commutation rules  and 
commute with  anyonic operators, while  the anyonic operators satisfy 
the  commutation relations
\be
 f^{(k)}_\alpha \ f^{+(j)}_\beta = e^{i\theta} \
 f^{+(j)}_\beta \ f^{(k)}_\alpha ,
%\ll{any1}\ee \be
~~~~ f^{(k)}_\alpha \ f^{(j)}_\beta = e^{-i\theta} \
 f^{(j)}_\beta \ f^{(k)}_\alpha ,\ll{any2}\ee
for $k>j$ and \be              [ p^{0(k)}_{\alpha },  \
 f^{(j)}_\beta]=0 ,~~~~
%\ll{any2a}\ee~~~\mbox{for $k\neq j$, ~~ while}~~
%\be
 f^{(l)}_\alpha \ f^{+(l)}_\beta - e^{i\theta} \
 f^{+(l)}_\beta \ f^{(l)}_\alpha = \delta_{\alpha \beta}
 p^{0(l)}_{\alpha }  \ll{any3}\ee
and
\be
p^{0(l)}_{\alpha }   f^{(l)}_\beta =  e^{-i\theta} \
 f^{(l)}_\beta p^{0(l)}_{\alpha }
, \quad [ f^{(l)}_\alpha ,\ ~~~~~~ f^{(l)}_\beta]=0,
  \ll{any4}\ee
at the same lattice points.
 Note that these relations are consistent
with the known anyonic algebra  \cite{anyon1},  though defined in
$1$-dimension. Even though
  explicit construction of integrable anyonic model  
has not been achieved yet, Kundu-Eckhaus equation considered below 
is however an interesting example of a similar realization.
For discussion on  various  physical aspects of the  anyon theory  we refer to 
\c{anyon0}.

By  choosing  trigonometric $R$-matrix solution 
and the same $Z$-matrix, 
one may obtain  a  $q$-deformed anyonic algebra \c{khijmp96}, which   
represents  a  generalization of the extended trigonometric
Sklyanin algebra \c {kbmpl92} to include  anyonic parameter
$\theta$ and
nonultralocal braiding.

%-------------MODULI
\subsubsection {Integrable model on moduli space}
In considering Hamiltonian Chern-Simons theory, moduli space of flat
connections on the Riemann surface with marked points appear.
 From
monodromies of flat connections along fundamental cycles of the Riemann
surface an  integrable model may be constructed 
(listed as IMMS in sect. 2), with   spectral parameter dependent
Lax operator   \c{alex} 
\be
L_n(\la) = K^n + \la I . \ll{Lmoduli}\ee 
The $K^n$ matrices are defined as 
\be
\K{n}{} = (\L-1)^{-1} \cdots (\L-{n-1})^{-1} K_n \L-{n-1} \cdots
\L-1 , \ll{Knmodu}\ee
by
 setting $K_i = (\L-i)^{-1} \L+i$, where $L^\pm$ satisfy the FRT relations
(\re{frt}).
  Using the
definition  (\re{Knmodu}) and the equations for $L^\pm$ 
the equations  satisfied by the Lax operator (\re{Lmoduli})
can be  shown  \c{alex,moduli}
to be 
\be 
S(\la,\mu) L_{1n}(\la) \Rt L_{2n}(\mu) = L_{2n}(\mu) R L_{1n}(\la)
\St(\la,\mu,)
\ll{bqybemodu}\ee
where $\St(\la,\mu) $ and $S(\la,\mu)=(\Rt)^{-1} \St(\la,\mu) R $
are  solutions of YBE (\re{ybe}), along with 
\be
   R L_{1m}(\la) \Ri L_{2n}(\mu) = L_{2n}(\mu) R L_{1m}(\la)
\Ri , \qquad m>n  \ll{braidmodu}\ee
These equations induce for the 
monodromy matrix $T(\la)=\prod_iL_i(\la)$  the   relation
\be
S(\la,\mu) T_1(\la) \Rt\, T_2(\mu) = T_2(\mu) R\, T_1(\la) \St(\la,\mu)
\ll{bybetmodu}\ee
giving commuting transfer  matrices
\be t(\la) = tr_q ( T(\la) ) = tr ( M T(\la) ) ,
\qquad M = diag(q^{-1} ,q) \ll{tracemodu}\ee
 for different values of spectral parameters 
 and thereby producing  a family of conserved operators in
involution.

This model can  be shown to be well described by 
 the general  nonultralocal
theory   presented here, 
 for which we identify $\St(\la,\mu) $ as the trigonometric $R_{trig}(\la-\mu)$
and  choose  $\ti Z_{12}=Z_{21}^{-1}=R_{12}$. This clearly 
 rewrites the general BQYBE
(\re{bqybel})  into (\re{bqybemodu}) and the braiding relation
(\re{zlzl2u}) for  different points into (\re{braidmodu}). The braided QYBE
(\re{rztzt}) for
the monodromy matrix related to nonperiodic models coincide therefore with 
(\re{bybetmodu}) recovering  the earlier  results. 
%_____________
We discuss below examples of models, which  unlike the  homogeneous
 long range
nonultralocality allows braiding 
 only  with  nearest-neighbors.
First we  look into  some models
described by  the spectral parameter independent
$R$-matrices   with equations like (\re{bqybe+}) and 
related to  the conformal
field theory,  without any demand on 
 trace factorization.
%---------------------+++3.8
\subsubsection { Current algebra in WZWN model}

The WZWN model is described by the unitary unimodular matrix-valued field
 $g(x,t): M \rightarrow su(2), \ \ M=S^1 \times R^1$.
Defining   the chiral left current as
$L= \frac {1} {2} (J_0+J_1)$, where $J_\mu= \partial_\mu gg^{-1},$
one gets the well known current algebra relation 
(\re{current}).
Introducing $\partial_xu = L u$ one can write the monodromy as
$M_L=u(x)^{-1}u(x+2\pi)$ and the similar relations for the right current.
With the aim of unraveling the quantum group structure in WZWN model
 a discrete  quantum
version of the current algebra
was formulated in \c{wzwn}.
We demonstrate that the basic equations found in \c{wzwn} can be reproduced
by choosing  $Z_{12}=  R^+_{q21}$ with $\ti Z=1$
in the spectral parameterless limit  (\re{bqybe+}) 
  of the 
braided equations as
\[
{R}^+_{q12}({R}^+_{q12})^{-1}L^+_{1j}  L_{2j}^+
= ({R}^+_{q12})^{-1}L_{2j}^+L_{1j}^+
({R}^+_{q12}), \]
which  
considering the relations 
 $ R^+_{q12}= (R^+_{{\bar q}12})^{-1}=
R^-_{{\bar q}21}$ with ${\bar q}=q^{-1}$ and $ Z^{-1}_{21}=
 R^+_{{\bar q}12}$ gives 
\be {R}^-_{{\bar q}12}L_{1j} L_{2j}
{R}^+_{{\bar q}12}=L_{2j}L_{1j},
\ll{fadcurrent0}\ee
where we have suppressed the $+$ symbol in $L_{aj}$.
Similarly the spectral independent limit of the braiding relations 
for the present choice of $Z, \ti Z$ reduce to 
\begin{equation}
 L_{2 j+1}
{R}^+_{{\bar q}12}L_{1 j}
=L_{1 j}L_{2 j+1}
,\quad
L_{2 k} L_{1 j}
=L_{1 j} L_{2 k}
\ll{fadcurrent2}\end{equation}
for $k>j+1$.
We observe 
 that the nonultralocal relations  (\re{fadcurrent0}-\re
 {fadcurrent2})
coincide exactly with the current algebras obtained in \c{wzwn}. 
Identifying further the chiral components as
 $~~u_j=T^{[j,1]},~~$ we get from (\re{rztzt}) a relation analogous to 
(\re{fadcurrent0}):
\be
{R}^+_{{\bar q}12}u_{1j} u_{2j}
{R}^-_{{\bar q}12}=u_{2j}u_{1j}
\ll{fadchiral}\end{equation}
Since $~~u_{n}=L_{n} \ldots L_{j+1}u_{j},~~$ multiplying 
(\re{fadchiral}) from the left successively by 
$~~L_{2j+1},L_{2j+2},\ldots L_{2n}~~ $ and using the braiding  
 relations (\re {fadcurrent2}) we can derive from (\re {fadchiral}) 
the algebra at $n>j$ as 
\be
u_{1n} u_{2j}
{R}^+_{{\bar q}12}=u_{2j}u_{1n}
\ll{fadchiral2}\end{equation}
Similar exchange relations can also be derived 
for $T_{1}^{[j,k]}, T^{[n,k]}
$.
For    monodromy matrix $M^L_{1} $ on the other hand,
 we rewrite  equation 
(\re{bqybet}) suitable  for the   
periodic models:
\be
M_{1}^L
({R}^-_{{\bar q}12})^{-1} M_{2}^L
{R}^-_{{\bar q}12}=
({R}^+_{{\bar q}12})^{-1}M_{2}^L
{R}^+_{{\bar q}12}M_{1}^L
\ll{fadmonodromy}\end{equation}
We notice again  
 that the  relations  (\re{fadchiral}-\re
 {fadmonodromy}) recover correctly the results of \c{wzwn}. 

It is checked easily,  that the present  $Z_{12}=R^+_{q21}$ 
satisfies  the conditions 
(\re{braid}) and (\re{bybe12}) for $ R_{12} = R^+_{q12}   $ and since it 
corresponds to the   trace factorization case d),
the model might serve as a candidate for  
 constructing  nonultralocal quantum integrable models.
Moreover,  if the long range braiding given by  
   $ \tilde Z$  is  chosen nontrivially, it might
also lead to a generalization of the current algebra 
of the WZWN model.
%-------------------------------------------------------
\subsubsection
 { Braided algebra}
 As described in sect. 3 the braided algebra (BA) may be given in the compact
form by the matrix relations (\re{braided1},\re{braided2}), where the elements
of the matrices $T$ and $T'$ are two copies 
of the generators of BA. It is  easily checked  that by a  homogeneous
choice for the   braiding as  
$Z_{12}=\ti Z_{12}=R_{21},$ one can  generate  from the same spectralfree
BQYBE (\re{bqybe+}) the braided algebra relations (\re{braided1},\re{braided2}).
\subsubsection
 { Quantum group structure in Coulomb gas picture of  CFT}
The Coulomb gas picture of CFT (CG-CFT) 
is based on the Drinfeld-Sokolov linear
system   \cite{babelon}
$\ \ \partial Q= L(x)Q,\ $ given by  the Lax operator
(\re{pbcft}) involving  periodic field
$P(x)$ with nonultralocal bracket. In
  \cite{babelon} a lattice regularized description was given
at the quantum level. We show again that the  same result can be obtained 
from the general relations formulated here for the nonultralocal models
at the spectralfree  limit considered in the above example.

We choose $\ti Z=1$ with  nontrivial braiding  
  $Z_{12}   = q^{ \sum_i H_i
\otimes H_i},~~ H_i=e_{ii}-e_{i+1 i+1},$ 
in agreement with  simplification   e), which reduces 
(\re{rzkzk})  to
$
[ R, C]=
0, \ \ C_{12}= Z^2_{12}k_1k_2.
$ Noticing that, $ H_{i1}+ H_{i2}$ commutes with
 the parameterfree $R^+_q$-matrix, 
 we may
 extract a solution  $k = q^{\sum_i H^2_i
} $, since   $ C_{12}
=Z^2_{12}k_1k_2= q^{\sum_i (H_{i1}+ H_{i2})^2}.$
Defining $\tilde L_{aj}=B_a^{-1} L_{aj},~~B=k^{\frac {1} {2}}$
we can eliminate  the  $Z$ dependence from the parameterless  BQYBE
(\re{bqybe+})
 yielding
\be
{R}_{12}^+\tilde L_{1j}\tilde L_{2j}
=\tilde L_{2j}\tilde L_{1j}
{R}_{12}^+
\ll{babcurrent10}\end{equation}
The braiding relations on the other hand give
\begin{equation}
\tilde L_{2 j+1}
{A}_{12}\tilde L_{1 j}
=\tilde L_{1 j} \tilde L_{2 j+1}
,\quad
\tilde L_{2 k}\tilde L_{1 j}
=\tilde L_{1 j}\tilde L_{2 k}, \ \ k>j+1
\ll{babcurrent2}\end{equation}
where ${A}_{12}\equiv {Z}_{12}^{-1}$.
 These equations recover  the related findings  of \c{babelon}, while
 their algebras of monodromies can be  found
from (\re{rztzt}) and (\re{bqybet}) at the parameterfree limit, by defining 
$ Q_j\equiv  B^{-1}T^{[j,1]}$ and   $ S\equiv  B^{-1}T$ as
\be
{R}^+_{12} Q_{1j}Q_{2j}
=Q_{2j}Q_{1j}
{R}^+_{12}
\ll{babmono1}\end{equation}
and 
\be
{R}^+_{12} S_{1}A_{12}S_{2}
=S_{2}A_{12}S_{1}
{R}^+_{12}.
\ll{babmono2}\end{equation}
%---------------------------------------------------

The following examples are of genuine integrable models with
 trigonometric or rational  $R(\la)$-matrix and spectral 
parameter dependent Lax operators.
\subsubsection
 { Nonabelian Toda chain }
NATC  is a one-dimensional evolution model on a periodic lattice
of $N$ sites described by
the nonabelian matrix-valued operator
$g_k \in GL(n)$. The model, representing  a discrete
analog of principal chiral field, was  set into the Yang-Baxter
formalism in \c{natoda} with  the nonultralocal Lax operator
  (\re{Lkor}).
The associated
 $R$-matrix is
 the rational solution  $R(\lambda)=ih P-\lambda$
 with $P= \Pi\otimes \pi  ,$
where $\Pi$
is a $4 \times 4$ and $\pi$   a $\ n^2 \times n^2$ permutation matrix.
We show that the  model can  be described well by the general  formalism
with the  choice   $\tilde Z=I$  and $Z_{12}=
 { 1} + i { h }(
 e_{22}\otimes e_{12})\otimes \pi$.
 This  is an example of  models with acquired nonultralocality
through gauge transformation.
 Using  (\re{Lkor})
 one  shows
 that $Z_{21}^{-1}L_{1N}=\hat l_2L_{1N}\hat l_2^{-1}
$ with $\hat l= diag({ I},g_N)$ and
 derives 
$ Z_{21}^{-1}T_1=\hat l_2T_1\hat l_2^{-1}$  and 
  other conditions of 
(\re{sym3}).
Therefore the quantum integrability of the model follows from
trace identity  (\re{trace3}).
 With this input  for $R,Z,\ti Z$
our main formulas
(\re{bqybel},
\re{bqybet}) recover  the QYBE's of \c{natoda}, while
 the  braiding relations are  given by
\begin{equation}
 L_{2 j+1}(\mu)Z_{21}^{-1}L_{1 j}(\la)
=L_{1 j}(\la)L_{2 j+1}(\mu)
,\quad
L_{2 k}(\mu)L_{1 j}(\la)
=L_{1 j}(\la) L_{2 k}(\mu)
\ll{korepin2}\end{equation}
for $k>j+1$.
%--------------------------+++++
\subsubsection { Nonultralocal quantum mapping }
The quantum mapping \c{Nijhof} $v_j \rw v'_j$ is
related to the
lattice version of the KdV type equation
\be v'_{2j-1}=v_{2j}, ~~v'_{2j}= v_{2j+1} {\ep \de \ov v_{2j}}
 -{\ep \de \ov v_{2j+2}}, ~~~~~\sum_{j=1}^{N+1}v_{2j}
=\sum_{j=0}^{N}v_{2j+1}=c .\ll{eqm}\ee
The generalization  of integrable  NQM is associated with   the Lax operator 
$L_n=V_{2n}V_{2n-1}$, with
$V_{n}(\la)$ (\re{Lqm}) involving 
hermitian nonultralocal operators $v_{i,j}$ (\re{algqm}).
 The  $R(\la)$-matrix is given by the rational solution
 \be R_{12}(\la_1-\la_2)= {I}+ \frac { h P_{12}}{\la_1-\la_2}
\ll{exR}\ee
with the braiding 
\be  Z_{12}(\la_2)= S_{12}(\la_2)^{-1}
=
 {\bf 1} + \frac { h }{\la_2}\sum_{\alpha}^{N-1}
 e_{N\alpha}\otimes e_{\alpha
 N}\ll{Zqm}\ee 
and similarly for $Z_{21}(\la_1)$
along with the choice $\ti Z=1$.
This  is an interesting   example for the nonultralocal scheme, since
it is the only model where 
 the braiding  matrices are spectral parameter dependent.
We can see that the  choice coincides with the simplifying ansatz 
c) presented above and therefore trace factorization is possible 
with $k=1$. Consequently  $tr T(\la)$  constitutes the commuting family
along with  $tr(sT)$, where  $s_a=diag(1,1+{h \ov \la_a})$
\c{Nijhof}.
We should stress here  on a  subtlety indicated  in sect. 2, 
that  the present $Z_{12}(\la_2)$ does not satisfy the standard YBE
(\re{ybe}), while the same equation  holds for  
the matrix $~~ \tilde R_{12}(\la_1,\la_2)=
 (S_{12}(\la_2))^{-1}  R_{12} (\la_1-\la_2) S_{21}(\la_1) ~.~$

Putting $\ti Z=1$ in the braided QYBE's (\re{bqybel})
 and introducing $\ti R(\la_1,\la_2)$,
 we derive therefore 
\begin{equation}
\tilde R_{12}(\la_1,\la_2)L_{1j}(\la_1)L_{2j}(\la_2)
= L_{2j}(\la_2)L_{1j}(\la_1){R}_{12}(\la_1-\la_2)
,\ll{nijhof1}\end{equation}
and from the braiding relations (\re{zlzl1u}-\re{zlzl2u}) the 
nonultralocal condition
\begin{equation}
 L_{2 j+1}(\la_2)S_{21}(\la_1)L_{1 j}(\la_1)
=L_{1 j}(\la_1)L_{2 j+1}(\la_2)
\ll{nijhof3}\end{equation} and the trivial commutation
$~~
 L_{2 k}(\la_2)L_{1 j}(\la_1)
=L_{1 j}(\la_1) L_{2 k}(\la_2), \ \
~~$
%\ll{nijhof4}\end{equation}
 {for} $  k>j+1.$

The equation for the monodromy matrix with the periodic
boundary condition can similarly be obtained from  
(\re{bqybet})
 in  the form
\begin{equation}
\tilde R_{12}(\la_1-\la_2)T_{1}(\la_1)S_{12}(\la_2)T_{2}(\la_2)
= T_{2}(\la_2)S_{21}(\la_1)T_{1}(\la_1){R}_{12}(\la_1-\la_2)
\ll{nijhof2}\end{equation}
Thus all the basic equations of NQM (\c{Nijhof}) can be 
 obtained   systematically  from
the general  nonultralocal scheme. 
\subsubsection { Reflection equation}
The reflection equation (RE) due to  \c{skly-r} can
 also be derived  from the
nonultralocal point of view, by considering a monodromy matrix 
arranged as $\ti T(\la)=\prod^{2N}_j
\tilde L_j(\la)$ such that 
\be \tilde L_j(\la)= \left \{ \begin{array}{ll} L_{j-N}(\la)& ~\mbox{for} ~ j
>N\\~~  K^-(\la) &~ \mbox{for} 
~j=N \\L^{-1}_{N-j}(-\la)&~ \mbox{for}~  j<N \ea \right.\ll{lre}\ee
 with $L_j(\la)$ being any ultralocal Lax operator. As a result the local QYBE
becomes the standard one: $
~~{R}_{12}(\la-\mu)\ti L_{1j}(\la)\ti L_{2j}(\mu)
=\ti L_{2j}(\mu) \ti L_{1j}(\la){R}_{12}(\la-\mu)
~~$ with the braiding
\begin{equation}
\ti L_{2 (2N-j)}(\mu)
R_{12}(\la+\mu)\ti L_{1 (j+1)}(\la)
=\ti L_{1 (j+1)}(\la)R_{12}(\la+\mu)\ti
 L_{2 (2N-j)}(\mu) .
\ll{braidre}\end{equation}
Due to the braiding  relation (\re{braidre}) 
and the equation for $  K^-(\la)  $:
\begin{equation}
 ~{R}_{12}(\la-\mu)K^-_{1}(\la)
R_{12}(\la+\mu)K^-_{2}(\mu)
=K^-_{2}(\mu)R_{12}(\la+\mu)K^-_{1}(\la){R}_{12}(\la-\mu), 
\ll{Kre}\end{equation}
the global QYBE  takes the form
\begin{equation}
 ~{R}_{12}(\la-\mu)\ti T_{1}(\la)
R_{12}(\la+\mu)\ti T_{2}(\mu)
=\ti T_{2}(\mu)R_{12}(\la+\mu)\ti T_{1}(\la){R}_{12}(\la-\mu), 
\ll{Tre}\end{equation}
and recovers the reflection equation of \c{skly-r}. 
It is interesting to note that, 
though it refers to a  
braiding different from that   considered here, the choice $Z=1, \tilde
Z_{12}=\ti Z_{21}=R_{12}(u_1+u_2)$, with symmetric $R$-matrices
derives formally  the reflection equation 
(\re{Tre}) from BQYBE (\re{rztzt}).
\subsubsection { Kundu-Eckhaus equation}
Higher order nonlinear equation KEE (\re{kuEk}) can be
 derived  from the NLS equation by  gauge transformation \c{kun84}.
However  the quantum  model becomes nonultralocal with 
the field $~~\ti \psi_n=  A_{n+1}\psi_n A_n,~~~A_n= e^{-i \th
 \De
 \sum_{i=1}^n    (\ti \psi^\dagger_i \ti \psi_i) }~
 $ exhibiting anyon like algebra
\be 
 \ti \psi_n 
\ti \psi_m= e^{2i \th} \ti \psi_m \ti \psi_n,~~\mbox{for}~~n>m+1,~
\ \ \  
\ti \psi_{m+1} \ti \psi_m =e^{i \th}\ti \psi_m 
\ti \psi_{m+1} 
\ll{nmKE}\ee
 along with the standard relation
$[\ti \psi_n ,\ti \psi^\dagger_n]={\hbar \ov \De}$ at the same point.
Note that the braiding relations are inhomogeneous and  different
 from the pure anyonic relations.  
The lattice regularization of the Lax operator (\re{lkuEk}) can be 
obtained as  $~~L_n=h_{n+1}L_n^{(lnls)}h_n^{-1}~~$
by considering the gauge transformation $~~h_n=A_n^{ \si^3 }
$ over the lattice NLS Lax  operator $~~L_n^{(lnls)}~~$ (\re{lnls}).

Using the commutation relations of $\ti \psi_n$ the braiding relations can
be
derived as 
\begin{equation}
 L_{2 j+1}(\mu)L_{1 j}(\la)
=L_{1 j}(\la)Z_{12}
 L_{2 j+1}(\mu) Z_{12}^{-1}
,~~
%\ll{zlzl1u}\end{equation}
 L_{2 k}(\mu)Z_{12}^{-1}L_{1 j}(\la)
= Z_{12}^{-1}L_{1 j}(\la) Z_{12}
 L_{2 k}(\mu) Z_{12}^{-1}
\ll{zlzlKE}
\end{equation}
with $k>j+1$ and 
 the braiding matrix 
$Z_{12}=Z_{21}=
 diag (e^{i\theta},1,1,e^{i\theta})$.
For  periodic BC on   $\psi_n,$ the KEE model does not show periodicity :$
L_{N+1}=h_{N+1}^{-1}L_{1}h_{N}$ and  we have
$~~\ti \psi_N 
\ti \psi_1= e^{3i \th} \ti \psi_1 \ti \psi_N~~$ giving
\be
 L_{2 N}(\mu)Z_{12}^{-1}L_{1 1}(\la)
= Z_{12}^{-2}L_{1 1}(\la) Z^2_{12}
 L_{2 N}(\mu) Z_{12}^{-1}.
\ll{zlzlKEn}
\end{equation}
 The braided QYBE with  rational $R(\la)$-matrix
has the simple form 
\begin{equation}
{R}_{12}(\la-\mu)L_{1j}(\la) L_{2j}(\mu)
= L_{2j}(\mu) L_{1j}(\la){R}_{12}(\la-\mu),
\ll{bqybelKE}\end{equation}
since $[R(\la),Z]=0$. Observe that the braiding for this model, 
 differs from the  types assumed 
for the  nonultralocal scheme.
Since the braiding in general can be widely different,  certain cases 
 naturally might go beyond  any particular class of choice and should be 
treated individually.
The KEE model due to peculiar inhomogeneous braiding  also gives trouble in
achieving trace factorization. By defining the monodromy matrix
in a   special way
$~~T(\la)=L_N(\la)\ldots L_2(\la)\ti L_1(\la)$, where $\ti L_1(\la) =Z^{-1}
 L_1(\la)Z~~$ one may derive  the global BQYBE 
\begin{equation}
{R}(\la-\mu)T(\la) Z T(\mu)
= T(\mu) ZT(\la){R}(\la-\mu).
\ll{bqybetKE}\end{equation}
 However, since $T\equiv T_{12}$ is defined now in both the  auxiliary spaces,
it is difficult to factorize the trace in $ tr(T(\la) Z T(\mu))
= tr (T(\mu) ZT(\la)), $ obtained from (\re{bqybetKE}).

\subsubsection { Quantum mKdV model  }
Modified KdV equation
is a well known classical integrable system  \c{soliton} with wide range
of applications \c{mkdv}. 
The  quantum mKdV (QMKD)  model is important for its relation with the 
conformal field theory \c{BLZ}. 
 However, 
due to its nonultralocal algebraic structure
$~~\{ v(x),v(y) \}=\mp 2\delta'(x-y)
~~$
 the  quantum generalization
of the model  could not be done  through standard QISM  \c{fadrev},
but  was achieved  
following the  scheme  for quantum nonultralocal models \c{kmpl95}
and presented here .

A lattice regularized Lax operator involving  
 a single field  \c{volkov} had  a problem of mismatching
the number of fields, which 
 was overcome   by  introducing   
the  Lax operator \c{kmpl95}
\be
L_{k}(\xi) = \left( \begin{array}{c}
(W_k^{-})^{-1}
\qquad  \ \
 i\De \xi W_k^{+}\\
-i\De \xi (W_k^{+})^{-1}
\qquad \ W_k^{-}
         \end{array}   \right),
\ll{Lmkdv}\ee
with two different quantum  operators $W_k^{\pm}=e^{iv_{j}^{\pm} }$.
 The  operators ${iv_{j}^{\pm} }$     satisfy
 nonultralocal   commutators
\be
[v^\pm_k, v^\pm_l]=\mp i \ga {\hbar \ov 2} (\delta_{k-1,l}- \delta_{k,l-1})
,\ll{cr1}\ee
and \be
[v^+_k, v^-_l]=i \ga {\hbar \ov 2}
(\delta_{k-1,l}- 2\delta_{k,l}+\delta_{k,l-1})
  \ll{cr2}\ee
which go at the continuum limit 
  $\De \rw 0,$
to 
\be
[v^\pm(x), v^\pm(y)]=\pm i \ga {\hbar \ov 2}
(\delta_{x}(x-y)- \delta_{y}(x-y)) =  \pm i \ga {\hbar } \de'(x-y)
   \ll{cr3}\ee
with
 discrete operators
$v^\pm_k$ turning  to the quantum field $\Delta v^\pm(x) $ and
 $ \partial_k {\de_{kl} \ov \De^2} \rw \de_x(x-y)$.
The  reduction of operators $ W_{j}^{\pm} \approx { I }
+i \De v^\pm (x) $
at the continuum limit yields from (\re{Lmkdv})
 $~~ L_{k}(v^-,v^+,\xi) \rightarrow { I} + \Delta {\cal L}(v^-,\xi)+
 O ({\Delta}^2 (v^+) ) , \ ~~$
where ${\cal L}(v^-,\xi)$ given by (\re{Lcmkdv}) is 
linked  to the well known mkdv  Lax operator 
 \c{soliton} as $~~\si^1
{\cal L}(v^-,\xi)= U(v(x), \xi)~~$.
It is remarkable that, though  the lattice Lax operator (\re{Lmkdv})
 contains  both
$v^\pm$ variables having  nontrivial commutation relation (\re{cr2}),
dependence of  one of 
them drops out from the continuum Lax operator (in the
present case it is $ v^+$).

Inserting  the lattice  Lax operator (\re
{Lmkdv}) in the braiding relations (\re{zlzl1u},\re{zlzl2u})
and  using  commutators (\re{cr1},\re{cr2}), 
we derive the matrices $~~ Z_{12}=Z_{21}=q^{-\si^3\otimes \si^3}~~$and$~~\ti
Z~=~I$.
 Plugging the 
  $Z$-matrices  
in  (\re{bqybel}) 
one finds that 
the $R(\la)$-matrix of the  lattice mKdV  model is the same 
   trigonometric solution
related to the $XXZ$-spin chain.
 Since this  $R(\la)$-matrix commutes with  the present  $Z$,
the braided QYBE's 
 simplify  to
\begin{equation}
{R}_{12}(\lambda-\mu) L_{1j}(\lambda)L_{2j}(\mu)
=  L_{2j}(\mu) L_{1j}(\lambda){R}_{12}(\lambda-\mu)
\ll{ybelmkdv}\end{equation}
and  
\begin{equation}
{R}_{12}(\lambda-\mu) T_{1}(\lambda)Z_{12}^{-1} T_{2}(\mu)
=  T_{2}(\mu) Z_{12}^{-1} T_{1}(\lambda){R}_{12}(\lambda-\mu)
,\ll{ybetmkdv}\end{equation}
for the periodic case.

For identifying  conserved quantities
from (\re{bqybet}),  one  finds easily that the $k=1$
is a trace factorization solution
giving 
$~~ [tr  T(\la),tr T(\mu)]= 0  \quad .$  Similarly 
  due to $[R_{12}, s_1s_2]=0$, the $s$-trace
$~~\tau_q (\la)= tr (s T(\la))= q^{-{\kap \ov 2}} A(\la) +
q^{{\kap \ov 2}} D(\la) ~~$
 with
$s= q^{-{\kap \ov 2} \si^3},$ also
  generates  the commuting family of conserved quantities.
Exact solution of  the eigenvalue problem 
of this model will be considered below.
Recently some other forms of quantum lattice KdV model have also been 
investigated in a different way \c{semenov95,volkov95}.

Thus we have seen that nonultralocal quantum models proposed in different
contexts at different times  and solved in individual ways, can be unified
under one scheme. All their basic equations can be derived
systematically starting from the same braided QYBE and the related braiding
relations for different choices of $Z, \ti Z$ and $R$-matrices
(see fig. 8.1). 
Therefore,
much in parallel to the QYBE for the ultralocal models, its braided
extensions can be considered to be some representing universal equations
for describing integrable nonultralocal models of certain class.
 
%%%%%%%%%%%%%%%%%%%%%%%%%%%%%%%%%%%%%%%%%%%%%%%%%%%%%%%%
%%%%%%%%%%%%%%%%%%%%%%%%%%%%%%%%%%%%%%%%%%%%
\subsection {Algebraic Bethe ansatz solution for the quantum mKdV model}
To demonstrate  that the Bethe ansatz technique described in sect. 7
is applicable also to  nonultralocal quantum models,
 we present here the ABA scheme 
for   the quantum mKdV model \c{kmpl95}. For solving
 exactly 
the related  eigenvalue problem (\re{evp})
including that of its energy spectrum, one has to follow  the standard ABA
formulation,
keeping simultaneously the  track of  nonultralocal signatures.
For calculating the eigenvalue of the transfer matrix
(\re {tev}), the eigenvectors are taken in the usual  form (\re{evm}), while 
the commutation relations
of $A (\la)$ and $ D(\la)$ with $B(\mu)
 $   should  be obtained now from the braided
QYBE (\re{ybetmkdv}) using $R,Z$-matrices as
\bea
A (\la) B(\mu) &=&q^{-2}  f(\mu-\la)  B(\mu)A (\la)-
q^{-2}\ti  f(\mu-\la) A(\mu)B (\la) \nonumber \\
D (\la) B(\mu) &=&q^{2}  { f(\la-\mu) } B(\mu)D (\la)-
q^{2}\ti  f(\la-\mu)  D(\mu)B (\la), \ll{ADB}
\eea 
where the factors coming from the $R$-matrix are
${ f(\la-\mu) }= { \sin(\la-\mu + \hbar \ga) \ov \sin(\la -\mu  )}$
and $\ti  f(\la-\mu) = { \sin \hbar \ga \ov \sin(\la -\mu  )}$.
Note that the multiplicative factors   $q^{\pm 2}$ appearing
in (\re{ADB}) are the contribution
of $Z$ due to  nonultralocality.

It is  evident from  the form of $L_k$  (\re{Lmkdv}),
that
like the case of SG or 
Liouville model (\re{l2}), instead 
 of a single   Lax operator 
one has  to use $\hat L_k= L_k L_{k+1}$ for  constructing the pseudovacuum.
 This gives the pseudovacuum \c{kmpl95} as 
\be \mid 0>=\prod_{({k+1 \ov 2}=1)}^N \mid 0>_k,~~~~ 
 \mid 0>_k=\delta ( W_{k+1}^+- q W_{k}^+)
\delta ( W_{k+1}^-+ (W_{k}^-)^{-1}).   
\ll{vac}\ee
and from the action of $\hat L_k\mid 0>_k,$
 \be
\al(\la)= 2i e^{i(\la+ {\ga \hbar \ov 2})} \sin(\la+ {\ga \hbar \ov 2})
,~~~~~~~\De \xi = e^{i \la}\ll{alpha}\ee
and
 \be
\beta(\la)= 2i e^{i(\la- {\ga \hbar \ov 2})} \sin(\la- {\ga \hbar \ov 2}).
\ll{beta}\ee
Similarly one derives the crucial triangular form  for 
 the  Lax operator acting on the pseudovacuum due to 
$\hat L_{k}\mid 0>_k  = 0$.
In analogy with the standard ABA,  
the eigenvalue problem (\re{tev}) for  the  mKdV
model  may therefore be solved by
following the relations (\re{a},\re{d}, \re{lambda}) and 
using  (\re{ADB},\re{albet}) as  the $s$-trace
 \bea
 \Lambda(\la)&=& q^{-{\kap \ov 2}}
\al(\la)^N \prod_j \l(q^{-2}  {f(\la_j-\la)} \r)
 +
  q^{{\kap \ov 2}} \beta(\la)^N \prod_j\l(q^2  {f(\la-\la_j)} \r) \nonumber \\
 &=& 2i e^{i \la} [q^{{-\kap \ov 2}+{N \ov 2}-2m} \sin^N(\la+ {\ga \hbar
\ov  2})\prod_{j=1}^m  {\sin  (\la-\la_j-{\ga  \hbar})  \ov \sin(\la-
\la_j)}  \nonumber \\
&+& 
 q^{-({-{\kap \ov 2}+{N \ov 2}-2m})} \sin^N(\la- {\ga \hbar
\ov  2})\prod_{j=1}^m  {\sin  (\la-\la_j+{\ga  \hbar})  \ov \sin(\la-
\la_j)} ].
\ll{spectrum} \eea
The energy eigenvalue can be obtained by expanding (\re{spectrum}) in the
spectral parameter and considering $C_{-3}=H$ as the Hamiltonian of the model.
The determining equation for the rapidity parameters  $\la_j$  may be
obtained   from   the   vanishing  residue  condition of   the   eigenvalue
equation
(\re{spectrum}) in analogy with the standard Bethe equation (\re{be}) as
\be
 q^{-\kap+{N }-4m} {\l({ \sin(\la+ {\ga \hbar \ov  2}) \ov
\sin(\la- {\ga \hbar \ov  2})} \right)}^N=
\prod_{j \neq  k}^m \l( {\sin  (\la-\la_j+{\ga  \hbar})  \ov
\sin  (\la-\la_j-{\ga  \hbar}}\right)
\ll{bethe} \ee
Remarkably,  these results 
for the quantum mKdV model, apart from an extra  $q$ factor, 
coincide exactly with  the $XXZ$  spin-$\ha$  chain.  It is true however
that, unlike $XXZ$ the Hamiltonian of the mKdV model is $H=C_{-3}$ and  
moreover
 one has to consider also  the continuum
limit for the mKdV field  model.

%%%%%%%%%%%%%%%%%%%%%%%%%%%%%%%%%%%%%%%%%%%%%%%

\vspace* {1cm} \fbox{fig.8.1} 

\vspace* {1cm}
  {\bf  Figure 8.1 } \ {\it  Classification of the nonultralocal
 models  of  
spectral parameter dependent or independent types 
belonging to  trigonometric
and  rational classes along with the
respective choices for $Z$ and $\ti Z$ matrices.}

%\input {prepfig.tex}
%______________________________________\input {preport9.tex}

\setcounter {section}{8}
\section{ Recent progress and unsolved problems}
\setcounter{equation}{0}
We briefly touch here on some recent progress in the fields related
to the  quantum integrable system or influenced by it, which can be
followed from the material  presented in the review. We also indicate for
each topic some unsolved problems, which are either long-standing problems
or have been opened up along with the recent development of the subject.  
%_________________

\subsection {Higher rank generalization of the ETSA 
and construction of multicomponent
models} 

We have seen in sect. 5 that through different realizations of ETSA
wider class of models can be reached, than those constructed 
 from the quantum algebra. The   $U_q(sl(n+1))$ algebra on the other 
hand has been shown to generate
 multicomponent discrete Toda field models. Therefore,
it may be hoped that the 
  higher rank generalization of the ETSA  would lead to 
 multicomponent models with richer structure of their Lax operators.   

Such a generalized ancestor Lax operator related to the trigonometric
$R_{trig}(\la)$-matrix (\re{rtrign}) was constructed in \c{kbjp92}
in the form 
\be L^{m-anc}(\xi)~
=~\sum _k^{\mathsf {N}} ({1 \ov \xi}~\tau_k^+ \xi ~\tau_k^- )
 ~E_{kk} ~+~\sum_{k,l}^{\mathsf {N}}{1 \ov \xi} ~\tau_{lk} ~E_{kl} 
+~\xi ~\tau_{kl} ~E_{lk}. ~\ll{nL-anc}\ee
 $~\tau_k^{\pm } ~,~\tau_{kl}~$ are generators of the higher rank ETSA
given by the relations
\[
\tau_k^{\pm }\tau_{kl} = e^{\pm i \alpha } ~ \tau_{kl} \tau_k^{\pm } ~,~~
\tau_k^{\pm } \tau_{lk} ~=~e^{\mp i \alpha } ~ \tau_{lk} \tau_k^{\pm } ,
\] \be [  \tau_{kl}~, \tau_{lk} ~ ] = 2i \sin  \alpha ~
(~ \tau_k^- \tau_l^+ ~-~ \tau_k^+ \tau_l^- ~) ~,  
\ll{netsa1}\ee
for all different $~k,l,m,n~$ indices.
The diagonally placed operators $~\tau^{\pm}_k~$ commute among
themselves in addition to
\[
 [ ~\tau_k^{\pm } ~,~\tau_{lm} ]  ~=~0 ~,~~
 \tau_{kl} \tau_{km} ~= ~ e^{i\epsilon \alpha }~\tau_{km} \tau_{kl} ~,~
 \tau_{kl} \tau_{ml} ~=~ e^{-i\epsilon \alpha }~\tau_{ml} \tau_{kl} ~, 
\] \be
    [~\tau_{mk}~,~\tau_{kl} ~ ] ~=~2i ~\epsilon ~ \sin \alpha ~
 {\tau_k}^{(\epsilon)} \tau_{ml} ~,~\ll{netsa2}\ee
where $~\epsilon ~=~
 {\rm sign }(k-l) + {\rm sign }(l-m) + {\rm sign }(m-k)~,~~$
 $~ \tau^{(\epsilon )}_k ~ 
=~ \tau^{\pm }_k  ~$ for $~\epsilon ~=~\pm 1~$  and
\bea
\left [~\tau_{kl}~,~\tau_{m n} ~\right ]
~=~ 2i ~\rho ~\sin \alpha ~\tau_{ml} \tau_{kn} ~,
\ll{netsa3}\eea
with $~\rho = +1 ~$ for $~ l>n>k>m ~$ ( and all its cyclic inequalities ),
while $~\rho = -1 ~$ for the reverse inequalities and $~\rho = 0 ~,~$
otherwise.

 It may be noted that the above relations
are the generalization of the  ETSA (\re{etsa}) and reduce to it
at $~\mathsf{N}=2~$.
On the other hand for a  symmetric reduction  through generators of the
quantum algebra, (\re{nL-anc}) reduces clearly to (\re{ltrign}) producing
generalized spin model and Toda field model. Therefore as descendants of  
(\re{nL-anc}) and its multiparameter 
generalization one can expect to generate
models with asymmetric form of their Lax operators like
 multicomponent DNLS, MTM, RTC,
ALM   etc.

At the  limit $q \rw 1$, (\re {nL-anc}) reduces to 
\be   L= \sum_l^{\mathsf {N}} (K_l^+ + {{i \lambda} \over \kappa} K_l^-) E_{ll} +
           \sum_{j \not= l}^{\mathsf {N}} K_{lj} E_{jl}, \ll{nL-anc0}\ee 
and  is  associated with the rational $(\mathsf{N}^2\times \mathsf{N}^2)$
 $R$-matrix (\re{rationalr}).
 The operators ${\bf K} $ satisfy the  algebra
\[
[K_{mk},K_{kl} ] = K_k^- K_{ml}
,~~ [K_{kl},K_{lk} ] =  K_k^+K_l^- -  K_k^-K_l^+   ,\] \be
     [ K_k^+ , K_{kl}] = K_{kl} K_k^- , [K_k^+,K_{lk}]= - K_{lk} K_k^- ,
 \ll{nk-alg}\ee
along with $~[K_k^{+},K_{lm}]= [K_{kl},K_{km}]=
K_{kl},K_{ml}]= [K_{kl},K_{mn}]=  0,
 $  where $K_k^- $ commute with all other generators and 
  $K_l^{\pm} $  form an abelian subalgebra.
 Notice
  that (\re{nk-alg}) is an higher rank generalization of (\re{k-alg})
and recovers the known  $sl(N)$ at  particular symmetric reduction. 
  
 Different bosonic 
  realizations of this algebra and its multiparameter extensions 
should be able to 
 generate from  (\re{nL-anc0}) quantum integrable models like 
multiparticle Toda chain or Toda chain like models, as well as the 
vector generalization of the SLNLS, generically  not exhibiting
  unitary symmetry.
Let us take up the last example \c{krjp94}, where 
 through a set of
 independent bosonic operators with  
$ [\psi_l,\phi_k] = \delta_{lk},~~  [\psi_l,\psi_k] = [\phi_l,\phi_k] = 0 ,$
 the algebra (\re{nk-alg}) may be realized as
\bea
K_1^-&=& -1, ~~ K_1^+ =\sum_j\phi_j\psi_j ,~~ K_i^+ = {\bf 1}_{ii},~
 K_i^- =0,~~(i=2,\cdots,N)     \nonumber \\
K_{1j} &=& \psi_j \equiv {\vec \psi},~~ K_{j1} =\phi_j\equiv {\vec \phi}^t
 ,~ K_{ij} = 0,~~~1<(i,j)\leq N.    
\ll{nk-npsi}\eea
The corresponding  Lax operator (\re{nL-anc0})  then reads
 \be
 L(\lambda) =  \l( \ba 
  - {{i \lambda} \over \kappa} +
   (\vec  \phi \cdot \vec \psi)   \qquad
{ \vec {\phi}
   }
 \\
{\vec {\psi}}^t
  \qquad  \qquad ~~~
{ {\bf I}  }  \ea \r),
 \ll{nL-vslnls}\ee
  which represents  a simple  lattice version of the 
vector NLS model,
generalizing (\re{slnls}) to the vector case.
 
It would be fruitful  to construct and analyze the algebraic structures 
of the various   
quantum integrable  multicomponent models through 
higher rank ETSA 
and solve them through the nested Bethe ansatz.  
 The quantum
Dodd--Bullough--Mikhailov model \c{dbm-kor} 
might be  a  good candidate for such analysis.

\subsection{ Integrable model with ladder symmetry}
The ladder symmetric models are of immense current interest due to their
possible connection with the high $T_C$ superconductivity \c{ladder}.
Theoretical models with ladder symmetry exhibiting quantum integrability
can also be introduced  \c{frahm}.
The idea of constructing $XXX$ spin-$\ha$  chain   with
inhomogeneity and its Bethe
ansatz solution (described earlier), can be extended  to the tensor product of 
similar Lax operators:  
\be L_{ab}(\la)= L_{b1}(\la+\eta) L_{b2}(\la) L_{a1}(\la) L_{a2}(\la-\eta),
\ll{rab}\ee
where $ L_{12}(\la)={\la -i P_{12} \ov \la -i}$ is 
equivalent to $ R_{rat}(\la)$ (\re{rationalr}) associated with  the $XXX$
chain.
Using now the method of Hamiltonian construction of sect. 6 and  noticing
 that 
$ L_{ab}(0)= R^+_{b2i} P_{b2i+1} P_{a2i} R^-_{a2i+1},$ where 
$ R^\pm_{b2i}= R_{b2i}(\pm \eta)$, we arrive at the 
Hamiltonian 
 \bea H&=&(i) \sum_{i,\ep=+,-}  R^{\ep '}_{2i,2i+\ep 1}
 (R^{\ep })^{-1}_{2i,2i+\ep 1} 
+
 (R^{\ep })_{2i,2i+\ep 1}
 R^{'}_{2i,2i+\ep 1}
 P_{2i,2i+\ep 1} (R^{\ep })^{-1}_{2i,2i+\ep 1}
\nonumber \\
&=& J_1 \sum_{i} ({\vec S}_{2i-1}{\vec S}_{2i}+{\vec S}_{2i}{\vec S}_{2i+1}
) +J_2 \sum_i (  {\vec S}_{2i-1}{\vec S}_{2i+1}+
{\vec S}_{2i}{\vec S}_{2i+2} ) \nonumber \\
&+& J_3 \sum_i (  {\vec S}_{2i}({\vec S}_{2i+1}\times {\vec S}_{2i+2})+
 {\vec S}_{2i-1}({\vec S}_{2i+1}\times {\vec S}_{2i}) 
\ll{hladder}\eea
where $J_1={1 \ov 1+\eta^2}, ~~J_2=\eta^2 J_1,~~ J_3=-2 \eta J_1 ~$.
Note that the lattice on which the above Hamiltonian is
defined has
ladder symmetric form, which has important physical consequences \c{ladder}
 (see fig. 9.1a,b).

The QYBE for  $T_{ab}= T_b(\la+\eta) T_a(\la),$
where $T_a(\la)$ is the monodromy matrix of the inhomogeneous model
constructed from the Lax  operators \be R_{ai}(\la)=
R_{a2i}(\la)R_{a2i+1}(\la-\eta) \ll{r2}\ee
 yields   $tr_{ab}
(T_{ab}(\la))=t(\la+\eta) t(\la)$ commuting for different $\la$,
where  $t(\la)=tr(T_a(\la))$    in
turn is the  commuting 
transfer matrix of the inhomogeneous $XXX$ chain.
 Therefore the eigenvalue of 
 $t_{ab}\mid m>=\Lambda(\la)\mid m>$ is given by $\Lambda(\la)=
\Lambda^0(\la+\eta)\Lambda^0(\la)$, where $\Lambda^0(\la)$ is the eigenvalue
of $t(\la).$  Due to the tensor product structure of the Lax operator
 (\re{r2})   the results  for the 
standard $XXX$ spin chain (\re {lamxxx}, \re{bexxx})  are generalized 
to the expression 
 \be \Lambda^0 (\la) =
 \pr_{j=1}^m { \la_j-\la+{i\ov 2} \ov  \la_j-\la-{i \ov 2}}
  +\bet^N(\la)
\pr_{j=1}^m {\la -\la_j+{3i \ov 2}  \ov \la-\la_j+{i \ov 2}}
\ll{lamlad}\ee
for the eigenvalue  along with  the Bethe equation
\be 
(\bet(\la_j-{i\ov 2}))^N= \pr_{j\neq k}^m {\la_j-\la_k-i
 \ov \la_j-\la_k +i},
\ll{belad}\ee
  where $~\bet (\la)=
\left(  {\la \ov \la+ {i }} \r)\left(  {\la -\eta \ov \la -{\eta }+i}
 \r).$ As a result similar to (\re{exxx}) we obtain the excitation spectrum 
$E_m= i {\partial \ov \partial \la } \ln \Lambda (\la) \mid _{\la =0}=
\sum_j e^0(\la_j)+
e^0(\la_j+ \eta),~ $
where $~e^0(\la_j)= - {1 \ov \la_j^2 +  {1 \ov 4}} $. The thermodynamic limit
of the result and other details can be found in \c{frahm}

This two-chain ladder model can be generalized to arbitrary k-chains.
It is conjectured that the ladder models with even (odd) number of chains
must show presence  (absence) of mass gap in their excited spectra. Whether
such integrable ladder models can throw any light on this conjecture is
 an important open problem.
\subsection {Elliptic quantum group}
The quantum group associated with the trigonometric $R$-matrix  and
generated by the QYBE (\re{qybel}) has been discussed in detail in this
review. In recent years there are some results  to extend
this notion to the elliptic case \c{eqg}. Such an algebra  ${\cal
A}_{k,q} (gl(2)) ~$ may be generated by an extension of the QYBE  as
\begin {equation}
R^+_{12}(\xi_1 /\xi_2) {L}_{1}(\xi_1 ) {L}_{2}(\xi_2 )
~ = ~  {L}_{2}(\xi_2 )~{L}_{1}(\xi_1 ) R_{12}^{*+}(\xi_1/ \xi_2 ),
\ll{eqgybe}
\end {equation}
where the quantum indices of $L$ are suppressed,
  $R^{*+}(\xi)= R^{+}(\xi, k^{* \ha}, q^\ha ),~
k^{* \ha}=k^{\ha}q^{-c}~$ with  $~~ R_{12}^{+}(\xi) ~$ being the elliptic
solutions like (\re{xyz}).  The complex parameters $  k^{* \ha}, q^\ha $ and $
\xi  $ may be interpreted as the elliptic nome, the deforming parameter  and
the spectral parameter, respectively.
Here for the central element $q^{-c}=q^{-l}~$, the algebra is called  level
$l$-algebra with its  simplest representation having $l=0$.
Imposing the constraint on quantum determinant $ det_q L(\xi)= 
q^{{c \ov 2}}$ the algebra  ${\cal
A}_{k,q} (sl(2)) ~$ may be obtained. Note that in contrast to (\re{qybel})
the relation
(\re{eqgybe}) contains two different $R, R^*$-matrices,  
which makes the introduction of the coproduct structure and hence the Hopf
algebra difficult. For obtaining the algebra from (\re{eqgybe}) (for details 
see \c{eqg}) the matrix elements $L_{\ep \ep'}(\xi),~\ep
\ep'=(++),(--),(+-),(-+) $ of $L(\xi)$ should be expanded as
$~~L_{\ep \ep'}(\xi)= \sum_{n\in {\bf Z}}L_{\ep \ep',n} \xi^{-n}~~$. Unlike the
trigonometric case this gives  expansion like
\be~~L^+(\xi)\equiv ~L(q^{{c \ov 2}}\xi)= \cdots +
 k^\ha~L^+_1\xi^{-1}+
~L^+_0\xi^0+
~L^+_{-1}\xi + \cdots ~
\ll{eqgl+} \ee
 mixing both $+ ve$ and $-ve$  powers of $\xi$ and similarly for 
$L^-(\xi).$ 

At $k \rw 0$, the elliptic $R_{ellip} \rw $ trigonometric $R_{trig}$ matrix
and $L^+(\xi)$ , as seen from (\re{eqgl+}), contains only $+ve$ powers of
$\xi$. (Similarly $L^-(\xi).$ contains only $-ve$ powers). Consequently the elliptic quantum group reduces to the well known
trigonometric quantum group.

 Discovering  the universal ${\cal R}$-matrix for the
elliptic quantum group, establishing 
 its Hopf algebra structure and finding  its relation
with the quadratic Sklyanin algebra discussed in sect. 3 are some
challenging   open problems. It would also be interesting to find physically
relevant integrable ultralocal and nonultralocal models related to such
elliptic algebra in the spirit of the present review.
  
\subsection { Coupled spin chain and extended Hubbard model} 
Using the idea of twisting   (\re{lgnl}), one can construct 
an integrable coupled  spin chain,  starting from 
noninteracting  spin chains
as
\be 
L(\si ,\tau \mid \la)= F(\si,\tau)
 \l(L_{xxz}(\si \mid \la)\otimes L_{xxz}(\tau \mid \la)\r)
F(\si,\tau).
\ll{coupxxz}\ee
Here  $L_{xxz}(\si \mid \la) $ and $L_{xxz}(\tau \mid \la)$ are  
related to the independent spin-$\ha$ operators $\si, \tau$ 
and are  equivalent to the Lax operator  (\re{XXZ1})
of the $XXZ$  model with Hamiltonian  (\re{hxxz1}). 
The twisting operator coupling the two spins is  
 given by
\be F(\si,\tau)= e^{ig (\si^3 \otimes \tau^3- \tau^3\otimes \si^3 )}
\ll{cxxztw}\ee  
with the needed property $F_{12}F_{21}=I.$ That 
makes the coupled system  quantum integrable with the Hamiltonian
\bea
H_{couple}&=& \sum_n^N F_{n~ n+1}\l({ H_{xxz}}^\si(n~ n+1)+
{ H_{xxz}}^\tau(n~ n+1)\r)F_{n~ n+1} \nonumber \\ 
&=& \sum_n^N {\ti  H_{xxz}}^{\si \tau}(n~ n+1)+
{\ti H_{xxz}}^{\tau \si}(n~ n+1).\ll{hcxxz}\eea
The interacting parts read as 
\bea
{\ti H_{xxz}}^{\si \tau}(n~ n+1)
&=& \Delta \si_n^3 \si_{n+1}^3
+f(\tau^3_n,\tau^3_{n+1})  
\si_n^+ \si_{n+1}^- + f^{-1}(\tau^3_n,\tau^3_{n+1})\si_n^- \si_{n+1}^+ \nonumber
\\
{\ti H_{xxz}}^{\tau\si}(n~ n+1)
&=& \Delta \tau_n^3 \tau_{n+1}^3
+f^{-1}(\si^3_n,\si^3_{n+1})  
\tau_n^+ \tau_{n+1}^- + f(\si^3_n,\si^3_{n+1})\tau_n^- \tau_{n+1}^+ 
\ll{hsitau}\eea
where
$~~
 f(\tau^3_n,\tau^3_{n+1})=
e^{ig (\tau_n^3+ \tau_{n+1}^3)}~~$ and similarly $~~
 f(\si^3_n,\si^3_{n+1})=
e^{ig (\si_n^3+ \si_{n+1}^3)}~~$  are responsible for cubic and quartic spin
interactions.

Recall that the Hubbard model may  be 
given by the Lax operator (\re{lhub}) involving two different
 spin-$\ha$ operators. Therefore  the same twist
 (\re{cxxztw})  can also be applied
 for constructing an extension of the Hubbard model. After
transforming  to the fermion operators $c_{(\pm)i}$
 by (\re{JWT},\re{JWT1}) one can write down the   
 Hamiltonian of such an integrable interacting fermion model as   
\bea
H &=& \sum_i^N t ( 
F^-_{i,i+1}
c^\da_{ (+) i} c_{ (+) i+1}+
(F^-)^{-1}_{i,i+1}c^\da_{ (+) i} c_{ (+) i-1} \nonumber \\
 & +&
(F^+)^{-1}_{i,i+1}c^\da_{ (-) i} c_{ (-) i+1}+
(F^+)_{i,i+1}c^\da_{ (-) i} c_{ (-) i-1} ) 
+~~ U~ n_{ (+) i}n_{ (-) i}     
\ll{exthhub} \eea
where $~F^\pm_{i,i+1}= e^{2ig (n_{(\pm)i}+n_{(\pm)i+1} -1)}. $ 

Note that in   the hopping part of the Hamiltonian,  one type of 
fermion is  influenced by the concentration 
of the other.
Exact eigenvalue solution of these novel integrable models through coordinate
 Bethe ansatz similar to the Hubbard model
 should  be an important  problem \c {kplan}. 

\subsection {Diffusion model as integrable system}
In recent years an important connection between integrable quantum spin
chains and some one dimensional reaction-diffusion processes has been
established \c{reacrit}. The reaction-diffusion models in the simplest case 
is  described by the 
probability distribution
$~~P_N(\tau_1, \tau_2, \ldots,\tau_N \mid t)~$   
 of finding certain configuration of molecules 
and  vacancies at time $t$ in an open chain of $N$ sites. The value 
 $\tau_k=1$ denotes occupied (molecule), while $\tau_k=0$ an empty (vacancy)
site $k$.   The master equation representing the evolution of probability
distribution $~ {~\partial  \ov \partial t}\mid P >=-H \mid P> ~$ is
equivalent to  an
imaginary-time Schr\"odinger equation, with $H$ as a quantum Hamiltonian.

	If we consider the asymmetric diffusion process involving two-body
interactions only, then the Hamiltonian may be given by 
\be H=\sum_{k=1}^{N-1} (H_{k k+1})^{\al \bet}_{\ga \de} E_k^{\ga \al}
E_{k+1}^{\de \bet}+(h_1)^\al_\ga E_1^{\ga \al}+
(h_N)^\al_\ga E_N^{\ga \al}\ll{hrde}\ee
where due to the existence of the stationary state:$~ <0\mid H=0 ~$
in the matrix representation of $H$ the sum of the elements of each column
must vanish. Due to this crucial property, for the process of 
partially {\it asymmetric diffusion}, we may have 
$~(H_{k k+1})^{01}_{10}=-q,~(H_{k k+1})^{10}_{01}=-p~$ denoting probability of
diffusion to the {\it left} and to the {\it right}, respectively along with 
$~(H_{k k+1})^{01}_{01}=q,~(H_{k k+1})^{10}_{10}=p~$ to ensure the
required stationary state property. The boundary contributions $h_1$ and
 $ h_N$
describe injection (extraction) of particles with rates $\al$ and $\de$
($\ga$ and $\bet$) at boundary sites $1$ and $N$, respectively. Therefore
    $~(h_1)^1_0=-\ga=-(h_1)^1_1, (h_1)^0_1=-\al=-(h_1)^0_0~$
and  $~(h_N)^1_0=-\bet=-(h_N)^1_1, (h_N)^0_1=-\de=-(h_N)^0_0~$.
After performing an unitary transformation, which does not change the 
average of the observables: $~<X>=\sum_{\{\tau\}} X(\{\tau\})
P_s(\{\tau\})~$, we may rewrite (\re{hrde}) as the integrable spin chain
model, where along with the $U_Q(su(2))$ invariant spin Hamiltonian 
 (\re{hqgroup}) additional  boundary
terms appear as
\bea
{ H}& =&- \sqrt{pq} \sum_{n=1}^N
\si_n^+ \si_{n+1}^- +\si_n^- \si_{n+1}^+ 
 +
 \qr (Q+Q^{-1})(\si_n^3 \si_{n+1}^3 -{\bf I}) \nonumber \\ &-&  
  \qr (Q-Q^{-1}) (\si_j^3 -\si_{j+1}^3) +B_1+B_N 
\ll{hadm}\eea
with $Q=\sqrt{{q \ov p}}$,
where 
the boundary contributions $B_1,B_N$  contain
 nondiagonal matrices $\si^\pm$ as
\bea
B_1&=&{1\ov 4 \sqrt{pq}} \l((\al-\ga ) \si^3_1-2 \al \si^-_1-2 \ga
 \si^+_1+(\al+\ga ) \r) \nonumber \\
B_N&=&{1\ov 4 \sqrt{pq}} \l((\de-\bet ) \si^3_N-2 \de Q^{N-1} \si^-_N
-2 \bet  Q^{1-N}
 \si^+_N+(\bet+\de ) \r)
\ll{b12}\eea

The physical consequences of this Hamiltonian related to the diffusion
process have been analyzed in \c{admrit}. However, though the spin model
(\re{hadm}) can be shown to be integrable \c{bvega}, due  to the lack of the
reference state its exact eigenvalue solution through  Bethe ansatz 
 remains as an  important unsolved
problem. 
\subsection{Spin Calogero-Sutherland model}
In recent years due to relevance in the high $T_c$ superconductivity,
interest towards models with long-range interactions has been enormously
increased.
An interesting class of integrable  models of this  kind is the 
 quantum Calogero-Sutherland (CS) model 
  and its spin extensions.
 The Hamiltonian of the spin CS 
 model may be given by \c{cs,hal-pas} 
\be
H_{cs}= \sum_{j=1}^{N} p^2_j+ 2 \sum_{1\geq j<k \geq N} (a^2-a
P_{jk})V(x_j-x_k)\ll{hscs}\ee
with 
$~[x_j,p_k]= i\de_{jk}~$, where the potential 
 $~V(x_j-x_k)= {1 \ov (x_j-x_k)^2}~$ for nonperiodic and 
$~ {1 \ov \sin ^2 (x_j-x_k)}~$
for the periodic model. 
 $~P_{jk}$ is  the permutation operator  responsible for exchanging the
spin states of the $j$-th and the $k$-th particles. In the absence of the
operator $P_{ij}$, (\re{hscs}) turns into the  original CS model without
 spin.

The spin CS model exhibits many fascinating features, namely its conserved
quantities including the Hamiltonian exhibit Yangian symmetry, the
eigenvalue problem  can be  solved  exactly using 
Dunkl operators,  
 the ground state is a solution of the Knizhnik-Zamolodchikov
equation, the system can be viewed  as the free anyonic  gas related to
the notion of fractional statistics etc. \c{hal-pas}. 
The quantum 
integrability  of the spin CS model may be proved  through an
alternative set of operators $L$ and $M$, 
much in common with  the classical case by assuming
$~~ [H_{cs}, L]=[L,M]
~~$, where  $\dot L$ is replaced by  $ [H_{cs}, L]$ in the quantum case
\c{wad-shas}. 
Explicitly,
\bea
L_{jk} &=& \de_{jk} p_j +i a(1-\de_{jk}) P_{jk} f(x_j-x_k), \nonumber \\
M_{jk} &=&  2a ( \de_{jk} \sum_{j \neq k}P_{jk} V(x_j-x_k)
 -(1-\de_{jk}) P_{jk} V(x_j-x_k)) \ll{lmcs}\eea
with $f(x_j-x_k)={1 \ov (x_j-x_k)} ~$ for the nonperiodic and 
$f(x_j-x_k)=a \cot (x_j-x_k) ~$ for periodic model.
The conserved set of operators can be constructed as 
$~ I_n= \sum_{j,k}  (L^n)_{jk} ,n \geq 1 $ or $~ 
 I^a_n= \sum_{j,k}  (X^a L^n)_{jk} ,n \geq 0 $, with $X^a$ being the generators of 
 $su(n)$ algebra. Using the crucial property $
\sum_{j}  M_{jk}=\sum_{k}  M_{jk}=0$, one can show that  both $I_n$ and
$I_n^a$
commute with the Hamiltonian $H_{cs}$, with $I_2=H_{cs}$.
These conserved quantities  commute  also with  the generators of the
Yangian algebra.

Remarkably, at $a \rw {\infty}$ the Hamiltonian of the CS model (\re{hscs})
for the periodic case reduces to the well known {\it Haldane-Shastry model}
\c{HS}
\be
H_{hs}=  \sum_{j<k } 
{P_{jk} \ov \sin^2(x_j-x_k)}.\ll{hhs}\ee
This discretized  long-range interacting spin chain like model seems to be
less well understood and its Lax operator description difficult to find.

Using the twisting transformation (\re{lgnl}) one can 
extend the  spin CS model in the line of \c{csyan} to have the deformed
Yangian symmetry. The Hamiltonian of such models can be   given  in the form
(\re{hscs}) with the $P_{jk}$ operator generalized to $Q_{jk}P_{jk}$, where 
$Q_{12}=F_{12}\Omega_{12}F_{21}^{-1} $  with     the twisting
operators $F_{12}$ and $\Omega=\sum_{\al \bet} \ep_\al e_{\al \al} \otimes
 e_{\al \al}+ e_{\al \al}\ot  e_{\bet \bet} ~$.
The parameters $~\ep_\al=\pm $ can be related to the parity of te
particles.
Another form of generalization can be made through
 the $q$-deformation leading to a 
relativistic spin CS model \c{relcsm}. A  QYBE type 
 formulation of this model 
involving  $R, L$-matrices also attempted in \c{csavan},
 though the related $R$-matrix  found was a  dynamical one. 

Possible description of the   spin CS model as a nonultralocal system 
and the formulation of its integrability through braided QYBE, introducing 
$Z$ and $R$-matrices
 following  the scheme of
  this    review,  would be a challenging problem.

\subsection{Finite-size corrections in  integrable systems and related CFT}
Statistical systems at critical points should possess conformal invariance.
Using this fact it is possible to extract the important information about
the conformal quantum field theory  in the scaling limit  of the
integrable lattice models. Interestingly, from the finite size correction
of the Bethe ansatz solutions, one can determine \c {karow} the CFT
characteristics like the central charge and the  conformal dimensions.

Recall that \c{BPZ} the central charge  $c$ appears in the central extension of 
the Virasoro algebra
\be [L_n,L_m]=(n-m)L_{n+m}+{c \ov  12}(n^3-n)\de_{n+m,0}, \ll{vir}\ee
 the algebra generating the conformal transformations
in two dimensions. $\Delta$ is the anomalous dimension of the {\it primary}
fields. For the minimal models ${\cal M}_{pp'}$ in 
 CFT, the central charge  is given by
 \[ c=1-{6 (p-p')\ov p p' } \]   
with positive integers $p,p'$ and finite number of primary fields.
For unitary theories one gets the restriction 
$p-p'=1$ yielding
\be c=1-{6 \ov \nu (\nu-1)} \ll{c} \ee
with integer
$~p\equiv \nu \geq 2 ~$ 
leading  to
$c \leq 1$.

One may analyze the finite size effect of the Bethe ansatz solutions
like (\re{exxz}) corresponding to the six-vertex model, but  with a seam 
given by  $\kappa.$ Considering the coupling parameter 
 $q=e^{i{ \pi \ov \nu+1}},$ one obtains
 at the large
$N$ limit  the expression
\[ E_0=Nf_{\infty}-{1 \ov N}{\pi \ov 6} c + O({1 \ov N^2}) \]
 for the ground state energy
and
\[ E_m -E_0 ={2 \pi \ov N}( \De+ \ti \De) + O({1 \ov N^2}) \qquad
 P_m -P_0 ={2 \pi \ov N}( \De- \ti \De) + O({1 \ov N^2}) \]
 for the excited states. Here 
  $\De,  \ti
\De$  are conformal weights of  unitary minimal
models  and 
$c=1-{6 \kappa^2 \ov \nu(\nu+1)}, \ \nu=2,3, \ldots$
is the central charge of the  corresponding conformal field theory.

The intriguing relation between
the  quantum mKdV  and the spin-$\ha$  $XXZ$  chain given by equations like 
(\re{spectrum},\re{bethe}) frames naturally an worthy problem of 
finding the conformal properties of the quantum mKdV model by
suitable choice of $\kappa$ and $\nu$.
\subsection{Integrable structure of CFT through quantum KdV}
To capture the integrable structure of CFT, an alternative approach 
 through {\it massless} $S$-matrix description 
and  based on the quantum KdV model has been proposed recently \c{BLZ},.

Note that the conformal symmetry of CFT is generated by its energy-momentum
tensor $~ T(u)=-{c \ov 24}+\sum_{-\infty}^{\infty} L_{-n}e^{inu}~,$ with
$L_n$ satisfying the Virasoro algebra (\re{vir}). The operators 
$I_{2k-1}= {1 \ov 2 \pi} \int_0^{2 \pi} du T_{2k}(u)~$ with $
 T_{2k}(u),$  depending on various
 powers and derivatives of $ T(u)$ represents an
infinite set of commuting integrals of motion. The idea is to solve their 
simultaneous diagonalization  problem, much in common to the QISM
 for the
integrable theory. Remarkably, this is equivalent  to  solving the quantum
KdV problem  \c{Gervais}, since  at the classical limit the field  $T(u)=-{c
\ov 6} U(u) $ with $U(u+2 \pi)=U(u)$  reduces the commutators of $T(u)$ to
\be \{U(u),U(v)\}=2(U(u)+U(v))\de'(u-v)+\de'''(u-v)\ll{pbkdv}\ee
which is the well known Poisson structure of the KdV.

In the quantum case, defining a Miura transformation  through a mKdV like
field $\phi(u),$ the monodromy matrix of the model can be expressed as
\c{BLZ}
\be M_j(\la)=\pi_j\left[ e^{i\pi PH} {\cal P}exp \l(\la\int_0^{2 \pi}
 du(:e^{-2
\phi(u)}:
q^{{H \ov 2}} E+:e^{2
\phi(u)}:
q^{-{H \ov 2}} F )\r) \right] \ll{moncft}\ee
with proper normal ordering :~ :. Here $E,F,H$ are generating elements of
$U_q(sl(2))$ and $ \pi_j$ is its $2j+1$ dimensional representation.
Interestingly, the operators $M_j(\la)$ satisfy now the QYBE like
(\re{qybet}) with the trigonometric $R(\la)$-matrix. The transfer matrix $
\tau_j(\la)=tr_{\pi_j} (e^{i\pi PH}
M_j(\la))$ commutes for different parameters $\la$  as well as with the
operators $I_{2k}$,
( which are related to the  expansion coefficients of $\ln
\tau_j(\la)$). 
For the values of the central charge 
\be c=1-3{(2n+1)^2 \ov 2n+3}, ~~n=1,2,3,\ldots \ll{ccft}\ee
the eigenvalues of $\tau_j(\la)$ was found to satisfy a closed system of
functional equations,
which is sufficient to derive the spectrum.
Such  equations are 
 obtained from the operator relations 
\be  
 \tau_j(q^{\ha}\la)\tau_j(q^{-\ha}\la)=1+\tau_{j-\ha}(\la)\tau_{j+\ha}.
(\la)\ll{talg}\ee
For the ground-state eigenvalues
the functional equations are equivalent to those of 
the thermodynamic Bethe ansatz
equations for the massless $S$-matrix
 theory associated with the minimal CFT
${\cal M}_{2,2k+1}$.

Since the quantum mKdV model plays  determining role in this approach,
application of its exact quantum Lax operator as presented in (\re{Lmkdv})
might be promising. Moreover, the finite-size corrections of its Bethe
ansatz solution (\re{spectrum}) might also be fruitful in  deriving 
the central charge  like (\re{ccft}).
\subsection {Other problems}
There are important  classically integrable nonultralocal models
like complex  sine-Gordon \c{comsg}, nonlinear $\si$-model \c{simodel},
 KdV model etc., whose
quantum generalizations remain a major  open problem. It is desirable 
to apply  the scheme for quantum nonultralocal models presented here 
or its generalizations to  these models.

\ni i) {\it KdV model}

The second Hamiltonian Poisson bracket structure of the  KdV  field is
given by   (\re{pbkdv}) and due to the appearance of higher derivative of
$\de$-function  the generalization of the scheme should include
more number of  braiding  matrices $Z$ \c{hlav94}, for
 incorporating the nonultralocality beyond nearest
neighbors. 

\ni ii) {\it Complex sine-Gordon model}

The complex SG model is a relativistic model given by the Lagrangian
density 
$~{\cal L}={1 \ov 2g}\l( {\partial_\mu \bar \psi\partial_\mu \psi \ov
(1-\bar \psi \psi)}-m^2 \bar \psi \psi \r) ~$, expressed through the complex
scalar field $ \psi $ with canonical bracket
like $~\{ \pi_{\psi} (x),\psi (y) \}=\de(x-y) ~$ 
 and similarly for its conjugate field $\bar \psi$. The
related
Lax operator is given in the form \c{comsg}   
\bea
L(x,\xi)=(m({1 \ov \xi} -\xi)-{m \ov \xi} \bar \psi \psi -i{\bar \psi
\partial_0 \psi - \psi \partial_0 \bar \psi \ov 1- \bar \psi \psi})\si^3
\nonumber \\ + 2i ( 1- \bar \psi \psi)^{-\ha}\l( \partial_0  \psi +
 \partial_x  \psi -{i m \ov \xi}( 1- \bar \psi \psi) \psi \r) \si^++ \mbox
 ~~~ {c.c.}
\ll{Lcsg}\eea
where ~c.c~ stands for the complex conjugate. Due to the 
$\partial_x  \psi$ term present in the Lax operator, the model exhibit 
nonultralocal property.

\ni iii) {\it Nonlinear $\si$-model}

It is a relativistic model given by the action $~{\it S}= \ha\int d^2x
(\partial_\mu n_a)^2, ~n^2_a=1, ~a=1,2,\ldots,N .~~$ The Lax operator of the
model may be given by \c{simodel}
\be L=-{\la \ov \la^2-1} (\la A_1+A_0), ~\vec A_\mu=2[\vec n\times
\partial_\mu \vec n], \ll{Lsi}\ee
where the Poisson brackets among the $n_a (x) $ are
\bea
\{n_a (x), n_b (y) \}=0,~~\{n_a (x), \pi_b (y)\}=(\de_{ab}-
n_a n_b) (x) \de(x-y), \nonumber \\
\{ \pi_a (x),\pi_b(y) \}=(\pi_a n_b- \pi_b
n_a)(x)\de(x-y),~~\pi_a=\partial_0  n_a. 
\ll{pbn}\eea 
The presence of the derivative in $x$   in the  Lax operator 
(\re{Lsi}) through  the term
 $A_1,$ 
makes the model nonultralocal and quantum generalization
difficult.

\vspace* {1cm} \fbox{fig.9.1} 

\vspace* {1cm}

  {\bf Figure 9.1}{ \it  Ladder symmetric  models}\\
a) {\it Quantum integrable theoretical model  (\re {hladder})\\
}b) {\it Ladder structure of Cu$_2$O$_3$  layers contained in the high
$T_c$ superconducting crystals} \c{ladder} 

%_______________________________________\input {preport10.tex}

\setcounter {section}{9}
\section{  Concluding remarks}
\setcounter{equation}{0}

Recent advancement  in the theory of quantum integrable systems, to which the
present review is dedicated, have shown a remarkable interpenetration between 
 formal algebraic structures appearing in pure mathematics   and the
integrable theory of physically interesting quantum models.
  The quantized algebra and the
quantized braided  algebra  are found to be the  important  
 structures  underlying
integrable systems, where their crucial 
 Hopf algebra property   plays a significant  role in the
transition from  local to  global form of the quantum Yang-Baxter
equation. This in turn leads
 to the integrability of both ultralocal and nonultralocal
systems.  
A major emphasis  is given here to highlight this interplay between 
 abstract algebras in one hand  and the construction 
of concrete quantum models on the other. We have tried to maintain
  an elementary level  stressing mainly  on  physical insights and
 avoiding mathematical rigor.
 The basic aim is to   understand the  
 algebraic  aspect inherent to  the quantum integrable systems and to see how
these formal structures can be fruitfully 
applied for constructing key  objects in the integrable theory for both
ultralocal and nonultralocal systems. 

A scheme for systematic generation of the representative Lax operators in 
ultralocal  models is presented.
The approach    in a sense  is { reverse }
to  the standard QISM and   instead of starting from
 concrete models or  known Lax operators and subsequently
deriving  quantum $R$-matrices, it  
constructs the $R$-matrix and the Lax operator by using abstract objects
like universal ${\cal R}$-matrix related to the  quantum algebra. 
Introducing spectral parameter through an algebraic
procedure
  one constructs for the Hecke algebra class the Lax operator of an ancestor model.
Through  different realizations of the quantized 
 algebra or its limiting Yangian,
one   builds up  known as well as new 
integrable models. A wide  range  of models as   descendant models
  are thus generated 
from the same algebra  and sharing the same quantum $R$-matrix.
The integrable systems are therefore classified 
into various   classes represented  by their $R$-matrices and ancestor 
Lax operators (see fig 8.1).
 Due to this 
relationship
at the algebraic level,
 the diverse models like sine-Gordon, $XXZ$ spin chain, derivative
Schr\"odinger equation, Liouville model, Ablowitz-Ladik model etc.
show remarkably common features at the {\it action-angle} level,
reflected  in their eigenvalue  solution through algebraic Bethe
ansatz.   
The intriguing question of finding 
  the criteria for  the 
 nonlinearity that makes a
system integrable, seems  also to be answered  by the specific form of
realization of  the
algebra related to the ancestor models.
 A   general realization
 in  bosonic field $\psi, \psi^\da$
 of such ancestor model  belonging to  the trigonometric 
class  may  be given by the $L$ operator 
  found in \c{tarasov}, while 
  the $L$ operator of 
\c{genlrat}  is a similar realization in 
 the rational case.

Lattice regularization of the field models  required
for 
tackling ultraviolate divergences
  often  becomes a nontrivial task,
even for  deriving integrable    models
 with  required
approximation.  The  present 
 construction  on the other hand yields,
 exploiting the underlying algebra determined by the QYBE,  
  integrable    lattice models exact 
 in all orders of  the lattice constant $\De$.
One of the drawbacks of this scheme however is that, for 
 wider class of  models with nonsymmetric Lax operators 
belonging to the 
     extended  trigonometric Sklyanin
algebra,
the universal  ${\cal R}$-matrix solution could not be found.
 This should therefore be an open problem to be solved.
Similarly
generalization 
 to include 
 other algebras like rotational,
 projective as well as
supersymmetric algebras 
would be important  directions.

Due to the recent development  in formulating   
 braided extensions of the QYBE,   
the integrable theory for quantum nonultralocal models,
which include important models like quantum mKdV, nonabelian Toda
chain, WZWN model, anyonic models etc. can now be  presented in a rather
unifying way (see fig. 9.1). The formulation is almost 
similar to the well established  theory for 
 ultralocal models, though many aspects in this regard need further 
refinement. 
Such derivation of known examples from a
  general scheme by particular
choices of braiding matrices  $Z$ and $\ti Z$ seems to suggest 
 also a possibility for their nontrivial
extensions. For example, choosing $\ti Z \neq I$ in the 
reduction for the  WZWN model, one is likely to obtain 
 generalizations of the known current algebra.

A distinguishing feature of this review perhaps is its parallel  
treatment  for 
 both ultralocal and nonultralocal quantum models
 emphasizing their algebraic aspects, integrability,  
 and interrelations. An extended list of concrete physical models 
of both these classes are provided with details of their constructions and
solutions. 
After  constructing systematically the representative   Lax operators 
from more fundamental level, the next task is to derive the Hamiltonian
 of the concrete models starting from the Lax operators and then to 
solve the eigenvalues of this Hamiltonian along with other conserved
quantities. In constructing Hamiltonian we have focused on  different 
 methods
adopted for fundamental and nonfundamental models in concrete examples.
Likewise, the specialization required 
 for  the Bethe ansatz solution  for
handling integrable field models, e.g. SG or DNLS models from that of the
lattice models like spin chains are    highlighted. 
Similarly, in presenting Bethe ansatz method for solving exactly
the eigenvalue problems,  
  both its algebraic as well as 
coordinate formulations are described on the same set of models with
comparison. This is done  to bring out the  deep relationship between these two 
 fascinating methods and  also  for
better understanding them.

The depth with which the ultralocal systems can be treated is still not
possible with nonultralocal models in the quantum case.
 For example we are not able to generate yet systematically the Lax operators 
for nonultralocal 
models  from more fundamental objects like universal {\cal
R}-matrices or ancestor models.
  Full theory should also be  able to describe  all  
     models with arbitrary braiding and in particular
should cover   well known models like quantum
 KdV,  chiral models, nonlinear $\si$-models, complex sine-Gordon model,
 Calogero-Sutherland models
etc.

Since  the subject is evergrowing, many interesting 
 recent developments are taking
place with yet new problems opening up. We have provided in the last section
a brief account of some of such recent progress, which can be understood
based on the material presented in the review. We also indicate in this
section 
various open
problems and  desirable directions stemming from the recent results.
\\ \\
{\bf Acknowledgement}

The author likes to thank Profs. Ladislav Hlavaty, Orlando
Ragnisco, Vladimir Rittenberg,  Piero Truini and also Drs. Indrani Bose and
Bireswar Basumallick for enlighting discussions on various topics of this
review. Sincere thanks are also extended to  the Alexander von
 Humboldt Foundation for
 financial and other  supports including that of a PC in which this
manuscript is prepared.  
\newpage
     
%_______________________________
%________________________________________\input {prepbib.tex}


\begin{thebibliography}{99}
%%%%%%%%%%%%%%%%%%Review QIS QG%%%%%%%%%%%%%%%%%
\bibitem{deVega}
 H. J. de Vega, 
{ Int. J. Mod. Phys. } A { 4} (1989) 2371
\bibitem{jimbo} M. Jimbo, Int. J. Mod. Phys. A 4 (1989) 3759
%3
\bibitem{selzub} H.
  Saleur  and J. B.  Zuber,  1990 { Saclay Preprint}  SPhT/90-071
%4
\bibitem{wadati} M. Wadati, T. Deguchi  and Y. Akutsu,  
 { Phys. Rep.}
 {180} (1989) 247
\bibitem{alvarez} L. Alvarez-Gaume, C. Gomes and G. Sierra, Nucl. Phys. B
319 (1989) 155
\bibitem{moor} G. Moor and N. Reshetikhin, Nucl. Phys. B328 (1989) 557
\bibitem{kauffman} L. Kauffman, Int. J. Mod. Phys. A5 (1990) 93
%----------------History%%%%%%%%%%%%%%%
\bibitem{russel} 
R. K. Bullough and P. J. Caudrey, in {\it Solitons} (ed.
R. K. Bullough and P. J. Caudrey, Springer-Verlag, 1980) 
\bibitem{fup} E. Fermi, J. R. Pasta and S. M. Ulam,
Collected Works of E. Fermi, Vol 2 (Univ. of Chicago Press, 1965) p. 978
%%%%%%%%%%%%%%%%class ISM%%%%%%%%%%%%%%%%
\bibitem{ggkm}
C S Gardner, J M Green, M D Kruskal and R H Miura, Phys. Rev. 
Lett, 19 (1967) 1095
\bibitem{krus65} N. Zabusky and M. D. Kruskal, Phys. Rev. Lett. 15 (1965)
240 
\bibitem{zakh72}V. E. Zakharov and A. B. Shabat, JETP 61 (1972) 118
\bibitem{soliton} S. Novikov, V. Manakov, L. Pitaevskii and V.  Zakharov,
Theory of Solitons (Plenum, N.Y., 1984).

\bibitem{ism}
M J Ablowitz, D J Kaup, A C Newell and H Segur, Phys Rev Lett, 31 (1973)
125
\bibitem{lakh93} M. Lakshmanan and R. Sahadeban, Phys. Rep. 224 (1993) 1 
%________________QISM__________________

\bibitem{fadrev} L. D. Faddeev, Sov. Sc. Rev. C1 (1980) 107.

\bibitem{qism}
H. B. Thacker, Lect. Notes in Phys. vol. 145 (Springer, 1981), 1

H B Thacker , Rev. Mod. Phys. 53 (1981) 253

J H Lowenstein in Les Houches Lect. Notes (ed. J B Zuber et al, 1984)p. 565
\bibitem{kulskly} P. Kulish and E. K. Sklyanin,
Lect. Notes in Phys. (ed. J. Hietarinta et al, Springer,Berlin, 1982) vol. 151
p. 61.
\bibitem{maillet} L. Freidel and J.M. Maillet, Phys. Lett. 262 B  (1991) 278.

 L. Freidel and J.M. Maillet, Phys. Lett. 263 B  (1991) 403
\bibitem{Mailletc}
S. A. Tsyplyaev, Teor. Mat. Fiz. 46 (1981) 24



M. Semenov-Tian-Shansky Funct. Anal. Appl. 17 (1983) 259
\bibitem{khijmp96}
L. Hlavaty  and Anjan Kundu, { Int J. Mod. Phys.}
  11 (1996) 2143
\bibitem{baxter} R. Baxter,
 {\it Exactly solved models in statistical mechanics}
(Acad. Press, 1981)
\bibitem{xyz} L.A. Takhtajan  and L.D. Faddeev, Russian Math. Surveys 34
(1979) 11
% Uspekhi Mat.Nauk 34 (1979) 13 

\bibitem{sgFad}
  E.K.Sklyanin , L.A. Takhtajan and L.D. Faddeev, Theor. Math. Phys.
40 (1979) 688
\bibitem{drinfeld} V. G. Drinfeld, Proc. Int. Cong. Mathematicians
(Berkeley, 1986) vol.1 p. 798.
\bibitem{fad95} L. D. Faddeev, Int. J. Mod. Phys. 10 (1995) 1845
\bibitem{kbmpl92}
     Anjan Kundu and B. Basumallick   {
     Mod. Phys. Lett. A} { 7} (1992) 61

\bibitem{kbmpl95} 
B. Basumallick and
  Anjan Kundu ,  Mod Phys Lett  A 10 (1995) 3113
\bibitem{dobrev} V. K. Dobrev, Introduction to Quantum Groups, Proc. of 22nd
Iranian Math. Conf. , Mashbad, 1991

\bibitem{qa} 
M.  Jimbo,   
 { Lett. Math. Phys. } { 10} (1985) 63


P. P.  Kulish 
 and N. Yu. Reshetikhin 
 { J. Sov. Math. } { 23} (1983) 2435


\bibitem{skalg}  E. K. Sklyanin 
 {  Funct. Anal. Appl.} { 16 } (1982) 27

\bibitem{kbplb92}
  B. Basumallick and      Anjan Kundu,  {
    Phys. Lett. B} { 287} (1992) 149
\bibitem{kbjp92} B. Basumallick and Anjan  Kundu, J. Phys. A 25 (1992) 4147
\bibitem{cheri} V Chari and A. Presley, {\it Introduction to quantum Groups},
(Cambridge), 1994

%%%%%%%%%%%%%%%%%%%%%%%%%%%%%%%%%%%Braid group%%%%%%%%%%%%%%
\bibitem{majid} S. Majid, J. Math. Phys. 32 (1991) 3246
\bibitem{majidq} S. Majid, Proc. Cambridge Philos. Soc. 113 (1993) 45,
 
S. Majid, J. Pure Appl. Algebra 86, (1993).
\bibitem{hlavqb}  L. Hlavat\'{y}, J. Math. Phys. 35 (1994) 2560

%-----------------------UNIV R MATRIX%%%%%%%%%%%%%%%
\bibitem{urslq2} M. Rosso, { Comm. Math. Phys.} { 124}  (1989) 307

A.N. Kirillov  and N. Reshetikhin, { Comm. Math. Phys.} { 134} 
           (1990) 421

\bibitem{reductive} P. Truini and V. S. Varadarajan, Lett. Math. Phys. 
26 (1992) 53
\bibitem{ktjp95}
Anjan Kundu
and P.Truini
, { J. Phys.A }  28 (1995) 4089


\bibitem{nls} E. K. Sklyanin, DAN SSSR, 244 (1979) 1337 
\bibitem{frt} N.Yu. Reshetikhin, L.A. Takhtajan and L.D. Faddeev,
Algebra and Analysis 1 (1989) 178

\bibitem{jones} V. F. R.  Jones, { Int. J. Mod. Phys. } B { 4} (1990) 701

\bibitem{babelontoda}
O. Babelon, Nucl. Phys. B 230 [FS 10] (1984) 241

\bibitem{twist} N.  Reshetikhin, { Lett. Math. Phys.} { 20}  (1990) 331

A. Schirrmacher, J. Phys. A24 (1991) L1249

\bibitem{liuFad}  L. D. Faddeev and O.  Tirkkonen, Nucl. Phys. B453
(1995) 647

  \bibitem{qrtoda} Anjan  Kundu, Phys. Lett. A 190 (1994) 73

\bibitem{ALM} M. Ablowitz 
 { Studies in Appl. Math.} { 58} (1978)  17

\bibitem{kb-dnls} 
 Anjan Kundu  and B. Basumallick, J. Math. Phys. 34 (1993) 1252
\bibitem{sglsg} A. G. Izergin  and
 V. E. Korepin, 
{ Nucl. Phys.} { B 205} [FS 5] (1982)
401


 A. G. Izergin  and
 V. E. Korepin, 
{ Sov. Phys. Dokl.} { 26}  (1981)
653.
\bibitem{Coker}  D. A. Coker, {\it Use of projectors for integrable models
of quantum field theory}, preprint ITP-SB-92-18 (1992).
\bibitem{krjp94} Anjan  Kundu and O. Ragnisco, J.  Phys. A 27 (1994) 6335.
\bibitem{izumov} Yu. A. Izumov and Yu. N. Skryabin,{\it
 Statistical Mechanics of
Magnetically Ordered Systems} (Consultant Bureau, N. Y. , 1988) p. 197
\bibitem{Bethe}
H. Bethe, Z. Phys. 71 (1931) 205

\bibitem{cba} C N Yang and C P Yang, J. Math. Phys., 10 (1969) 1115

B Sutherland, Phys Rev Lett, 19 (1967) 103

M. Gaudin, Phys. Rev. Lett. 26 (1971) 1301

 D. C. Mattis ed. 
 {\it Encyclopedia of exactly solved  models in one dimension}
 (World Sc., 1993)

\bibitem{korbook} V. E. Korepin, N. M. Bogoliubov, A. G. Izergin, {\it QISM
and Correlation Functions}, {Cambridge Univ. Press , 1993)
\bibitem{nba} P. P.  Kulish and N. Yu. Reshetikhin, J. Phys. A 16 (1983)
L591

 \bibitem{fba} B. T.
 Batchelor, R. J. Baxter, M. J. O'Rourke and C. M. Yung
preprint hep-th/9502040

\bibitem{toda}  E. K. Sklyanin,
Lect. Notes in Phys vol 226 (1985) 196

\bibitem{simodel}  J.M. Maillet,  Phys. Lett. B162  (1985) 137
\bibitem{wzwn}
L. D. Faddeev,  Comm. Math. Phys.  132 (1990) 131

A. Alekseev, L.D. Faddeev, M. Semenov-Tian-Shansky  and
A. Volkov,{\it The unraveling of the quantum group structure in the
WZWN theory},  preprint CERN-TH-5981/91 (1991)


A. Alekseev, S. Statashvili, Comm. Math. Phys.  133 (1990) 353

B. Blok,  Phys. Lett. 233B (1989) 359


\bibitem{comsg} J.M. Maillet, Nucl. Phys. B269 (1986) 54

\bibitem{skly-r} E. Sklyanin, J. Phys. A 21 (1988) 2375

\bibitem{hlavopsc} L. Hlavat\'{y}, J.Phys. A 27 (1994) 5645
\bibitem{kmpl95} 
 Anjan Kundu,  Mod. Phys. Lett.  A 10 (1995) 2955

\bibitem{natoda} V. E. Korepin, J. Sov. Math. 23 (1983) 2429

\bibitem{Nijhof} F.W. Nijhoff, H.W. Capel and V.G. Papageorgiou,
Phys. Rev. A 46 (1992) 2155

\bibitem{babelon} O. Babelon and L. Bonora, Phys. Lett. 253 B (1991) 365

 O. Babelon, Comm. Math. Phys.  139 (1991) 619


 L. Bonora and V. Bonservizi, Nucl. Phys. B 390  (1993) 205
\bibitem{SUSY} P. Kulish  J. Sov. Math. 35 (1986) 2648

%\bibitem{}
  P. Kulish Sov. Phys. Dokl. 25 (1980) 912.
%\bibitem{}

  A. Forester and M. Karowski, Nucl. Phys. B408 (1993) 512
%%%%%%%%%%%%%%%%%%%%%%%Anyon%%%%%%%%%%%%%%%%%%%%%%%
\bibitem{anyon1} A. Lerda and S. Sciuto,   Nucl. Phys.B 401 (1993) 613.

M. Frau and S. Sciuto, Proc. Int. School of Phys. 'Enrico Fermi' Course,
(Eds. L. Castellani and J. Wess, IOS Press, Amsterdam,1996), p. 215

\bibitem{cs} F. Calogero, J. Math. Phys. 12 (1971) 418

J. Moser,  Adv. Math. 16 (1975) 197

B. Sutherland, Phys. Rev. A5 (1972) 1372

\bibitem{csavan} J. Avan, O. Babelon and E. Billey, {\it The Gervais- Neveu
-Fedler equation and quantum Calogero-Sutherland systems}. preprint
hep-th/9505091

\bibitem{csyan} B. Basumallick, {\it  Spin dependent extension of
Calogero-Sutherland model through anyon like representation of permutation
operators }, preprint hep-th/9602107

%%%%%%%%%%%%%%ISM---CFT   Chapt 9%%%%%%%%%%%%%%%%%%%%

\bibitem{karow} A. Karowski, Nucl. Phys. B 300 [FS 22] (1988) 479

H. J. de Vega and  A. Karowski, Nucl. Phys. B 285 [FS 19] (1987) 619

\bibitem{BLZ} V. V. Bazhanov, S. L. Lukyanov and A. B. Zamolodchikov,
Comm. Math. Phys. 177 (1996) 381



V. A. Fateev and S. Lukyanov, Int. J. Mod. Phys. A7 (1992) 853, 1325

\bibitem{reacrit}
F. C. Alcaraz, M. Droz, M. Henkel, V. Rittenberg, Ann. Phys. 230 (1994) 667

\bibitem{frahm} H. Frahm and C. R\"odenbeck, {\it
 Integrable models of coupled
Heisenberg chains}, preprint cond-mat/9502090 (1995)

\bibitem{eqg} O. Foda, K. Iohara, M. Jimbo, R. Kedem, T. Miwa and H. Yan,
Lett. Math. Phys. 32 (1994) 259

 D. B. Uglov, Lett. Mat. Phys. 268 (1993) 139


\bibitem{kplan} Anjan Kundu, under preparation

\bibitem{WM} B. M. McCoy and T.T. Wu, Nuovo Cim. B (1968) 311

\bibitem{6v1}  
K. Sogo, M. Uchinami, Y.  Akutsu and M. Wadati, 
{ Prog. Theor. Phys.} { 68 } (1982)  508

\bibitem{llm} L. D.  Faddeev  and  L. A. Takhtajan   
{ Lect. Notes Phys.}
 {vol. 246} (ed. H.  de Vega et al, Sringer, 1986) 166

\bibitem {closetc} P.L. Christiansen, M. F. Jorgensen and V.B. Kuznetsov
,{Lett. Math. Phys.} {29  } (1993) 165

\bibitem{semenov95} M. Semenov Tian-Shansky and A. Sevostyanov
{\it Classical and quantum nonultralocal systems on the lattice}
preprint hep-th/ 9509029

\bibitem{goddard} M. Chu, P. Goddard, I. Halliday, D. Olive and
A. Schwimmer,
 Phys. Lett. 266 B  (1991) 71

\bibitem{volkov95} A. Yu. Volkov, Comm. Math. Phys. 177 (1996) 381

\bibitem{kun84} Anjan  Kundu J. Math. Phys. 25 (1984) 3433

F. Calogero, Inverse Prob. 3 (1987) 229

L. Y. Shen, in {\it Symmetries and Singularity Structures} (ed. M.
Lakshmanan, Springer Verlag, NY, 1990) p. 27

\bibitem{qalg-gr} C. Fonsdal and A. Galindo, Lett. Math. Phys. 27 (1993) 59

F. Bonechi, E. Cheleghini, R. Giachetti, C. M. Perena, E. Sorace and M.
Tarlini, J. Phys. A 27 (1994) 1307

J. V. der Jeugt and R. Jagannathan, preprint q-alg/9507009 (1995)

\bibitem{openc} L. Mezincescu  and R. Nepomechie, J. Phys. A24 (1991) L17; 
A 25 (1992) 2533
\bibitem{tolstoy} S.M. Khoroshkin  and V.N. Tolstoy,
     { Comm. Math. Phys.} { 141}  (1991) 599

\bibitem{mlg} M. Couture, M. L. Ge, H. C. Lee N. C. Scheming, J. Phys. A 23
(1990) 4751

 Y. Cheng, M. L. Ge and K. Xue, { Comm. Math. Phys. } { 136} (1991) 195

\bibitem{hlav94} L. Hlavat\'y, J. Math. Phys. 36 (1995) 4882

 %%%%%%%%%%%%%%%%%%Qmodels%%%%%%%%%%%%%%
\bibitem{lecber} D. Bernard, Quantum symmetries in 2D massive field
theories, preprint SPhT -91-124 (1991)
%%%%%%%%%%%%%%%%%%%%QG-formal%%%%%%%%%%%%%%%%%%%%%%%
\bibitem{qoscl}
A. J. Macfarlane,
{  J.Phys. }  { A  22 }( 1989)  4581
%33

 L. C.  Biederharn, 1989
 { J.Phys.}  { A 22 } (1989) L873

\bibitem{tarasov} V. O. Tarasov,Teor. Mat. Fiz. 63 (1985) 175

\bibitem{sun}
O. Babelon, H. J. de Vega and C. M. Viallet, Nucl. Phys. B 140 [FS 3] (1981)
542; B 200 [FS 54] (1982) 266

\bibitem{perks} J.H.H. Perk, C.L. Schultz, {\it Nonlinear Integrable
Systems -- Classical Theory and Quantum Theory} in Proc. of RIMS Symposium
ed. M. Jimbo and T. Miwa, World Scientific 1983.  

\bibitem{dnls1} 
D.  J. Kaup and A. C. Newell, J. Math. Phys. 19( 1978) 798

Anjan Kundu, J. Phys A 21 (1988) 945

\bibitem{rel2} Y.K. Zhou, Phys. Lett. B 276 (1992) 135

\bibitem{sgKS} Anjan Kundu and S. Ghosh, J. Phys. A 21 (1988) 3951
\bibitem {suris} Yu. B. Suris, 
{ Phys.Lett.} { A 145 } (1990)  113


\bibitem{ruij} S. N. M. Ruijsenaars, Comm. Math. Phys. 133 (1990) 217

\bibitem{hqg} V. Pasquier and H.  Saleur, Nucl. Phys. B330 (1990) 523

\bibitem{smodel} H. M. Babujian, Nucl. Phys. [FS 7] (1983) 317

L. A. Takhtajan, Phys. Lett. 87 A (1982) 479

\bibitem{spin1} A. B. Zamolodchikov and V. A. Fateev, Sov. J. Nucl. Phys. 32
(1980) 293


\bibitem{alters} H. J. de Vega and F. Woynarovich, 
{  J.  Phys. } A 25 (1992) 4499

\bibitem{twistbc} A. Kl\"umper, M. Batchelor and P.  Pearce,
J. Phys. A 24 (1991) 3111;
 A. Kl\"umper, T. Wehner and J. Zittarz,
J. Phys. A 26 (1993) 2815

\bibitem{TTF}  V. O. Tarasov, L. A.  Takhtajan  and L. D. Faddeev,  
 Teor. Mat. Fiz. { 57} (1983)  163

 \bibitem{fba1}  C. M. Yung and B. T.
 Batchelor, Nulc. Phys. B 446 [FS] (1995) 461

\bibitem{dbose} E Lieb and W Liniger, Phys. Rev. 130 (1963) 1605

J. B. McGuire,  J. Math. Phys. 5 (1964) 622


\bibitem{sgothers}
R. Rajaraman, Solitons and Instantons (North Holland publ., 1982)

S. Coleman, Phys. Rev. D11 (1975) 2088 

R.P. Dashen, B. Hasslacher and A. Neveu, Phys. Rev. D11 (1975) 3424

 
J. Rubinstein, J. Math. Phys. 11 (1970) 258

\bibitem{sgGS} G. Bhattacharya and S. Ghosh, Int. J. Mod. Phys. 4 (1998) 627

\bibitem{fabian}

  F. Essler and V. Korepin, Phys. Rev. B46 (1992) 9147

\bibitem{vnls} P. P.  Kulish and N. Yu. Reshetikhin, Zap. Nauch. Sem. LOMI
120 (1981) 92

 P. P.  Kulish, Physica D 3(1981) 246

\bibitem{nwave} M. Wadati and K. Ohkuma, J. Phys. Soc. Jpn. 53 (1984)
1229,2899

P. P. Kulish, Physica D18 (1986) 360

\bibitem{schultz} L. Schultz, Physica 122 A (1983) 711

\bibitem{lopez} E. Lopes Nucl. Phys. B 370 (1992) 636

\bibitem{BaxterHC} R. J. Baxter, Ann. Phys. 10 (1972) 323
\bibitem{zamfac} 
A. B. Zamolodchikov and Al. B. Zamolodchikov, Ann. Phys. 120 (1979) 253 
\bibitem{abaHub}  P. B. Ramos and M. J.  Martins,
{\it Algebraic Bethe ansatz approach for the one-dimensional Hubberd model},
 preprint  hep-th/9605141 (1996)


%--------------Hubberd------------

\bibitem{wadatiH} S. Shiroishi and M. Wadati, J. Phys. Soc. Jpn. 64 (1995)
57\bibitem{shastryH} B. S. Shastry, Phys. Rev. Lett. 56 (1986) 1529
 B. S. Shastry, J. Stat. Phys.  50 (1988) 57
\bibitem{LWHub} E. H. Lieb and F. Y. Wu, Phys. Rev. Lett. 20 (1968) 1445


\bibitem{SUSYq}
 M. Chaichian and P. Kulish, Phys. Lett. 234 B (1990) 72


\bibitem{anyon0} F. Wilczek in {\it Fractional statistics and anyon
superconductivity} (ed. F. Wilczek, World Sc., Singapore, 1990)

Y. S. Wu,   Phys. Rev. Lett. 52 (1984) 2103

D. Eliezer and G. W. Semenoff,  Phys. Lett. 286 B (1992) 118


 F. D. M. Haldane,   Phys. Rev. Lett. 67 (1991) 937.

 E. Fradkin,   Phys. Rev. Lett. 63 (1989) 322

%%%%%%%%%%%%%%%%%%%Nonultralocal%%%%%%%%%%%%%%%%%%%%%%%%%%%%

\bibitem{alex} A. Yu. Alexeev, {\it Integrability in the Hamiltonian
Chern-Simons theory}, preprint hep-th/9311074 (1993).
\bibitem{moduli} C. Schwiebert, {\it Generalized quantum inverse scattering
method}, RIMS preprint 1003, hep-th/9412237 (1994)
\bibitem{mkdv} {\it Solitons in Action } (Acad Press, eds. K. Lonngren and
A. Scott, 1978)

\bibitem{volkov} A. Yu. Volkov, Phys. Lett. A167 (1992) 345

\bibitem{dbm-kor} A. G. Izergin and V. E. Korepin, Comm. Math. Phys., 79
(1981) 303
\bibitem{admrit}
F. Essler and V. Rittenberg, {\it Representation of the quadratic algebra
and partially asymmetric diffusion with open boundaries}, preprint
cond-mat/9506131

\bibitem{hal-pas} D. Bernard, M. Gaudin, F. Haldane, V. Pasquier, J. Phys.
A 21 ? (1993) 5219

\bibitem{wad-shas} K. Hikami and M. Wadati, J. Phys. Soc. Japn 62 (1993) 469

B. Sutherland and S, Shastry, Phys. Rev. Lett. 71 (1993) 5

\bibitem{relcsm} Hitoshi Konno, preprint hep-th/9508016 (1995)

\bibitem{BPZ} A. A. Belavin, A. M. Polyakov and A. B. Zamolodchikov,
Nucl. Phys. B241 (1984) 333
 \bibitem{Gervais}B. A. Kuppershmidt, P. Mathieu, Phys. Lett. B 227 (1989)
245, 250  

J. L. Gervais Comm. Math. Phys. 138 (1991) 301
\bibitem{fadq} L. D.  Faddeev  in {\it Fields and Particles, } eds.
H. Mitter et. al. ( Springer-Verlag, Berlin, 1990 ) p. 89
 
 

\bibitem{bwm} J. Birman and H, Wenzl, Trans. Am. Math. Soc. 313 (1989) 249

J. Murakami, Osaka J. Math. 24 (1987) 745 
%6
%8
\bibitem{ncgeom} D. D.   Demidov , Yu. I. Manin, E. E. Mukhin 
 and D. V. Zhdanovich,
{ Kyoto Univ. preprint } RIMS-70, 1990

A. Sudbey, J. Phys. A 23 (1990) L697

A. Schirrmacher, J. Wess and B. Zumino, Z. Phys. C 49 (1991) 317

D. B. Fairlie and C. K. Zachos, Phys. Lett. B 256 (1991) 43


\bibitem{genlrat} A. G. Izergin and V. E. Korepin, Lett. Math. Phys. 8
(1984) 259 

\bibitem{urslqn} S.Z. Levendorsky  and  Ya.S. Soibelman, RGU preprint (1990)

%\item{7.} Khoroshkin S.M. and Tolstoy V.N., Wroclaw preprint ITP UWr
%          800/92
 
\bibitem{nstR} L. H. Kauffman, in {\it Knots, Topology and
Quantum Field Theory} (World Sc., 1989) 
%%%%%%%%%%%%%%%%%%%%%%%Coordinate Bethe%%%%%%%%%%%%%%%

\bibitem{TDqb} I. Tamm, J. Fiz. (USSR) 9 (1945) 499

S. Dancoff, Phys. Rev. 78 (1950) 382
\bibitem{asym6} I. M. Nolden, J. Stat. Phys. 67 (1992) 155

%1
\bibitem {toda0}  S.V.~Manakov,   { JETP }  { 67} (1974) 269

 H.~Flaschka and D.W.~McLaughlin, 
 { Progr.Theor.Phys. } { 55}  (1976) 438

  M.A. Olshanetsky and A.M. Perelomov,
 { Lett.Math.Phys.} { 2 }  (1977) 7

   M.C. Gutzwiller,   { Ann. Phys. } {133} (1981) 304

 V. Pasquier and M. Gaudin,  { J. Phys.} { A 25} (1992) 5243
%------------------------
%\bibitem{spinchains} L. A. Takhtajan and L. D. Faddeev,
% Russian Math. Survey 34 (1979) 11.
% P. P.  Kulish and E. K. Sklyanin, Phys. Lett. A70 (1979) 461
%%%%%%%%%%%%%%%%%twistbc

%%%%%%%%%%%%%%%%%%%%%%%NBA%%%%%%%%%%%%%%%%%%%%%%%%%%%%%%%%

  
%%%%%%%%%%%Functional Bethe ansatz%%%%%%%%%%%5%%%%%%%%%%%%%%%%%
  

%%%%%%%%%%%%%%%%%%%%OUr work%%%%%%%%%%%%

\bibitem{qhpt} Anjan Kundu and B. Basumallick, Phys. Lett. A 156 (1991) 175
  
%________________________________NOT TO BE INCLUDED__________
%\bibitem{kbneed91}  Anjan  Kundu  and B. Basumallick, 
%{ Proc. of Int.conf.NEEDS'91,Gallipoli (Italy),1991}
% (Ed.M.Boiti,L.Martina and F.Pompinelli ,World Sc.,1992),p.357:
%\bibitem{ksoliton95} Anjan Kundu,  {\it Chaos Soliton and Fractals}
%( Pergamon  publ.,, 1995), vol. 5 no. 12 (1995), p. 2329-2344
%_______________________________________________________

\bibitem{Liao} Li Liao, X-Ch. Song, Mod. Phys.  Lett.  6  (1991)
959


\bibitem{reshet} N. Yu. Reshetikhin and  M. Semenov-Tian-Shansky ,
 Lett. Math. Phys. 19  (1990) 133


\bibitem{ladder}
E. Dagotto and T. M. Rice, Science 271 (1996) 618

 S. Maekawa, Science 273 (1996) 1515
 
\bibitem{wadatiIH}  M. Wadati, T. Nagao and K. Hikami,  Physica D68 (1993)
163


\bibitem{HS} F.D.M. Haldane, Phys. Rev. Lett. 60 (1988) 635

B.S. Shastry, Phys. Rev. Lett.   60 (1988) 639

 \bibitem{bvega} H. J. de Vega, A. Gonzales-Ruiz, J. Phys.  A 27 (1994) 6129
}
\end{thebibliography}
 \end{document}